\documentclass[11pt, amssymb]{article}
\usepackage{amsmath}
\usepackage{amsfonts}
\usepackage{amscd}
\usepackage{amssymb}
\usepackage{graphicx}
\usepackage{color}
\input{colordvi.tex}

\def\kesu#1{}

\newcommand{\vev}[1]{ \left\langle {#1} \right\rangle }

\newcommand{\nin}{\mbox{\ooalign{\hfil/\hfil\crcr$\in$}}}

\def\diag{\mathop{\rm diag}\nolimits}
\def\tr{\mathop{\rm tr}}

\def\SO{\mathop{\rm SO}}
\def\O{\mathop{\rm O}}
\def\SU{\mathop{\rm SU}}

\def\Z{\mathbb{Z}}
\def\Q{\mathbb{Q}}
\def\R{\mathbb{R}}
\def\C{\mathbb{C}}
\def\P{\mathbb{P}}

\begin{document}

\begin{titlepage}
 
\begin{flushright}
KCL-mth-13-12 \\
YITP-13-129 \\
IPMU13-0237 
\end{flushright}
 
\vskip 1cm
\begin{center}
 
{\large \bf The Noether-Lefschetz Problem and Gauge-Group-Resolved Landscapes: \\
F-Theory on  ${\rm K3} \times {\rm K3}$ as a Test Case } 
 
 \vskip 1.2cm
 
 Andreas P. Braun$^1$, Yusuke Kimura$^2$ and Taizan Watari$^{3}$

 \vskip 0.4cm
 
 {\it $^1$Department of Mathematics, King's College, London WC2R 2LS, UK
\\[2mm]

 $^2$Yukawa Institute for Theoretical Physics, Kyoto University, 
Kyoto 606-8502, Japan  \\[2mm]
 
 $^3$Kavli Institute for the Physics and Mathematics of the Universe, 
University of Tokyo, Kashiwano-ha 5-1-5, 277-8583, Japan
 }
 \vskip 1.5cm
 
\abstract{
Four-form flux in F-theory compactifications not only
stabilizes moduli, but gives rise to ensembles of string vacua,
providing a scientific basis for a stringy notion of naturalness.
Of particular interest in this context is the ability to keep 
track of algebraic information (such as the gauge group) associated with individual vacua
while dealing with statistics. In the present work, we aim to clarify conceptual issues and sharpen
methods for this purpose, using compactification on ${\rm K3} \times {\rm K3}$ as a test case.
Our first approach exploits the connection between the stabilization of
complex structure moduli and the Noether--Lefschetz problem.
Compactification data for F-theory, however, involve not only a
four-fold (with a given complex structure) $Y_4$ and a flux on it, but
also an elliptic fibration morphism $Y_4 \longrightarrow B_3$, which
makes this problem complicated. The heterotic--F-theory duality indicates that elliptic fibration
morphisms should be identified modulo isomorphism. Based on this
principle, we explain how to count F-theory vacua on ${\rm K3} \times {\rm K3}$ while
keeping the gauge group information. Mathematical results
reviewed/developed in our companion paper are exploited heavily.
With applications to more general four-folds in mind, we also
clarify how to use Ashok--Denef--Douglas' theory of the distribution
of flux vacua in order to deal with statistics of sub-ensembles
tagged by a given set of algebraic/topological information. 
As a side remark, we extend the heterotic/F-theory duality dictionary on flux quanta and elaborate on its connection 
to the semistable degeneration of a K3 surface. } 
 
\end{center}
\end{titlepage}

\tableofcontents

\newpage 

\section{Introduction}

Flux compactifications of type IIB string theory / F-theory can generate 
large supersymmetric masses for moduli, so that the moduli particles 
decay well before the period of big-bang nucleosynthesis. In addition to 
this phenomenological advantage, the discretum of vacua in this class of
compactifications provides an ensemble of vacua (or landscape), 
which gives rise to a scientific/stringy basis for a notion of naturalness
\cite{Denef:2007pq, Denef:2008wq}.
Certainly the (geometric phase) Calabi--Yau compactifications of 
type IIB / F-theory are not more than a small subset of all possible
vacua in string theory. However, one can still think of some use in such a
restricted ensemble of vacua because supersymmetric extensions of 
the Standard Model and grand unification can be naturally accommodated in 
this framework.  

Given a topological configuration of three-form fluxes in type IIB string 
theory compactified on a Calabi--Yau threefold $M_3$, the dilaton 
and complex structure moduli of $M_3$ are stabilized and their
vacuum expectation values (vevs) are determined. 
By this mechanism, however, not only the moduli vevs (= the coupling constants of the low-energy 
effective theories), but also a configuration of D7-branes  
(= (a part of) the gauge group of the effective theories) is determined. 

Low-energy effective theories in particle physics are usually crudely classified by such information as gauge
groups, matter representations, types of non-vanishing interactions and matter 
multiplicity, and then the effective theories sharing this information 
are distinguished by the values of coupling constants. In order to fit 
this natural framework of thought, ensembles of string flux vacua 
should also be crudely classified by algebraic and topological
information. After this, statistics should be presented in the form of distributions 
over the moduli space of compactifications sharing the same set of 
algebraic and topological information. With the statistics of flux vacua 
presented in this way, we begin to be able to ask such
naturalness-related questions as the ratio of the number of vacua having 
various algebraic and/or topological data or the distribution of various 
coupling constants (moduli parameters) in a class of theories having a
give set of algebraic/topological data. This article aims at taking one step 
further in this program. The distribution of gauge groups in effective four-dimensional
theories derived from string compactifications has been studied from 
several perspectives in the literature, see 
\cite{Kumar:2004pv,Blumenhagen:2004xx,Dienes:2006ut,Anderson:2008uw}
for examples.

Stabilization / determination of D7-brane configurations can be
understood purely in type IIB language in terms of calibration conditions 
\cite{Cascales:2004qp,Gomis:2005wc}. Another option is to use F-theory, where the 7-brane 
configuration, dilaton vev and complex structure moduli of $M_3$ are 
all treated as part of the complex structure moduli of a Calabi--Yau fourfold $Y_4$.
In F-theory language, there are two ways to understand the mechanism of 
determination of the moduli vevs. One is to specify the four-form flux 
on an elliptic fibred Calabi--Yau fourfold $Y$ topologically,\footnote{
The four-form flux should be in the Abelian group 
$\left[H^4(Y; \Z) + \frac{1}{2}[c_2(TY)] \right]\cap H^{2,2}(Y; \R)$
\cite{Witten:1996md}, with a possibly half-integral shift $(1/2)[c_2(TY)]$. 
When $Y$ is a smooth Weierstrass model, however, $c_2(TY)$ is always even,
and the four-form flux takes its value in (\ref{eq:Hodge-22}) \cite{Collinucci:2010gz,Collinucci:2012as}.
In the present case, where we work with  ${\rm K3} \times {\rm K3}$, $c_2(TY)$ is manifestly even and no such issue arises. Hence we ignore this point
from now on.} 
\begin{equation}
 [G^{(4)}] \in H^4(Y; \Z)\, .
\end{equation}
The Gukov--Vafa--Witten superpotential 
\begin{equation}
 W_{\rm GVW} \propto \int_Y \Omega_Y \wedge G^{(4)}
\label{eq:GVW}
\end{equation}
gives rise to an F-term scalar potential that depends on the complex structure 
moduli of $Y$. The minimization of this potential determines the 
vevs of those moduli. Once the moduli vevs arrive at the minimum of 
the potential (and if the cosmological constant happens to vanish), 
the four-form flux $G^{(4)}$ is guaranteed to only have a $(2,2)$ component 
in the Hodge decomposition under the complex structure corresponding to 
the vevs \cite{Giddings:2001yu, Kachru:2002he}. In the presence of the 
four-form flux, the moduli fields slide down the potential to find a vacuum 
complex structure, so that $[G^{(4)}]$ only has the $(2,2)$ component. 
If we are to allow large vacuum expectation values of $W_{\rm GVW}$ 
$[G^{(4)}]$ may also have $(4,0)$ and $(0,4)$ components.

An alternative way to characterize the vacuum choice of the complex 
structure of $Y$ is available by focussing on the finitely generated 
Abelian group
\begin{equation}
 H^4(Y; \Z) \cap H^{2,2}(Y; \R) \, \qquad \qquad 
  \left({\rm or~} H^4(Y; \Z) \cap 
    H^{2,2}(Y;\R) \oplus [ H^{4,0}(Y; \C) + {\rm h.c} ] \right)\, .
\label{eq:Hodge-22}
\end{equation}
The rank of this Abelian group remains constant almost everywhere 
on the moduli space of the complex structure of $Y$, but it jumps
at special loci. In mathematics, 
this problem---at which loci in the moduli space the rank of this Abelian group 
jumps, and how it changes there---is known as the Noether--Lefschetz problem. Once we find 
a point in the Noether--Lefschetz locus and insert four-form flux in the 
enhanced part of the Abelian group (\ref{eq:Hodge-22}), we can 
no longer go away continuously from the Noether--Lefschetz loci 
in the moduli space while keeping the flux purely of type $(2,2)$. 
The higher the codimension of a Noether--Lefschetz locus is in the 
complex structure moduli space, the more moduli are given masses and 
stabilized. Therefore, the problem of determination of vacuum complex 
structure is equivalent to the Noether--Lefschetz problem
(e.g., \cite{Moore:2004fg}). \footnote{We can also draw an analogy with the
attractor mechanism \cite{Moore:2004fg}, although the analogy is particularly good 
in the case with $G_1 \neq 0$ and $G_0 = 0$ in 
(\ref{eq:G0-def}, \ref{eq:G1-def-W=0}).}

This article begins, in Sections \ref{ssec:extended-list}, 
\ref{ssec:J1-J2-phys}, \ref{ssec:sample-stats} and \ref{ssec:rank-16} 
in particular, with exploiting this equivalence\footnote{This is
an obvious continuation of a program initiated a decade earlier.
This idea is already evident in pioneering works such 
as \cite{Moore:2004fg, Gorlich:2004qm, DeWolfe:2004ns, AK}, to name a few, 
and has also been reflected in recent articles such 
as \cite{DW-3, BCV}.}
to see how the Noether--Lefschetz problem in F-theory determines
statistics of such things as gauge group, discrete symmetry and 
moduli vevs. We focus our attention on $K3 \times K3$ compactifications of 
F-theory, as in \cite{Becker:2002sx,Tripathy:2002qw,Becker:2003yv, Gorlich:2004qm, Lust:2005bd,AK, Heidelberg-BHLV}. 
This compactification cannot be considered realistic enough for an immediate use in particle 
physics (e.g. there are no matter curves), but a sufficient complexity is involved
in this toy model of landscape to make it suitable for the purpose of 
clarifying various concepts as well as sharpening technical tools.

In the process of deriving statistics, one cannot avoid asking about the modular group (e.g., \cite{DeWolfe:2004ns}). 
In other words, we have to understand when a pair of seemingly different 
compactification data actually correspond to the same vacuum in physics. 
In F-theory we have to introduce some equivalence
relation among the space of elliptic fibrations that are admitted by $Y_4$, so 
that the quotient space corresponds to the set of physically distinct
vacua. In Section \ref{ssec:Het-F-AutX}, we use the duality between heterotic string theory and F-theory, 
and find that the modulo-isomorphism classification of elliptic fibrations should be adopted. 
This observation yields two problems. One is purely mathematical: how can we work out 
the modulo-isomorphism classification of elliptic fibration for a given $Y_4$ ? 
A companion paper by the present authors \cite{BKW-math} is dedicated to this problem, 
with a Calabi--Yau fourfold $Y_4$ replaced by a K3 surface, and the results in \cite{BKW-math} are reviewed mainly in 
Section \ref{ssec:J1-J2-phys} in this article. The other problem is how 
to use such results in mathematics to carry out vacuum counting in
physics. We take on this issue in Section \ref{ssec:sbtl-stat}.

Sample statistics, which give us some feeling of what string landscapes 
can do to answer statistical/naturalness questions in particle physics, 
are obtained in Sections \ref{ssec:sample-stats} and \ref{ssec:rank-16}.

At the same time, though, the study running up to Section \ref{ssec:rank-16} in 
this article along with \cite{BKW-math} also hints that it may not be easy to pursue 
the Noether--Lefschetz problem approach for Calabi--Yau fourfolds which are not as 
simple as ${\rm K3} \times {\rm K3}$ or K3-fibration over some complex surface. One may
of course always use the strategy of computing periods, as done e.g. in \cite{Grimm:2009ef} 
in the present context. Of course, this approach has its own technical challenges.

The theory of \cite{Ashok:2003gk, Denef:2004ze, Denef:2008wq} is 
a promising direction to go beyond a case-by-case study for 
different choices of fourfolds $Y$. Therefore, the second theme begins 
to dominate in Section \ref{ssec:orientifold-beyond} toward the end of this
article. Since articles as \cite{Denef:2004ze} and
\cite{DeWolfe:2004ns} seemed to have had applications in Type IIB 
orientifold compactifications in mind primarily, we clarify how to use 
the Ashok--Denef--Douglas theory to study statistics of flux vacua 
in F-theory, with the total ensemble resolved into sub-ensembles
according to their algebraic and/or topological data such as gauge
groups and matter multiplicity.
The presentation in \cite{DeWolfe:2004ns} sits in the middle 
between our discussion up to Section \ref{ssec:rank-16} and that of 
\cite{Ashok:2003gk, Denef:2004ze}, and makes it easier to understand
how the conceptual issues discussed in the sections 
up to \ref{ssec:MW-singl-fbr-flux} fit into the Ashok--Denef--Douglas 
theory. Although the presentation in Section
\ref{ssec:orientifold-beyond} and Appendix \ref{sec:ADD} uses $K3\times K3$
compactification as an example, we tried to phrase it in a way ready 
for generalization at least to cases with K3-fibred Calabi--Yau
fourfolds, and possibly to general F-theory compactifications. 

There is also the third theme behind Sections \ref{sssec:FMW}, 
\ref{ssec:MW-singl-fbr-flux} and Appendix \ref{sec:Km-E6E6} in 
this article. In the duality between heterotic string theory and F-theory, 
the dictionary of flux data has been mostly phrased by using the stable 
degeneration limit of \cite{Friedman:1997yq, Morrison:1996pp}.
This was for good reasons, because \cite{Curio:1998bva, DW-1,
Hayashi-1} focused on fluxes in F-theory that are directly responsible 
for the chirality of non-Abelian (GUT gauge group) charged matter fields 
on the matter curves. There is an extra algebraic curve in the singular
fibre over the matter curve in the F-theory geometry, and a flux can be 
introduced in this algebraic cycle \cite{DW-1, Hayashi-1}.
For more general flux configurations, however, it is not a priori clear
to what extent we can use the $dP_9 \cup dP_9$ limit in the
duality dictionary, because $dP_9 \cup dP_9$ is quite different from a K3 
surface when it comes to whether two-cycles are algebraic or not.  
There is the work of \cite{BCV}, indicating that ${\rm U}(1)$ flux 
associated with an elliptically fibred geometry with an extra section
stabilizes some complex structure moduli. Furthermore,
the spectral cover description of vector bundles in heterotic string theory 
\cite{Friedman:1997yq} did not rule out twisting information $\gamma$ 
which is more general than (\ref{eq:FMW-flux}) for special choices of complex
structure. Section \ref{ssec:MW-singl-fbr-flux} and the 
Appendix \ref{sec:Km-E6E6} provide a comprehensive understanding of
this material, generalizing the duality dictionary of the flux data in
the literature without relying on the $dP_9 \cup dP_9$ limit. 
Appendix \ref{sec:Km-E6E6} also makes a trial
attempt of studying how much information of such fluxes can be captured 
by the $dP_9 \cup dP_9$ limit. 

A similar theme has already been studied extensively in the series of 
papers \cite{Caltech-0906, Caltech-0912, Caltech-1006, Caltech-1102,
Caltech-1107, Caltech-1211}. It will be interesting to clarify the relation between the 
logical construction given there and the presentation in this article, but this task is 
beyond the scope of this present article.

All $K3$ surfaces which appear as solutions in this article have Picard number $20$, which
fixes the rank of the total gauge group to be $18$ (this is the `geometric' gauge group, which 
can still be further broken by fluxes). 
Whenever the non-abelian part of the gauge group has rank less than $18$, there are ${\rm U}(1)$ factors which are
geometrically realized as extra sections of the elliptic fibration. The explicit construction 
of extra section has been an active research program in recent years. As discussed in
\cite{Grimm:2010ez,BCV,Mayrhofer:2012zy}, sections can be realized by demanding appropriate 
factorization conditions in the Weierstrass model. A study of fluxes
in (a resolution of) the scenario of \cite{Grimm:2010ez} appeared in \cite{Krause:2011xj}. 
As already discussed in \cite{Grimm:2010ez}, extra sections can equivalently be obtained by realizing 
the elliptic fibre as a hypersurface in ambient spaces with more than a single toric divisor.
This strategy is systematically exploited in 
\cite{Cvetic:2013nia, Cvetic:2013uta, Cvetic:2013jta, Cvetic:2013qsa}. 
Using toric techniques, in particular the classification of tops, models with $I_5$ fibres 
and extra sections were constructed in 
\cite{Braun:2013yti, Borchmann:2013jwa, Braun:2013nqa, Borchmann:2013hta}.
Given a specific embedding of the fibre, one can also use a similar approach as 
\cite{Bershadsky:1996nh}, i.e. use Tate's algorithm, to find all possible degenerations 
leading to a prescribed gauge group \cite{sakuracraigmoritz}. 
F-theory compactifications with ${\rm U}(1)$ symmetries also give rise to an interesting interplay between
geometry and anomalies of the effective field theory, see 
\cite{Park:2011ji, Morrison:2012ei, Cvetic:2012xn, Grimm:2013oga} for some recent works in this
direction.

We regret that we use some mathematical jargon and notations, which are non-standard 
in the physics literature, without explanations. Sections 2--4 of the mathematical companion paper of the
present article \cite{BKW-math} should contain the necessary background.

\section{Four-Form Flux in M/F-Theory on ${\rm K3} \times {\rm K3}$}

\subsection{Review of Known Results}
\label{ssec:review-AK}

Compactification of F-theory on $Y = {\rm K3} \times {\rm K3}$ has 
been studied from various perspectives in the literature. To start 
off, we begin this section with a review of a result in \cite{AK}. \footnote{See also 
Sections \ref{ssec:extended-list}, \ref{ssec:remark-G0} 
and \ref{ssec:rank-16} in this article, 
where the material reviewed in this section is extended.} 
Their results are immediate for M-theory compactification 
down to 2+1-dimensions, but it is clear that we can build a study of F-theory compactification 
down to 3+1-dimensions by adding extra structure and imposing conditions on top of the discussion 
for M-theory \cite{AK}.

When the Calabi--Yau fourfold $Y$ is a product of two K3 surfaces, $S_1$ and 
$S_2$, the complex structure moduli space of $Y$, 
${\cal M}_{\rm cpx}(Y)$, 
is the product of the complex structure moduli space of $S_1$ and $S_2$, 
${\cal M}_{\rm cpx}(S_1) \times {\cal M}_{\rm cpx}(S_2)$.
A discussion of the modular group is postponed to later sections. 
Over the moduli space of $[h^{3,1}(Y) = h^{1,1}(S_1) + h^{1,1}(S_2) = 40]$ 
dimensions, the Hodge decomposition of $H^4(Y; \C)$ varies, because the decompositions 
of $H^2(S_1; \C)$ and $H^2(S_2; \C)$ vary on the moduli spaces of the two K3 surfaces.\footnote{
The $H^4(S_1; \Z) \otimes H^0(S_2;\Z) \oplus 
H^4(S_1; \Z) \otimes H^4(S_2;\Z)$ components in $H^4(Y;\Z)$ (and their 
$\R$-coefficient versions) are ignored here, because fluxes in these 
components do not preserve the $\SO(3,1)$ symmetry in the application 
to F-theory compactifications. This extra assumption is made implicitly everywhere in this article.} 
\begin{align}
 H^{2,2}(Y; \R) & =   H^{1,1}(S_1; \R) \otimes H^{1,1}(S_2; \R) + 
[H^{2,0}(S_1; \C) \otimes H^{0,2}(S_2; \C) + {\rm h.c.}]\, ,   
\label{eq:H22-Y-decomp} \\   
 [ H^{4,0}(Y; \C) + {\rm h.c.} ] & =       
[H^{2,0} (S_1 ; \C) \otimes H^{2,0} (S_2; \C) + {\rm h.c.}]\, ,
    \label{eq:H40-Y-decomp}
\end{align}
where $[V + {\rm h.c.}]$ for a complex vector space $V$ 
with ${\rm dim}_\C V = 1$ denotes the corresponding 2-dimensional 
vector space over $\R$. The Hodge components 
$[H^{4,0}(Y;\C) + {\rm h.c}]$ are also included here for now, partly 
because the four-form flux with non-vanishing $(4,0)$ and $(0,4)$ components 
still preserves AdS supersymmetry. The overlap between
$H^2(S_1; \Z) \otimes H^2(S_2; \Z) \subset H^4(Y; \Z)$ and 
$H^{2,2} \oplus [H^{4,0} + {\rm h.c.}]$ has the maximal 
rank, 404, when 
\begin{equation}\label{attrK3cond}
 {\rm rank} \left[ H^2(S_1; \Z) \cap H^{1,1}(S_1; \R) \right] = 20, \qquad 
 {\rm rank} \left[ H^2(S_2; \Z) \cap H^{1,1}(S_2; \R) \right] = 20 \, .
\end{equation}
The loci satisfying these conditions have complex codimension $40$ in 
the moduli space ${\cal M}_{\rm cpx}(S_1) \times {\cal M}_{\rm cpx}(S_2)$, 
and hence are isolated points. Once plenty of fluxes are introduced 
in this rank 404 free Abelian group, all the complex structure moduli are 
stabilized. 

The Abelian group 
\begin{equation}
  S_X = \left[ H^{1,1}(X; \R) \cap H^2(X; \Z) \right] \subset H^2(X; \Z)  
\end{equation}
for a K3 surface $X$ is called Neron--Severi lattice (or group), and 
the rank of $S_X$---denoted by $\rho_X$ or $\rho(X)$---is called the Picard number of $X$. 
$S_X$ is empty for $X$ with a generic complex structure, but its rank 
can be as large as 20, which is possible only in points of ${\cal M}_{\rm cpx}(X)$. 
K3 surfaces with $\rho_X = 20$ are called {\it attractive K3 surfaces} in \cite{Moore:1998pn}.\footnote{
In the mathematics literature, a K3 surface with $\rho_X = 20$ 
is sometimes called a {\it singular K3 surface}, although the word 
``singular'' only means ``very special'' in this case, and does not imply that 
the surface has a singularity. Ref. \cite{Moore:1998pn} 
introduced the term {\it attractive K3 surface} for K3 surfaces satisfying 
the same condition, which allows us to avoid confusing terminology. 
This terminology is a natural choice: just like the complex structure 
of Calabi--Yau threefolds for type IIB compactifications is attracted 
towards special loci in ${\cal M}_{\rm cpx}$ near the horizon of a BPS 
black hole in 4D ${\cal N}=2$ effective theory of IIB/$CY_3$ in the
attractor mechanism \cite{Ferrara:1995ih}, the complex structure of fourfold 
for F-theory/M-theory should be driven towards special loci 
in ${\cal M}_{\rm cpx}$ in a cosmological evolution in the presence of
($G_1$-type) flux due to the F-term potential from (\ref{eq:GVW}). 
In both cases, special loci are characterized by the condition 
that some topological flux falls into some particular Hodge component.
\cite{Moore:1998pn}. 
In this article, we follow \cite{Moore:1998pn} and use the word 
attractive K3 surface.} Thus, the ensemble of flux vacua of M-theory/F-theory 
compactifications on $Y= {\rm K3} \times {\rm K3}$ are mapped into 
a subset of ${\cal M}_{\rm cpx}(S_1) \times {\cal M}_{\rm cpx}(S_2)$ where 
both $S_1$ and $S_2$ are attractive K3 surfaces. 

It is convenient for the classification of K3 surfaces with large Picard number 
to use the transcendental lattice. For a K3 surface $X$, it is defined as 
the orthogonal complement of $S_X$ under the intersection form in 
$H^2(X; \Z)$:
\begin{equation}
  T_X := \left[ (S_X)^\perp \subset H^2(X; \Z) \right]. 
\end{equation}
For a K3 surface with Picard number $\rho_X$, ${\rm rank}(T_X)=22-\rho_X$.
K3 surfaces with a given transcendental lattice form a 
$(20-\rho_X)$-dimensional subspace of ${\cal M}_{\rm cpx}({\rm K3})$, and 
in particular, attractive K3 surfaces are in one-to-one 
correspondence\footnote{A pair of K3 surfaces $X$ and $X'$ are regarded 
equivalent iff there is a holomorphic bijection between them.} 
with rank-2 transcendental lattices (modulo orientation-preserving 
basis change).

For an attractive K3 surface $X$, its rank-2 transcendental lattice
has to be even and positive definite. This is equivalent to the condition 
that, for a set of generators $\{p,q\}$ of $T_X$, the intersection from 
is given by\footnote{A set of generators $\{q,p\}$ of $T_X$ with the ordering 
between $q$ and $q$ specified is called an {\it oriented basis} of an 
attractive K3 surface $X$, if 
${\rm Im}[ \langle \Omega_X, q \rangle / \langle \Omega_X, p\rangle ] > 0$. 
Choosing ${\rm Im}(\tau) > 0$ as in (\ref{eq:TX-OmegaX}), $\{q,p\}$ is 
indeed an oriented basis. 
We follow the convention of \cite{AK}, and present and parametrize the 
intersection form of $T_X$ as in (\ref{eq:def-of-abc-1}) in this article. 
But it looks more common in math literatures (such as \cite{SI}) and also 
in \cite{Moore:1998pn} to parametrize the intersection form in this way: 
\begin{equation}
 \left[ \begin{array}{cc} (q,q) & (q,p) \\ (p,q) & (p,p) \end{array} \right]
 = \left[ \begin{array}{cc} 2a & b \\ b & 2c \end{array} \right].
\end{equation}
Thus, $[a~b~c]$ here (and in \cite{AK}) correspond to $[c~b~a]$ in 
\cite{SI}. } 
\begin{equation}
 \left[ \begin{array}{cc} (p,p) & (p,q) \\ (q,p) & (q,q) 
        \end{array} \right] = 
 \left[ \begin{array}{cc} 2a & b \\ b & 2c \end{array} \right], 
\label{eq:def-of-abc-1}
\end{equation}
where $a,b,c$ are all integers, $Q := 4ac-b^2$ is positive, and 
$a,c>0$. The 2-dimensional vector space $T_X \otimes \C$ over $\C$
(resp. $T_X \otimes \R$ over $\R$) agrees precisely with 
the vector space $H^{2,0}(X; \C) \oplus H^{0,2}(X; \C)$
(resp. $[H^{2,0}(X; \C) + {\rm h.c.}]$), and the complex vector subspace 
$H^{2,0}(X; \C) \subset T_X \otimes \C$ is identified with
$\C \cdot \Omega_X \subset T_X \otimes \C$, where
\begin{equation}
 T_X \otimes \C \ni \Omega_X := p + \tau q, \qquad \qquad 
  \tau := \frac{-b + i\sqrt{Q}}{2c}\, .
\label{eq:TX-OmegaX}
\end{equation}
With orientation-preserving basis changes of $T_X$, one can always 
choose the integers $a, b, c$ such that 
\begin{equation}
 0 \leq |b| \leq c \leq a \qquad ({\rm but} \quad 0\leq b \quad 
{\rm if~} c = a) \qquad {\rm and~} Q > 0\, .
\label{eq:def-of-abc-2}
\end{equation}
An attractive K3 surface characterized by integers $a, b, c$ in the way explained above
is denoted by $X_{[a~b~c]}$ in this article. 

For a pair of attractive K3 surfaces $S_1$ and $S_2$, let $\{q_1, p_1\}$
and $\{q_2,p_2\}$ be the oriented basis of $T_{S_1}$ and $T_{S_2}$,
respectively. The intersection form in this basis is denoted by 
\begin{equation}
 \left[ \begin{array}{cc} (p_1, p_1) & (p_1,q_1) \\ (q_1,p_1) & (q_1,q_1) 
        \end{array} \right] = 
 \left[ \begin{array}{cc} 2a & b \\ b & 2c \end{array} \right]\, , \qquad \qquad 
 \left[ \begin{array}{cc} (p_2, p_2) & (p_2,q_2) \\ (q_2,p_2) & (q_2,q_2) 
        \end{array} \right] = 
 \left[ \begin{array}{cc} 2d & e \\ e & 2f \end{array} \right]\, ,  
\label{eq:def-of-abcdef}
\end{equation}
where $a,b,c,d,e$ and $f$ are all integers. The positive
definiteness implies that 
\begin{eqnarray}
&& 0 \leq |b| \leq c \leq a, \qquad ({\rm but~}0 \leq b {\rm ~if~}c =
 a), \qquad 
     Q_1:=4ac - b^2 > 0,  \label{eq:cond-abc} \\
&& 0 \leq |e| \leq f \leq d, \qquad ({\rm but~}0 \leq e {\rm ~if~}f = 
 d), \qquad 
      Q_2:= 4df - c^2 > 0, 
\end{eqnarray}
where all $a, \cdots, f$ are integers. 
The holomorphic (2,0)-forms on the K3 surfaces $S_1 = X_{[a~b~c]}$ and 
$S_2 = X_{[d~e~f]}$ can then be written as
\begin{equation}\label{omega_ifromlattice}
 \Omega_{S_1} = p_1 + \tau_1 q_1, \quad 
     \tau_1 = \frac{-b + i \sqrt{Q_1}}{2c},  \qquad 
 \Omega_{S_2} = p_2 + \tau_2 q_2, \quad 
     \tau_2 = \frac{-e + i \sqrt{Q_2}}{2f} \, .
\end{equation}
Note that $\tau_1 \in \Q[\sqrt{Q_1}]$ and $\tau_2 \in \Q[\sqrt{Q_2}]$.
Both $\Q[\sqrt{Q_1}]$ and $\Q[\sqrt{Q_2}]$ are degree 2 (${\cal D}=2$) 
algebraic extensions of $\Q$ (see Ref. \cite{Moore:1998pn}).
 
In physics applications, we would rather like to impose one more condition.
The M2/D3-brane tadpole cancellation is equivalent to 
\begin{equation}
\frac{1}{2} \int_Y (G \wedge G) + N_{M2/D3} =  \frac{\chi(Y)}{24} = 24\, , 
\label{eq:D3-tadpole}
\end{equation}
where $N_{M2/D3}$ is the number of M2/D3-branes minus anti-M2/D3-branes 
(that are point like) in $Y$. When we exclude anti M2/D3-branes on $Y$, 
$N_{M2/D3} \geq 0$, and thus not all the pairs of attractive K3 surfaces in 
${\cal M}_{\rm cpx}(S_1) \times {\cal M}_{\rm cpx}(S_2)$ qualify 
for the landscape of flux vacua of M-theory compactified on 
$Y = {\rm K3} \times {\rm K3}$.

Aspinwall and Kallosh carried out an explicit study of which pairs of 
attractive K3 surfaces can satisfy the condition
(\ref{eq:D3-tadpole}), within a couple of constraints that make 
the analysis easier \cite{AK}. In order to state one of the constraints
introduced in \cite{AK}, we need the following definition. 
Let us focus on a pair of attractive K3 surfaces $S_1$ and $S_2$.
Any $G^{(4)}$ in (\ref{eq:Hodge-22}) on $Y = S_1 \times S_2$ 
can be decomposed, under the Hodge structure of $Y$, into  
$[G] = [G_1] + [G_0]$, where
\begin{align}
  [G_0] & \in  \left[ H^{1,1}(S_1; \R) \otimes H^{1,1}(S_2; \R) \right]
      = S_{S_1} \otimes S_{S_2} \otimes \R, 
      \label{eq:G0-def} \\
   [G_1] & \in \left[ H^{2,0}(S_1; \C) \otimes H^{0,2}(S_2; \C) + {\rm h.c.}
     \right],  \label{eq:G1-def-W=0} \\
   & \qquad \left({\rm or} \qquad 
    \in \left[ H^{2,0}(S_1 ; \C) + {\rm h.c.} \right] \otimes 
        \left[ H^{2,0}(S_2 ; \C) + {\rm h.c.} \right] 
     = T_{S_1} \otimes T_{S_2} \otimes \R \right)\, .
       \label{eq:G1-def-Wnot=0}
\end{align}
The explicit study in Ref. \cite{AK} assumes that 
\begin{equation}
 [G_0] = 0, \label{eq:no-G0}
\end{equation}
and\footnote{
If the $[G_1]$ component were to vanish, then there would be no 
interaction in the effective theory violating ${\cal N}=2$ supersymmetry 
in 3+1 dimensions \cite{Dasgupta:1999ss, Tripathy:2002qw}. Moduli mass terms purely from the $[G_0]$ component are consistent with ${\cal N}=2$ supersymmetry.} 
$[G_1] \neq 0$ is that of (\ref{eq:G1-def-W=0}) rather 
than (\ref{eq:G1-def-Wnot=0}), so that $G^{(4)} = G_1$ is 
purely of $(2,2)$ type in the Hodge structure of $Y=S_1 \times S_2$, 
and the vev of $W_{\rm GVW}$ vanishes. 
Another assumption is to set 
\begin{equation}
 N_{M2/D3} = 0\, , 
\end{equation}
so that 
\begin{equation}
 \frac{1}{2} [G_1] \cdot [G_1] = 24\, . 
\label{eq:G1-totflux=24}
\end{equation}
Under these constraints, $[G_1]$ has to be an integral element of $T_{S_1} \otimes T_{S_2}$:
\begin{equation}
 [G_1] \in [H^{2,0}(S_1; \C) \otimes H^{0,2}(S_2; \C) + {\rm h.c.}] \cap 
   (T_{S_1} \otimes T_{S_2}).
\label{eq:2dimR-in-4dimR}
\end{equation}
It is not always guaranteed for any pair of attractive K3 surfaces 
$S_1$ and $S_2$ that there can be $[G_1] \neq 0$. The Abelian group 
on the right hand side of (\ref{eq:2dimR-in-4dimR}) can be empty. 
Writing down $[G_1]$ as 
\begin{equation}
 [G_1] = {\rm Re} 
   \left(\gamma \Omega_{S_1}\wedge\overline{\Omega}_{S_2} \right)\, 
\label{eq:G1-gamma-def}
\end{equation} 
for some $\gamma \in \C$ and expanding this in the integral basis 
$\{p_1 \otimes p_2, \; q_1 \otimes p_2, \; p_1 \otimes q_2, \; 
q_1 \otimes q_2 \}$
of $T_{S_1} \otimes T_{S_2}$, Aspinwall and Kallosh found that 
(\ref{eq:2dimR-in-4dimR}) is non-empty if and only if 
\begin{equation}
{}^\exists m \in \Z \quad {\rm s.t.} \quad Q_1 Q_2 = m^2\, . 
\label{eq:square-integer}
\end{equation}
This implies that the two algebraic number fields 
$\Q[\sqrt{Q_1}]$ and $\Q[\sqrt{Q_2}]$ are the same. 
All the period integrals of the holomorphic (4,0) form 
$\Omega_Y = \Omega_{S_1} \wedge \Omega_{S_2}$ take values in 
the common degree-2 (${\cal D}=2$) algebraic extension field 
$\Q[\sqrt{Q_1}]$ of $\Q$ (c.f. \cite{DeWolfe:2004ns}).
\begin{table}[tbp]
 \begin{center}
  \begin{tabular}{||c|c|c||c||c|c|c||}
\hline
 [a b c] & [d e f] & $\gamma$ & $\qquad \qquad$ &
 [a b c] & [d e f] & $\gamma$ \\
\hline
\hline
 [8 8 8] & [1 1 1] & $\gamma^{(6)}$ & &
 [6 0 6] & [1 0 1] & $\gamma^{(4)}$ \\
{} [6 0 3] & [2 0 1] & $\pm i/\sqrt{2}$ & &
{} [6 0 2] & [3 0 1] & $\pm i / \sqrt{3}$ \\
{} [6 0 2] & [1 1 1] & $\gamma^{(6)}$ & & 
{} [6 0 1] & [6 0 1] & $ \pm i / \sqrt{6}$ \\
{} [4 4 4] & [2 2 2] & $\gamma^{(6)}$ & & 
{} [3 0 3] & [2 0 2] & $\gamma^{(4)}$ \\
{} [3 0 3] & [1 0 1] & $(1+i)\gamma^{(4)}$ & & 
{} [3 0 2] & [3 0 2] & $\pm i \sqrt{2/3}$ \\
{} [3 0 1] & [2 2 2] & $\gamma^{(6)}$ & & 
{} [2 2 2] & [1 1 1] & $2 \times \gamma^{(6)}$ \\
{} [2 0 1] & [2 0 1] & $\pm 1 \pm i/\sqrt{2}$ & & & & \\
\hline
  \end{tabular}
\caption{\label{tab:AK-flux24} Table 1 of Ref. \cite{AK} is reproduced
here (with minor modifications) for the convenience of the reader. 
Intersection forms (\ref{eq:def-of-abcdef}) of $T_{S_1}$ and $T_{S_2}$ 
are simply denoted by [a b c] and [d e f] in this table. 
All the possible choices of $\gamma \in \C$ are listed; 
$\gamma^{(6)} := \left\{\pm 2i/\sqrt{3}, \pm 1 \pm i/\sqrt{3}\right\}
  = 2i/\sqrt{3} \times \{e^{2 \pi i k /6}| k=0,1,2,3,4,5\}$, 
$\gamma^{(4)} := \left\{ \pm 1, \pm i \right\}
 = \{ e^{2 \pi i k/4} | k=0,1,2,3\}$.}
 \end{center}
\end{table}

Imposing the condition (\ref{eq:G1-totflux=24}) on $[G_1]$ 
in (\ref{eq:2dimR-in-4dimR}), \cite{AK} worked out the complete 
list of pairs of attractive K3 surfaces where there exists a flux 
$G^{(4)}$ satisfying (\ref{eq:no-G0}, \ref{eq:2dimR-in-4dimR}, 
\ref{eq:G1-totflux=24}).
The result are 13 pairs of attractive K3 surfaces \cite{AK}, which are  
listed in Table \ref{tab:AK-flux24} along with all possible values 
of $\gamma \in \C$. 

Two remarks are in order here. First note that for a supersymmetric 
compactification of M-theory on 
$Y = S_1 \times S_2$ with a four-form flux $G^{(4)} = G_1$, 
we could think of a compactification on $\overline{Y} = \overline{S_1} \times
\overline{S}_2$ obtained by simply declaring that the holomorphic local
coordinates on $Y$ are anti-holomorphic coordinates on $\overline{Y}$,
keeping the underlying real-8-dimensional manifold the same.
The flux $[G_1] \in H^4(Y; \Z)$ remains the same. This new compactification,
however, should not be regarded as a vacuum physically different from the
original one, only the role of the superpotential and its hermitian conjugate, 
and that of chiral multiplets and anti-chiral multiplets in the
low-energy effective theory, are exchanged. The physics remains the same.

The transcendental lattice of the K3 surface $S'_1 = \overline{S}_1$ 
can be regarded as $T_{S'_1} \cong {\rm Span}_{\Z} \{ p'_1, q'_1\} := 
{\rm Span}_{\Z} \{ p_1, -q_1\}$, where the symmetric pairing 
is described by [a' b' c'] = [a -b c]. The holomorphic $(2,0)$-form 
is given by\footnote{Here, $\{q'_1, p'_1\}$ is still an oriented basis
of $T_{S'_1}$.} 
\begin{equation}
 \Omega_{S'_1} = p'_1 + \tau'_1 q'_1 := 
p_1 + \bar{\tau}_1 q_1 = \overline{\Omega}_{S_1}. 
\end{equation}
Thus, the four-form flux 
$G_1 = {\rm Re}[ \gamma \Omega_{S_1} \wedge \overline{\Omega}_{S_2}]$ is 
rewritten in terms of $S'_1 \times S'_2 = 
\overline{S}_1 \times \overline{S}_2$ as 
$G_1 = {\rm Re}[ \gamma^* \Omega_{S'_1} \wedge \overline{\Omega}_{S'_2}]$.
Thus, $Y = S_1 \times S_2$ compactification with [a b c], [d e f] and $\gamma$
and another compactification with [a -b c] and [d -e f] and $\gamma^*$
are completely equivalent, and should not be regarded as different 
compactifications (or different vacua). 
For this reason, only one of each such pairs is shown in Table \ref{tab:AK-flux24}.

Secondly, as for M-theory compactification, $Y = S_1 \times S_2$ with 
$G_1 = {\rm Re}(\gamma \Omega_{S_1} \wedge \overline{\Omega}_{S_2})$ and 
$Y= S_2 \times S_1$ with 
$G_1 = {\rm Re}(\gamma^* \Omega_{S_2} \wedge \overline{\Omega}_{S_1})$ 
should also be regarded equivalent. Thus, 
Table \ref{tab:AK-flux24} only shows cases where $a \geq d$, and
furthermore, in cases with $a=d$, we impose $c \geq f$.

\subsection{Extending the List}
\label{ssec:extended-list}

Before proceeding to the next section, it is worthwhile to extend the 
list so that the condition (\ref{eq:G1-totflux=24}) is relaxed to 
\begin{equation}
 \frac{1}{2} [G_1] \cdot [G_1] \leq  24.
\label{eq:G1-partflux<=24}
\end{equation}
Certainly for any compactification of M-theory over 
$Y = S_1 \times S_2$ with a four-form flux $[G_1]$ 
in (\ref{eq:2dimR-in-4dimR}) satisfying the inequality above, at least 
we can introduce an appropriate number of M2-branes 
to satisfy the tadpole condition (\ref{eq:D3-tadpole}). One might even be 
able to find a flux $[G_0] \in (S_{S_1} \otimes S_{S_2})$ to 
satisfy (\ref{eq:D3-tadpole}). See Section \ref{sec:G0-flux} for more 
about the case with $[G_0] \neq 0$, however.

Straightforward analysis allows us to extend Table
\ref{tab:AK-flux24}, so that it contains all pairs of attractive 
K3 surfaces $S_1$ and $S_2$ and a choice of $\gamma \in \C$ satisfying 
(\ref{eq:no-G0}, \ref{eq:2dimR-in-4dimR}) and
(\ref{eq:G1-partflux<=24}), rather than 
(\ref{eq:no-G0}, \ref{eq:2dimR-in-4dimR}) and (\ref{eq:G1-totflux=24}). 
The result of our analysis is presented in Table \ref{tab:AK-notmorethan24}.
\begin{table}[htbp]
\begin{center}
 \begin{tabular}{||c|c|c|c|r|r||}
\hline
 flux & [a b c] & [d e f] & $\gamma$ & M & F \\
\hline
\hline
24 & [8 8 8] & [1 1 1] & $\gamma^{(6)}$ & 6 & 6 \\
 & [6 0 6] & [1 0 1] & $\gamma^{(4)}$ & 3 & 3 \\
 & [6 0 3] & [2 0 1] & $\pm i/\sqrt{2}$ & 1 & 1 \\
 & [6 0 2] & [3 0 1] & $\pm i / \sqrt{3}$ & 1 & 1 \\
 & [6 0 2] & [1 1 1] & $\gamma^{(6)}$ & 6 & 6 \\
 & [6 0 1] & [6 0 1] & $ \pm i / \sqrt{6}$ & 1 & 1 \\
 & [4 4 4] & [2 2 2] & $\gamma^{(6)}$ & 6 & 6 \\
 & [3 0 3] & [2 0 2] & $\gamma^{(4)}$ & 3 & 3 \\
 & [3 0 3] & [1 0 1] & $(1+i)\gamma^{(4)}$ & 2 & 2 \\
 & [3 0 2] & [3 0 2] & $\pm i \sqrt{2/3}$ & 1 & 1 \\
 & [3 0 1] & [2 2 2] & $\gamma^{(6)}$ & 6 & 6 \\
 & [2 2 2] & [1 1 1] & $2\gamma^{(6)}$ & 6 & 6 \\
 & [2 0 1] & [2 0 1] & $\pm 1 \pm i/\sqrt{2}$ & 2 & 2 \\
\hline
 23 & [6 1 1] & [6 1 1] & $\pm 2 i/\sqrt{23}$ & 1 & 2 \\
 & [3 1 2] & [3 1 2] & $\pm 4i / \sqrt{23}$ & 1 & 2 \\
\hline
 22 & [6 2 2] & [3 1 1] & $\pm 2 i/\sqrt{11}$ & 2 & 2 \\
\hline
 21 & [7 7 7] & [1 1 1] & $\gamma^{(6)}$ & 6 & 6 \\
 & [6 3 3] & [2 1 1] & $\pm 2 i/\sqrt{7}$ & 2 & 2 \\
 & [1 1 1] & [1 1 1] & $(2 \pm \sqrt{3}i) \gamma^{(6)}$ & 
		      6 & 12 \\
\hline
 20 & [5 0 5] & [1 0 1] & $\gamma^{(4)}$ & 3 & 3 \\
 & [5 0 1] & [5 0 1] & $\pm i/\sqrt{5}$ & 1 & 1 \\
 & [3 2 2] & [3 2 2] & $\pm 2 i/\sqrt{5}$ & 1 & 2 \\
 & [1 0 1] & [1 0 1] & $(1 \pm 2i) \gamma^{(4)}$ & 4 & 4 \\
\hline 
 19 & [5 1 1] & [5 1 1] & $\pm 2 i /\sqrt{19}$ & 1 & 2 \\
\hline
 18 & [6 6 6] & [1 1 1] & $\gamma^{(6)}$ & 6 & 6 \\
 & [3 3 3] & [2 2 2] & $\gamma^{(6)}$ & 6 & 6 \\
 & [2 2 2] & [1 1 1] & $2 \gamma^{(6)}$ & 6 & 6 \\
\hline
 17 & & & & & \\
\hline
 16 & [4 0 4] & [1 0 1] & $\gamma^{(4)}$ & 3 & 3 \\
 & [4 0 2] & [2 0 1] & $\pm i/\sqrt{2}$ & 1 & 1 \\
 & [4 0 1] & [4 0 1] & $\pm i/2$ & 1 & 1 \\
 & [4 0 1] & [1 0 1] & $\gamma^{(4)}$ & 3 & 3 \\
 & [2 0 2] & [2 0 2] & $\gamma^{(4)}$ & 3 & 3 \\
 & [2 0 2] & [1 0 1] & $(1+i)\gamma^{(4)}$ & 2 & 2 \\
 & [2 0 1] & [2 0 1] & $\pm 1$ & 2 & 2 \\
 & [1 0 1] & [1 0 1] & $2 \gamma^{(4)}$ & 3 & 3 \\ 
\hline
\end{tabular}
\begin{tabular}{||c|c|c|c|c|r|r||}
\hline
 flux & [a b c] & [d e f] & $\gamma$ & M & F \\
\hline
\hline
 15 & [5 5 5] & [1 1 1] & $\gamma^{(6)}$ & 6 & 6 \\
 & [4 1 1] & [4 1 1] & $\pm 2 i/\sqrt{15}$ & 1 & 2 \\
 & [2 1 2] & [2 1 2] & $\pm 4 i /\sqrt{15}$ & 1 & 2 \\ 
\hline
 14 & [4 2 2] & [2 1 1] & $\pm 2i / \sqrt{7}$ & 2 & 2 \\
 & [2 1 1] & [2 1 1] & $\pm 1 \pm i /\sqrt{7}$ & *2 & 4 \\
\hline
 13 & & & & & \\
\hline
 12 & [4 4 4] & [1 1 1] & $\gamma^{(6)}$ & 6 & 6 \\
 & [3 0 3] & [1 0 1] & $\gamma^{(4)}$ & 3 & 3 \\
 & [3 0 1] & [3 0 1] & $\pm i/\sqrt{3}$ & 1 & 1 \\
 & [3 0 1] & [1 1 1] & $\gamma^{(6)}$ & 6 & 6 \\
 & [2 2 2] & [2 2 2] & $\gamma^{(6)}$ & 3 & 6 \\
 & [1 1 1] & [1 1 1] & $2 \gamma^{(6)}$ & 3 & 6 \\
\hline
 11 & [3 1 1] & [3 1 1] & $\pm 2 i /\sqrt{11}$ & 1 & 2 \\
\hline 
 10 & & & & & \\
\hline
 9 & [3 3 3] & [1 1 1] & $\gamma^{(6)}$ & 6 & 6 \\
 & [1 1 1] & [1 1 1] & $\sqrt{3}i \gamma^{(6)}$ & 3 & 6 \\
\hline
 8 & [2 0 2] & [1 0 1] & $\gamma^{(4)}$ & 3 & 3 \\
 & [2 0 1] & [2 0 1] & $\pm i/\sqrt{2}$ & 1 & 1 \\
 & [1 0 1] & [1 0 1] & $(1+i)\gamma^{(4)}$ & 2 & 2 \\
\hline
 7 & [2 1 1] & [2 1 1] & $\pm 2 i /\sqrt{7}$ & 1 & 2 \\
\hline 
 6 & [2 2 2] & [1 1 1] & $\gamma^{(6)}$ & 6 & 6 \\
\hline
 5 & & & & & \\
\hline
 4 & [1 0 1] & [1 0 1] & $\gamma^{(4)}$ & 3 & 3 \\
\hline
 3 & [1 1 1] & [1 1 1] & $\gamma^{(6)}$ & 3 & 6 \\
\hline
\hline
\hline
\hline
\hline
 flux & [a b c] & [d e f] & $\gamma$ & M & F \\
\hline
\hline
 23 & [6 -1 1] & [6 1 1] &  $\pm 2i/\sqrt{23}$ & 2 & 2 \\
\hline
 22 & [3 -1 1] & [6 2 2] & $\pm 2i \sqrt{11}$ & 2 & 2 \\
\hline
 21 & [2 -1 1] & [6 3 3] & $\pm 2i /\sqrt{7}$ & 2 & 2 \\
\hline
 20 & [3 -2 2] & [3 2 2] & $\pm 2i/\sqrt{5}$ & 2 & 2 \\
\hline
 19 & [5 -1 1] & [5 1 1] & $\pm 2i /\sqrt{19}$ & 2 & 2 \\
\hline 
 15 & [4 -1 1] & [4 1 1] & $\pm 2i/\sqrt{15}$ & 2 & 2 \\
\hline 
 14 & [2 -1 1] & [2 1 1] & $\pm 1 \pm i/\sqrt{7}$ & *4 & 4 \\
 & [2 -1 1] & [4 2 2] & $\pm 2i /\sqrt{7}$ & 2 & 2 \\
\hline
 11 & [3 -1 1] & [3 1 1] & $\pm 2 i/\sqrt{11}$ & 2 & 2 \\
\hline 
 7 & [2 -1 1] & [2 1 1] & $\pm 2i/\sqrt{7}$ & 2 & 2 \\
\hline  
\end{tabular}
\caption{\label{tab:AK-notmorethan24}
This table reproduces all the 13 cases with the total flux of $G_1$ type 
being 24 in \cite{AK}. The first 5/6 of this table covers the cases 
with $be \geq 0$; see comments at the end of Section \ref{ssec:review-AK}.
The last 1/6 of this table is the list of cases with $(be) < 0$; 
this table only shows cases with $b < 0$ (so $e > 0$), in order to
 reduce the redundant information associated with $S_2
 \longleftrightarrow S_1$ and $\gamma \longleftrightarrow \gamma^*$.
See Section \ref{ssec:sbtl-stat} for the meaning of the last two 
columns.}
 \end{center}
\end{table}

\begin{figure}[tbp]
 \begin{center}
  \begin{tabular}{ccc}
  \includegraphics[width=.3\linewidth]{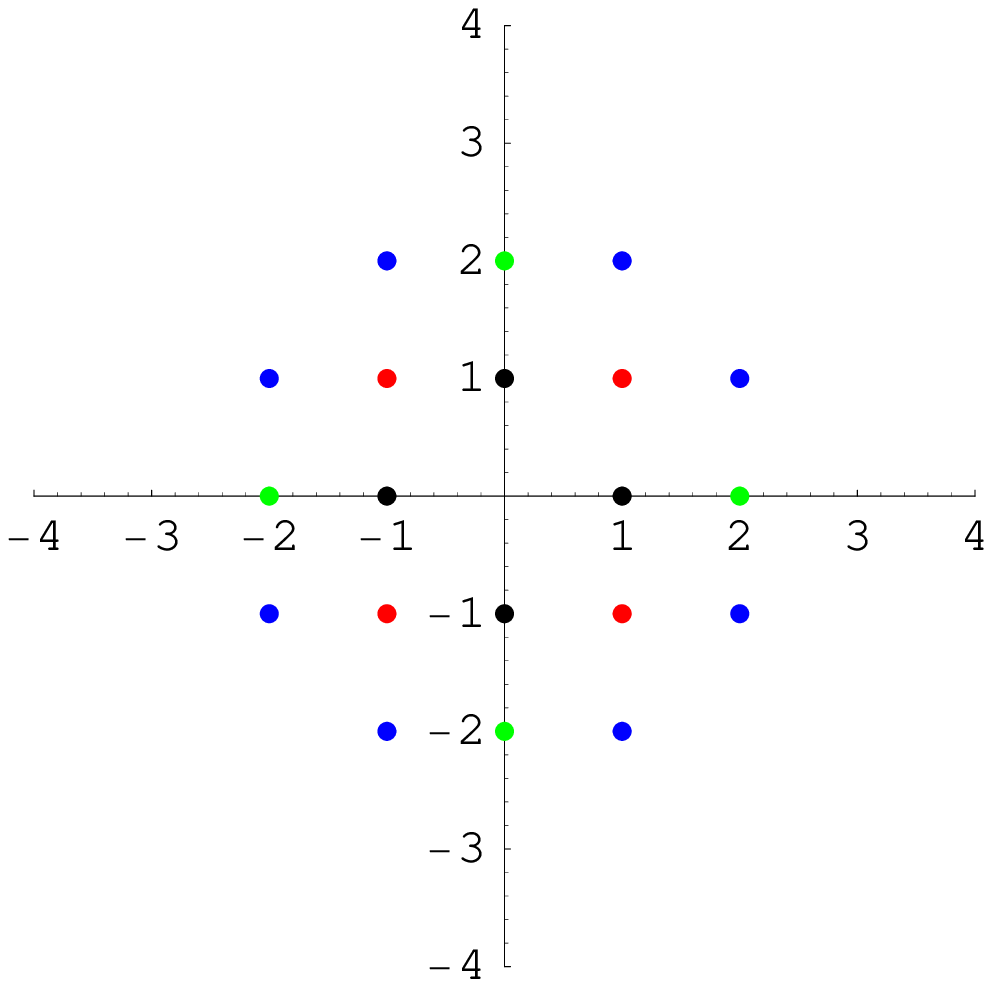}   & & 
  \includegraphics[width=.3\linewidth]{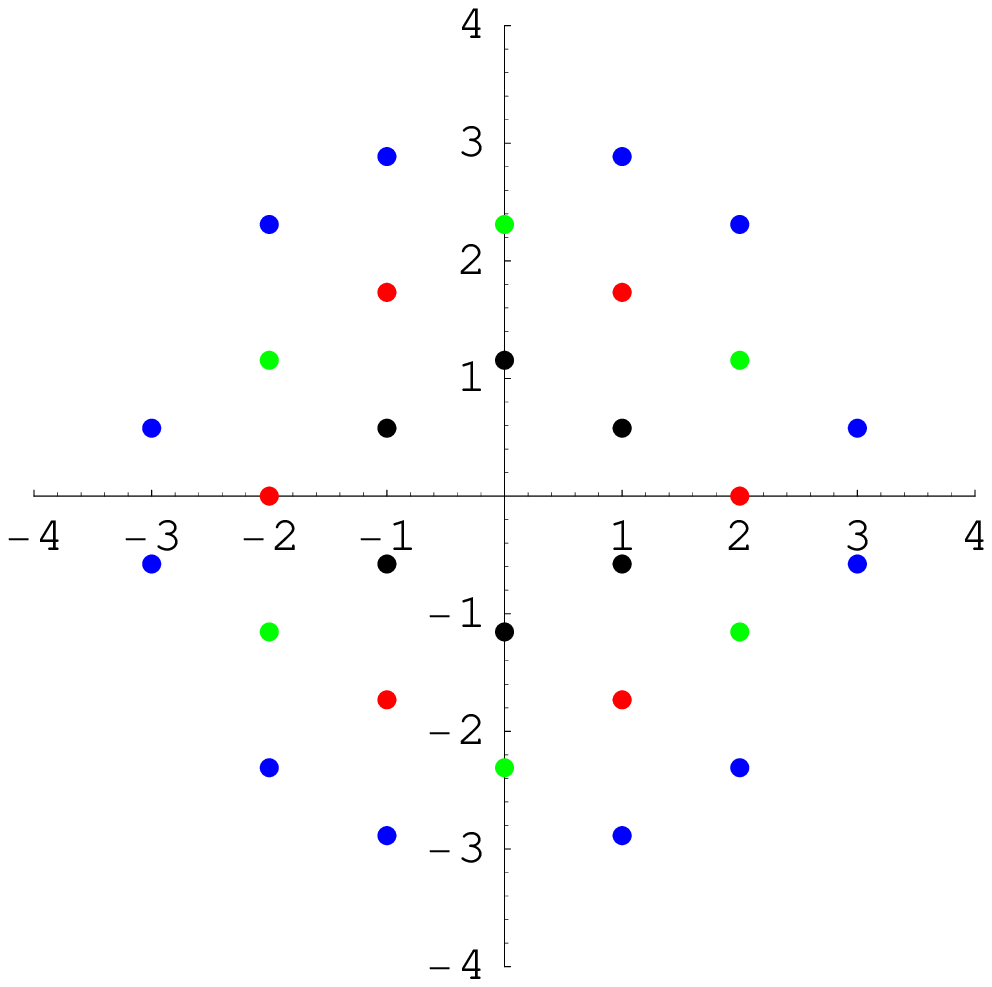} \\
 (a)  for $X_{[1~0~1]} \times X_{[1~0~1]}$ & &
 (b)  for $X_{[1~1~1]} \times X_{[1~1~1]}$
  \end{tabular}
\caption{\label{fig:gamma-lattice}Possible values of 
$\gamma \in \C$ appearing in Table \ref{tab:AK-notmorethan24} for 
$X_{[1~0~1]} \times X_{[1~0~1]}$ and $X_{[1~1~1]} \times X_{[1~1~1]}$, 
respectively. In (a), black, red, green and blue points 
(from inside to outside) correspond to $[G_1] \cdot [G_1]/2 = 4, 8, 16$
  and $20$, respectively, while in (b), the black, red, green and blue
  points give rise to the contributions $3, 9, 12$ and $21$, respectively.}
 \end{center}
\end{figure}

Out of the 66 entries in Table \ref{tab:AK-notmorethan24}, some pairs of
K3 surfaces appear more than once. In these cases, there are different possible choices 
of $\gamma$ which give rise to different contributions to the tadpole that are less than $24$.
For the pairs $S_1 \times S_2 = X_{[1~0~1]} \times X_{[1~0~1]}$
and $X_{[1~1~1]} \times X_{[1~1~1]}$, all the possible values of 
$\gamma \in \C$ (and the corresponding $[G_1] \cdot [G_1]/2$
contribution to the M2/D3 tadpole) are shown in Figure~\ref{fig:gamma-lattice}.
They form a lattice within $\C$, and so do $[G_1]$ 
in (\ref{eq:2dimR-in-4dimR}) \cite{Moore:1998pn}. In fact, one
can show that that the possible values for $\gamma$ form a lattice
for any pair of K3 surfaces satisfying \eqref{eq:square-integer}. 
The upper bound (\ref{eq:G1-partflux<=24}), however, allows 
only finitely many choices of $G^{(4)} = G_1$ for a given pair 
of attractive K3 surfaces satisfying (\ref{eq:square-integer}).

\section{F-theory Classification of Elliptic Fibrations on a K3 Surface}
\label{sec:F-classify-elliptic-fib}

The study reviewed in the previous section implies that
a pair of K3 surfaces $S_1$ and $S_2$ corresponding to 
a pair of transcendental lattices 
$T_{S_1} = \left[ \begin{array}{cc}2a & b\\b & 2c \end{array}\right]$
and 
$T_{S_2} = \left[ \begin{array}{cc}2d & e\\e & 2f \end{array}\right]$
in Tables \ref{tab:AK-flux24} and \ref{tab:AK-notmorethan24} is realized 
in the landscape of flux compactifications of M-theory on 
$Y = S_1 \times S_2$ down to 2+1-dimensions, with all the 40 complex 
structure moduli stabilised. In order to translate this result to 
the landscape of F-theory compactifications to 3+1-dimensions, however, 
we have to impose a couple of extra conditions \cite{AK}. 

One of the conditions to be imposed, of course, is that either one 
of the K3 surfaces $S_1$ or $S_2$ admits an elliptic fibration 
with a section (in its vacuum complex structure), and the vev of the  
K\"{a}hler moduli should be such that the volume of the elliptic fibre 
vanishes. Let $X$ be this elliptic fibred K3 surface,\footnote{
In this article, we always imply by ``elliptic fibration'' that 
it is accompanied by a section. 
} and the other one 
of $S_1$ and $S_2$ be denoted by $S$;
\begin{equation}
 Y = X \times S,  \label{eq:Y=XxS} \qquad 
 \pi_X: X \rightarrow \P^1,   
  \qquad 
 \sigma: \P^1 \rightarrow X.
\end{equation}
The authors of \cite{AK} pointed out that 
$S_1$ (resp. $S_2$) can be identified with a K3 surface of the form 
$X = T^4/\Z_2$, if and only if all of a, b and c 
(resp. d, e and f) in Table \ref{tab:AK-flux24} are even (the same 
rule applies also to Table \ref{tab:AK-notmorethan24}), based on 
a known fact on Kummer surfaces (see \cite{PSS, SI, Barthetal}). 
Projecting down to $T^2/\Z_2 \simeq \P^1$, we obtain an elliptic fibration 
with four singular fibres of type $I^*_0$ (namely, $D_4 = \SO(8)$ 
gauge groups on 7-branes); this is the F-theory/Type IIB orientifold 
model in \cite{Sen-SO8}.
%

This class of F-theory vacua, which admits a type IIB orientifold
interpretation without any approximation or ambiguity, is only a subset 
of all possible vacua of F-theory, however. In fact, it is known 
that any K3 surface with $\rho\geq 13$ admits an elliptic fibration with a
section (Lemma 12.22 of \cite{Sch}). Hence all the K3 surfaces in Tables \ref{tab:AK-flux24} and \ref{tab:AK-notmorethan24} admit 
an elliptic fibration with a section, so that they all have an interpretation 
in terms of F-theory if the vev of the K\"{a}hler moduli is chosen 
appropriately. 
It should be noted, however, that there can be more than one elliptic 
fibration morphism $\pi_X : X \longrightarrow \P^1$ for
a given K3 surface $X$, and furthermore, the type of singular fibres 
(type = collection of some of $I_n$, $I^*_n$, II, III, IV$^*$, III$^*$ and
II$^*$) will in general be different for each of the fibrations. 
We are thus facing at least two questions: 
\begin{itemize}
 \item How do we find out the list of all possible elliptic fibrations,
       when the transcendental lattice $T_X \subset H^2(X; \Z)$ of a K3
       surface $X$ is given?  
 \item Suppose that there are two elliptic fibrations 
       $\pi_X: X \longrightarrow \P^1$ and 
       $\pi'_X: X \longrightarrow \P^1$ available; how do we find out
       whether the two fibrations correspond to the same vacuum in
       physics or not?  
\end{itemize}
The former question is purely mathematical in nature, while the latter 
is a question in physics. 
Our companion paper \cite{BKW-math} is dedicated to a study of the first 
question, while the latter question is addressed in this section. 
The primary conclusion in this section 
is (\ref{eq:classify-Uembed-fibre-p}) and the discussion that follows 
immediately after.\footnote{It is an option to skip this
section and proceed to the next section, if one is happy to accept 
this statement.} We begin by reviewing Torelli
theorem for K3 surface in Section \ref{ssec:Torelli}, as it plays 
a crucial role in our discussion in Section \ref{ssec:Het-F-AutX}.  

\subsection{On the Torelli theorem for K3 surfaces}
\label{ssec:Torelli}

In this we discuss the relation between 
the moduli space of complex structures of K3 surfaces and periods
of the holomorphic $(2,0)$ form. Statements of this type are generally 
referred to as `Torelli theorems'. For K3 surfaces, there exist several 
powerful versions, which are closely related to each other, yet shed
light on the subject from slightly different angles. Those Torelli
theorems combined allow many questions on K3 moduli to be reformulated 
in terms of lattice theory. The following review on Torelli theorems for 
K3 surfaces is designed to serve as preparation for 
Section \ref{ssec:Het-F-AutX}. This review together with Section
2 of \cite{BKW-math} is designed to be self-contained, all the jargon and
notation without definition or explanation in this section should 
be explained in Section 2 of \cite{BKW-math}.
See also \cite{Huybrechts, Barthetal} for a concise mathematical exposition.

We begin with defining such words as ``(moduli space of) marked K3
surface'' and ``period domain'', and proceed to explain the period map.
A {\it marked K3 surface} is a K3 surface $X$ for which we have fixed a
set of generators for $H^{2}(X; \Z)$, i.e. we consider
a pair $(X, \varphi)$ which consists of the K3 surface $X$ and an isometry between lattices 
$\varphi: H_2(X; \Z) \longrightarrow \Lambda_{\rm K3}$, where 
\begin{equation}
 \Lambda_{\rm K3} = U \oplus U \oplus U \oplus E_8 \oplus E_8\, .
\end{equation}
The map $\varphi$ is called the {\it marking}. Note that we use conventions 
in which the $A$--$D$--$E$ lattices have a negative definite inner product, as is natural
in the present context.

Two marked K3 surfaces $(X, \varphi)$ and $(X', \varphi')$ are said 
to be {\it equivalent}, if and only if there is an isomorphism 
$f: X \rightarrow X'$ such that $\varphi' \cdot f_* = \varphi$ as an
isometry from $H_2(X; \Z)$ to $\Lambda_{\rm K3}$. Each point of 
the {\it moduli space of marked K3 surface} $N$ corresponds to such an equivalence 
class of marked K3 surfaces. 


The {\it period domain} $D$, on the other hand, 
is a subspace of $\P [ \Lambda_{\rm K3} \otimes \C]$ given by 
\begin{equation}
 D := \{ [\omega] \in \P[ \Lambda_{\rm K3} \otimes \C] \; | \; 
     \omega \cdot \omega = 0, \; \omega \cdot \overline{\omega} > 0 \;
     \} \subset \P [ \Lambda_{\rm K3} \otimes \C]\, .
\label{eq:def-period-domain-K3}
\end{equation}
The global structure of $D$ is given by 
\begin{equation}
 D \cong O(\Lambda_{\rm K3} \otimes \R)/ {\rm SO}(2) \times O(1,19) =
  {\rm Gr}^{\rm po}(2; \Lambda_{\rm K3} \otimes \R)\, , 
\end{equation}
where the superscript ``po'' stands for ``positive and oriented'', 
in the sense that we consider the Grassmannian of oriented 2-dimensional 
subspaces in $\Lambda_{\rm K3} \otimes \R$ with signature $(2,0)$.\footnote{
By forgetting the
orientation, a twofold cover $D = {\rm Gr}^{\rm po}(2; \Lambda_{\rm
K3}\otimes \R) \rightarrow {\rm Gr}^{\rm p}(2; \Lambda_{\rm K3} \otimes
\R)$ can be constructed; 
${\rm Gr}^{\rm p}(2; \Lambda_{\rm K3}\otimes \R) \cong 
O(\Lambda_{\rm K3} \otimes \R)/ O(2) \times O(1,19)$.}  

Using the holomorphic $(2,0)$-form $\Omega_X\in H_2(X; \C)$ and the intersection form on $H_2(X;\C)$,
we may map a point in the moduli space of marked K3 surface $N$ to 
a point in the period domain $D$:
\begin{equation}
 {\cal P}: N \ni [(X,\varphi)] \longmapsto \varphi([\Omega_X]) 
  \in D \subset \P[\Lambda_{\rm K3} \otimes \C]\, .
\end{equation}
${\cal P}$ is called the {\it period map}. Here, $[\Omega_X]$ stands for both the complex line 
$\C \Omega_X \subset H_2(X; \C)$ as well as its image in $\P[H_2(X;
\C)]$; the same notation has already been used 
in (\ref{eq:def-period-domain-K3}).

\vspace{5mm}

The classic {\bf local Torelli theorem} for K3 surface states that the 
period map is locally an isomorphism between $N$ and $D$.  
For any point in $N$ and its image under the period map in $D$, we can
always take an appropriate open set in $N$ and $D$ so that the period
map becomes an isomorphism between the two open sets. 
Thus, locally in the moduli space(s), K3 surfaces are uniquely 
determined by their periods. 

In the following, we will turn to global aspects of the moduli spaces 
$N$ and $D$, and the period map ${\cal P}$ between them. 
While it turns out that 
complex deformations can be used to introduce local coordinates 
on $N$, and to give it the structure of a complex manifold, 
the moduli space $N$ fails to be Hausdorff. 
The way the period maps glue globally is expressed by the


\begin{quote}
 {\bf Global Torelli Theorem, version 1} (see e.g. \cite{Huybrechts}): 
the moduli space of marked K3 surfaces $N$ consists of two connected 
components $N^o$ and $N^{o'}$, and the period map ${\cal P}$ maps each 
one of them surjectively, and also generically injectively 
to the period domain $D$.
\end{quote}


To elaborate more on this, note first that  
the group of all the isometries of the lattice 
$\Lambda_{\rm K3}$---${\rm Isom}(\Lambda_{\rm K3})$---acts naturally 
on $D$ (from the left), and it also acts on $N$ through the marking; 
$g \in {\rm Isom}(\Lambda_{\rm K3})$ maps $[(X,\varphi)] \in N$ to 
$[(X,g \cdot \varphi)] \in N$. The action of this symmetry 
group on $N$ and $D$ commutes with the period map ${\cal P}: N
\longrightarrow D$. This isometry group has a structure 
${\rm Isom}(\Lambda_{\rm K3}) \cong 
\{ \pm {\rm id.}\} \times {\rm Isom}^+(\Lambda_{\rm K3})$.
The ${\rm Isom}^+(\Lambda_{\rm K3})$ subgroup is such that 
the orientation of the 3-dimensional positive definite subspace of 
$\Lambda_{\rm K3} \otimes \R$ is preserved. 

A pair of elements $[(X,\varphi)]$ and $[(X,-\varphi)]$ 
are different points in $N$ because automorphisms of $X$
cannot induce $(-{\rm id.})$ on $H_2(X; \C)$. The period map 
${\cal P}: N \longrightarrow D$ sends these two elements to the same 
point in $D$, $[\varphi(\Omega_X)] = [-\varphi(\Omega_X)] \in
\P[\Lambda_{\rm K3} \otimes \C]$.
In the Torelli theorem above, such a pair of points in $N$ corresponds 
to two inverse images\footnote{In this version of the Torelli theorem, 
we included a statement that there are just {\it two} and not more than 
two connected components in the moduli space of marked K3 surfaces $N$.
This statement comes from the fact that the full ${\rm Isom}^+(H_2(X;
\Z))$ subgroup of ${\rm Isom}(H_2(X; \Z))$ is realized as the monodromy 
group on $H_2(X; \Z)$ for a given K3 surface $X$ through continuous 
complex structure deformations \cite{matumoto,borcea,donaldson}.
See \cite{Huybrechts} Chapt.10 for more detailed list of references.} 
of a given point in $D$; one is in $N^o$, and 
the other is in the other connected component $N^{'o}$. 
The action of $(- {\rm id.}) \in {\rm Isom}(\Lambda_{\rm K3})$ maps 
these two elements in $N$ to each other, and hence the subgroup 
$\{ \pm {\rm id.} \}$ of ${\rm Isom}(\Lambda_{\rm K3})$ acts trivially 
on $D$, while it exchanges the two connected components of $N$.
Next we discuss the classical form of the global Torelli theorem.
First, the homomorphism ${\rm Aut}(X) \longrightarrow {\rm Isom}(H_2(X;
\Z))$ is injective (Prop. 2 of \S 2 in \cite{PSS}), 
where ${\rm Aut}(X)$ is the group of automorphisms of a K3 surface $X$, 
and ${\rm Isom}(H_2(X; \Z))$ is the isometry group of $H_2(X; \Z)$ 
endowed with a symmetric pairing from intersection number. 
Note that this means that there cannot be any non-trivial automorphism 
which acts as the identity on $H_2(X;\Z)$. 
Furthermore, 

%
%
\begin{quote}
 {\bf Global Torelli Theorem, version 2}: 
(Prop. of \S 7 and Thm. 1 of \S 6 in \cite{PSS})
the image of ${\rm Aut}(X)$ under the injective homomorphism is 
${\rm Isom}(H_2(X; \Z))^{({\rm Hodge~eff})}$, the group  
of isometries that are both Hodge and effective. 
This means that for a given 
$\varphi \in {\rm Isom}(H_2(X; \Z))^{({\rm Hodge~eff})}$, there exists 
a unique automorphism $f \in {\rm Aut}(X)$ such that 
$\varphi = f_*$. Furthermore, the subgroup 
${\rm Isom}(H_2(X; \Z))^{({\rm Hodge~eff.})} \cong {\rm Aut}(X)$ 
sits inside the group of Hodge isometries, which have the form  
\begin{equation}
 {\rm Isom}\left( H_2(X; \Z) \right)^{({\rm Hodge})} \cong 
   \{ \pm {\rm id.} \} \times \left[ W^{(2)}(S_X) \rtimes {\rm Aut}(X)
			      \right]\, .
\label{eq:str-Isom-Hodge-H2}
\end{equation}
\hspace{5mm} For K3 surfaces $X$ and $X'$ there is an isomorphism of 
surfaces $f: X \longrightarrow X'$ if and only if there is a Hodge 
and effective isometry $\varphi: H_2(X; \Z) \longrightarrow H_2(X'; \Z)$. In
this case, $f_* = \varphi$.
\end{quote}

%
See the mathematics literature (such as \cite{PSS}) or \cite{BKW-math} for the
definition of Hodge and effective isometries. 
$W^{(2)}(S_X)$ is the group generated by reflections associated with 
algebraic curves of self-intersection $(-2)$ in the
Neron--Severi lattice $S_X$. 
Section 2.2 of \cite{BKW-math} explains the structure of 
the group (\ref{eq:str-Isom-Hodge-H2}) in more detail. 
Another version of the theorem, which is equivalent to version 2, 
is also useful:

%
%
\begin{quote}
{\bf Global Torelli Theorem, version 3} (e.g., Chapt. 10 of \cite{Huybrechts}): 
For a pair of K3 surface $X$ and $X'$, there is an automorphism 
$f: X \longrightarrow X'$, if and only if there is a Hodge isometry 
$\varphi: H_2(X; \Z) \longrightarrow H_2(X'; \Z)$. Furthermore, 
in this case, 
$\varphi^{-1} \cdot f_* \in {\rm Isom}(H_2(X; \Z))^{({\rm Hodge})}$.
When it is known that $\varphi$ maps ${\rm Pos}^+_X$ to 
${\rm Pos}^+_{X'}$, then 
$\varphi^{-1} \cdot f_* \in {\rm Isom}^+(H_2(X; \Z))^{({\rm Hodge})}$,
the index 2 subgroup of ${\rm Isom}(H_2(X; \Z))^{({\rm Hodge})}$ obtained
by dropping $\{ \pm {\rm id.}\}$ from (\ref{eq:str-Isom-Hodge-H2}).
\end{quote}

Having seen these three versions of the Torelli theorem, we may now
use the perspective of version 3 of the global Torelli theorem 
to elucidate the meaning of the expression `generically injective' used in version 1.
Suppose that the period map ${\cal P}: N^o \longrightarrow D$ restricted 
to one of the two connected components maps two points 
$[(X, \varphi)]$ and $[(X',\varphi')]$ in $N^o$ 
to one and the same point $[\omega] \in D$. That is, both 
$[(X, \varphi)]$ and $[(X', \varphi')]$ are contained in 
${\cal P}^{-1}([\omega]) \cap N^o$.
It then follows from version 3 of the global Torelli theorem that there exists 
an isomorphism of surfaces $f: X \longrightarrow X'$, because 
$\varphi^{'-1} \cdot \varphi: H_2(X; \Z) \longrightarrow 
\Lambda_{\rm K3} \longrightarrow H_2(X'; \Z)$ is a Hodge isometry. 
This means that a marked K3 surface $(X,\varphi' \cdot f)$ is equivalent 
to the marked K3 surface $[(X', \varphi')]$ in $N^o$. Thus, if ${\cal P}: [(X_0, \varphi_0)] \longmapsto [\omega_0]
= \varphi_0([\Omega_X])$, then all elements of 
${\cal P}^{-1}([\omega_0]) \cap N^o$ can be written in the form of 
$[(X_0,\varphi)]$ with some marking $\varphi$.

Deviation from the injectiveness of the period map ${\cal P}|_{N^o}$
therefore corresponds to the variety of marking $\varphi$ allowed 
for ${\cal P}^{-1}([\omega_0]) \cap N^o$. The remaining variety 
for $\varphi$ can also be read out from the version 3 of the global 
Torelli theorem. Since $[(X, \varphi)]$ belongs to the same 
connected component as $[(X, \varphi_0)]$, 
$\varphi_0^{-1} \cdot \varphi \in W^{(2)}(S_X) \rtimes {\rm Aut}(X)$,
and conversely, any $\varphi$ satisfying this condition gives rise 
to $[(X, \varphi)] \in {\cal P}^{-1}([\omega_0]) \cap N^o$. Therefore, 
reminding ourselves of the definition of the equivalence relation between 
$(X,\varphi)$ and $(X,\varphi_0)$ in the moduli space $N$, we see that 
\begin{equation}
 {\cal P}^{-1} ([\omega_0]) \cap N^o = \left\{ 
   [(X,\varphi)] \; \left| \; \varphi \in \varphi_0 \cdot
       \left[ W^{(2)}(S_X) \rtimes {\rm Aut}(X) \right] / {\rm Aut}(X) 
                   \right.             \right\}. 
\label{eq:period-map-fibr}
\end{equation}

For a general (non algebraic) complex K3 surface $X$, the Neron--Severi 
lattice $S_X$ is trivial, $\rho_X = 0$, so that $W^{(2)}(S_X)$ is the 
trivial group. In this case, there is only one element 
$[(X, \varphi)] \in N^o$ that is mapped to a given point $[\omega] \in
D$, that is, the period map ${\cal P}: N^o \longrightarrow D$ is 
injective there. Since only a measure-zero subspace of $N$ is occupied by 
algebraic K3 surfaces, the period map is indeed generically injective. 
For an algebraic K3 surface $X$, however, the group $W^{(2)}(S_X)$ 
can be non-trivial, and there can be multiple points in the inverse 
image of the period map, as in (\ref{eq:period-map-fibr}).
%
Since our interest in this article is primarily in K3 surfaces $X$ 
with large Picard number, $\rho_X = {\rm rank}(S_X)$, this non-injective 
behaviour of the period map is of particular importance.

Although we have seen above that any two points in $N$ that are mapped to the same 
point in $D$ are represented by a common K3 surface $X$, there are 
more points in $N$ that share the same K3 surface $X$.
To see this, remember that the ${\rm Isom}^+(\Lambda_{\rm K3})$ subgroup 
of ${\rm Isom}(\Lambda_{\rm K3})$ acts on individual connected
components of $N$, that is, $N^o$ and $N^{o'}$, as well as on the period 
domain $D$.  
If there is an isometry $g \in {\rm Isom}^+(\Lambda_{\rm K3})$
mapping $[\omega] \in D$ to $[\omega'] \in D$, then it also maps 
${\cal P}^{-1}([\omega]) \cap N^o$ to ${\cal P}^{-1}([\omega']) \cap N^o$.
For any element in these inverse images, 
$[(X, \varphi)] \in {\cal P}^{-1}([\omega]) \cap N^o$ and 
$[(X',\varphi')] \in {\cal P}^{-1}([\omega']) \cap N^o$, 
$\varphi^{'-1} \cdot g \cdot \varphi: H_2(X; \Z) 
\longrightarrow H_2(X'; \Z)$ is a Hodge isometry, and hence the version
3 of the global Torelli theorem implies that there is an isomorphism of
surfaces $f: X \longrightarrow X'$. 
Thus, for all the points $[\omega] \in D$ in a given orbit of 
${\rm Isom}(\Lambda_{\rm K3})$, all the points in $N$ mapped to this 
orbit can be represented by a common K3 surface $X$ and some markings. 
Conversely, if two points $[{X, \varphi}]$ and $[(X, \varphi')]$ in $N$ 
share the same K3 surface, then $\varphi' \cdot \varphi^{-1}$ is an
isometry of $\Lambda_{\rm K3}$ mapping the image of $[(X,\varphi)]$ to 
that of $[(X,\varphi')]$. Therefore, the ${\rm Isom}(\Lambda_{\rm
K3})$-orbit decomposition of the period domain, 
${\rm Isom}^+(\Lambda_{\rm K3}) \backslash D$ is equivalent to the 
classification of K3 surfaces modulo isomorphism of surfaces. 

Finally, let us take a closer look at how 
the ${\rm Isom}(\Lambda_{\rm K3})$ symmetry group acts on the moduli
space $N$ or $N^o$. Its action on $D$ is quite simple, but its action 
on $N^o$ has a more interesting structure, and we will need that 
in Section \ref{ssec:Het-F-AutX}.
When an element $g \in {\rm Isom}^+(\Lambda_{\rm K3})$ maps $[\omega]
\in D$ to another element $[\omega'] \neq [\omega]$, the fibres of those
two points under the period map can be described by 
$ 
  \{[(X,\varphi)] \; | \; \varphi \in
  \varphi_0 \cdot [W^{(2)}(S_X) \rtimes {\rm Aut}(X)] / {\rm Aut}(X)\}$
and 
$ 
  \{[(X,\varphi')] \; | \; \varphi' \in
\varphi'_0 \cdot [W^{(2)}(S_X) \rtimes {\rm Aut}(X)] / {\rm Aut}(X)\}$ 
for some $\varphi_0$ and $\varphi'_0$, respectively.\footnote{
Here, $\varphi'_0 \nin \varphi_0 \cdot [ W^{(2)}(S_X) \rtimes {\rm Aut}(X)]$.} 
%
The action of $g$ establishes a one-to-one correspondence between 
the two fibres by setting $\varphi' = g \cdot \varphi$.

The stabilizer subgroup of $[\omega] \in D$ in 
${\rm Isom}^+(\Lambda_{\rm K3})$ is 
%
\begin{equation}
 G_{[\omega]} = \varphi_0 \cdot
   \left[ W^{(2)}(S_X) \rtimes {\rm Aut}(X) \right] 
   \cdot \varphi^{-1}_0 \subset {\rm Isom}^+(\Lambda_{\rm K3}).
\label{eq:stabl-grp-omega-in-D}
\end{equation}
%
This stabiliser group acts naturally on the inverse image of 
$[\omega]$, given in (\ref{eq:period-map-fibr}).

\subsection{When Are Two Elliptic Fibrations Considered ``Different'' 
in F-theory?}
\label{ssec:Het-F-AutX}

In the description of complex structure moduli of K3 surfaces, 
one can think of several different moduli spaces in mathematics, 
such as $N$ (or $N^o$) (the moduli space of marked K3 surfaces), $D$ (the period 
domain) and ${\rm Isom}^+(\Lambda_{\rm K3}) \backslash D$ (the moduli space
of K3 surfaces modulo automorphism). These different moduli spaces
contain different information and are mutually related in the way we
have reviewed above. When we refer to ``the moduli space'' in string
theory applications, however, we want it to parametrize vacua (hopefully 
with less redundancy in the parametrization, and at least with
information on the redundancy), and use it as the target space of a non-linear 
sigma model to describe light degrees of freedom. 

It is considered that, in M-theory compactification on 
$Y = S_1 \times S_2$ with both $S_1$ and $S_2$ being K3 surfaces, 
the moduli space (in the absence of four-form flux) is given
by\footnote{We postpone a slightly more refined argument for the choice 
of the quotient group to Section \ref{sssec:moduli-flux-quotient-MF},
for M-theory moduli space as well as for F-theory. The essence of the
argument in this section remains valid after Section \ref{sssec:moduli-flux-quotient-MF}.} 
\begin{equation}
\left[ {\rm Isom}^+(\Lambda^{(S_1)}_{\rm K3}) \times 
       {\rm Isom}^+(\Lambda^{(S_2)}_{\rm K3}) \right] \backslash  
\left[D^{(S_1)} \times D^{(S_2)}\right]\, .  
\label{eq:classify-M-cpx-noflux}
\end{equation}
It makes perfect sense to take a quotient by the symmetry group 
${\rm Isom}^+(\Lambda_{\rm K3})$, because two marked K3 surfaces 
$[(X,\varphi)]$ and $[(X,\varphi')]$ in $N^o$ which differ only in the markings 
$\varphi$ and $\varphi'$ should not be considered as different 
compactifications in 11-dimensional supergravity.\footnote{
Homogeneous coordinates 
can be introduced to the period domain $D$ by taking a basis 
$\{\Sigma_I \}_{I=1, \cdots, 22}$ in the lattice $\Lambda_{\rm K3}$. 
The period integrals $\Pi_I := \int_{\Sigma_I} \omega$
can be used as the coordinates. With these coordinates, 
the K\"{a}hler potential (obtained by dimensional reduction) is given by 
\begin{equation}
 K \propto - \ln \left[ \int_Y (\Omega_{S_1} \wedge \Omega_{S_2}) \wedge 
                             \left( \overline{\Omega}_{S_1} \wedge
			      \overline{\Omega}_{S_2} \right)
               \right] = -
   \ln \left[ \Pi^{(S_1)}_I C^{IJ} \overline{\Pi}^{(S_1)}_J \right] - 
   \ln \left[ \Pi^{(S_2)}_I C^{IJ} \overline{\Pi}^{(S_2)}_J \right]\, ,  
\end{equation} 
where $C^{IJ}$ is the inverse of the intersection form of $\Lambda_{\rm
K3}$ in the basis $\{ \Sigma_I \}_{I=1,\cdots,22}$.
The action of the ${\rm Isom}^+(\Lambda^{(S_1)}_{\rm K3}) \times 
{\rm Isom}^+(\Lambda^{(S_2)}_{\rm K3})$ group on the $\Pi^{(S_{1,2})}_I$ leaves
the K\"{a}hler potential unchanged, because it preserves the
intersection form.}  Since F-theory compactification on an
elliptic-fibred Calabi--Yau fourfold is regarded as a special case of 
M-theory compactification on a Calabi--Yau fourfold, this moduli space can be 
regarded as a reliable place to start for F-theory as well. 

The moduli space of F-theory compactification on a K3 surface $X$ (where 
we require that there is an elliptic fibration $\pi_X: X \longrightarrow
\P^1$ and a section $\sigma: \P^1 \longrightarrow X$) without any
flux is given by 
\begin{equation}
 {\cal M}^{\rm cpx}_{F; {\rm K3}} := 
 \left[ {\rm Isom}^+(\Lambda_{\rm K3}) \right] \backslash 
  \left\{ (\phi_U, [\omega]) \; | \; [\omega]|_{\phi_U(U)} = 0 \right\} / 
  \left\{ \pm {\rm id.}_U \right\}\, ,
\label{eq:classify-F-K3-cpx-noflux}
\end{equation}
where $[\omega] \in D$ as before, and 
$\phi_U: U \hookrightarrow \Lambda_{\rm K3}$ is an embedding of 
the hyperbolic plane lattice $U$. It is a popular way to make sure 
that there is an elliptic fibration by specifying a sublattice (which is
isomorphic to $U$) 
generated by algebraic cycles corresponding to the elliptic fibre and 
the section (e.g. \cite{Aspinwall:1996mn, BKW-math}).
A remaining subtlety can arise in the choice of the
quotient group. The group ${\rm Isom}^+(\Lambda_{\rm K3})$ acts on 
$D$ and embeddings of $U$ (while preserving $[\omega]|_{\phi_U(U)} =
0$), while $\{ \pm {\rm id.}_U\}$ is a subgroup of 
${\rm Isom}(U) \cong \Z_2 \times \Z_2$ (see Section 2 of \cite{BKW-math}), 
and acts on embeddings $\phi_U: U \hookrightarrow \Lambda_{\rm K3}$ 
from the right by changing $\phi_U$ to $\pm \phi_U$. 
We will see shortly that this is the right choice of the quotient group. 
Once this statement is accepted, then 
it wouldn't be difficult to accept the following: the moduli space of
F-theory compactification on $Y = X \times S$ (where $X$ has an elliptic
fibration as before) is given by 
\begin{equation}
{\cal M}^{\rm cpx}_{F; {\rm K3}} \times
 \left( {\rm Isom}^+(\Lambda^{(S)}_{\rm K3}) \backslash D^{(S)}\right)\, .
\label{eq:classify-F-cpx-noflux}
\end{equation}
%

Now, in order to justify the statement (\ref{eq:classify-F-K3-cpx-noflux}), 
we need to understand the space ${\cal M}^{\rm cpx}_{F; {\rm K3}}$
better. It is often a good strategy in understanding a space $M$ 
to construct a map $f: M \longrightarrow B$ to some simple space $B$, 
and study how the ``fibres'' $f^{-1}(b) \subset M$ change with $b \in B$. 
First consider a projection 
\begin{equation}
{\rm fgt}_{[\omega]}: {\cal M}^{\rm cpx}_{F; {\rm K3}} \longrightarrow 
  {\rm Isom}^+(\Lambda_{\rm K3}) \backslash 
   \left\{ \phi_U : U \hookrightarrow \Lambda_{\rm K3} \right\} /
  \left\{ \pm {\rm id.}_U \right\} 
\label{eq:fgt-omega-base}
\end{equation}
by throwing away the information of $[\omega]$ from 
${\cal M}^{\rm cpx}_{F; {\rm K3}}$.
The base of this forgetful map, ``$B$'', 
consists of only one point: due to the uniqueness (modulo isometry) of
even unimodular lattices of signature $(n,n+16)$, there always exists an
isometry $\varphi \in {\rm Isom}^+(\Lambda_{\rm K3})$ such that either 
$\phi'_U = \varphi \cdot \phi_U$ or 
$-\phi'_U = \varphi \cdot \phi_U$ holds for any two embeddings $\phi_U$
and $\phi_U'$ of the hyperbolic plane lattice $U$. 
Thus, the whole set ${\cal M}^{\rm cpx}_{F; {\rm K3}}$
can be studied by looking at the fibre over just one point in the base; 
that is, we can take an arbitrary embedding 
$\phi_{U0}: U \hookrightarrow \Lambda_{\rm K3}$, and just study 
the fibre ${\rm fgt}_{[\omega]}^{-1}( [\phi_{U0}])$.
The fibre over this one point is
\begin{eqnarray}
&& {\rm Isom}(U^{\oplus 2} \oplus E_8^{\oplus 2}) \backslash
 \left\{ [\omega] \in D_{U^\perp} \right\}, 
   \label{eq:classify-F-K3-cpx-noflux-2} \\
&&\quad  D_{U^\perp} : =  \{ [\omega] \in 
\P[(\phi_{U0}(U))^\perp \subset \Lambda_{\rm K3}] \; | \; \omega \cdot \omega = 0,
\; \omega \cdot \overline{\omega} > 0 \}\, . 
\end{eqnarray}
Therefore, the set ${\cal M}^{\rm cpx}_{F; {\rm K3}}$
is equivalent to (\ref{eq:classify-F-K3-cpx-noflux-2}), 
the global structure of which is 
\begin{equation}
 {\rm Isom}(U^{\oplus 2} \oplus E_8^{\oplus 2}) \backslash 
  {\rm Gr}^{\rm po}(2; (U^{\oplus 2} \oplus E_8^{\oplus 2}) \otimes \R)
\cong
 {\rm Isom}(U^{\oplus 2} \oplus E_8^{\oplus 2}) \backslash 
  \O(2,18; \R) / \SO(2) \times \O(18)\, .
\label{eq:double-Het-T2-moduli}
\end{equation}
This is a double cover over what we know as the moduli space of 
heterotic string compactifications on $T^2$ \cite{Vafa-evidence} 
(see e.g., also \S 5 of \cite{Bershadsky:1998vn}), 
\begin{equation}
{\rm Isom}(U^{\oplus 2} \oplus E_8^{\oplus 2}) \backslash 
 {\rm Gr}^{\rm p}(2; (U^{\oplus 2}\oplus E_8^{\oplus 2})\otimes \R ) 
\cong
 {\rm Isom}(U^{\oplus 2} \oplus E_8^{\oplus 2}) \backslash 
  \O(2,18; \R) / \O(2) \times \O(18)\, . 
\label{eq:Het-T2-moduli}
\end{equation}
This argument almost proves\footnote{Without microscopic foundation of
F-theory, it is hard to make any precise statement about F-theory
physics directly. Here, this problem is overcome by relying 
on heterotic--F-theory duality.} that we can 
take ${\cal M}^{\rm cpx}_{F;{\rm K3}}$ in (\ref{eq:classify-F-K3-cpx-noflux}) 
as the classification scheme of F-theory vacua when an elliptic 
fibred K3 surface is involved as part of the compactification data. 

We understand 
that the remaining subtlety---double cover---corresponds, 
in F-theory language, to a pair of (elliptic fibred) K3 surfaces $X$ and 
$X'= \overline{X}$ where $H^{2,0}(X; \C)$ and $H^{0,2}(X; \C)$ 
in the Hodge decomposition of $H^2(X; \C)$ are identified with 
$H^{0,2}(X'; \C)$ and $H^{2,0}(X'; \C)$. $X$ and $X'$ are a mutually 
complex conjugate pair. 
The difference between them is only in declaring a complex coordinate 
as holomorphic or anti-holomorphic, and that should not make a
difference in physics in 7+1-dimensions. Thus, even in F-theory, 
the moduli space of K3 compactification to 7+1-dimensions 
should be (\ref{eq:Het-T2-moduli}), rather 
than (\ref{eq:double-Het-T2-moduli}). As we proceed to consider
compactifications of F-theory on $Y = X \times S$ 
along with a four-form flux on $Y$, it {\it does} make a difference 
in low-energy physics in 3+1-dimensions to take complex conjugation of 
$X$, while keeping the complex structure of $S$ and the flux. 
We therefore take 
(\ref{eq:classify-F-cpx-noflux}) as the classification scheme for  ${\rm K3} \times {\rm K3}$
compactification of F-theory for now; the $\Z_2$ quotient associated
with unphysical holomorphic--anti-holomorphic distinction will be
implemented in Section \ref{ssec:sbtl-stat} after introducing fluxes.

\vspace{5mm}

Let us now start from ${\cal M}^{\rm cpx}_{F;{\rm K3}}$ in 
(\ref{eq:classify-F-K3-cpx-noflux}) again, and derive a useful way to look 
at it in order to address the second one of the two questions raised 
at the beginning of this section. 
Consider a projection 
%
\begin{eqnarray}
{\rm fgt}_{\phi_U}: \; 
{\cal M}^{\rm cpx}_{F; {\rm K3}} & \longrightarrow  &
{\rm Isom}^+(\Lambda_{\rm K3}) \backslash D\, , 
\label{eq:forget-quot} 
\end{eqnarray}
This time, we throw away the information on the embedding $\phi_U$ from 
${\cal M}^{\rm cpx}_{F; {\rm K3}}$. As we have seen in 
Section \ref{ssec:Torelli}, the ``base'' space of this projection, 
${\rm Isom}^+(\Lambda_{\rm K3}) \backslash D$,  
corresponds to the classification of K3 surfaces modulo surface isomorphism.
Thus, by studying the ``fibre'' of this projection, we can find 
the variety of F-theory vacua that (the surface-isomorphism class of) a K3
surface admits.

We begin this study by looking at the fibration structure of 
the following projection map instead:
\begin{eqnarray}
{\rm fgt}_{\phi_U}: \; 
\left\{ (\phi_U, [\omega]) \; | \; [\omega] \in D, \;
   [\omega]|_{\phi_U(U)} = 0 \right\} / \left\{ \pm {\rm id.}_U \right\}
 & \longrightarrow  & \left\{ [\omega] \; | \; [\omega] \in D
\right\} = D\, . \label{eq:forget-no-quot}
\end{eqnarray}
Before and after the projection, we are not taking a quotient by the 
symmetry group ${\rm Isom}^+(\Lambda_{\rm K3})$ action here, which 
makes the problem easier to get started.
For a given $[\omega] \in D$, and for any $\phi_U$ satisfying 
the condition $[\omega]|_{\phi_U(U)} = 0$, 
$\phi_U$ embeds the hyperbolic plane lattice $U$ into 
$S_{[\omega]} := \left[ [\omega]^\perp \subset \Lambda_{\rm K3}
\right]$. That is, 
\begin{equation}
  {\rm fgt}_{\phi_U} ^{-1}([\omega]) =
   \left\{ \phi_U: U \hookrightarrow S_{[\omega]} \right\} /
   \left\{ \pm {\rm id.}_U \right\}\, .
\label{eq:forget-map-fibr-noquot}
\end{equation}
This means geometrically that for any 
one of the inverse images $[(X_{[\omega]}, \varphi_{[\omega]})] \in 
{\cal P}^{-1}([\omega]) \cap N^o$ under the period map, 
an embedding of the hyperbolic plane lattice into the Neron--Severi 
lattice of the K3 surface $X_{[\omega]}$, $S_{X_{[\omega]}}$, is defined:
\begin{equation}
 \varphi_{[\omega]}^{-1} \cdot \phi_U: \; U \hookrightarrow 
  S_{X_{[\omega]}} \subset H^2(X_{[\omega]}; \Z)\, . 
\end{equation}
The inverse image ${\cal P}^{-1}([\omega]) \cap N^o$ of 
any given element $[\omega] \in D$ is described 
in (\ref{eq:period-map-fibr}); we can choose an appropriate $\varphi$ 
in (\ref{eq:period-map-fibr}) so that either 
$\varphi^{-1} \cdot \phi_U$ or $\varphi^{-1} \cdot (-\phi_U)$ 
defines a {\it canonical embedding} of hyperbolic plane lattice into 
$S_{X_{[\omega]}}$. This is a sufficient condition to 
construct an elliptic fibration $\pi_X: X\longrightarrow \P^1$ along
with a zero section $\sigma: \P^1\longrightarrow X$,
see Section 3.1 of \cite{BKW-math} for a more detailed explanation. 
Each element $\phi_U$ in ${\rm fgt}_{\phi_U}^{-1}([\omega])$ 
in (\ref{eq:forget-map-fibr-noquot}) therefore defines an 
elliptic fibration on $X$.
%

Let us now go back to the study of the fibre of the projection 
(\ref{eq:forget-quot}), bringing back the quotient by 
${\rm Isom}^+(\Lambda_{\rm K3})$. First of all, when an element 
$g \in {\rm Isom}^+(\Lambda_{\rm K3})$ maps $[\omega] \in D$ to 
$[\omega'] := g \cdot [\omega] \neq [\omega]$, taking a quotient does
not change the fibre of the projection map. It only establishes a
one-to-one identification between the elements in the fibre 
${\rm fgt}_{\phi_U}^{-1}([\omega])$ and ${\rm fgt}_{\phi_U}^{-1}([\omega'])$.
%
%
%
The stabilizer subgroup $G_{[\omega]}$ of 
${\rm Isom}^+(\Lambda_{\rm K3})$ for $[\omega] \in D$, however, 
can be non-trivial, as we have seen in (\ref{eq:stabl-grp-omega-in-D}).
We have to take a quotient of (\ref{eq:forget-map-fibr-noquot}) by
the stabilizer group $G_{[\omega]}$ in order to obtain the fibre of 
the projection map in (\ref{eq:forget-quot}) at 
$[[\omega]] \in {\rm Isom}^+(\Lambda_{\rm K3}) \backslash D$. 
Therefore, we conclude that 
\begin{eqnarray}
 {\rm fgt}_{\phi_U}^{-1} \left( [[\omega]] \right) & = &
 \left[ W^{(2)}(S_{[\omega]}) \rtimes \varphi_0({\rm Aut}(X_{[\omega]})) \right]
 \backslash 
 \left\{\phi_U : U \hookrightarrow S_{[\omega]} \subset \Lambda_{\rm K3} 
 \right\} / \left\{ \pm {\rm id.}_U \right\}  \\
 & = &
 \left[ W^{(2)}(S_{X_{[\omega]}}) \rtimes {\rm Aut}(X_{[\omega]})
 \right] 
 \backslash  
 \left\{ \varphi_0^{-1} \cdot \phi_U: U \hookrightarrow 
  S_{X_{[\omega]}} \subset H^2(X_{[\omega]}; \Z) \right\}
 / \left\{ \pm {\rm id.}_U \right\}  \nonumber \\
 & = & 
 \left[ W^{(2)}(S_{X_{[\omega]}}) \rtimes 
    {\rm Isom}(S_{X_{[\omega]}})^{({\rm Amp~Hodge})} \right] 
 \backslash  
 \left\{ \varphi_0^{-1} \cdot \phi_U: U \hookrightarrow 
  S_{X_{[\omega]}} \right\} 
 / \left\{ \pm {\rm id.}_U \right\}  \nonumber \\
 & = & {\cal J}_{1} (X_{[\omega]})\, .
\label{eq:classify-Uembed-fibre-p}
\end{eqnarray}
As we have explained in Section 3 of \cite{BKW-math}, 
this ${\cal J}_{1}(X_{[\omega]})$ corresponds to the classification of 
elliptic fibrations $(\pi_X, \sigma)$ for a K3 surface $X$ 
($\pi_X: X \longrightarrow \P^1$ along with $\sigma: \P^1
\longrightarrow X$ so that $\pi_X \cdot \sigma = {\rm id.}_{\P^1}$)
modulo 
${\rm Aut}(X) \times {\rm Aut}(\P^1) = {\rm Aut}(X) \times PGL(2;\C)$.
Therefore, the projection map ${\rm fgt}_{\phi_U}$ in
(\ref{eq:forget-quot}) enables us to apprehend the moduli space 
(vacuum classification scheme) of F-theory compactifications on K3
surface ${\cal M}^{\rm cpx}_{F; {\rm K3}}$ as a fibration over 
${\rm Isom}^+(\Lambda_{\rm K3}) \backslash D$ 
(i.e., complex structure moduli space of K3 surface modulo surface 
isomorphism), with the fibre given by the ${\cal J}_{1}$ classification 
(i.e., modulo automorphism) of elliptic fibrations.

\section{A Miniature Landscape: F-theory on  ${\rm K3} \times {\rm K3}$ with $G_0 = 0$}
\label{sec:only-G1}

\subsection{${\cal J}_1(X)$ and ${\cal J}_2(X)$ Classification}
\label{ssec:J1-J2-phys}

When we classify low-energy effective theories, we normally group
together theories with the same gauge groups and matter representations 
first, and then pay attention to the values of various coupling constants. 
Although two elliptic fibrations $(\pi_X, \sigma)$ and $(\pi'_X,
\sigma')$ for a K3 surface $X$ are not regarded as the same vacuum 
(or the same low-energy effective theory) in the absence of an
appropriate automorphism in ${\rm Aut}(X) \times PGL(2; \C)$,  
they might still have the same gauge groups and matter presentations. 

Corresponding to the coarse classification in terms of gauge groups and 
matter representations is the ${\cal J}_2(X)$ classification of elliptic 
fibrations on a K3 surface $X$. This is close to the classification of  
singular fibre types, but slightly different and more suited for 
physicists' needs. As reviewed in detail in \cite{BKW-math}, 
\begin{eqnarray}
{\cal J}_1(X) & = & \left[ W^{(2)}(S_{X}) \rtimes 
    {\rm Isom}(S_{X})^{({\rm Amp~Hodge})} \right] 
 \backslash  
 \left\{ U \hookrightarrow  S_{X} \right\} 
 / \left\{ \pm {\rm id.}_U \right\},   \nonumber \\
{\cal J}_2(X) & = & \left[ W^{(2)}(S_{X}) \rtimes 
    {\rm Isom}(S_{X})^{({\rm Amp})} \right] 
 \backslash  
 \left\{ U \hookrightarrow  S_{X} \right\} 
 / \left\{ \pm {\rm id.}_U \right\}\, .   
\end{eqnarray}
Here, the group ${\rm Isom}(S_X)^{({\rm Amp~Hodge})}$ is a subgroup 
of ${\rm Isom}(S_X)^{({\rm Amp})}$, and hence the ${\cal J}_2(X)$
classification is obviously more coarse than the ${\cal J}_1(X)$ 
classification.\footnote{The quotient group $[W^{(2)}(S_X) \rtimes {\rm
Isom}(S_X)^{({\rm Amp})}]$ for the ${\cal J}_2(X)$ classification 
is equivalent to ${\rm Isom}^+(S_X)$ (an index 2 subgroup of the entire
isometry group of the Neron--Severi lattice).} The ${\cal J}_2(X)$
classification is equivalent to the classification of frame lattices 
of elliptic fibrations modulo isometry. For an elliptic fibration 
$\pi: X \longrightarrow \P^1$ with a fibre class $[F] \in S_X$, 
the frame lattice is given by 
\begin{equation}
 W_{\rm frame} = \left[ [F]^\perp \subset S_X \right] / \vev{[F]}\, .
\label{eq:def-frame}
\end{equation}
Readers are referred to \cite{BKW-math} for more mathematical aspects of 
this discussion. The frame lattice $W_{\rm frame}$ (modulo isometry) 
contains all the information of 7-brane gauge groups and representations 
of charged matters.
Individual equivalence classes in ${\cal J}_2(X)$ are referred to as 
{\it types}, and those in ${\cal J}_1(X)$ as {\it isomorphism classes}.

There is a systematic procedure to study the ${\cal J}_2(X)$
classification for a given K3 surface $X$ with large Picard number 
$\rho_X$ (see \cite{Nish01} or \S 4 of \cite{BKW-math}). 
The ${\cal J}_2(X)$ classification of elliptic fibrations has already 
been studied for some K3 surfaces (i.e., for some particular choices of 
complex structures of K3 surface). For most generic Kummer surfaces\footnote{
The Picard number of this family is $\rho_X = 17$, so that there are 3 complex structure parameters.} $X={\rm Km}(A)$, for example, 
there are 25 different types in the ${\cal J}_2(X)$ 
classification \cite{Kumar}. Roughly speaking, this means that the 
compactifications of F-theory on $Y=X \times S$ with $X = {\rm Km}(A)$ 
admits 25 different choices of 7-brane gauge groups and matter representations. 
A slightly more special class (2-parameter family) of Kummer surfaces, 
$X = {\rm Km}(E \times F)$, admits 11 different types of elliptic fibrations 
in the ${\cal J}_2(X)$ classification. Reference \cite{Nish01} worked out 
the ${\cal J}_2(X)$ classification for four attractive K3 surfaces, 
$X_{[1~0~1]}$, $X_{[1~1~1]}$, $X_{[2~0~2]}$ and $X_{[2~2~2]}$ among others, 
and found that there are ${\cal O}(10 \sim 100)$ inequivalent types of 
elliptic fibrations in the ${\cal J}_2(X)$ classification 
(Table \ref{tab:X3-frame} in this article contains detailed information of the 
${\cal J}_2(X)$ classification of $X_{[1~1~1]}$). We also worked out 
the ${\cal J}_2(X)$ classification partially for another attractive K3 
surface $X_{[3~0~2]}$ (see \S4.4 of \cite{BKW-math}) and found that there are 
at least 54 inequivalent types in ${\cal J}_2(X)$.
Based on such an experience, it may not be too far off the mark to guess 
that the attractive K3 surfaces in Table \ref{tab:AK-notmorethan24} have 
${\cal O}(10 \sim 100)$ inequivalent types of elliptic fibrations in 
the ${\cal J}_2(X)$ classification.\footnote{A brute force calculation 
(or automatized/computerized calculation) following the procedure 
reviewed in \S4 of \cite{BKW-math} should be able to verify or correct 
this statement, but this task is beyond the scope of this article 
and \cite{BKW-math}.} 

Let us now focus on a given type of elliptic fibration in ${\cal J}_2(X)$ 
(i.e., we focus on a particular choice of 7-brane gauge group and matter 
representation) for some K3 surface $X$. There can be more than one 
isomorphism class of elliptic fibrations in the ${\cal J}_1(X)$ classification 
(fine classification) that corresponds to the same type. The number of 
such mutually non-isomorphic elliptic fibrations is referred to as the 
``number of isomorphism classes'', or simply ``multiplicity'' of that type 
in this article. Reference \cite{Oguiso} worked out the multiplicity for 
each one of the types in ${\cal J}_2(X)$ for $X = {\rm Km}(E \times F)$.
There is no theory known (at least to the authors) that computes multiplicities for 
any K3 surface, and the authors of this article made an attempt at 
generalizing the study of \cite{Oguiso} so the multiplicities are estimated, 
if not computed, for a broader class of K3 surfaces with large Picard number. 
The primary goal of $\S 5$ of \cite{BKW-math} is to develop a theory for 
this purpose.

One of the solid results obtained in \cite{BKW-math} is that 
the multiplicities are at most 16 for any type and for any one of 
the 34 attractive K3 surfaces that appear in Table \ref{tab:AK-notmorethan24}. 
For individual attractive K3 surfaces (or for individual types of elliptic fibrations of a given 
attractive K3 surface), stronger upper bounds on the multiplicity are obtained.
For example, the multiplicity is at most 2 for all types of 20 out of 
the 34 attractive K3 surfaces in Table \ref{tab:AK-notmorethan24}, 
and furthermore, the multiplicity is 1---any two elliptic fibrations 
of a given type must be mutually isomorphic---for 10 attractive K3
surfaces among them.\footnote{\label{fn:10-K3s} They are 
$X_{[1~0~1]}$, $X_{[1~1~1]}$, $X_{[2~0~1]}$, 
$X_{[2~1~1]}$, $X_{[3~0~1]}$, $X_{[3~1~1]}$, $X_{[4~0~1]}$,
$X_{[5~1~1]}$, $X_{[6~1~1]}$ and $X_{[3~1~2]}$.} 
See Corollary D in \cite{BKW-math} for more information.

There are two remarks to be made: first, it is not guaranteed that 
${\rm Isom}(S_X)^{({\rm Amp~Hodge})}$ is always a normal subgroup of 
${\rm Isom}(S_X)^{({\rm Amp})}$. If it is, then the map from ${\cal J}_1(X)$ 
to ${\cal J}_2(X)$ is regarded as the quotient map under the 
action of the quotient group ${\rm Isom}(S_X)^{({\rm Amp~Hodge})} \backslash
{\rm Isom}(S_X)^{({\rm Amp})}$. The multiplicity of a given type is the number 
of elements of the orbit under this group. When the quotient group is not 
a normal subgroup, however, mutually non-isomorphic elliptic fibres do not 
necessarily form an orbit of a group action.

There seems to be a correlation between the 
multiplicity of a given type and the Picard number, at least among the 
examples that have been looked at in \cite{BKW-math}. 
The multiplicities for various types range in 
${\cal O}(10)$--${\cal O}(100)$ for a 3-parameter family of K3 surfaces 
$X = {\rm Km}(A)$ (where $\rho_X=17$), while they range in 
a few--10 for a 2-parameter family of K3 surfaces 
$X = {\rm Km}(E \times F)$ (where $\rho_X = 18$), and 
the multiplicities often become a few or even less for many attractive 
K3 surfaces ($\rho_X = 20$) appearing in Table \ref{tab:AK-notmorethan24}. 
This is far from a rigorous mathematical statement, and in particular, 
it is conceivable that the physics-motivated 
condition (\ref{eq:G1-partflux<=24}) has extracted biased samples 
from all the attractive K3 surfaces. For a study of supersymmetric landscapes, 
however, it is mandatory to set upper bounds like (\ref{eq:G1-partflux<=24}) 
on the flux quanta. The 34 attractive K3 surfaces are then our sample of 
interest (see also Section \ref{sec:G0-flux} for a related discussion), 
and this bias is not a problem at all. 

\subsubsection{Frame Lattice, Mordell--Weil Group and {\rm U}(1) Charges}
\label{sssec:frame-MW}

Before proceeding to Section \ref{ssec:sbtl-stat}, we take 
a moment to give a detailed account of how physics information 
is read out from the frame lattice (\ref{eq:def-frame}).
This is largely a well-known subject, and this section 
is primarily meant to be a review or reading guide 
for \S 4 of \cite{BKW-math}. The details of the following presentation 
are not directly relevant to the rest of this article. 
However, this section also contains a generalization
of the discussion in \cite{Aspinwall:1998xj} in a way applicable to K3 
surfaces away from the stable degeneration limit.

The Cartan (maximal torus) part of 7-brane gauge fields in F-theory 
originates from the three-form field of 11-dimensional supergravity.
These fields correspond to fluctuations of the three-form field in the form of 
$A^a \wedge \omega^a$, where $A^a$ is a vector field in the low-energy
effective theory, and $\omega^a$ is chosen from 
\begin{equation}
 F^1 / F^0 \cong H^1(B_3; R^1\pi_{Y*} \Z)\, ;  
\end{equation}
$H^2(Y; \Z)$ for an elliptic fibred Calabi--Yau fourfold $Y$ 
with $\pi_Y: Y \longrightarrow B_3$ has a filtration 
\begin{equation}
 H^2(Y; \R) = F^2 \supset F^1 \supset F^0 
\end{equation}
and 
\begin{eqnarray}
 F^2/ F^1 & \cong & H^0(B_3; R^2\pi_{Y*}\R) = H^0(B_3; \R), \\ 
 F^1/ F^0 & \cong & H^1(B_3; R^1\pi_{Y*} \R), \\ 
 F^0 & \cong & H^2(B_3; R^0\pi_{Y*}\R) = H^2(B_3; \R).   
\end{eqnarray}
This---choosing $\omega$ from $F^1/F^0$---is because two-forms purely in 
the base, $F^0$, correspond to scalars (or two-forms) in the effective
theory in 3+1-dimensions, and those containing two-forms in the elliptic 
fibre, $F^2/F^1$, to a part of metric in 3+1-dimensions \cite{Dasgupta:1999ss}. 
The total rank of the 7-brane gauge group in the effective theory is therefore 
$h^{2}(Y) - h^{2}(B_3)-1$ \cite{Morrison:1996pp}. In the case of 
$\pi_Y: Y = {\rm K3} \times {\rm K3} \longrightarrow 
\P^1 \times {\rm K3}$, the rank is $44-23-1=20$. 

In the case of $Y = X \times S$ with an elliptic K3 surface $X$, 
$F^1/F^0$ can simply be identified with 
\begin{equation}
F^1/F^0 \cong H^1(\P^1; R^1\pi_{X*}\R)\, . 
\end{equation}
The condition that $\omega^a$ be within $F^1 \subset F^2 = H^2(X; \R)$
corresponds to $\omega^a \in [[F]^\perp \subset H^2(X; \R)]$.
One can see that $(F^1/F^0) \cong (T_X \oplus W_{\rm frame}) \otimes \R$, 
because i) $H^2(X; \R) = (T_X \oplus S_X) \otimes \R$, and 
ii) the generator of $F^0 \cong H^2(\P^1; \R) = \R$ is Poincar\'e dual 
to the fibre class $[F]$ of the elliptic K3 surface $X$, and 
iii) also because of the definition of the frame lattice (\ref{eq:def-frame}). 
For a K3 surface $X$ with $\rho_X = 20$, rank-2 ${\rm U}(1)$ gauge fields are 
associated with $T_X \otimes \R$, while the remaining 18 Cartan ${\rm U}(1)$'s are
related to $W_{\rm frame}\otimes \R$.

In the presence of four-form flux purely of $G_1$ type, the two ${\rm U}(1)$ 
vector fields associated with $T_X \otimes \R$ become massive by a 
St\"uckelberg mechanism. At the level of analysis in this article (where 
non-perturbative effects are not considered, and stabilization of 
K\"{a}hler moduli is also ignored), those two ${\rm U}(1)$ symmetries remain 
in the effective theory as global symmetries. 

The frame lattice is negative definite. As we always assume that 
the elliptic fibration $\pi_X: X \longrightarrow \P^1$ has a section 
$\sigma: \P^1 \longrightarrow X$, we can identify a sublattice of 
$S_X$ isomorphic to $W_{{\rm frame}}$ in the case of K3 surface $X$; 
it is characterized as 
\begin{equation}
 W_{{\rm frame}*} := \left[ \left({\rm Span}_\Z\{ [F], [\sigma] \} \cong
			     U \right)^\perp \subset S_X \right]\, ,
\end{equation}
the orthogonal complement of a sublattice generated by the fibre class 
$[F]$ and the section $[\sigma]$, and we call this sublattice 
$W_{{\rm frame}*}$ the canonical frame lattice of a given elliptic 
fibration $(\pi_X, \sigma; X, \P^1)$.

The non-Abelian part of the gauge group in F-theory is associated 
with the (Poincar\'e dual of the) irreducible $(-2)$ curves in the singular 
fibres of $X$ that do not meet the zero section $[\sigma]$. 
They are contained in $W_{{\rm frame}*}$ and are linearly independent. 
The sublattice generated by these $(-2)$-curves is contained in 
\begin{equation}
 W_{\rm root} := {\rm Span}_{\Z} \left\{ D \in W_{{\rm frame}*} \; 
   | \; D^2 = -2 \right\}\, ,
\end{equation}
the sublattice generated by norm-$(-2)$ elements of the canonical 
frame lattice. But this $W_{\rm root}$---called the root lattice of 
$W_{{\rm frame}*}$---is also known to be the same as the sublattice
generated by the $(-2)$-curves (not meeting the section) in the singular 
fibres of $X$.\footnote{To see this, suppose that $D$ is a generator of $W_{\rm
root}$, i.e., $D \in W$ and $D^2 = -2$. Then either $D$ or $-D$ 
corresponds to a class containing an effective divisor (curve) due 
to the Riemann--Roch theorem (Lemma 2.2 in \S 1 of \cite{PSS}), and secondly, 
it should be mapped down to a point  
in the base space $\P^1$ of the elliptic fibration, 
because the effective divisor in $W_{{\rm frame}*}$ does not intersect 
with the fibre class. Therefore, it has to be contained in some 
singular fibres. The $W_{\rm root}$ lattice is attributed purely to 
singular fibres, not to any other sort of non-trivial sections
of the elliptic fibration.} 
Therefore, once an elliptic fibration is specified in the form of 
an embedding of the lattice $U$ into $S_X$, the non-Abelian part of 
the gauge group can be read out by calculating the $W_{\rm root}$
lattice from $W_{\rm frame}$ without dealing with defining equations 
(or the fibration map) of the K3 surface.

When the rank of the frame lattice 
$W_{{\rm frame}*} \cong W_{\rm frame}$ is larger than $W_{\rm root}$, 
there is a massless ${\rm U}(1)$ vector field in the effective theory 
(if there is only $G_1$ component of the flux). 
Since ``W-bosons'' in the non-Abelian gauge groups should not be charged 
under such a ${\rm U}(1)$ symmetry, the two-form $\omega^a$ for such a ${\rm U}(1)$ vector 
field should be in the sublattice 
\begin{equation}
W_{{\rm U}(1)} := \left[ W_{{\rm root}}^\perp \subset W_{{\rm frame}}\right]\, .
\label{eq:U(1)-characterization}
\end{equation}
This is equivalent to an object known as the {\it essential lattice} 
of an elliptic surface $X$ in the mathematics literature \cite{Shioda, Sch}, 
and may also be denoted by $L(X)$. 
Let $\{ \omega^a \}$ be an independent set of generators of 
$W_{{\rm U}(1)} = L(X)$. The massless ${\rm U}(1)$ vector fields in the effective theory are 
obtained from 
\begin{equation}
  C^{(3)} = \sum_{a=1}^{\rho_X-2-{\rm rk}(W_{\rm root})} 
  A^a \omega^a\, ,
\label{eq:3form-expansion}
\end{equation}
where $\rho_X-2 = {\rm rk}(W_{{\rm frame}*})$ \footnote{The $\omega^a$ are not 
necessarily Poincar\'e dual to effective curves. This does not pose a problem
as we only have to carry out a dimensional reduction to obtain their 
physics properties.}.  Theorem 1.3 in \cite{Shioda} states that the relation between 
the Mordell--Weil group $MW(X)$ and $W_{\rm frame}$ of an elliptically fibred 
surface is as follows:
\begin{equation}
 {\rm MW}(X) \cong {\rm NS}(X) / [ U \oplus W_{\rm root}] \cong
   W_{{\rm frame}}/W_{\rm root}\, .   
\end{equation}
Thus, the rank of Mordell--Weil lattice is the same as 
${\rm rank}(W_{{\rm U}(1)})$, the number of massless ${\rm U}(1)$ vector 
fields in the effective theory (when $G_1 \neq 0$, and $G_0=0$), and 
serves the purpose of counting degrees of freedom \cite{Morrison:1996pp}.
It should be remembered, though, that the ${\rm U}(1)$ vector fields are directly
associated  with two-forms in $F^1$, and hence in $W_{\rm frame}$, in 
physics. The connection with the Mordell--Weil group is only through 
an extra theorem in mathematics \cite{Shioda, Sch}.\footnote{
If we are to exploit this connection, the narrow Mordell--Weil lattice 
$MW(X)^0$ will be a more appropriate object than $MW(X)$. $MW(X)^0$ is 
defined as the subgroup of $MW(X)$ that consists of sections of an elliptic 
fibration $(\pi_X, \sigma; X, \P^1)$ that cross singular fibres only 
through the $(-2)$ curves meeting the zero section $\sigma$, rather than 
through $(-2)$ curves generating $A$--$D$--$E$ root lattices in $W_{\rm root}$. 
Theorem 8.9 in \cite{Shioda} states that the narrow Mordell--Weil lattice 
is isomorphic to $W_{{\rm U}(1)} = L(X)$ as an Abelian group, and the height pairing 
of $MW(X)^0$ (positive definite) is precisely the intersection form of 
$L(X)$ (negative definite) times $(-1)$.}
To go beyond the degree-of-freedom counting in \cite{Morrison:1996pp}, 
and extract more physics information, $W_{{\rm U}(1)}$ lattice is the right object
to deal with, as will be clear in the following discussion.

A preceding attempt of extracting more physics data, matter representations 
in F-theory compactifications on K3 surfaces in particular, 
has been made in \cite{Aspinwall:1998xj}. The discussion in \cite{Aspinwall:1998xj} 
leaves room for further sophistication in that 
\begin{itemize}
 \item only the stable degeneration limit of K3 surface was considered and,
 instead of a K3 surface, rational elliptic surfaces ($X=$dP$_9$) were used for the analysis. 
 This means that that $H^2(X;\Z)=S_X$, and the transcendental 
 lattice is trivial. That is now different for a K3 surface.
 \item The primary interest in \cite{Aspinwall:1998xj} was to keep track of 
matter representations under the non-Abelian part of the gauge group. 
But one may also be interested in classifying matter representations 
using not just non-Abelian charges but also massless (as well as global) 
${\rm U}(1)$ charges. As we will see in Section \ref{ssec:sample-stats}, 
it is not rare among attractive K3 surfaces that $W_{{\rm U}(1)}$ is non-empty.
\end{itemize}
Thus, a revised version of the discussion in \cite{Aspinwall:1998xj} 
is provided in the following, using the lattice-theory language that has
already been explaining in this section.

Obviously we can think of (not necessarily light) matter fields
originating from ``somehow quantizing'' an M2-brane wrapped 
on a cycle in $U^{\oplus 2} \oplus E_8^{\oplus 2} \cong 
[U_*^\perp \subset H_2(X; \Z)]$. 
Their representations under the massless gauge group associated with 
two-forms $W_{\rm gauge}:= W_{\rm root} \oplus W_{{\rm U}(1)}$ (resp. under the 
symmetry group associated with $W_{\rm gauge} \oplus T_X$)
should be specified by their weights, elements in the dual space 
$W_{\rm gauge}^* := {\rm Hom}(W_{\rm gauge}, \Z)$ 
(resp. $W_{\rm gauge}^* \oplus T_X^*$). Any quantized states arising from an 
M2-brane wrapped on a two-cycle in $[U_*^\perp \subset H_2(X; \Z)] \cong 
U^{\oplus 2} \oplus E_8^{\oplus 2}$ are in the same weight, and the 
weight is determined by the pairing between the divisors in $W_{\rm gauge}$ 
(resp. $W_{\rm gauge} \oplus T_X$) and the two-cycle. 
The collection of weights realized in this way forms a sublattice 
of the weight lattice $W_{\rm gauge}^*$ (resp. $W_{\rm gauge}^* \oplus T_X^*$). 
Let $G_{\rm matter}$ (resp. $\widetilde{G}_{\rm matter}$) be the image of 
this sublattice in the quotient space $G_{W_{\rm gauge}} = W_{\rm gauge}^*/W_{\rm gauge}$ 
(resp. $(W_{\rm gauge}^*/W_{\rm gauge}) \times G_{T_X}$). $G_{\rm matter}$ (resp.  
$\widetilde{G}_{\rm matter}$) is referred to as the $N$-ality of a given effective theory. 
Remembering that the unimodular lattice $U^{\oplus 2} \oplus E_8^{\oplus 2}$
is an overlattice of $W_{{\rm frame}*} \oplus T_X$, and that 
$W_{\rm gauge} \subset W_{{\rm frame}*} \subset W_{{\rm frame}*}^* \subset W_{\rm gauge}$, 
one finds an exact sequence 
\begin{equation}
 0 \longrightarrow (W_{{\rm frame}*} / W_{\rm gauge}) \longrightarrow
  \widetilde{G}_{\rm matter} \longrightarrow \Delta \longrightarrow 0\, ,
\end{equation}
where $\Delta$ is the diagonal subgroup of $G_{S_X} \times G_{T_X} \cong
G_{W_{{\rm frame}*}} \times G_{T_X}$. This characterizes the $N$-ality 
of matter representations $\widetilde{G}_{\rm matter}$ under the symmetry group.
For definitions of lattice theory jargon as well as reviews 
on background material, see e.g., \cite{BKW-math}. 
If we are to ignore the ${\rm U}(1)$ symmetry charges associated with the vector 
fields from $T_x$ (which are not massless in the presence of $G_1$ type flux), 
then the $N$-ality is given by 
\begin{equation}
 0 \longrightarrow \left(W_{{\rm frame}*}/W_{{\rm gauge}} \right)
  \longrightarrow G_{\rm matter} \longrightarrow 
  \left[\left(W_{{\rm frame}*}^*/W_{{\rm frame}*}\right)
         \cong G_{S_X} \right]
  \longrightarrow 0\, .
\end{equation} 
The matter fields in $W_{{\rm frame}*}/W_{{\rm gauge}}$
form a subgroup in $G_{\rm matter}$, which means that interactions among 
these fields must be closed within themselves.
Techniques to calculate $W_{{\rm frame}*}/W_{\rm gauge}$ as well as 
$G_{S_X} \cong \Delta$ are presented in \cite{BKW-math}, Section 4. 
Note that $W_{{\rm frame}*}^* \subset W_{{\rm gauge}}^*$ is now regarded
as the kernel of 
\begin{equation}
[W_{{\rm gauge}}^* = {\rm Ext}^0(W_{{\rm gauge}}, \Z)] 
\longrightarrow [{\rm Ext}^1(W_{{\rm frame}*}/W_{{\rm gauge}}, \Z) = 
W_{{\rm gauge}}^*/W_{{\rm frame}*}^*]\, , 
\end{equation}
rather than the kernel of 
\begin{equation}
[W_{\rm root}^* = {\rm Ext}^0(W_{\rm root}, \Z)] \longrightarrow
[{\rm Ext}^1(W_{{\rm frame}*}/W_{\rm gauge}, \Z) \cong{\rm Tor}(MW(X))] 
\end{equation}
as presented in \cite{Aspinwall:1998xj}. This difference 
from \cite{Aspinwall:1998xj} is due to the generalization from the 
stable degeneration limit (rational elliptic surface) to K3 surfaces and 
the inclusion of information on Abelian charges of the matter fields.
It is thus best for physics purposes to extract the information of 
an elliptic fibration in the form of the sublattice $W_{\rm gauge}$ and 
the quotient $W_{{\rm frame}*}/W_{\rm gauge}$. Consequently. computation results 
in \cite{BKW-math} are presented in this way.

Explicit examples will help understand the abstract theory above. 
In this article, we only show Table \ref{tab:X3-frame}, more examples are 
found in \cite{BKW-math}. For an attractive K3 surface $X = X_{[1~1~1]}$ (often 
denoted also by $X_3$), which has 6 different types of elliptic fibrations,  
the Mordell--Weil group has been computed for any one of these types 
(see Table 1.1 of \cite{Nish01}).
It is certainly well-motivated to study Mordell--Weil groups of elliptic 
fibrations in mathematics, to begin with, and decompose them into their free 
part and torsion part. However, more suitable for physicists' needs 
is to extract information from $W_{{\rm frame}*}$ in the form of 
$W_{\rm gauge}$, $(W_{{\rm frame}*}/W_{\rm gauge})$ and $G_{\rm matter}$. 
The subtle differences between them should be visible in the examples 
in Table \ref{tab:X3-frame}.
\begin{table}[tbp]
 \begin{center}
  \begin{tabular}{|c||c|c|c|c|}
\hline
  $W_{\rm gauge}$ & MW & $W_{{\rm frame}*}/W_{\rm gauge}$ \\
   &  $G_{W_{\rm gauge}}:= W^*_{\rm gauge}/W_{\rm gauge}$
     & $G_{\rm matter} := W^*_{{\rm frane}*}/W_{\rm gauge}$ \\
\hline
\hline
   $A_2E_8E_8$ 
     & $\{ 1 \}$ & $\{1 \}$ \\
     &  $\Z_3\vev{a_3}$ & $\Z_3\vev{a_3}$ \\
\hline
   $A_2D_{16}$ 
     & $\Z_2\vev{{\bf sp}}$ & $\Z_2\vev{{\bf sp}}$ \\
     &  $\Z_3\vev{a_3} \times \Z_2\vev{{\bf sp}} \times 
       \Z_2\vev{\overline{\bf sp}}$ &
      $\Z_3\vev{a_3} \times \Z_2\vev{{\bf sp}}$  \\
\hline
   $D_{10}E_7(-6)$
     & $\Z\vev{({\bf sp},0,{\bf 56})} \times
       \Z_2\vev{(\overline{\bf sp},{\bf 56},0)}$ &
       $\Z_2\vev{({\bf sp},0,{\bf 56})} \times
	   \Z_2\vev{(\overline{\bf sp},{\bf 56},0)}$ \\
     & $(\Z_2\vev{{\bf sp}} \times \Z_2\vev{\overline{\bf sp}}) \times
       \Z_2\vev{{\bf 56}} \times \Z_6\vev{{\bf 56}/3}$  
       & $\Z_2\vev{(\overline{\bf sp},{\bf 56},0)} \times \Z_6\vev{({\bf
	       sp},0,{\bf 56}/3)}$ \\
\hline 
   $A_{17}(-6)$ 
     & $\Z\vev{(3a_{18},{\bf 56})} \times \Z_3\vev{(0,2\cdot {\bf 56})}$
       & $\Z_6\vev{(3a_{18},{\bf 56})}$ \\
     & $\Z_{18}\vev{a_{18}} \times \Z_6\vev{{\bf 56}/3}$
       & $\Z_3\vev{(6,0)} \times \Z_3\vev{(0,2)} \times \Z_2\vev{(9,3)}$ \\ 
\hline        
   $E_6E_6E_6$ & $\Z_3\vev{(1,1,1)}$ & $\Z_3\vev{(1,1,1)}$ \\
       & $\Z_3\vev{{\bf 27}} \times \Z_3\vev{{\bf 27}} \times
         \Z_3\vev{{\bf 27}}$ 
         & $\Z_3\vev{(1,1,1)} \times \Z_3\vev{(0,2,1)}$ \\
\hline 
   $A_{11}D_7$ & $\Z_4\vev{(3a_{12},{\bf sp})}$ &
	   $\Z_4\vev{(3a_{12},{\bf sp})}$ \\
        & $\Z_{12}\vev{a_{12}} \times \Z_4\vev{{\bf sp}}$  
          & $\Z_4\vev{(3a_{12},{\bf sp})} \times \Z_3\vev{(4a_{12},0)}$ \\
\hline
  \end{tabular}
\caption{\label{tab:X3-frame}
The $N$-ality $G_{\rm matter} \subset G_{W_{\rm gauge}}$ of the 
six different types of elliptic fibrations in ${\cal J}_2(X)$ for an 
attractive K3 surface $X = X_3$. 
In this table, one can confirm that i) 
$(W_{\rm frame}/W_{\rm gauge}) \subset G_{\rm matter} \subset 
G_{W_{\rm gauge}}$, 
ii) $MW \cong W_{{\rm frame}*}/W_{\rm root} \longrightarrow 
 (W_{{\rm frame}*}/W_{\rm gauge})$ is a quotient, and 
iii) the quotient of $(W_{{\rm frame}*}/W_{\rm gauge}) \hookrightarrow
  G_{\rm matter}$ is always $G_{T_X} \cong \Z_3$ of the attractive K3
  surface $X_3$. 
In the 3rd and 4th entries, the generators of the rank-1 lattice 
$(-6)$ are denoted by $2 \cdot{\bf 56}$, because the generator is that of
  the weight $2 \cdot {\bf 56}$ in $E_7^*$, when this rank-1 lattice 
is regarded as $[E_6^\perp \subset E_7]$. 
}
 \end{center}
\end{table}

When we employ the expansion in the form of (\ref{eq:3form-expansion}), 
the gauge kinetic term of the vector fields on $S \times \R^{3,1}$ is 
given by 
\begin{equation}
\propto - \int_{\R^{3,1}} d^4x \int_{S} d^4y \sqrt{g(y)} \; 
   M_*^4 \left(
    T_R^{-1} {\rm tr}_R \left[ F_{mn} F^{mn} \right]
  -2 (\omega_a, \omega_b) F_{mn}^a F_{mn}^b
  \right)\, ;
\end{equation}
the normalization of the second term is set relatively to that of the 
first term, so that the maximal torus part of the non-Abelian components 
also have the same normalization\footnote{When the maximal torus part of
a non-Abelian $A_{n-1}$ component is expanded as 
$C^{(3)} = \sum_{a=1}^{n-1} C_a A^a$, with $A^a$ being $\R$-valued
vector fields, then this corresponds to 
$\diag(-A^1, A^1-A^2,\cdots,A^{n-2}-A^{n-1}, A^{n-1})$ in the
fundamental ($n$-dimensional) representation of $\SU(n)$, since an
M2-brane wrapped on $C_b$---usually assigned to the $(b,b+1)$ entry of
the $n \times n$ matrix representation---should have a covariant
derivative involving $A^{b-1}-2A^b+A^{b+1}$. 
Then $T_n^{-1} \tr_n\left[(\diag(-F^1,F^1-F^2, \cdots,
F^{n-1}))^2\right] = 4 [(F^1)^2 + \cdots + (F^{n-1})^2] 
- 4 [F^1 F^2+\cdots+F^{n-2}F^{n-1}] = -2 (C_a,C_b) F^a_{mn} F^b_{mn}$.} 
as the Abelian components given by the intersection form on the K3 surface $X$.
$T_R^{-1} \tr_R [ \cdots ]$ is the ordinary convention adopted 
for non-Abelian gauge theories.
The gauge coupling constant\footnote{Note that this kinetic term most 
likely corresponds to the one renormalized at the Kaluza--Klein scale, 
since this is obtained from a simple dimensional reduction (truncation) 
of 11-dimensional supergravity on a flat spacetime, whose infrared 
physics (especially when it comes to renormalization) is quite different 
from what we are really interested in.} 
of the massless ${\rm U}(1)$ vector fields is given by the opposite of the 
intersection form on the essential lattice $L(X)=W_{{\rm U}(1)}$, 
$- (\omega_a, \omega_b) = - \int_X \omega_a \wedge \omega_b$, which is 
equivalent to the (positive definite) height pairing of the narrow 
Mordell-Weil lattice $MW(X)^0$.

\subsection{Moduli Space of F-theory with Flux}
\label{ssec:sbtl-stat}

\subsubsection{Subspace of K3 Moduli Space with a Given Picard Number}

The discussion in Sections \ref{ssec:review-AK} and \ref{ssec:extended-list} 
centres on (pairs of) attractive ($\rho = 20$) K3 surfaces, 
while that of Sections \ref{sec:F-classify-elliptic-fib} 
and \ref{ssec:J1-J2-phys} is applicable for K3 surfaces with any 
Picard number $\rho$. Thus all the statements 
in Sections \ref{sec:F-classify-elliptic-fib} and \ref{ssec:J1-J2-phys} 
are applicable to the special cases treated in Sections \ref{ssec:review-AK} and 
\ref{ssec:extended-list}. As a warming up for the discussion in 
Section \ref{sssec:moduli-flux-quotient-MF} and later, however, 
let us first elaborate a little more about the relation between the 
characterization of attractive K3 surfaces in terms 
of (\ref{eq:def-of-abc-1}, \ref{eq:TX-OmegaX}, \ref{eq:def-of-abc-2}) 
and the complex structure moduli space 
${\rm Isom}^+(\Lambda_{\rm K3}) \backslash D$. This is only to repeat 
material presented in \cite{Mor, Moore:1998pn, AK}, apart from 
the purpose of setting up notations that we need later.
%

Let us first define a pair of sublattices 
$(T_{[\omega]},S_{[\omega]})$ for $[\omega] \in D$ as 
\begin{equation}
 S_{[\omega]} := \left[\omega^\perp \subset \Lambda_{\rm K3} \right], \qquad 
 T_{[\omega]} := \left[ S_{[\omega]}^\perp \subset \Lambda_{\rm K3} \right]\, .
\end{equation}
These two sublattices are mutually orthogonal complements in $\Lambda_{\rm K3}$ (which also means that they are primitive 
sublattices of $\Lambda_{\rm K3}$). Thus, one can define a map 
\begin{equation}
D \ni [\omega] \mapsto (T_{[\omega]}, S_{[\omega]}) \in \left\{ (T, S) |\; {\rm
				mutually~orthog.~sublattices~of~}\Lambda_{\rm
				K3} \right\} =: {\cal C}\, .
\end{equation}
${\cal C}$ is further decomposed into ${\cal C}_\rho$ with 
$\rho=0,1,\cdots,20$ where $T_{[\omega]}$ and $S_{[\omega]}$ have
signature $(2,20-\rho)$ and $(1,\rho - 1)$, respectively, 
and others which we are not interested in.\footnote{
Either $T_{[\omega]}$ or $S_{[\omega]}$ contains all three positive 
directions in such cases. They are not in the image of $D$, however.}
$D$ is also decomposed into $D_\rho$ with $\rho=0,\cdots,20$, where 
$D_\rho$ is the fibres over ${\cal C}_\rho$. Each irreducible component
of the fibres is of complex dimension $20-\rho$. 
The group ${\rm Isom}^+(\Lambda_{\rm K3})$ acts also 
on the ${\cal C}_\rho$, and the action on $D_\rho$ and ${\cal C}_\rho$ 
commutes with the map introduced above.

The Theorem 2.10 of \cite{Mor} states that there is a map that is both
injective and surjective between 
${\rm Isom}^{+}(\Lambda_{\rm K3}) \backslash {\cal C}_{\rho}$ and 
the classification of even lattice $T$ of signature $(2,20-\rho)$ 
modulo isometry, if $\rho \geq 12$
(which comes from a condition ${\rm rank}(T) \leq {\rm rank}(S)-2$). 
In the case of $\rho=20$, 
${\rm Isom}^+(\Lambda_{\rm K3}) \backslash D_{\rho=20}
\longrightarrow {\rm Isom}^+(\Lambda_{\rm K3}) \backslash {\cal C}_{\rho=20}$
(or, equivalently, $D_{\rho=20} \longrightarrow {\cal C}_{\rho=20}$)
is surjective\footnote{The original proof of surjectivity of 
the map from ${\rm Isom}^+(\Lambda_{\rm K3}) \backslash D_{\rho=20}$ 
to the set of even lattices of signature $(2,0)$ with orientation 
in \cite{SI} was to show that, for any even $(2,0)$ lattice with
orientation, $T^{\rm or}$, a K3 surface can be constructed whose 
transcendental lattice with an oriented basis becomes the even $(2,0)$ 
lattice $T^{\rm or}$.} 
and the fibre consists of 2 elements; they correspond to 
the two different choices of an orientation in 
$T_{[\omega]}\otimes \R$ that turns it into a complex line 
$[\omega] \in \Lambda_{\rm K3} \otimes \C$.
Thus, the scan over even lattices of signature $(2,0)$ with orientation 
in the basis---the scanning in \cite{AK} and in 
Sections \ref{ssec:review-AK} and \ref{ssec:extended-list}---is in
one-to-one correspondence with ${\rm Isom}^+(\Lambda_{\rm K3})
\backslash D_{\rho=20}$ \cite{AK}.
Therefore, the entries in Table \ref{tab:AK-notmorethan24} are regarded as 
a subset of 
\begin{equation}
\left[ {\rm Isom}^+(\Lambda_{\rm K3}^{(X)}) \backslash
        D^{(X)}_{\rho=20} \right] 
\times
\left[ {\rm Isom}^+(\Lambda_{\rm K3}^{(S)}) \backslash
        D^{(S)}_{\rho=20} \right]  \, ,
\end{equation}
specified by the condition (\ref{eq:G1-partflux<=24}). 


\subsubsection{Moduli Space in the Presence of Flux}
\label{sssec:moduli-flux-quotient-MF}

Moduli spaces such as (\ref{eq:classify-M-cpx-noflux}, 
\ref{eq:classify-F-K3-cpx-noflux}, \ref{eq:classify-F-cpx-noflux}) 
arise from compactifications of M/F-theory without flux. 
Let us now move on formulate the moduli spaces for compactifications 
including fluxes, paying close attention to the choice of the quotient group
which should tell us when a pair of vacua should be regarded the same 
in physics and when as distinct.

To get started, let us return to M-theory compactification on $Y=S_1
\times S_2$ down to 2+1-dimensions. Remembering that the moduli space 
was (\ref{eq:classify-M-cpx-noflux}) because we take a quotient by 
${\rm Isom}^+(\Lambda_{\rm K3}^{(S_1)}) \times
 {\rm Isom}^+(\Lambda_{\rm K3}^{(S_2)})$ in order to reduce the 
unphysical difference in the choice of marking, we claim that 
the complex structure moduli space\footnote{
There is nothing wrong to introduce the flux 
$G^{(4)}$ also in $H^4(S_1;\Z)\otimes H^0(S_2;\Z) 
\oplus H^0(S_1;\Z) \otimes H^4(S_2; \Z)$ in M-theory compactifications down 
to 2+1-dimensions, where we do not have to preserve $\SO(3,1)$ 
Lorentz symmetry. Strictly speaking, $H^{2,2}(Y; \R)$ in this equation should be 
replaced by its image under the marking. We do not try to be precise beyond our need. } 
of compactifications on $Y=S_1 \times S_2$ in the presence of 4-from flux should be given by 
the quotient space of 
\begin{equation}
 \left\{\left. \left( [\omega_1], [\omega_2], G^{(4)} \right) \; \right|
    \; [\omega_i] \in D^{(S_i)}, \;
      [G^{(4)}] \in \left( \Lambda_{\rm K3}^{(S_1)} \otimes 
                           \Lambda_{\rm K3}^{(S_2)} \right) \cap 
                     H^{2,2}(Y; \R)  
\right\} 
\label{eq:classify-M-cpx-wflux}
\end{equation}
by\footnote{Note that $\Z_2\vev{{\rm exch}_{12}}$ in the modular group 
$\Gamma$ also acts on the K\"{a}hler
moduli. }\raisebox{5pt}{,}\footnote{
To be more precise, we only know that the true modular group should
contain this $\Gamma$ in (\ref{eq:classify-M-cpx-wflux-grp}) as a
subgroup. To draw an analogy, $T^2 \times T^2 \times \cdots$
compactifications of type II string theory has a larger duality group
than just ${\rm SL}(2; \Z) \times {\rm SL}(2; \Z) \times \cdots $. The same 
comment applies also to the choice of modular group (\ref{eq:modular-group-F}) 
for F-theory.} 
\begin{equation}
\Gamma =  \Z_2\vev{{\rm c.c.}} \times \Z_2\vev{{\rm exch}_{12}} \ltimes
 \left[ 
     {\rm Isom}^+(\Lambda_{\rm K3}^{(S_1)}) \times 
     {\rm Isom}^+(\Lambda_{\rm K3}^{(S_2)}) 
  \right]\, ,
\label{eq:classify-M-cpx-wflux-grp}
\end{equation}
where ${\rm exch}_{12}$ exchanges $S_1$ and $S_2$, and 
${\rm c.c.}$ denotes complex conjugation of the entire $Y = S_1 \times
S_2$. As stated at the end of Section \ref{ssec:review-AK}, a pair of 
descriptions related by $\Z_2\vev{{\rm c.c.}} \times \Z_2\vev{{\rm exch}_{12}}$ 
should not be regarded distinct vacua in physics. 


This moduli space has a number of disconnected components corresponding 
to topological choices of the four-form flux. For non-trivial fluxes, 
some moduli have masses, and such connected components of the moduli
space have reduced dimensions. Thus, this moduli space should be that 
of effective theories below the mass scale of stabilized moduli,\footnote{ 
In the case of type IIB/F-theory compactifications, the mass scale is
typically $M_{\rm KK}^3/M_{\rm str}^2$ \cite{Kachru:2002he}.}
and can be used at least for the purpose of parametrizing/counting
vacua.\footnote{In order to use this as the target space of a non-linear 
sigma model below the mass scale of the stabilized moduli, one has 
to study corrections to the metric (K\"{a}hler potential) on moduli space.
Note that the classification of matter representations in Section \ref{sssec:frame-MW} 
includes information on stringy states, and hence is not a classification of effective 
field theories below the scale of moduli masses. Note also that the restricted moduli 
space ${\cal M}_*$ to be introduced in Section \ref{ssec:orientifold-beyond} should be 
regarded more as a mathematical (rather than physical) object on which the 
$\rho_{\rm ind}$ distribution is presented.} 
%

It is instructive to use the landscape of vacua already shown in 
Table \ref{tab:AK-notmorethan24}, where $G_0 = 0$, to see what 
the isolated (completely stabilized) components of this moduli is like. 
Already the table serves as the list of quotient of 
$D^{(S_1)}_{\rho=20} \times D^{(S_2)}_{\rho=20}$ by the group 
(\ref{eq:classify-M-cpx-wflux-grp}). The rest is to work out 
the number of different choices of fluxes $G^{(4)}=G_1$ (or equivalently
the number of different choices of $\gamma$) modulo the action of 
the residual symmetry in the group (\ref{eq:classify-M-cpx-wflux-grp}).
Written in the second to last column of 
Table~\ref{tab:AK-notmorethan24} is the number of different $\gamma$ modulo 
the residual symmetry in 
$\Z_2\vev{{\rm c.c.}} \times \Z_2\vev{{\rm exch}_{12}}$. Assuming further that 
all of the ${\rm Isom}(T_{S_1})^{({\rm Hodge})} \times 
{\rm Isom}(T_{S_2})^{({\rm Hodge})} \cong \Z_{m_1} \times \Z_{m_2}$ 
($m_{1,2} = 2,4,6$) symmetries of the transcendental lattices 
$T_{S_1}$ and $T_{S_2}$ can be lifted to isometries of the entire 
lattice $\Lambda_{\rm K3}$, however,\footnote{
This assumption is satisfied, if  
$p_S: {\rm Isom}(S_{S_{1,2}}) \longrightarrow {\rm Isom}(q_{1,2})$ 
is surjective. It is known that this is the case for some K3 surfaces
with large Picard number. See \cite{BKW-math} for more information.} 
all the $\gamma$'s are equivalent for all the entries, except in 
two entries marked by $*$ in the table, where there are two inequivalent
values of $\gamma$. Thus, we conclude---under this assumption---that 
the landscape of M-theory compactification on $Y=S_1 \times S_2$ 
with a four-form flux purely of type $G_1$ and completely stabilized 
complex structure moduli consists $1 \times 64 + 2 \times 2=68$ vacua.

Let us now turn to F-theory and try to figure out the moduli space 
for F-theory compactifications on elliptically fibred $X \times S$, with 
a four-form flux preserving $\SO(3,1)$ symmetry. 
From the experience so far, it is natural to 
consider that the moduli space is given by 
\begin{equation}
 \Gamma \backslash 
 \left\{ \left. \left([\omega_X], [\omega_S], G^{(4)}, \phi_U \right) \;
         \right| \;	\; G^{(4)} \in L, \; 
    \right\} / \{ \pm {\rm id.}_U \}\,, 
\label{eq:classify-F-cpx-wflux}
\end{equation}
where $\left([\omega_X],[\omega_S]\right) \in D^{(X)} \times D^{(S)}$, 
$\phi_U: U \hookrightarrow
  \left[ [\omega_X]^\perp \subset \Lambda_{\rm K3}\right]$, and 
the four-form flux $G^{(4)}$ is in 
\begin{eqnarray}
L & := & \left(
\left[ \phi_U(U)^\perp \subset \Lambda^{(X)}_{\rm K3} \right]
 \otimes \Lambda^{(S)}_{\rm K3} \right) \cap \\
& & \quad  \left[
  \varphi_{[\omega_X]}(H^{2,0}(X_{[\omega_X]}; \C)) \otimes
  \varphi_{[\omega_S]} (H^{0,2}(S_{[\omega_S]}; \C)) + {\rm h.c.}
           \right. \nonumber \\
& & \qquad \left. + 
  \varphi_{[\omega_X]}(H^{1,1}(X_{[\omega_X]}; \R)) \otimes \varphi_{[\omega_S]}(H^{1,1}(S_{[\omega_S]}; \R)) 
  \right]\, , \nonumber 
\end{eqnarray}
where $(X_{[\omega_X]},\varphi_{[\omega_X]})$ and 
$(S_{[\omega_S]}, \varphi_{[\omega_S]})$ are either one of two inverse 
images of $[\omega_X]$ and $[\omega_S]$, respectively, under the period map.
When only flux of $G_1$ type is introduced, the last line in $L$ is
dropped. The quotient group is given by
\begin{equation}
 \Gamma = \Z_2\vev{{\rm c.c.}} \times 
{\rm Isom}^+(\Lambda^{(X)}_{\rm K3}) \times 
{\rm Isom}^+(\Lambda^{(S)}_{\rm K3})\, .
\label{eq:modular-group-F}
\end{equation}
The $\Z_2\vev{{\rm exch}_{12}}$ is gone at this point, because 
we have already set up a convention that it is $X$, rather than $S$, 
whose vev of the volume of elliptic fibre goes to zero. 
If we are to focus on vacua with $\rho_X = \rho_S = 20$, then 
simply the condition that 
$\left([\omega_X], [\omega_S]\right) \in D^{(X)} \times D^{(S)}$ is replaced by 
$\left([\omega_X], [\omega_S]\right) \in 
D^{(X)}_{\rho=20} \times D^{(S)}_{\rho=20}$.

\subsubsection{How to Carry out the Vacuum Counting for F-theory on ${\rm K3} \times {\rm K3}$ 
in Practice}
\label{sssec:counting-F-practice}

As long as we consider compactifications on 
$Y= {\rm K3} \times {\rm K3} = X \times S$, with the elliptic fibration 
implemented as $\pi_X: X \longrightarrow \P^1$, 
all the 7-branes are in the form $\{{\rm point} \} \times S$;
in particular, there are no matter curves. 
Thus, all algebraic information (such as gauge groups and matter 
representations) of low-energy effective theories is captured by 
the frame lattice $W_{\rm frame}(X)$ and $T_X$. This means that 
\begin{equation}
 \amalg_{[a~b~c]} \; {\cal J}_2(X_{[a~b~c]})
\label{eq:theory-classify-alg}
\end{equation}
serves as the classification of effective theories by their 
algebraic information.\footnote{It may be possible that the difference 
between a pair of non-equivalent embedding of 
$T_X \oplus W_{\rm frame} \oplus U$ into $H_2(X; \Z)$ is absorbed by
rescaling of ${\rm U}(1)$ charges, only to result in different gauge coupling 
constants (gauge kinetic terms). We are not paying attention at this
level of detail in this article, however.} 
Here, $[a~b~c]$ runs over the thirty-four 
choices of the three integers characterizing the transcendental 
lattice of either $S_1$ or $S_2$ in Table \ref{tab:AK-notmorethan24}.

Let us take $X_{[1~0~1]}$ (also denoted by $X_4$ in the mathematics literature) 
as the first example. There are 13 different types of elliptic fibrations 
for this K3 surface \cite{Nish01}, i.e., $\# [{\cal J}_2(X_{[1~0~1]})] = 13$. 
When this K3 surface $X_{[1~0~1]}$ is to be used for the $X$ of 
$Y = X \times S$ in (\ref{eq:Y=XxS}), one can use 
Table \ref{tab:AK-notmorethan24} to see that the other K3 surface $S$ 
can be 
$X_{[6~0~6]}$ or $X_{[3~0~3]}$ (when $N_{D3}=0$), 
$X_{[5~0~5]}$ or $X_{[1~0~1]}$ (when $N_{D3}=4$), 
$X_{[4~0~4]}$, $X_{[4~0~1]}$, $X_{[2~0~2]}$ and $X_{[1~0~1]}$ 
(when $N_{D3}=8$), $X_{[3~0~3]}$ (when $N_{D3}=12$), 
$X_{[2~0~2]}$, $X_{[1~0~1]}$ (when $N_{D3}=16$) and finally 
$X_{[1~0~1]}$ (when $N_{D3}=20$).
There are 12 options for the choice of $(S, N_{D3})$. 
For any one of these 12 choices of $(Y = X_{[1~0~1]} \times S, N_{D3})$, 
the stabilizer subgroup of $\Gamma$ (i.e., the residual modular group) is 
\begin{equation}
 \Z_2\vev{{\rm c.c.}} \times 
 \left[ W^{(2)}(S_X) \rtimes {\rm Aut}(X) \right] \times 
 \left[ W^{(2)}(S_S) \rtimes {\rm Aut}(S) \right]\, , 
\end{equation}
which acts on the possible choices of elliptic fibrations 
($\phi_U: U \hookrightarrow S_X$) and flux of $G_1$ type 
($\gamma$ in Table \ref{tab:AK-notmorethan24}).
This is quite a complicated problem to work out. 
If we are to first exploit this remaining symmetry in $\Gamma$ 
in (\ref{eq:modular-group-F}) to eliminate a redundant description 
of elliptic fibrations, we can use the Corollary D 
of \cite{BKW-math}), which states that any one of the 13 types of elliptic fibrations 
of $X_{[1~0~1]}$ consists of a unique isomorphism class. There is no 
extra multiplicity coming from the difference between the ${\cal J}_1(X)$ 
and the ${\cal J}_2(X)$ classifications. The $\Z_2\vev{{\rm c.c.}}$ action 
in $\Gamma$ is not necessary in eliminating redundant descriptions of 
elliptic fibrations on $X_{[1~0~1]}$, and we can exploit this to see that 
the number of inequivalent choices of the flux $G^{(4)} = G_1$ is not more than 
the numbers presented in the last column of Table \ref{tab:AK-notmorethan24}.
Furthermore, in the cases $S=X_{[1~0~1]}$ or $X_{[2~0~2]}$, we can also see 
that the combined choice of flux and elliptic fibration is unique under
the action of the whole group $\Gamma$ because 
${\rm Isom}(T_S)^{({\rm Hodge})} \cong \Z_4$ and the generator 
of this group can be extended\footnote{This is because 
$p_S: {\rm Isom}(S_S) \longrightarrow {\rm Isom}(q)$ is known to be 
surjective for $S=X_{[1~0~1]}$ and $X_{[2~0~2]}$.} to an isometry of $H_2(S; \Z)$ for 
$S=X_{[1~0~1]}$ and $X_{[2~0~2]}$. For other $S$, the number of non-equivalent 
choices of flux and elliptic fibration combined cannot be determined 
without more information. We thus conclude that for any one of the 13 
types of elliptic fibrations in ${\cal J}_2(X_{[1~0~1]})$, the total 
number of inequivalent choices of $(S, N_{D3}, \gamma, \phi_U)$, and hence 
the number of inequivalent choices of vacua, is somewhere in 
between 12 and 23. 

The attractive K3 surface $X_{[2~1~1]}$ is another example for which
there is a unique isomorphism class in each type of elliptic fibration
(see Corollary D of \cite{BKW-math} or footnote \ref{fn:10-K3s} 
in this article). Thus, for theories 
in the classification of ${\cal J}_2(X_{[2~1~1]})$ 
in (\ref{eq:theory-classify-alg}), the counting of inequivalent vacua arises
only from the choice of fluxes ($\gamma$), not in the isomorphism
classes of elliptic fibrations. Thus, for any type of elliptic fibration 
in ${\cal J}_2(X_{[2~1~1]})$, the number of inequivalent vacua lies 
somewhere in between 9 and 18. These statistics originate from 
$(S, N_{D3})$ being $(X_{[4~2~2]}, 10)$, $(X_{[2~1~1]}, 10)$, 
$(X_{[2~1~1]}, 17)$, $(X_{[6~-3~3]}, 3)$, $(X_{[4~-2~2]}, 10)$, 
$(X_{[2~-1~1]}, 10)$, $(X_{[2~-1~1]}, 17)$ in Table \ref{tab:AK-notmorethan24}.
Note that we have exploited $\Z_2\vev{{\rm c.c.}}$ to set 
$X = X_{[2~1~1]}$ rather than $X_{[2~-1~1]}$. 

As an example of attractive K3 surfaces where there can be multiple 
isomorphism classes of elliptic fibrations of the same type, let us 
first consider $X = X_{[2~2~2]}$.
This K3 surface admits 30 different types of elliptic fibrations, 
$\# \left[ {\cal J}_2(X_{[2~2~2]}) \right] = 30$ \cite{Nish01}.
The number of isomorphism classes of each type can be either one or two, 
and it turns out (Example J of \cite{BKW-math}) that there is a unique 
isomorphism class in at least 15 out of the 30 different types.  
The number of remaining inequivalent choices of flux $G_1 \propto \gamma$ 
can be estimated as above, and it falls within 7--22, using the information 
in the last column of Table \ref{tab:AK-notmorethan24}. 
Thus, in conclusion, at least 15 classes of effective theories 
in ${\cal J}_2(X_{[2~0~2]})$ consist of 7--22 inequivalent vacua
individually, and there may be $2 \times$ (7--22) inequivalent 
effective theories of a given algebraic information corresponding to 
any one of the remaining 15 types in ${\cal J}_2(X_{[2~0~2]})$ 

Finally, let us take a look at the cases $X = X_{[6~0~6]}$ and
$X = X_{[6~6~6]}$. For these two attractive K3 surfaces, there is only 
one possible choice of $(S, N_{D3})$; $(S, N_{D3}) = 
(X_{[1~0~1]}, 0)$ and $(X_{[1~1~1]}, 6)$, respectively. 
All the choices of the flux $G_1 \propto \gamma$ turn out to be
equivalent under the residual $W^{(2)}(S_S) \rtimes {\rm Aut}(S) \subset 
{\rm Isom}^+(\Lambda^{(S)}_{\rm K3})$ symmetry in $\Gamma$, because of 
the surjectiveness of $p_S: {\rm Isom}^+(S_S) \longrightarrow {\rm Isom}(q)$
for $S=X_{[1~0~1]}$ and $X_{[1~1~1]}$. The number of distinct 
isomorphism classes of elliptic fibrations is not more than 
16 and 12 for $X=X_{[6~0~6]}$ and $X=X_{[6~6~6]}$, respectively, for any 
types in ${\cal J}_2(X)$ (Corollary D of \cite{BKW-math}). Thus, 
for these two attractive K3 surfaces chosen as $X$, the number of 
inequivalent vacua is bounded from above by 16 and 12, respectively. 

\subsection{Sample Statistics}
\label{ssec:sample-stats}

The example-based study in Section \ref{sssec:counting-F-practice} indicates 
that each class of theories in (\ref{eq:theory-classify-alg}) consists 
of ${\cal O}(10)$ vacua inequivalent under the modular group $\Gamma$ 
in (\ref{eq:modular-group-F}). Although the study only covers 
five attractive K3 surfaces
$X_{[a~b~c]}$ out of thirty-four, small as well as large $a, c$ are 
covered in the five examples. We expect that an estimate of
the vacuum counting would not be different so much for the other
twenty-nine attractive K3 surfaces. 

This fact---the numbers of vacua in individual classes of effective
theories in (\ref{eq:theory-classify-alg}) are much the same---allows 
us to take a short-cut approach in studying statistical distributions 
of more inclusive classifications of effective theories. 
By more inclusive classifications, we mean classifications of low-energy
effective theories coarser than in (\ref{eq:theory-classify-alg}).
One might be interested, for example, in the number of effective theories 
that contain a certain gauge group $G$ (such as 
$\SU(3)_C \times \SU(2)_L \times {\rm U}(1)_Y$, $\SU(5)$ or $\SO(10)$), 
and compare the numbers for various choices of $G$.
When we ask this question, we have to include all the vacua
from (\ref{eq:theory-classify-alg}) containing the specified gauge
group, regardless of the gauge groups in the hidden sector. 
Given the fact that the number of vacua in each class of theories 
in (\ref{eq:theory-classify-alg}) are much the same, we can
simply count the number of classes of effective theories contained 
in inclusive classes of theories, because more or less ``the same'' 
multiplicity ${\cal O}(10)$ factors out in the ratio.
In this section, we take this short-cut approach in order to address 
three questions of interest.

\subsubsection{Statistics on 7-brane Gauge Groups and CP Violation}

{\bf 7-brane Gauge Groups}

It is one of the most important questions we can address by using 
a toy/miniature supersymmetric landscape whether or not there are 
more vacua with an $\SU(5)$ unified gauge group than those with 
$\SU(3)_C \times \SU(2)_L \times {\rm U}(1)_Y$ gauge group that just happens 
to satisfy gauge coupling unification ``by accident''. 
As is well-known, it makes sense in the context of unified theories 
to focus on vacua of string theory realized as
compactifications for which the volume of internal space is
parametrically larger than the string length. This is because in 
$\SU(5)$ unified gauge theories, for example, the doublet-triplet
splitting problem will be too difficult to solve within string
theory\footnote{Imagine implementing the missing partner mechanism in
string theory, for example.} in a form other than implementing a
non-trivial line bundle (flat or non-flat) in the hypercharge direction.
The Kaluza--Klein scale has to be set at the scale of gauge coupling 
unification then. If it turns out, however, that there are more vacua 
with the $\SU(3)_C \times \SU(2)_L \times {\rm U}(1)_Y$ gauge group and
accidental gauge coupling unification than those with $\SU(5)$ gauge
group in the landscape obtained by assuming geometric
compactification, then there must be much more $\SU(3)_C \times \SU(2)_L
\times {\rm U}(1)_Y$ vacua when we include string vacua with non-geometric
(just CFT-based) ``internal space''. Therefore, it is a necessary
condition that there are more $\SU(5)$ vacua than $\SU(3)_C \times
\SU(2)_L \times {\rm U}(1)_Y$ vacua in landscapes based on geometric 
compactification for the study of $\SU(5)$ unification in string 
compactification. It is this necessary condition that we intend to 
test below.

Instead of carrying out this test itself, we consider a similar (and a
little easier) test in this article. Instead of studying the ratio of
vacua with $\SU(5) \times ({\rm any~non\mbox{-}Abelian})$ and those with 
$\SU(3) \times \SU(2) \times {\rm U}(1) \times ({\rm
any~non\mbox{-}Abelian})$, 
we study the ratio\footnote{It must be a reasonable assumption that such 
gauge groups originate from 7-branes rather than D3-branes; D3-branes 
(rather than fractional 3-branes) only give rise to ${\cal N}=4$ super 
Yang--Mills theory, and furthermore, their gauge couplings are
parametrically larger than those of 7-branes because of the volume of 
four-cycles that the 7-branes are wrapped.} of vacua with 
$E_n \times {\rm U}(1)^{r} \times ({\rm any~non\mbox{-}Abelian})$ and 
those with $E_{n-1} \times {\rm U}(1)^{r+1} \times 
({\rm any~non\mbox{-}Abelian})$ ($n=6,7,8$).

The ${\cal J}_2(X)$ classification has been worked out completely for 
four attractive K3 surfaces, $X=X_{[1~0~1]}$, $X_{[1~1~1]}$, 
$X_{[2~0~2]}$ and $X_{[2~2~2]}$ \cite{Nish01}. Let us first use these
statistics---a subset of (\ref{eq:theory-classify-alg})---and further 
use the short-cut approach we explained above to see the ratio of 
vacua with a $E_{6,7,8}$ gauge group on 7-branes along with how many 
${\rm U}(1)$ vector fields they are accompanied. There are 112 different
types of elliptic fibrations in ${\cal J}_2(X)$ for the four K3 surfaces 
combined, and 32 among them contain one of $E_{6,7,8}$ as a part of the 
7-brane gauge group.\footnote{Here, 23/112 is the fraction of the types of 
elliptic fibration containing ${\rm IV}^*$, ${\rm III}^*$ or ${\rm II}^*$ type 
singular fibres in $X_{[1~0~1]}$, $X_{[1~1~1]}$, $X_{[2~0~2]}$ and $X_{[2~2~2]}$. 
Such a fraction for an attractive K3 surface $X_{[a~b~c]}$ may, however, have 
some correlation with whether $a$, $c$ are large or small. } 
Using the information in Table 1.1--1.4 of \cite{Nish01}, the statistics 
turns out to be the following (Table \ref{tab:nmbr-U1-4K3s-Nishiyama}):
\begin{table}[htbp]
 \begin{center}
  \begin{tabular}{|c||c|c|c|c|c|}
\hline
   group (non-Abelian) & rk 0 & rk 1 & rk 2 & rk 3 & tot. \\
\hline
\hline
 $E_8+$ any other & 5 & 2 & 3 & 0 & 10 \\
 $E_7+$ any other & 3 & 5 & 2 & 0 & 10 \\
 $E_6+$ any other & 3 & 5 & 5 & 0 & 13 \\
\hline 
  \end{tabular}
\caption{\label{tab:nmbr-U1-4K3s-Nishiyama}
This table shows the number of different types of elliptic fibrations 
in $X_{[1~0~1]}$, $X_{[1~1~1]}$, $X_{[2~0~2]}$ and $X_{[2~2~2]}$ that 
contain one of ${\rm IV}^*$, ${\rm III}^*$ and ${\rm II}^*$ type
singular fibres, and have Mordell--Weil lattice of a given rank.
One type, where $W_{\rm frame} = E_8\oplus E_6 \oplus D_4$ for $X_{[2~2~2]}$, 
is counted twice in this table. }
 \end{center}
\end{table}

In addition to the four attractive K3 surfaces, the authors have
partially carried out the ${\cal J}_2(X)$ classification for another 
attractive K3 surface\footnote{There is no particular reason to choose
this attractive K3 surface (rather than thirty-three others) from the
perspective of physics.} $X_{[3~0~2]}$ in \cite{BKW-math}, so that the same 
statistics as above can be extracted. There are 43 different types of 
elliptic fibrations on $X_{[3~0~2]}$ which contain 
either one of the ${\rm IV}^*$, ${\rm III}^*$ or ${\rm II}^*$ type
singular fibres. The distribution of the rank of the
Mordell--Weil lattice turns out be the following
(Table \ref{tab:nmbr-U1-X302}):
\begin{table}[htbp]
 \begin{center}
  \begin{tabular}{|c||c|c|c|c|c|}
\hline
   group (non-Abelian) & rk 0 & rk 1 & rk 2 & rk 3 & tot. \\
\hline
\hline
 $E_8+$ any other & 2 & 5 & 2 & 0 & 9 \\
 $E_7+$ any other & 2 & 9 & 7 & 0 & 18 \\
 $E_6+$ any other & 0 & 8 & 8 & 3 & 19 \\
\hline 
  \end{tabular}
\caption{\label{tab:nmbr-U1-X302} The number of different types of
elliptic fibrations of ${\cal J}_2(X_{[3~0~2]})$ containing either one
of ${\rm IV}^*$, ${\rm III}^*$ or ${\rm II}^*$ type singular fibres and 
having Mordell--Weil lattice with various ranks. Three types 
with $W_{\rm root} = A_2E_7E_8$, $A_1E_7E_8$ and $A_3E_6E_7$ contribute 
twice in this table, so that the total number is summed up to 46, 
rather than 43.}
 \end{center}
\end{table}

When we compare the numbers in the two tables 
(Tables \ref{tab:nmbr-U1-4K3s-Nishiyama} and \ref{tab:nmbr-U1-X302}) 
for $E_n$ with ${\rm rank}(MW)=r$ 
($E_n \times {\rm U}(1)^{r} \times ({\rm any~non~Abelian})$ massless gauge fields 
on 7-branes) and for $E_{n-1}$ with ${\rm rank}(MW)=r+1$ 
($E_{n-1} \times {\rm U}(1)^{r+1} \times ({\rm any~non~Abelian})$ massless gauge 
fields on 7-branes), we cannot observe a clear tendency. There are 
as many $E_n \times {\rm U}(1)^{r}$ (i.e., more unified) vacua as 
$E_{n-1} \times {\rm U}(1)^{r+1}$ (i.e., less unified) vacua in this sample.  
Although we cannot hope to extract too many lessons from this study 
based on a miniature landscape, it may not be too outrageous to say 
that this statistics does not indicate that it is nonsense to study 
unified theories in compactifications.
 

{\bf CP Violation}

Just like in earlier work such as \cite{DeWolfe:2004ns}, 
the miniature landscape in this section can also be used to study 
the fraction of vacua preserving CP symmetry. To be more specific, 
we study the fraction of CP preserving vacua in the class of effective 
theories containing $E_8 \times E_8$ gauge group from 7-branes. Such a 
formulation of the problem is meaningful, because we would eventually 
like to ask the fraction of CP-preserving/violating vacua in effective 
theories with the Standard Model gauge group. The choice of the gauge 
group---$E_8 \times E_8$---is entirely for a technical reason. For most
other choices of the gauge group, we would have to work out the
${\cal J}_2(X)$ classification for all the 34 attractive K3 surfaces 
appearing in Table \ref{tab:AK-notmorethan24}. 
It is known, however, (see \cite{Moore:1998pn, Gorlich:2004qm}, \cite{BKW-math} or Footnote \ref{fn:E8E8} for an
explanation) that any attractive K3 surface admits an elliptic fibration 
whose frame lattice contains the $E_8 \oplus E_8$ lattice, so that there must
be two fibres of type $II^*$.

In Table \ref{tab:AK-notmorethan24}, there are 66 pairs of attractive K3 
surfaces $Y = S_1 \times S_2$ for M-theory compactifications. From these, 
one can find 98 choices of 
$Y= X \times S = X_{[a~b~c]} \times X_{[a'~b'~c']}$ for F-theory 
compactifications, where elliptic fibrations (with the vanishing volume 
of the fibre) are implemented in $X_{[a~b~c]}$. Exploiting the 
$\Z_2\vev{{\rm c.c.}}$ in the modular group $\Gamma$ 
in (\ref{eq:modular-group-F}), we can always take $b \geq 0$. 
We take these 98 different choices as the denominator (whole ensemble)
of the statistics (see the cautionary remark at the end of this CP study).

In order to see when the low-energy effective theory possesses CP symmetry,
let us write down the Gukov--Vafa--Witten superpotential 
(\ref{eq:GVW}, \ref{eq:G1-gamma-def}) explicitly in terms of local 
coordinates of the moduli space 
${\cal M}_{K3}^{(S_1)} \times {\cal M}_{K3}^{(S_2)}$.
Let $\Omega_{S_i}^{\rm tot}$ for $i = 1,2$ be the total holomorphic 
$(2,0)$-form on the K3 surface $S_i$ for $i=1,2$, including both 
the vacuum value and fluctuation around it:
\begin{equation}
 \Omega_{S_i}^{\rm tot} = \vev{\Omega_{S_i}} + \delta \Omega_{S_i} 
 =  p_i + q_i (\tau_i + \widetilde{\Pi}_i) + C^{(i)}_I \delta\Pi^{(i)}_I,
   \qquad  i=1,2\, .
\label{eq:Omega-fluct-expansion};  
\end{equation}
The $\Omega_{S_i}$ which appeared in Section \ref{ssec:review-AK} corresponds 
to the vacuum value $\vev{\Omega_{S_i}}$ here, and 
$\{C_I^{(i)}\}_{I=1,\cdots,20}$ is a basis of the Neron--Severi lattice 
$S_{S_i}$.
%
$\delta\Pi_I^{(i)}$ for $i=1,2$ and $I = 1,\cdots, 20$ combined are
the independent local coordinates of 
${\cal M}_{K3}^{(S_1)} \times {\cal M}_{K3}^{(S_2)}$, 
and 
$\widetilde{\Pi}_i$ are determined by the condition 
$(\Omega_{S_i} + \delta \Omega_{S_i})^2 = 0$. In practice, 
\begin{eqnarray}
 \tau_1 + \widetilde{\Pi}_1 & = &
    \frac{-b + i \sqrt{Q_1 + 2c (\delta\Pi^{(1)})^2  }}{2c} 
 =
   \tau_1 + \frac{i\sqrt{Q_1}}{2c}
   \left( \frac{c (\delta \Pi^{(1)})^2}{Q_1} - 
          \frac{1}{2}\left(\frac{c(\delta\Pi^{(1)})^2}{Q_1}\right)^2
        + \cdots 
   \right), \\
 \tau_2 + \widetilde{\Pi}_2 & = & 
    \frac{-e + i \sqrt{Q_2 + 2f (\delta\Pi^{(2)})^2  }}{2f} 
=
  \tau_2 + \frac{i \sqrt{Q_2}}{2f} 
    \left( \frac{f (\delta \Pi^{(2)})^2}{Q_2} - 
          \frac{1}{2}\left(\frac{f(\delta\Pi^{(2)})^2}{Q_2}\right)^2 +
	  \cdots
    \right)\, , 
\end{eqnarray}
where $(\delta\Pi^{(i)})^2$ is the norm of $C_I^{(i)} \delta\Pi_I^{(i)}$ 
under the symmetric pairing of the Neron--Severi lattices $S_{S_i}$.
Substituting (\ref{eq:Omega-fluct-expansion}) 
into (\ref{eq:GVW}, \ref{eq:G1-gamma-def}), we obtain 
\begin{eqnarray}
W & \propto &
   \left(G_1, (\Omega^{\rm tot}_{S_1} \otimes \Omega^{\rm tot}_{S_2})
   \right) \\
& = & \sqrt{Q_1Q_2} \left[
   {\rm Re}(\gamma\tau_1\bar{\tau}_2) 
 + {\rm Re}(\gamma) (\tau_1 + \widetilde{\Pi}_1)(\tau_2 + \widetilde{\Pi}_2) 
   \right.   \nonumber \\
 & & \qquad \qquad \left.
    - {\rm Re}(\gamma \bar{\tau}_2) (\tau_1+\widetilde{\Pi}_1)
    - {\rm Re}(\gamma \tau_1) (\tau_2+\widetilde{\Pi}_2)
 \right]\, , \label{eq:flux-eff-super} 
\end{eqnarray}
This potential contains mass terms of all the fluctuations,\footnote{
Chiral multiplets in the adjoint representation of the 
non-Abelian 7-brane gauge groups (in $W_{\rm root}$), i.e., transverse 
fluctuations of the 7-branes, also become massive because of this.
This mass term, due to the $G_1$ type four-form flux, has nothing to do 
with the level-2 differential 
\begin{equation}
 d_2: [E^{0,1}_2 = H^0(S; K_S)] \rightarrow [E^{2,0}_2 = H^2(S; {\cal O}_S)]
\end{equation}
that we encounter in the spectral sequence calculation in heterotic 
language \cite{Bershadsky:1997zs, Hayashi-1}. It is only the kernel of 
this $d_2$ that remains massless (in the absence of the $G_1$ type
flux), but this $d_2$ is always trivial for the case of our interest, 
for reasons that are explained in \cite{Katz-Sharpe-02}. 
} as expected (moduli stabilization), and furthermore quartic and higher 
order interactions, as is known very well \cite{Tripathy:2002qw,Gorlich:2004qm,Lust:2005bd,AK}.

Note first the vacuum expectation value of this superpotential vanishes.
One can see this by expanding (\ref{eq:flux-eff-super}) in a power
series of $(\delta \Pi^{(i)})^2$ and evaluating the zero-th order term.
The vanishing vev of $W_{\rm GVW}$, however, is a straightforward 
consequence of choosing the four-form flux to be purely of $(2,2)$ type 
in the Hodge decomposition. It will be difficult to find a symmetry
reason for this vanishing vev of $W_{\rm GVW}$ covering all the 
66 pairs of $S_1 \times S_2$ for M-theory compactification, or all 
the 98 choices of $X \times S$ for F-theory. It seems more appropriate 
to consider that the ${\cal D}=2$ condition (\ref{eq:square-integer}) 
is essential, see the discussion right after (\ref{eq:square-integer}) and 
(\ref{eq:40-04-components}), and also \cite{DeWolfe:2004ns}.
The integral structure of the flux quanta plays an essential role in 
determining the moduli vev through $W_{\rm GVW}$. Hence it is not 
appropriate to apply naive naturalness arguments of bottom up
phenomenology. 

In a subset of vacua where $b = e = 0$ for M-theory 
($b = b' = 0$ for F-theory), both $\tau_1$ and $\tau_2$ are purely
imaginary and the GVW superpotential becomes CP invariant when 
we choose $\gamma$ to be either purely real or pure imaginary. 
In the first case, the two terms in the second line of 
(\ref{eq:flux-eff-super}) vanish, and all the terms in the superpotential 
have real valued coefficients. In the latter case, the two terms in the 
first line of (\ref{eq:flux-eff-super}) vanish, and all the remaining 
terms---those in the second line---have purely imaginary coefficients. 
With an appropriate phase redefinition of fermion fields 
(R-symmetry transformation), those coefficients can be made real valued. 
Thus, for these two case, all of the coefficients in the effective superpotential can be made 
real valued (by field redefinitions, if necessary), and CP symmetry 
is preserved. Following \cite{DeWolfe:2004ns}, we understand that 
the CP invariance in the former case is due to the compactification 
data $([a~b~c], [d~e~f], \gamma)$ invariant under 
the $\Z_2\vev{{\rm c.c.}}$ subgroup of the modular group $\Gamma$, and 
is due, in the latter case, to the compactification data being invariant under 
the combination of the ${\rm c.c.}$ operation followed by a non-modular 
group symmetry operation (that somehow becomes an R-symmetry transformation).
 
Among the 66 pairs of attractive K3 surfaces in 
Table \ref{tab:AK-notmorethan24}, there are 20 pairs where $b = e = 0$
and $\gamma$ is either purely real or purely imaginary.
To turn this statistics into that of F-theory compactifications, 
note that among the 98 different ways of identifying the 66 pairs with 
$X \times S$ for F-theory, 30 different ways correspond to $b = b' = 0$
and a purely real/imaginary choice of $\gamma$. Although such precise
values as $30/98$ do not have much importance,\footnote{
It should be noted that we ignore the possibility that an attractive K3
surface may admit more than one type of elliptic fibration 
where $E_8 \times E_8 \subset W_{\rm root} \subset W_{\rm frame}$. 
In fact, one can find an example of this in \S4.4 of \cite{BKW-math}:  
the attractive K3 surface $X_{[3~0~2]}$ admits a pair of elliptic fibrations 
where $W_{\rm root}$ are the same, but their $W_{\rm frame}$ are not 
isometric. We also ignore multiple 
inequivalent choices of $\gamma$ ($G_1$ type flux) and the number of isomorphism
classes of elliptic fibrations. These factors should, in principle, be 
treated as a non-trivial weight on the 98 different choices in the main
text. Thus, the precise value of the fraction of CP-invariant vacua does 
not have much importance.}\raisebox{5pt}{,}\footnote{ Complex conjugation 
$\Z_2\vev{{\rm c.c.}}$ of the M-theory real 8-dimensional manifold is included 
as a part of the modular group (\ref{eq:classify-M-cpx-wflux-grp}, 
\ref{eq:modular-group-F}) in this article, while it is not 
in \cite{DeWolfe:2004ns}. This subtle differences, however, only leads to 
at most a factor of 2 difference in the fraction of CP-preserving vacua.
Given the other factors that we did not try to bring under control, 
this issue would not be particularly important from practical perspectives. } 
it will not be too outrageous to conclude that a non-negligible fraction of vacua possesses CP symmetry 
in the landscape of  ${\rm K3} \times {\rm K3}$ compactifications of F-theory with all the 40
complex structure moduli stabilized by the $G_1$ type four-form flux. 
See also related discussion in Section \ref{sec:G0-flux}.

\subsubsection{Stable Degeneration ``Limit''}
\label{sssec:FMW}

F-theory compactifications on $Y= {\rm K3} \times {\rm K3} = X \times S$ 
(without a four-form background) are dual to heterotic string compactifications 
on $Z = T^2 \times {\rm K3}$ 
\cite{Kachru:1995wm,Aldazabal:1995yw,Vafa-evidence, Morrison:1996na, Morrison:1996pp,Bershadsky:1996nh, Friedman:1997yq, 
Bershadsky:1997zs}. From the heterotic string
picture, one would naively expect that the Kaluza--Klein reduction of metric and $B$-field on
$T^2$ gives rise to 4 massless ${\rm U}(1)$ vector fields in 7+1-dimensions, 
in addition to the at most rank-16 gauge group from either 
$E_8 \times E_8$ or $\SO(32)$.
Since F-theory compactifications on an attractive K3 surface $X$ have 
rank 20 gauge groups on 7-branes (two ${\rm U}(1)$'s from $T_X$ and rank 18
from $W_{\rm frame}$), there is no mismatch in the rank of the gauge
groups. It often happens in the miniature landscape we studied, however, 
that there are not more than four ${\rm U}(1)$ factors in the 7-brane gauge group. 
In Tables \ref{tab:nmbr-U1-4K3s-Nishiyama} and \ref{tab:nmbr-U1-X302},
for example, the rank of Mordell--Weil lattice is less than two (meaning
that the number of ${\rm U}(1)$ gauge group is less than four) in a large
fraction of the types of elliptic fibrations. By looking at the 
Tables in \cite{Nish01}, one can confirm that this phenomenon is not an artefact of 
requiring either one of $E_6$, $E_7$ or $E_8$ in the 7-brane gauge group. 
An appropriate interpretation of this phenomenon should be that 
the large fraction of attractive K3 surfaces in 
${\rm Isom}^+(\Lambda_{\rm K3}) \backslash D^{(X)}_{\rho=20}$ satisfying 
both the D3-tadpole condition (\ref{eq:G1-partflux<=24}) and 
the pure $G_1$-type assumption (\ref{eq:no-G0}) does not correspond to the large 
${\rm vol}(T^2)/\ell_s^2$ region of the heterotic string moduli space. 
In this case, the supergravity approximation is not valid and some of the non-Abelian 
7-brane gauge groups should be understood as stringy effects in heterotic language.  

Corresponding to the supergravity (large ${\rm vol}(T^2)/\ell_s^2$) ``limit''
in heterotic string theory is the stable degeneration ``limit'' of a K3 surface 
in F-theory \cite{Morrison:1996pp, Friedman:1997yq}.
In this section, we discuss a couple of issues 
associated with this supergravity/stable degeneration ``limit'', based
on the statistics in the miniature landscape. In this article, we mean by 
large/small ``limit'' of a moduli parameter xxx the region of moduli 
space where xxx is parametrically large/small, i.e., 
${\rm xxx} \gg 1$ or ${\rm xxx} \ll 1$. However, we still assume xxx to be 
different from literally being $\infty$ or $0$.

Let us begin with reminding ourselves of the following.
Suppose that the elliptic fibration of a K3 surface 
$\pi_X: X \longrightarrow \P^1$ is given by the generalized Weierstrass 
form (or Tate form)
\begin{eqnarray}
y^2 & = & x^3 + x f_0 z^4 + g_0 z^6  \nonumber \\
 & & \qquad  +  \epsilon_\eta
   (a^{\rm v}_{0*} z^5 + \epsilon_K^2 a^{\rm v}_{2*} x z^3 + \epsilon_K^3
 a^{\rm v}_{3*}y z^2 + \epsilon_K^4 a^{\rm v}_{4*} x^2 z + \cdots )  \nonumber \\
 & & \qquad + \epsilon_\eta
   (a^{\rm h}_{0*} z^7 + \epsilon_K^2 a^{\rm h}_{2*} x z^5 + \epsilon_K^3 a^{\rm h}_{3*} y z^4 + \cdots )\, ,
\label{eq:Weierstrass-degen-lim}
\end{eqnarray}
where $(x,y)$ are the coordinates of the elliptic fibre, $z$ the
inhomogeneous coordinate of the base $\P^1$, and $f_0$, $g_0$, 
$a^{\rm v}_{r*}$'s $(r=0,2,\cdots)$ and $a^{\rm h}_{r*}$'s are complex 
numbers of order unity. In the heterotic dual, ${\rm vol}(T^2)/\ell_s^2$ 
is parametrically large when\footnote{The Wilson lines on $T^2$ 
are small, so the 8D field theory approximation is valid 
when one more condition, $|\epsilon_K|\ll 1$, is also satisfied.} 
$|\epsilon_\eta| \ll 1$, $|\epsilon_K| \lesssim {\cal O}(1)$.  
K3 surfaces with an elliptic fibration with small $\epsilon_\eta$ are said to be 
in the stable degeneration ``limit'' \cite{Morrison:1996pp, 
Friedman:1997yq}.\footnote{\label{fn:degeneration-limit-A}
A family of K3 surfaces $\pi': {\cal X}'
\longrightarrow D$ was introduced in \cite{Morrison:1996pp}, where 
$D := \{t \in \C \; | \; |t| \leq 1\}$ is the unit disc,
${\cal X}'$ is given by $y^2=x^3 + f_0 x z^4 + 
(g_0 z^6 + \epsilon_\eta a^{\rm v} z^5 + \epsilon_\eta a^{\rm h} z^7)$ defined
as a subspace of $(y,x,z,\epsilon_\eta)$ for some complex valued parameters 
$f_0$, $g_0$, $a^{\rm v}$ and $a^{\rm h}$. The morphism $\pi'$ is defined by 
$\pi': (y,x,z,\epsilon_\eta) \mapsto t = \epsilon_\eta$. 
Instead of this family of K3 surfaces, one can also consider another
family $\pi: {\cal X} \longrightarrow D$ given by 
${\cal X} := \left\{ (\eta,\xi,u,v,t) \in \C^5 \; | \; 
\eta^2 = \xi^3 + f_0 \xi + (g_0 + a^{\rm v} u + a^{\rm h} v), \; 
u v = t \right\}$ (to be more precise, $(\eta, \xi)$ and $(u,t)$ are
affine coordinates of projective space). 
The original family $\pi': {\cal X}' \longmapsto D$ is regarded as 
the base change of order 2 of the second family $\pi: {\cal X}
\longmapsto D$, $D \ni \epsilon_\eta \mapsto \epsilon_\eta^2 = t \in D$.
Other coordinates are mapped by $(u,v)=(\epsilon_\eta / z, \epsilon_\eta z)$, 
$\eta = y/z^3$ and $\xi = x/z^2$. 
The second family shows a semistable degeneration, in that 
i) the threefold ${\cal X}$ is non-singular, and 
ii) the central fibre $X_0 := \pi^{-1}(t = 0)$ consists of two rational 
elliptic surfaces (a.k.a ``$dP_9$'') crossing normally along an elliptic 
curve at $u=v=0$. See \cite{FM} for more information.
Provided that $f_0$, $g_0$, $a^{\rm v}$ and $a^{\rm h}$ are generic, 
$X_t := \pi^{-1}(t)$ for any $t \in D$ has ${\rm II}^* + {\rm
II}^*$-type singular fibre in the non-singular model at 
$(u,v)=(\infty,0)$ and $=(0,\infty)$. The Weierstrass model version 
of this family has two ordinary $E_8 + E_8$ singularities. }

When $|\epsilon_{\eta}| \ll 1$, and the heterotic dual (in the supergravity 
approximation) corresponds to $E^{\rm v}_8 \times E^{\rm h}_8$ theory with 
the structure group $G^{\rm v}_{\rm str} \times G^{\rm h}_{\rm str} 
\subset E^{\rm v}_8 \times E^{\rm h}_8$, 
so that the unbroken symmetry group in the visible and hidden sector are 
$H^{\rm v} \times H^{\rm h}$, the non-$U_*$ part of the cohomology group of K3 surface $X$, 
\begin{equation}
[U_*^\perp \subset H_2(X; \Z)] \cong [U^{\oplus 2} \oplus E_8^{\oplus 2}] 
  = {\rm II}_{2,18}\, , 
\label{eq:U2-E82}
\end{equation}
contains $H^{\rm v} + G^{\rm v}_{\rm str} + U + U + 
G^{\rm h}_{\rm str} + H^{\rm h}$. 
The moduli spaces on both sides of the duality are identified 
by identifying the right-moving momenta (see Appendix \ref{sec:Het-Narain} 
for conventions on the description of $T^2$ compactification of heterotic string theory)
\begin{equation}
 {\cal Z}^R : = \sqrt{\frac{\alpha'}{2}} (k^R_8+ik^R_9) \in {\rm Hom}({\rm II}_{2,18}, \C)
\end{equation}
in heterotic string theory with the period integral 
$\Omega^{\rm norm.} \in 
[U_*^\perp \subset \Lambda_{\rm K3}]^* \otimes \C$ satisfying 
the normalization\footnote{For the period integral $\Omega$, we can always take 
$\Omega^{\rm norm.} = \sqrt{2/(\Omega,\overline{\Omega})} \times \Omega$.} 
\begin{equation}
 ({\cal Z}^R, \overline{{\cal Z}^R}) = 2 \leftrightarrow 
 (\Omega^{\rm norm.}, \overline{\Omega^{\rm norm.}} ) = 2 \, .
\end{equation}
Parametrically large volume ${\rm vol}(T^2)/\ell_s^2$ with 
$\ell_s^2 := (2\pi)^2\alpha'$ in heterotic string theory corresponds 
to the behaviour 
\begin{equation}
 {\cal Z}^R|_{U \oplus U}  \sim  \frac{1}{\sqrt{2}} 
   \left( \frac{\sqrt{\alpha'}}{R_8}, - \frac{R_8}{\sqrt{\alpha'}}, 
          i \frac{\sqrt{\alpha'}}{R_9}, -i \frac{R_9}{\sqrt{\alpha'}} 
   \right), \qquad 
 {\cal Z}^R|_{G^{\rm v}_{\rm str} \oplus G^{\rm h}_{\rm str}}  \sim  
    \frac{\sqrt{\alpha'}}{R}\, , 
\end{equation}
while the period integral becomes 
\begin{equation}
 \Omega^{\rm norm.}|_{U \oplus U} \sim
 \frac{1}{\sqrt{2}} \left(
   \frac{1}{\sqrt{\ln(1/\epsilon_\eta)}}, \sqrt{\ln(1/\epsilon_\eta)},  
   \frac{1}{\sqrt{\ln(1/\epsilon_\eta)}}, \sqrt{\ln(1/\epsilon_\eta)}  
  \right)\, .
\label{eq:degen-limit-U+U}
\end{equation}
for small $\epsilon_{\eta}$ in F-theory. Hence the relation is
${\rm vol}(T^2)/\ell_s^2 \sim (R_8R_9)/\alpha' 
\sim \ln (1/\epsilon_\eta)$ (see e.g. Appendix B of \cite{Hayashi-4}).\footnote{
\label{fn:degeneration-limit-B}
In physics, it is an interesting question whether or not the
$t=0$ point should be included in the string-theory moduli space. 
Along a one-dimensional subspace parametrized by 
$\epsilon_\eta$, the distance from a point with finite $\epsilon_\eta$ 
to the $\epsilon_\eta \longrightarrow 0$ limit diverges, when 
the metric from the K\"{a}hler
potential $K = -\ln [\int \Omega \wedge \overline{\Omega} ]$ is used 
(\S 5 \cite{Bershadsky:1998vn}). It is also impossible to stay 
within the period domain $D$ while setting $\epsilon_\eta = 0$ 
(see (\ref{eq:degen-limit-U+U})).
Note also, however, that $\epsilon_\eta$ can always be absorbed by 
redefinition of the coordinates $(y,x,z)$ in a special locus 
$a^{\rm h}_{0} = a^{\rm h}_2 = \cdots = 0$. In this special locus, 
the K3 surface has a bad singularity in the fibre over the $z = \infty$ point 
in the base $\P^1$. Although this singularity can be removed by a birational
transformation, the new geometry is a rational elliptic 
surface (where $c_1 \neq 0$), rather than a K3 surface. 
Physics implications of this mathematical facts should be considered 
separately. 
}

Let us now study the distribution of vacua in the miniature landscape, 
focusing on whether the vacua are close to the stable degeneration ``limit'' 
of a K3 surface, or to the large large ${\rm vol}(T^2)/\ell_s^2$ ``limit'' 
in heterotic language. For this purpose, the volume in heterotic string theory 
can be defined easily and unambiguously by using the Narain moduli (see 
below), and we use this $[{\rm vol}(T^2)/\ell_s^2]_{\rm Het}$ as the 
parameter of distribution. 

It is then more direct and convenient to deal with the F-theory data in terms 
of period integrals, rather than the defining equation(s) of K3, since we use 
the heterotic--F-theory duality. Let us follow \cite{Moore:1998pn,Gorlich:2004qm} 
and consider (for simplicity and concreteness) 
only the case in which the gauge groups in the visible and hidden sector $E_8$ 
both remain unbroken. 
For any attractive K3 surface $X=X_{[a~b~c]}$ with 
$T_X = \left[ \begin{array}{cc} 2a & b \\ b & 2c  \end{array}\right]$ there always exists such an elliptic fibration.\footnote{
\label{fn:E8E8} Indeed, 
we can always embed $T_0$ into one of the Niemeier lattices, $T_0 \hookrightarrow E_8 \subset E_8+E_8+E_8$, precisely in the same way $T_0$ is obtained, 
$T_0:= [T_X[-1]^\perp \subset E_8]$. The frame lattice then becomes simply 
$T_X[-1]+E_8+E_8$. See \cite{Nish01} or \S 4 of \cite{BKW-math} for a more detailed 
explanation. No special condition on $T_X$ needs to be satisfied 
for this fibration to exist in the ${\cal J}_2(X)$ classification. } 
$T_X \oplus T_X[-1]$ forms a sublattice of $U \oplus U$, and 
$W_{\rm frame} = E_8 \oplus E_8 \oplus T_X[-1]$.
To be more explicit, let the oriented basis of $T_X$ be
$\{q, p\}$ and $T_X[-1] = {\rm Span}_\Z \{Q, P\}$. 
Then we can embed $T_X$ and $T_X[-1]$ primitively into 
$U \oplus U = {\rm Span}_\Z \{ v, V \} 
\oplus {\rm Span}_\Z \{ v', V' \}$ as\footnote{The symmetric pairing 
is given by $(v, V) = (v', V') = 1$, zero otherwise.} 
\begin{equation}
 \left(p,q,P,Q \right) = 
 \left( v, V, v', V' \right) 
 \left( \begin{array}{cccc}
  1 &   & -1 &     \\
  a & b & a  &     \\
    & 1 &    & -1  \\
    & c & b  & c   \\
	\end{array}\right)\, .
\end{equation}
The period vector $\vev{\Omega_X} = (p + \tau_X q, \bullet)$ 
(with $\tau_X$ and $Q_X$ given by $\tau$ in (\ref{eq:TX-OmegaX}))
is written as 
\begin{equation}
 \vev{\Omega_X}|_{U \oplus U} = 
(v + (a+b\tau_X) V + \tau_X v' + c\tau_X V' , \bullet)
\Longrightarrow 
 \vev{\Omega_X^{\rm norm.}} = 
  \sqrt{\frac{2c}{Q_X}}((a+b \tau_X), 1, c \tau_X, \tau_X) 
\end{equation}
in the component description of $[U \oplus U]^* \otimes \C$.
This moduli data for F-theory is to be identified with 
\begin{equation}
 Z: \left( \begin{array}{c} n_1 \\ -w^1 \\ n_2 \\ -w^2 \end{array} \right)
  \longrightarrow 
   \frac{i}{\sqrt{2 ({\rm Im}\tau_H) ({\rm Im}\rho_H)}}
      (-\tau_H, -\rho_H, 1, -\rho_H \tau_H)
    \left( \begin{array}{c} n_1 \\ - w^1 \\ n_2 \\ - w^2 \end{array} \right)
\end{equation}
modulo ${\rm Isom}(U \oplus U)$. Here $\tau_H$ is the complex structure modulus 
of $T^2$ for heterotic string compactification and $\rho_H = (B+iJ)/\ell_s^2$ is 
the complexified K\"{a}hler modulus of $T^2$.
With the condition (\ref{eq:cond-abc}), the best identification is \cite{Moore:1998pn}
\begin{equation}
 \tau_H \leftrightarrow \tau_X, \qquad \rho_H \leftrightarrow c\tau_X\, , 
\qquad \tau_H \rho_H \leftrightarrow - (a+b\tau_X)\, .
\end{equation}
From this, we can read off that 
\begin{equation}
 \left[\frac{{\rm vol}(T^2)}{\ell_s^2} = {\rm Im}(\rho_H) \right]_{\rm
  Het} 
= \left[\frac{\sqrt{Q_X}}{2} = \frac{\sqrt{4ac - b^2}}{2} \right]_F,
   \qquad   |\rho_H| = \sqrt{ac}\, .
\end{equation}
Vacua parametrically close to the stable degeneration limit (i.e.,
parametrically large ${\rm vol}(T^2)/\ell_s^2$ for heterotic string theory) 
are realized when the attractive K3 surface is characterized by 
large $ac$ or $Q_X$. There is an upper limit on the value of $a$ and $c$, 
which comes from the D3-brane tadpole condition (\ref{eq:G1-partflux<=24}). 
For moderate choices of $a, b$ and $c$, neither $|(a+b\tau)|$ nor $|c|$ 
are large, and hence the F-theory flux vacua are not attracted towards the region 
where the heterotic dual has a good supergravity approximation. 

\begin{figure}[tbp]
 \begin{center}
\begin{tabular}{cc}
  \includegraphics[width=.4\linewidth]{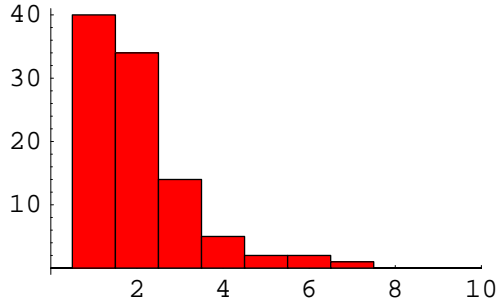} &
\end{tabular}
\caption{\label{fig:Het-vol-distr}
Distribution of $\sqrt{Q}/2 = [{\rm vol}(T^2)/\ell_s^2]_{\rm Het}$ 
in the miniature landscape 
(conditions (\ref{eq:no-G0}, \ref{eq:G1-partflux<=24})) from 
F-theory on ${\rm K3} \times {\rm K3}$. 
}
 \end{center}
\end{figure}

The distribution of $[{\rm vol}(T^2)/\ell_s^2]_{\rm Het}$, and hence 
of $\sqrt{Q_X}/2$ in F-theory, is determined and presented 
in Figure \ref{fig:Het-vol-distr}, using the 98 different
identifications of $S_1 \times S_2$ for M-theory 
in Table \ref{tab:AK-notmorethan24} with 
$X \times S = X_{[a~b~c]} \times X_{[a'~b'~c']}$ for F-theory.
It is clear that only a small fraction of vacua in the tail of the 
distribution in Figure~\ref{fig:Het-vol-distr} are in the moduli space 
of K3 surfaces $X$ in the stable degeneration ``limit'', or in the 
large ${\rm vol}(T^2)/\ell_s^2$ region of heterotic $E_8 \times E_8$ 
compactification.

This distribution can be regarded as that of an ensemble of effective
theories containing $E_8 \times E_8$ 7-brane gauge groups (which are left
unbroken), but it can also be regarded as that of theories with 
$\SO(32)$ 7-brane gauge group.\footnote{See footnote \ref{fn:E8E8}, 
and replace the lattice $L^{(\gamma)} = E_8 \oplus E_8 \oplus E_8$ 
with $L^{(\beta)} = E_8 \oplus D_{16}; \Z_2$. See \cite{BKW-math} 
for more explanations.}

For moderate choices of a, b and c, we should not expect that $\epsilon_K$ 
is small, either. In other words, all the period integrals in such a K3 
surface either vanish or are of order unity. This means that there is no 
extra singular fibre located closely to a singular fibre supporting 7-brane 
non-Abelian gauge groups, like $H^{\rm v}$ and $H^{\rm h}$ in the
visible and hidden sector.
Correspondingly, the appropriate field theory local model is to take $H^{\rm v}$ and $H^{\rm h}$ as the 
gauge group everywhere 
\footnote{When $Y$ is a non-trivial 
K3 fibration over some surface $S$, then there will usually be matter 
curves and possibly Yukawa points. That is where ``singular fibres 
collide'' and the field theory local model should be such that the rank 
of the gauge group is higher than that of $H^{\rm v}$ or $H^{\rm h}$ 
by (at least) 1 or 2, respectively \cite{Katz:1996xe, 
DW-1, Beasley:2008dc, Hayashi-2}. In the class of models studied 
in this article, however, such loci are absent 
because of the direct product structure $X \times S$.} 
on the other K3 surface $S$. 

Let us now turn our attention back to the defining 
equation (\ref{eq:Weierstrass-degen-lim}) of the K3 surface. 
Certainly it is a better and more direct approach for a systematic search for 
the Noether--Lefschetz locus of $Y$ to parametrize the moduli space 
by period integrals and deal with the lattices $T_X$, $S_X$ etc., but it 
is also nice if we can figure out what kind of defining equations such 
vacua correspond to. Precisely the same problem has been addressed by 
\cite{KuwShio, Kumar} for a specific class of K3 surfaces. 
In the following, we provide a simplified summary of their 
results so it fits to the context in this article. 

Consider an attractive K3 surface $X_{[2A~2B~2C]}$ 
(i.e., special case $a=2A$, $b=2B$ and $c = 2C$). Such a K3 surface 
corresponds to 
${\rm Km}(E_{\rho_1} \times E_{\rho_2}) = 
(E_{\rho_1} \times E_{\rho_2})/\Z_2$ with 
\begin{equation}
\rho_1 = \frac{-B + i \sqrt{4AC-B^2}}{2C}, \qquad 
\rho_2 = \frac{-B+i\sqrt{4AC-B^2}}{2} = C \rho_1\, . \label{rho12ABC}
\end{equation}
$E_{\rho_i}$ ($i=1,2$) is an elliptic curve with the complex structure 
$\rho_i$, and we let its defining equation\footnote{
From the definition given in the main text, it follows that 
\begin{equation}
 \frac{2^{11} (\lambda^2_i - \lambda_i + 1)^3}{(\lambda_i - 1)^2 \lambda^2_i} = 
 j(\rho_i) = q^{-1}_i + 744 + {\cal O}(q_i), \qquad q_i = \exp (2\pi i \rho_i)\, .
\end{equation}
Thus, for large ${\rm Im}(\rho_i)$, small $|q_i|$ and large $\lambda_i$, 
$2^{11} \lambda_i^2 \sim e^{- 2 \pi i \rho_i}$.
}  be $y^2 = x(x-1)(x-\lambda_i)$.
Kummer surfaces associated with a product type 
Abelian surface, which have $\rho_X = 18$, are known to have at least 11 
different types of elliptic fibrations \cite{Oguiso}. One of them 
(type 3 in \cite{Oguiso}) has $2{\rm IV}^*+8I_1$ type singular fibres 
(including two ``$E_6$-type'' fibres) in the Kodaira classification. 
Let us take this type of elliptic fibration as an example.
Ref. \cite{KuwShio} found that the Weierstrass model for this type of 
fibration (Inose's pencil) is given by 
\begin{equation}
Y^2 = (X + 4u^2)(X+4u^2 (\lambda_1+\lambda_2))
      (X + 4u^2 \lambda_1\lambda_2) 
    + \frac{(4u^2)^3}{4} 
      \left\{ \frac{\lambda_2(\lambda_2-1)}{u}
            + \lambda_1(\lambda_1-1) u
      \right\}^2\, , 
\label{eq:Inose-pencil}
\end{equation}
where $u$ is the inhomogeneous coordinate on the base $\P^1$, or equivalently, 
\begin{equation}
 \tilde{Y} \left\{ \tilde{Y} - (2u)^3
      \left(\frac{\lambda_2(\lambda_2-1)}{u}
    + \lambda_1(\lambda_1-1) u \right)
    \right\} = (X + 4u^2)(X+4u^2 (\lambda_1+\lambda_2))
      (X + 4u^2 \lambda_1\lambda_2)\, . 
\label{eq:Inose-pencil-2}
\end{equation}
Tate's condition for a ${\rm IV}^*$ fibre is satisfied at $u=0$ and
$u=\infty$. To turn this $\rho_X = 18$ K3 surface into 
an attractive K3 surface ($\rho_X = 20$), we only have to make a specific choice of 
$\rho_1$ and $\rho_2$ (and hence $\lambda_{1,2}$) in terms of integers $A$, $B$ and $C$
as in \eqref{rho12ABC}. The second equation above is already in the form of 
the generalized Weierstrass form (Tate form,
(\ref{eq:Weierstrass-degen-lim})), apart from a shift in $X$ that does not
play an essential role.\footnote{\label{fn:Inose-rescale}
The stable degeneration 
``limit'' corresponds to large $A$ and $C$, and hence to large 
${\rm Im}(\rho_2)$ and large $|\lambda_2|$. 
With the coordinate redefinition 
$\tilde{Y} \rightarrow \lambda_2^{9/2} \tilde{Y}'$, 
$X \rightarrow \lambda_2^3 X'$ and $u \rightarrow \lambda_2 u'$, 
one finds that $\epsilon_\eta \sim 1/\sqrt{\lambda_2}$, and hence 
$t = \epsilon_\eta^2 = 1/\lambda_2$.}
The defining equation of the spectral surface can be read off from 
the Tate form: $a^{\rm v}_0 = a^{\rm v}_2 = 0$, and $a^{\rm v}_3 y = 0$.
Thus, in the language of the supergravity approximation of heterotic string
theory, the $\rho_X = 20$ vacua of F-theory are realized due to spectral
surfaces (gauge field configurations) which are far from generic and imply an
intricate conspiracy among $f_0$, $g_0$, $a^{\rm v}_3$ and $a^{\rm h}_3$. 
We will come back to this issue in Section \ref{ssec:MW-singl-fbr-flux} and Appendix \ref{sec:Km-E6E6}.

\section{Landscapes of F-theory on  ${\rm K3} \times {\rm K3}$ with $G_0 \neq 0$ Flux}
\label{sec:G0-flux}

The study in the previous section is for a landscape with a limited choice 
of four-form flux; the condition (\ref{eq:no-G0}) has been imposed. 
Consequently, the number of vacua is limited and the level of complexity of 
the analysis remains (barely) manageable.
One can maintain full control of details, if one wishes, which enables 
us to understand various subtleties associated with multiplicity counting 
and the role played by the modular group. The $G_0=0$ assumption on the flux, 
however, has much to do with this advantage.  

Suppose that $Y= X \times S$ for F-theory compactification is 
given by a pair of attractive K3 surfaces $X$ and $S$, 
as in Section \ref{ssec:review-AK}.
The $G_0$ type flux is then in $S_X \otimes S_S \otimes \R$
and can be written in the form  
\begin{equation}
 [G_0] = \sum_{I=1}^{18} C_I \otimes F_I\, ,
\label{eq:G0-Wx-Ss-expansion}
\end{equation}
where $\{C_I\}_{I=1,\cdots, 18}$ is chosen as a basis of 
$W_{\rm root} \oplus W_{\rm {\rm U}(1)} \subset W_{\rm frame} \subset S_X$ and 
$F_I \in S_S \otimes \Q \subset H^{1,1}(S; \R) $. 
The gauge bosons corresponding to the Cartan part of the 7-brane gauge fields remain massless for 
a rank-$rk_7$ group when the $\{F_I \}$s span an $(18-rk_7)$-dimensional 
subspace of $H^{1,1}(S; \R)$. The rest of the Cartan gauge fields become 
massive through the St\"uckelberg mechanism. There will be $(18-rk_7)$ global 
${\rm U}(1)$ symmetries left in the effective theory. Non-perturbative 
corrections break these symmetries, but they may still look like 
approximate symmetries if the non-perturbative corrections break them 
only weakly. 

Therefore it is not a complete nonsense to focus on a sub-ensemble of 
vacua of F-theory on  ${\rm K3} \times {\rm K3}$ with $G_{\rm tot} = G_1+G_0$ flux of 
a given value of $(18-rk_7)$ and to study the statistics of this sub-ensemble, 
as those compactifications share a common property of the resulting effective theories. 
In this context, the statistical results of the $G_0= 0$ landscape 
in Section \ref{ssec:sample-stats} are also of some value. They are 
regarded as the statistics in the sub-ensemble characterized by 
the rank $rk_7 = 18$ massless gauge group on 7-branes and ${\cal N}=1$ 
supersymmetry. 

There is hence a physical motivation to study the statistics of the sub-landscape 
characterized by some properties of the $G_0$ flux, such as $rk_7$ 
(see Sections \ref{ssec:MW-singl-fbr-flux} and \ref{ssec:orientifold-beyond} 
for more), and also to compare the numbers of vacua that have 
$G_0$ with different properties. These are the kind of problems we address in this
section.
 
\subsection{General Remarks}
\label{ssec:remark-G0}

\subsubsection{Primitivity}

Let us begin with a brief remark on the K\"{a}hler moduli. 
For supersymmetric compactifications, $G_{\rm tot} = G_1 + G_0$ not only 
has to be purely of Hodge type $(2, 2)$,  but also primitive: 
\begin{equation}
 J_Y \wedge G_{\rm tot} = J_Y \wedge G_0 = 0\, .
\label{eq:primitive-cond}
\end{equation}
The K\"{a}hler form $J_Y$ on $Y = X \times S$ has the shape 
\begin{equation}
 J_Y = t_X [F_X] + J_S\, ,
\label{eq:Kahler-4-F-theory}
\end{equation}
where $t_X \in \R_{0<}$, $[F_X]$ is the fibre class (elliptic divisor)
associated with $\pi_X: X \rightarrow \P^1$, and $J_S$ a K\"{a}hler form 
on $S$ in the positive cone of $S_S$ (i.e., $J_S \in S_S \otimes \R$,
and $J_S^2  > 0$). 
%
The primitivity condition (\ref{eq:primitive-cond}) implies that all of
the $F_I$ ($I = 1,\cdots, 18$) are orthogonal to $J_S$ in the inner product 
in $S_S \otimes \R \cong H^{1,1}(S; \R)$. For a given K\"{a}hler form 
$J_S$, the $G_0$ type flux in 
$W_{{\rm frame}*} \otimes [J_S^\perp \subset (S_S \otimes \R)]$ always gives 
rise to a positive contribution---$1/2 [G_0] \cdot [G_0]$---to the 
D3-tadpole, since both $W_{\rm frame}$ and $[J_S^\perp \subset S_S \otimes \R]$ 
are negative definite.\footnote{It is not impossible to think of a case with 
some $F_I$ not orthogonal to $J_S^\perp$. 
Such an $F_I$ associated with $C_I$ in $W_{\rm root}$  
corresponds to non-anti-self-dual flux on a 7-brane wrapped on 
a K3 surface $S$, and is known to lead to a de-Sitter vacuum (if K\"{a}hler 
moduli is stabilized properly). 
If this negative contribution to the D3-brane tadpole---virtually the 
presence of an anti-D3-brane---coexists with a positive $N_{D3}$, 
we expect D3--D7 hybrid inflation to occur \cite{Dasgupta:2002ew}.} 

Because of the primitivity condition of the four-form flux, the complex 
structure moduli and K\"{a}hler moduli talk to each other. There are 
two equivalent ways to see how they ``talk''. One way is, 
to think of the K\"{a}hler moduli $(t_X, J_S)$ as being given 
first, after which the $F_I$ describing $G_0$ are forced to be in the subspace 
$[J_S^\perp \subset S_S \otimes \Q]$, as above. If the ratio of components of
$J_S$ are in $\Q$---this situation is denoted by $[J_S] \in \Q\P[S_S]$ in this article---then 
$[J_S^\perp \subset S_S \otimes \Q]$ is of dimension [20-1=19]. If the ratio is not just 
in $\Q$, but in an extension field of $\Q$, then the dimension of the space for $G_0$ is smaller 
than 19. In the rest of this article, we assume that $J_S \in \Q\P[S_S]$. 
The other way is to think of the $F_I$ as being given first. In this case, $J_S$ 
is forced to be in the subspace of $S_S \otimes \R$ orthogonal to all of the $F_I$ 
due to the D-term potential (e.g., \cite{Kachru:2002he, 
Tripathy:2002qw}). When the $F_I$ span a $(18-rk_7)$-dimensional subspace of 
$S_S \otimes \Q$, the remaining moduli space of $J_S$ is of 
$(2+rk_7)$-dimensions (in $\R$).\footnote{
$t_X \in \R_{0<}$ also remains unconstrained under the primitivity condition.
}\raisebox{5pt}{,}\footnote{\label{fn:massless-g33}
As long as we keep the flux in the form of $G_0 = \sum_I C_I \otimes F_I$, 
a perturbation $\delta J = t_b [\sigma_X]$ of the K\"{a}hler form 
does not violate the primitivity condition. Thus, there is no mass given 
to this degree of freedom either. This mode, however, corresponds to 
a part of the metric, $g_{33}$, of the effective field theory on $\R^{3,1}$.
Phenomenologically, we are interested in cases where the vacuum value of 
$t_b$ is zero (the small elliptic fibre volume limit of M-theory), yet 
the fluctuation in that direction---$g_{33}$ remains massless. Thus, there 
is no problem at all that the fluctuation $t_b$ does not have a mass.}
Eventually one has to think of stabilization of both complex structure
moduli and K\"{a}hler moduli, so that these two perspectives are
equivalent. Common to these two is the idea that the stabilization of the
two groups of moduli can be dealt with separately, which is true as long as there is a 
separation of scales between those two stabilization mechanisms, such as in the 
KKLT scenario \cite{Kachru:2003aw}.

In this article, we focus on aspects of complex structure moduli
stabilization in flux compactifications, using  ${\rm K3} \times {\rm K3}$ as an
example. This article is not committed to a particular mechanism of 
K\"{a}hler moduli stabilization available on  ${\rm K3} \times {\rm K3}$, except that 
we implicitly assume this separation of scales. It is thus impossible 
to determine the full landscape distribution of completely discrete vacua 
in the product of complex structure and K\"{a}hler moduli spaces 
${\cal M}_{\rm cpx} \times {\cal M}_{\rm Kahler}$ for  ${\rm K3} \times {\rm K3}$, or 
projections of these distributions to ${\cal M}_{\rm cpx}$, without 
introducing extra assumptions (on K\"{a}hler moduli stabilization).  
Our statements in the rest of this article are often presented in the 
form of distributions of flux vacua on ${\cal M}_{\rm cpx}$ for a fixed 
choice of K\"{a}hler moduli.

\subsubsection{Integrality}

Let us now switch the subject, and discuss the ``integrality condition'' on 
$[G_1]$ and $[G_0]$. When both $[G_1]$ and $[G_0]$ are non-zero, each 
one of them does not have to be integral by itself; only the total flux 
$[G_{\rm tot}] = [G_1] + [G_0]$ needs to be integral, i.e. an element of 
$H^4(Y; \Z)$. Once we find that $[G_1]$ does not have to be 
integral---an element of $T_X \otimes T_S$---on its own, 
the condition (\ref{eq:2dimR-in-4dimR}) is too restrictive. 
Contributions to the D3-brane tadpole, $[G_1] \cdot[G_1]/2$ and 
$[G_0] \cdot [G_0]/2$, are not necessarily integers separately either, 
though their sum is always integral, when $G_{\rm tot}$ is integral.
This means that the search for pairs of attractive K3 surface in 
Sections \ref{ssec:review-AK} and \ref{ssec:extended-list} needs to be 
carried out once again for cases with $G_0 \neq 0$. How large 
is the impact of this generalization ?

In cases where both $X$ and $S$ are attractive K3 surfaces, we can decompose 
the $22 \times 22$-dimensional vector space 
$H^2(X,\mathbb{R})\times H^2(S,\mathbb{R})$ into
\begin{eqnarray}
[(T_{X} \otimes T_{S}) \otimes \R]\,  \oplus\,  [(T_{X} \otimes S_{S}) \otimes \R]\, \oplus\, [(S_{X} \otimes T_{S}) \otimes \R]\,  \oplus \, [(S_{X} \otimes S_{S}) \otimes \R].
\end{eqnarray}
%
%
Such a decomposition is not necessarily possible with integer coefficients,
not all of the elements of $H^2(X; \Z)$ (resp. $H^2(S; \Z)$) can be written 
in the form of a sum of integral elements in $T_X$ and $S_X$ (resp. 
$T_S$ and $S_S$), and not all the elements in 
$H^2(X,\mathbb{Z})\times H^2(S,\mathbb{Z})$ can be written as a sum of 
integral elements in 
\begin{eqnarray}
[(T_{X} \otimes T_{S})]\,  \oplus\,  [(T_{X} \otimes S_{S}) ]\, \oplus\, [(S_{X} \otimes T_{S}) ]\,  \oplus \, [(S_{X} \otimes S_{S})].
\end{eqnarray}
%
The integrality of the total flux $[G]=[G_1] + [G_0]$, however, 
implies that for all the generators of $T_X \otimes T_S$ and 
$S_X \otimes S_S$, which are all integral four-cycles, the flux quanta 
evaluated on these cycles are integers. 
Hence $[G_1]$ and $[G_0]$ are contained within 
$(T_{X} \otimes T_{S})^* =(T_{X}^* \otimes T_{S}^*) \subset 
 (T_{X} \otimes T_{S})\otimes \Q$ and 
$(S_{X} \otimes S_{S})^* =(S_{X}^* \otimes S_{S}^*) \subset 
 (S_{X} \otimes S_{S})\otimes \Q$, respectively.\footnote{ 
$T_X \otimes T_S$, $T_X \otimes S_S$, $S_X \otimes T_S$ and $S_X \otimes S_S$ 
form sublattices of the lattice $H^2(X; \Z) \otimes H^2(S; \Z)$}
In particular, it follows that $[G_1]$ should be within $T_{X}^* \otimes T_{S}^*$, 
but may not necessarily be within $T_{X} \otimes T_{S}$. 
 
In fact, there is a stronger necessary condition than this; 
although the projection of $H^2(S; \Z) \subset T_S^* \oplus S_S^*$ to $T_S^*$
for an attractive K3 surface is always surjective, the projection image of 
\begin{equation}
(H^2(S_1; \Z) \otimes H^2(S_2; \Z)) \cap 
[H^{4,0}(Y;\C) \oplus H^{2,2}(Y;\R) \oplus H^{0,4}(Y;\C)]
\label{eq:integral-4cycle}
\end{equation}
to $(T_{S_1} \otimes T_{S_2})^*$, for $Y = S_1 \times S_2$ with 
a pair of attractive K3 surfaces $S_1$ and $S_2$, is not.
In order to state a condition on the image of this projection, note that, 
for a pair of attractive K3 surfaces $S_1$ and $S_2$, the transcendental 
lattices, Neron--Severi lattices and their dual lattices have the following 
properties: as Abelian groups, 
\begin{equation}
 T_{S_i}^*/T_{S_i} \cong \Z_{m_i} \times \Z_{n_i}, \qquad 
 S_{S_i}^*/S_{S_i} \cong \Z_{m_i} \times \Z_{n_i}, \qquad 
  {}^\exists \gamma_i: T_{S_i}^*/T_{S_i} \cong S_{S_i}^*/S_{S_i} \qquad 
  (i=1,2)
\end{equation}
for some positive integers $m_i, n_i$. It then follows that 
\begin{equation}
  (T_{S_1} \otimes T_{S_2})^*/(T_{S_1} \otimes T_{S_2}) \cong 
   \Z_{m_1m_2} \times \Z_{m_1n_2} \times \Z_{n_1m_1} \times \Z_{n_1n_2}\, .
\end{equation}
One can prove\footnote{We are not presenting the proof here because it is 
just technical, and is not particularly illuminating. After all, this is not 
a sufficient condition. It is not guaranteed that an appropriate 
$G_0 \in W_{\rm frame} \otimes [J_S^\perp \subset S_S] \otimes \Q$ exists and 
$G_{1} + G_0 =G_{\rm tot}$ becomes integral, even when $G_1$ satisfies this 
criterion.} that the image of the integral four-cycles 
in (\ref{eq:integral-4cycle}) has to be within the subgroup of 
$(T_{S_1} \otimes T_{S_2})^*$ characterized by 
\begin{equation}
{\rm LCD}(m_1,m_2)\Z_{m_1m_2} \times \cdots \times 
{\rm LCD}(n_1,n_2)\Z_{n_1n_2} \subset 
 \Z_{m_1m_2} \times \cdots \times \Z_{n_1n_2}\, .
\label{eq:subgroup-LCD}
\end{equation}

Because of this necessary condition on $[G_1]$, one can see that 
$[G_1]$ still has to be an integral element of $T_X \otimes T_S$ if 
all the four pairs $(m_1, m_2)$, $(m_1,n_2)$, $(n_1,m_2)$ and $(n_1,n_2)$ 
are mutually coprime. In this case, all the possible forms of the $[G_1]$ component 
on a pair of attractive K3 surfaces $X \times S = S_1 \times S_2$ 
remain the same as in Sections \ref{ssec:review-AK} and 
\ref{ssec:extended-list}. 
When the four pairs of integers are not coprime, however, there are more 
chances for the $[G_1]$ component available in the (sublattice of) the
$(T_X \otimes T_S)^*$ lattice than in the $T_X \otimes T_S$ lattice. 
That makes it easier to pass the D3-tadpole 
constraint (\ref{eq:G1-partflux<=24}), even for a pair of attractive K3 
surfaces with relatively large values of $a, c$ and $d, f$.
Section \ref{ssec:rank-16} is devoted to an explicit enumerative study
in order to see more of the consequences of this possibility of non-integral choice 
of $G_1$. 

\subsubsection{$\vev{W_{\rm GVW}} = 0$}

When we write the $[G_1]$ component as $G_1 = {\rm Re}
[\gamma \Omega_{S_1} \wedge \overline{\Omega}_{S_2}]$, as in \cite{AK} 
or in (\ref{eq:G1-gamma-def}), it manifestly has only a 
$(2,2)$ Hodge component, without $(4,0)$ or $(0,4)$ component. 
However, we may also ask what is the statistical cost of requiring 
the absence of the $(4,0)+(0,4)$ components, or equivalently, a vanishingly 
small cosmological constant. For this purpose, the $G_0$ component is 
irrelevant and $[G_1] \in (T_{S_1} \otimes T_{S_2}) \otimes \Q$, which we can write as:
\begin{equation}
 [G_1] = k_1 (p_1 \otimes p_2) + k_2 (p_1 \otimes q_2) + k_3 (q_1 \otimes p_2)
  + k_4 (q_1 \otimes q_2)\, , 
\end{equation}
where $k_1, k_2, k_3, k_4 \in \Q$ are not necessarily integers.
The $(4,0)$ and $(0,4)$ Hodge components of $[G_1]$ are
\begin{equation} 
\frac{\bar{\tau}_1 \bar{\tau}_2 k_1 - \bar{\tau}_1 k_2 - \bar{\tau}_2 k_3 + k_4}
{(\bar{\tau}_1 - \tau_1)(\bar{\tau}_2 - \tau_2)} \; 
 \Omega_{S_1} \wedge \Omega_{S_2}
 + 
\frac{\tau_1 \tau_2 k_1 - \tau_1 k_2 - \tau_2 k_3 + k_4}
{(\bar{\tau}_1 - \tau_1)(\bar{\tau}_2 - \tau_2)} \; 
\overline{\Omega}_{S_1} \wedge \overline{\Omega}_{S_2}\, .
\label{eq:40-04-components}
\end{equation}
Thus, the absence of the $(4,0)+(0,4)$ component is equivalent to the condition 
that either $(\tau_1\tau_2)$, $\tau_1$, $\tau_2$ and $1$ are not linear 
independent over the field $\Q$, or all of $k_{1,2,3,4}$ vanish 
(and $G_1 = 0$). 
For an arbitrary pair of attractive K3 surfaces $S_1$ and $S_2$, the period 
integrals take their values in $\Q[\tau_1, \tau_2]$, which is a degree 
${\cal D}=4$ algebraic extension field over $\Q$. The condition for this 
pair $S_1$ and $S_2$ to admit a flux with vanishing cosmological 
constant\footnote{A closely resembling phenomenon is found in \S 4.2.3 
of \cite{DeWolfe:2004ns}. The extension degree changes from ${\cal D}=4$ 
to ${\cal D}=2$ for $W=0$ vacua.} (while $G_1 \neq 0$), however, has 
turned out to be that $\tau_2$ is already contained in $\Q[\tau_1]$, and 
$\Q[\tau_1, \tau_2]$ is a degree ${\cal D} = 2$ extension field of $\Q$. 
The condition (\ref{eq:square-integer}) is for $\vev{W} = 0$, 
for a general non-vanishing $[G_1]$ in $(T_X \otimes T_S) \otimes \Q$, 
rather than just for an integral $[G_1] \in (T_X \otimes T_S)$.

\subsection{A Landscape of Vacua with a Rank-16 7-brane Gauge Group}
\label{ssec:rank-16}

Just like the $G_0 = 0$ landscape in Section \ref{ssec:sample-stats},
which is interpreted as the ensemble of vacua with rank-18 massless
gauge fields on 7-branes, let us also consider the ensemble of
F-theory compactifications on  ${\rm K3} \times {\rm K3}$ with rank-16 massless gauge fields 
on 7-branes. This is realized by allowing a $G_0 \neq 0$ with only two
terms non-zero in the expansion (\ref{eq:G0-Wx-Ss-expansion}).
With this study, we hope to elucidate various aspects of the landscape 
involving $G_0 \neq 0$. In the context of F-theory, there is not particular 
importance to the specific choice of having just two terms 
in (\ref{eq:G0-Wx-Ss-expansion}), or equivalently $rk_7 = 16$. However, 
this choice makes it easier to link the present discussion to earlier results 
on type IIB orientifold landscapes, as discussed in Section \ref{ssec:orientifold-beyond}.

Suppose that $G_0 \neq 0$ flux is introduced by using some 
$C_{I=1}$ and $C_{I=2}$ in (\ref{eq:G0-Wx-Ss-expansion}), where 
$C_{I=1,2}$ are elements of the frame lattice $W_{\rm frame}(X_{[a~b~c]})$ of 
some attractive K3 surface $X_{[a~b~c]}$. Then the gauge fields along the 
direction of $C_{I=1,2}$ become massive, and only the gauge fields
within $[\{C_{I=1,2}\}^\perp \subset W_{\rm frame}]$ remain massless. 
Since it is motivated from the perspective of physics to collect vacua
with a given unbroken gauge group, it will be
interesting to consider, for a given rank-$16 = rk_7$ lattice 
$W_{\rm unbroken}$, the vacuum ensemble associated with 
\begin{eqnarray}
&& 
\left\{\left(W_{\rm frame}, \{C_{I=1,2} \} \right) \; | \; 
   W_{\rm frame} \in 
  \amalg_{
     X_{[a~b~c]} \in {\rm Isom}(\Lambda_{\rm K3}) \backslash D_{\rho=20}}
  {\cal J}_2(X_{[a~b~c]}),   \right.  \\
&& \qquad \qquad \qquad \qquad \qquad \qquad \left. 
\; C_{I=1,2} \in W_{\rm frame}, \;
 [\{C_{1,2}\}^\perp \subset W_{\rm frame}] = W_{\rm unbroken}
\right\}\, . 
 \nonumber 
\end{eqnarray} 
For this reason, it also makes sense to use the notation $W_{\rm noscan}$
instead of $W_{\rm unbroken}$. These two are used interchangeably in 
Section \ref{ssec:rank-16}, but they are distinguished in 
Section \ref{ssec:orientifold-beyond}.

If we are to take the rank-16 lattice to be e.g.
$W_{\rm unbroken} = E_8 \oplus E_8$ or $D_{16}; \Z_2\vev{\bf sp}$, 
then for all the attractive K3 surfaces $X$, there is an embedding\footnote{
For an embedding of $T_X$ into $U \oplus U$, 
$[T_X^\perp \subset U \oplus U]$ is not necessarily the same as $T_X[-1]$.
For two even rank-2 lattices $T_1$ (positive definite) and $T_2$ 
(negative definite) with an isometric discriminant group, $U \oplus U$ 
can be constructed as an overlattice of $T_1 \oplus T_2$ 
\cite{HLOY}. } of the transcendental lattice $T_X$, 
the hyperbolic plane lattice $U_{*} \cong {\rm Span}_\Z\{[F_X], \sigma\}$ 
associated with elliptic fibration, and $W_{\rm unbroken}$ (which is 
an even unimodular lattice of signature $(0,16)$) into the 
cohomology lattice $\Lambda_{\rm K3}^{(X)}$:
%
\begin{equation}
 U_* \oplus \left(T_X \oplus T_X[-1] \right) \oplus W_{\rm unbroken} 
    \hookrightarrow 
 U_* \oplus \left( U \oplus U \right) \oplus W_{\rm unbroken} 
 \cong \Lambda_{\rm K3}^{(X)}\, .
\label{eq:TxTx[-]-into-UU}
\end{equation}
The Neron--Severi lattice is 
$S_X \cong U_* \oplus T_X[-1] \oplus W_{\rm unbroken}$, and the frame lattice 
is $W_{\rm frame} \cong T_X[-1] \oplus W_{\rm unbroken}$. Hence we choose the generators of the 
rank-2 lattice $T_X[-1] = [W_{\rm unbroken}^\perp \subset W_{\rm frame}]$ as 
$\{C_{I=1,2} \}$ \cite{Gorlich:2004qm,Lust:2005bd}. 

As another example of $W_{\rm unbroken}$, one can also think of 
$(D_4 \oplus D_4 \oplus D_4 \oplus D_4); (\Z_2 \times \Z_2)$, 
in which case a type IIB orientifold interpretation is given.
It is known that there is an embedding\footnote{For Kummer surfaces 
associated with product type Abelian surfaces ($\rho_X=18$), 
$X = {\rm Km}(E_\rho \times E_{\rho'}) = (E_{\rho} \times E_{\rho'})/\Z_2$, 
the Neron--Severi lattice is 
$S_X^{\rho=18} \cong U_* \oplus (D_4^{\oplus 4}); (\Z_2 \times \Z_2)$ and 
the transcendental lattice is $T_X^{\rho=18} \cong U[2] \oplus U[2]$.}
\begin{equation}
 \left(U[2] \oplus U[2]\right) \oplus (U_* \oplus W_{\rm unbroken}) 
\hookrightarrow \Lambda_{\rm K3}^{(X)}\, .
\end{equation}
Furthermore, it is known that for attractive K3 surfaces $X_{[a~b~c]}$ with 
$a,b,c$ even, there is an embedding 
$T_X \oplus T_X[-1] \hookrightarrow U[2] \oplus U[2]$ (\cite{PSS, SI, Barthetal, AK}).
Thus, $W_{\rm frame} \cong T_X[-1] \oplus W_{\rm unbroken}$, and the generators of 
the rank-2 lattice $T_X[-1] = [W_{\rm unbroken}^\perp \subset W_{\rm frame}]$ 
are chosen as $\{C_{I=1,2}\}$s. \cite{Gorlich:2004qm,Lust:2005bd}
Hence all the attractive K3 surfaces with even $a,b,c$ contribute
to the ensemble of vacua characterized by $W_{\rm unbroken} \cong (D_4^{\oplus 4}); (\Z_2 \times \Z_2)$.

Let $T_S$ be the transcendental lattice of the attractive K3 surface $S$.
Modulo ${\rm Isom}(\Lambda_{\rm K3}^{(S)})$, it is always possible to embed 
$T_S$ into $\Lambda_{\rm K3}^{(S)}$ as 
\begin{equation}
 T_S \hookrightarrow (T_S \oplus T_S[-1]) \hookrightarrow 
   (U \oplus U) \oplus U \oplus E_8 \oplus E_8 \cong \Lambda_{\rm K3}^{(S)}\, .
\label{eq:TsTs[-]-into-UU}
\end{equation} 
This is not only always possible, but also unique in that any embedding 
to $\Lambda_{\rm K3}^{(S)} \cong {\rm II}_{3,19}$ (not to ${\rm II}_{2,18}$) 
can be brought in the form above modulo ${\rm Isom}(\Lambda_{\rm K3}^{(S)})$ 
\cite{Mor}. Thus, $S_S \cong T_S[-1] \oplus U \oplus E_8 \oplus E_8$.
Therefore, the flux $G_{\rm tot} = G_1 + G_0$ is introduced to the 
space\footnote{The $T_X[-1]$ part is meant to be $[W_{\rm unbroken}^\perp \subset 
W_{\rm frame}]$, which is the same as $T_X[-1]$ for all the three choices of 
$W_{\rm unbroken}$ mentioned in the text.} 
\begin{equation}
 \left[ (T_X \otimes T_S)^* \oplus 
 (T_X[-1] \otimes (T_S[-1] \oplus U \oplus E_8 \oplus E_8))^* \right] \cap 
   H^2(X; \Z) \otimes H^2(S; \Z)\, .
\label{eq:tot-flux-options}
\end{equation}

Now, let us take the case $W_{\rm unbroken} = E_8 \oplus E_8$ as an 
example, and work out the details.
In order to avoid inessential complexity, we restrict our attention 
to $Y=X \times S$ of the form $X_{[a~b~c]} \times X_{[a'~b'~c']}$ with 
$b=b'=0$. Embeddings of (\ref{eq:TxTx[-]-into-UU},
\ref{eq:TsTs[-]-into-UU}) are given by\footnote{This embedding can be 
used also for the case $W_{\rm unbroken} \cong D_{16}; \Z_2$, another 
even unimodular lattice of signature $(0, 16)$. For the case 
$W_{\rm unbroken} \cong (D_4^{\oplus 4}); (\Z_2 \times \Z_2)$, the
embedding of $T_X = \diag[4A, 4C]$ and $T_X[-1]$ to $U[2] \oplus U[2]$
is given by the same expression, except that $a$ and $c$ are replaced by
$A$ and $C$.} 
\begin{eqnarray}
 (p_1, P_1, q_1, Q_1) & = & (v_1, V_1, v'_1, V'_1)
 \left(\begin{array}{cccc}
 1 & -1 & & \\
 a &  a & & \\
   & & 1 & -1 \\
   & & c & c 
   \end{array}\right)\, , \\ 
 (p_2, P_2, q_2, Q_2) & = & (v_2, V_2, v'_2, V'_2)
 \left(\begin{array}{cccc}
 1 & -1 & & \\
 a' &  a' & & \\
   & & 1 & -1 \\
   & & c' & c' 
   \end{array}\right)\, , 
\end{eqnarray}
where $\{p_1, q_1\}$ (resp. $\{p_2, q_2\}$) are generators of 
$T_X$ (resp. $T_S$), and $\{P_1, Q_1\}$ (resp. $\{P_2, Q_2\}$) are generators of 
$T_X[-1]$ (resp. $T_S[-1]$). The $(U \oplus U)$ sublattice of
$\Lambda_{\rm K3}^{(X)}$ is generated by $\{v_1, V_1\}$ and $\{v'_1,
V'_1\}$, and that of $\Lambda_{\rm K3}^{(S)}$ by $\{v_2, V_2\}$ and 
$\{v'_2, V'_2\}$. The intersection forms on the $U$s are given by 
$(v,v) = (V, V) = 0$ and $(v, V) = (V,v) = 1$.

It is now straightforward to enumerate the flux 
$G_{\rm tot} = G_1 + G_0$ in (\ref{eq:tot-flux-options}), using the
basis give above:
\begin{eqnarray}
 G_1 & = &
 \frac{n_{11}{\rm GCD}(a,a')+m_{11}}{2{\rm GCD}(a,a')}(p_1 \otimes p_2)+
 \frac{n_{22}{\rm GCD}(c,c')+m_{22}}{2{\rm GCD}(c,c')}(q_1 \otimes q_2)
   \nonumber  \\
& & \quad +  
 \frac{n_{12}{\rm GCD}(a,c')+m_{12}}{2{\rm GCD}(a,c')}(p_1 \otimes q_2)+
 \frac{n_{21}{\rm GCD}(c,a')+m_{21}}{2{\rm GCD}(c,a')}(q_1 \otimes p_2), 
\label{eq:G1-nm-parametrized}
\\
 G_0 & = &
 \frac{n_{11}{\rm GCD}(a,a')-m_{11}}{2{\rm GCD}(a,a')}(P_1 \otimes P_2)+
 \frac{n_{22}{\rm GCD}(c,c')-m_{22}}{2{\rm GCD}(c,c')}(Q_1 \otimes Q_2)
   + (P_1 \otimes F_P) + (Q_1 \otimes F_Q)    \nonumber  \\
& & \quad + 
 \frac{n_{12}{\rm GCD}(a,c')-m_{12}}{2{\rm GCD}(a,c')}(P_1 \otimes Q_2)+
 \frac{n_{21}{\rm GCD}(c,a')-m_{21}}{2{\rm GCD}(c,a')}(Q_1 \otimes P_2), 
\label{eq:G0-nm-parametrized} 
\\
G_{\rm tot} & = & n_{11} (v_1 \otimes v_2 + aa' V_1 \otimes V_2) + 
  m_{11} \frac{a' v_1 \otimes V_2 + a V_1 \otimes v_2}{{\rm GCD}(a,a')} +
  \cdots + (P_1 \otimes F_P) + (Q_1 \otimes F_Q)\, .
\end{eqnarray}
Here, $n_{11}, m_{11}, \cdots, n_{22}, m_{22}$ are integers and 
$F_P, F_Q \in U \oplus E_8 \oplus E_8 \subset \Lambda_{\rm K3}^{(S)}$.
The denominators ${\rm GCD}(2a, 2a')$, ${\rm GCD}(2c, 2c')$ etc. in 
(\ref{eq:G1-nm-parametrized}, \ref{eq:G0-nm-parametrized}) correspond 
to our earlier discussion around (\ref{eq:subgroup-LCD}).
$G_1$ and $G_0$ are not necessarily integral separately, but 
$G_{\rm tot}$ is. 
At this moment, it is not guaranteed that the $(4,0)+(0,4)$ Hodge
components vanish. After imposing the $\vev{W}=0$ condition, one is left 
with the following cases
\begin{itemize}
 \item [i)] ${\cal D} := \dim_\Q \Q[\tau_1, \tau_2] = 2$, i.e., 
the condition (\ref{eq:square-integer}) is satisfied when $G_1 \neq 0$
and there are $(8-{\cal D})=6$ independent integers out of $\{n_{11}, \cdots,
m_{22}\}$, in addition to $F_P, F_Q$. In this case there are ${\cal D} = 2$ 
less scanning integers just like in Sections \ref{ssec:review-AK}, \ref{ssec:extended-list} and 
\cite{AK}.
\item [ii)]  ${\cal D} = 4$, when $G_1 = 0$, and
there are $(8-{\cal D})=4$ independent integers as well as $F_P, F_Q$ (the ``second branch''
in \S 5 of \cite{Tripathy:2002qw}). 
\end{itemize}
The latter ii) cases leave 16 moduli of $X$ and 18 of $S$ unstabilized, and we 
restrict our attention only to the cases i) ,as \cite{AK} did, in the following.

Note, however, that the K\"{a}hler form and the primitivity condition on
the flux has been ignored completely in the argument above. The
K\"{a}hler form $J_S$ on $S$ needs to be introduced in the positive 
definite part of $[T_S^\perp \subset \Lambda_{\rm K3}] \otimes \R$, and 
the flux $G_0$ component has to be orthogonal to this $J_S$. We provide
the following presentation for a fixed choice of $J_S$ (rather than
scanning over all possible $J_S$), and in particular, we choose 
\begin{equation}
 J_S = t_S (v''_2 + V''_2)\, ,  
\label{eq:assume-Js}
\end{equation}
where $v''_2$ and $V''_2$ are the generators of the third copy of the
hyperbolic plane lattice $U$ in $\Lambda_{\rm K3}^{(S)}$ 
in (\ref{eq:TsTs[-]-into-UU}), so that the computation becomes as easy
as possible.  This means that the primitivity condition does not
impose a constraint on the integers $\{n_{11}, \cdots, m_{22}\}$, and 
$F_P, F_Q$ are chosen from the negative definite lattice 
$(-2) \oplus E_8 \oplus E_8$. Here, $(-2)$ is the lattice generated by 
$(v''_2-V''_2)$, where $(v''_2-V''_2)^2 = -2$.

Assuming that $W_{\rm unbroken} = E_8 \oplus E_8$ (or $D_{16};
\Z_2\vev{\bf sp}$), we scan all possible pairs of attractive K3 surfaces 
of the form $X_{[a~0~c]} \times X_{[a'~0~c']}$ and list up all the
possible fluxes (\ref{eq:G1-nm-parametrized},
\ref{eq:G0-nm-parametrized}) satisfying (i) the condition that $\vev{W}=0$, 
(ii) the primitivity condition with respect to the K\"{a}hler form
(\ref{eq:assume-Js}) and (iii) the D3-tadpole condition 
\begin{equation}
 \frac{1}{2} G_{\rm tot} \cdot G_{\rm tot} \leq \frac{\chi(X \times
  S)}{24} = 24\, ;
\label{eq:total-tadpole-<=24}
\end{equation}
the remaining D3-brane charge is supplied by placing an appropriate number of D3-branes.

\begin{figure}[tbp]
 \begin{center}
    \includegraphics[width=.4\linewidth]{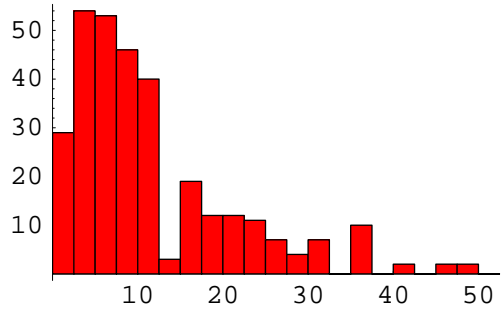} 
 \caption{\label{fig:a-distr} Variety in the choice of pairs 
of attractive K3 surfaces, $X \times S = X_{[a~0~c]} \times X_{[a'~0~c']}$ 
is shown in the form of histograms of $a \leq 50$. This distribution is 
not weighted by the number of flux choices available.
}
 \end{center}
\end{figure}

Scanning within the range of $0<c\leq a \leq 50$ and $0<c' \leq a' \leq
50$ for $Y = X \times S = X_{[a~0~c]} \times X_{[a'~0~c']}$, we found 
that there are 313 different choices of $X \times S$ admitting the 
flux satisfying all the three conditions above. The distribution of the 
value of $a$ in these 313 choices are shown in
Figure~\ref{fig:a-distr}. If both $G_1$ and $G_0$ were required to be
integral, there would only be 28 different
choices,\footnote{There are 8 pairs of $[a~0~c]$ and $[d~0~f]$ where 
$a=d$ and $c=f$ in Table \ref{tab:AK-notmorethan24}, and there are 10
pairs where either $a \neq d$ or $c \neq f$. Thus, there are
$28=8+2\times 10$ different choices of ($[a~0~c]$, $[a'~0~c']$).} 
and the largest possible value of $a$ would be 6.

\begin{figure}[tbp]
 \begin{center}
\begin{tabular}{ccc}
     \includegraphics[width=.3\linewidth]{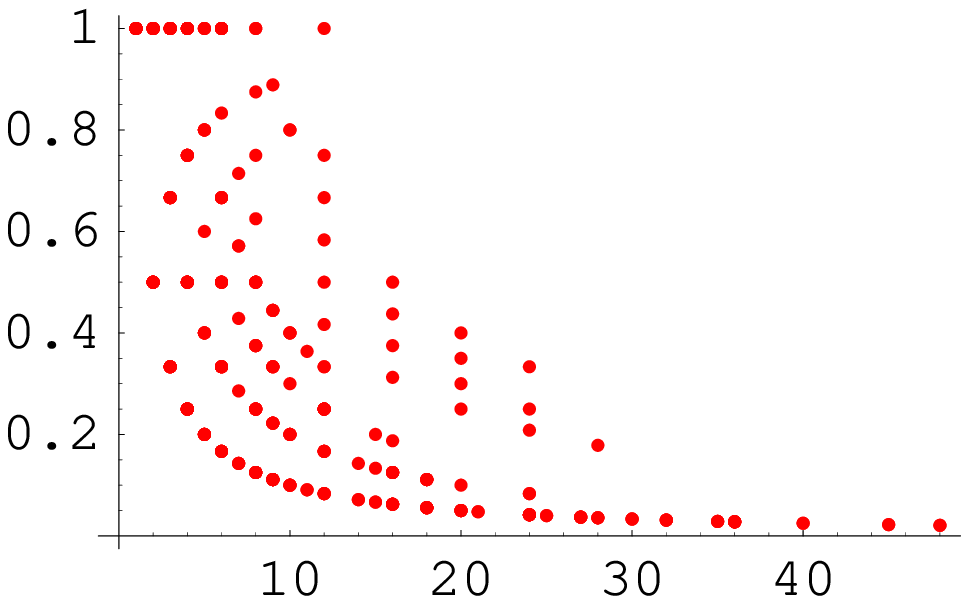} & 
     \includegraphics[width=.3\linewidth]{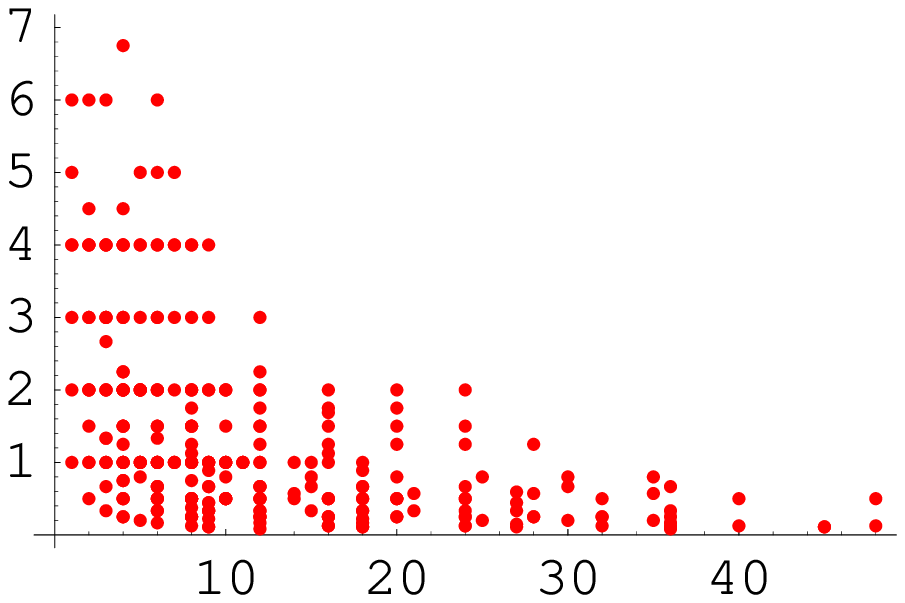} & 
     \includegraphics[width=.3\linewidth]{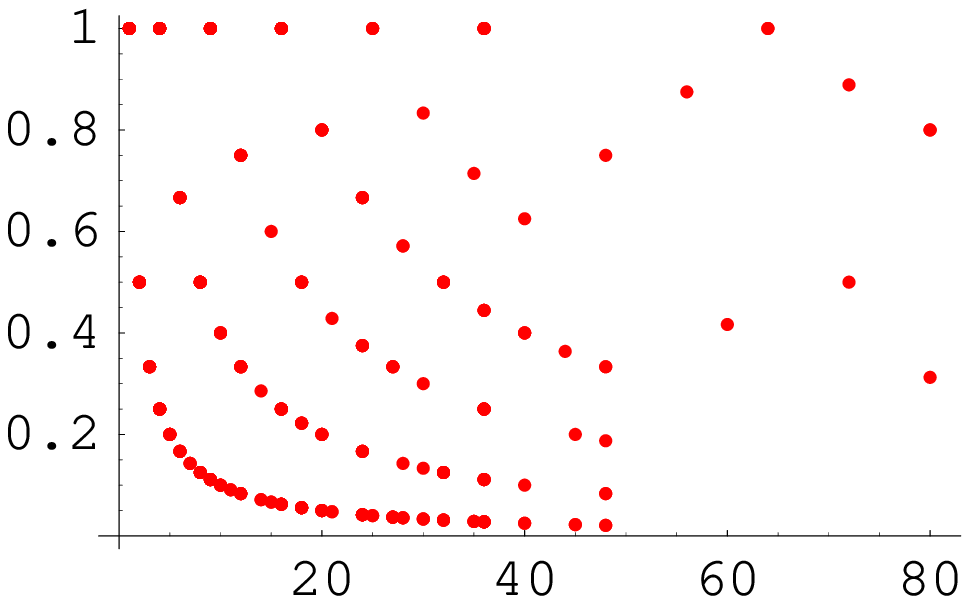}   \\  
 (i) $a$ vs $c/a$ & (ii) $a$ vs $a'/a$ & (iii) $ac$ vs $c/a$
\end{tabular}
 \caption{\label{fig:vsA} Scatter plots showing correlation between
  various modulus parameters $X$ (horizontal axis) vs $Y$ (vertical
  axis) for the 313 pairs of
attractive K3 surfaces $X \times S = X_{[a~0~c]} \times X_{[a'~0~c']}$.
No weight proportional to the number of flux choices is included. 
(the blank region in the lower right corner of (iii) is an artefact of 
cutting the scan at $a, a'\leq 50$)}
 \end{center}
\end{figure}

Figure \ref{fig:vsA} shows the correlation among moduli parameters 
for the 313 pairs of attractive K3 surfaces admitting a flux satisfying 
all the three conditions above. For the first two scatter plots (i) and (ii) 
of Figure~\ref{fig:vsA}, we can see clear correlations. When $a$ is
very large, there is presumably not much freedom to choose $a'$ other
than setting it to be comparable to $a$ itself (see (ii)), so that 
${\rm GCD}(a,a')$ is large, and the D3-tadpole contribution remains 
below the bound. 
On the other hand, there is no clear correlation to be read out from 
the plot (iii), we will have a comment on this in 
Section \ref{ssec:orientifold-beyond}. 
%

There is a tremendous amount of combined choices of $X \times S = X_{[a~0~c]} \times
X_{[a'~0~c']}$ and the fluxes on it. We found about 
$795 \times 10^{15}$ choices satisfying all the three conditions on the
flux by naively scanning $\{n_{ij}, m_{ij}, F_P, F_Q \}$ (apart from 
identification $\{n_{ij}, m_{ij}, F_P, F_Q \} \longleftrightarrow 
(-1) \times \{n_{ij}, m_{ij}, F_P, F_Q\}$). This large ensemble is
dominated by the flux choices on a very small group of possibilities 
of $X_{[a~0~c]} \times X_{[a'~0~c']}$. 
There are about $777 \times 10^{15}$ choices of flux for 
$X_{[1~0~1]} \times X_{[1~0~1]}$, 98\% of the ensemble. 
Almost 99.6\% of the ensemble is accounted for when we combine 
all the flux choices on $X_{[1~0~1]} \times X_{[1~0~1]}$, 
$X_{[2~0~1]} \times X_{[2~0~1]}$, 
$X_{[2~0~2]} \times X_{[1~0~1]}$, 
$X_{[1~0~1]} \times X_{[2~0~2]}$ and 
$X_{[2~0~2]} \times X_{[2~0~2]}$. 

It is a fluke, though, that the statistics is dominated by a small number 
of pairs of attractive K3 surfaces. The large number of flux choices for 
$X_{[a~0~c]} \times X_{[a'~0~c']}$ with small values of $a,c,a',c'$ is 
primarily due to\footnote{The 240 roots of $E_8$ are the norm $(-2)$ points 
on the $E_8$ root lattice; there are 2160 points of norm $(-4)$ 
in the $E_8$ lattice, and the number of norm $(-2m)$ points scale as $m^3$
There are 490560 points of norm $(-24)$ in the $E_8$ lattice \cite{CS-book}. 
These numbers were used in obtaining the flux choices of order 
$795 \times 10^{15}$. } $F_P, F_Q$ in 
$(-2) \oplus E_8 \oplus E_8 \subset \Lambda_{\rm K3}^{(S)}$.
It is possible for such pairs of K3 surfaces 
to find a flux $(G_1 + G_0)$ on 
$(T_X \oplus T_X[-1]) \otimes (T_S \oplus T_S[-1])$ so that the contribution 
to the D3-tadpole is much smaller than 24. There is a lot of room left in the 
tadpole condition for an extra $G_0$-type flux in 
$T_X[-1] \otimes ((-2) \oplus E_8 \oplus E_8)$. It should be remembered that 
we should take a quotient of flux configurations on 
$X \times S = X_{[a~0~c]} \times X_{[a'~0~c']}$ (and elliptic fibration morphisms) 
by the subgroup of $\Gamma$ that fixes the embeddings of 
$U_* \oplus T_X$ and $T_S$ in 
(\ref{eq:TxTx[-]-into-UU}, \ref{eq:TsTs[-]-into-UU}) and 
$J_S$ in (\ref{eq:assume-Js}). Included in this stabilizer subgroup is 
$W_{E_8} \times W_{E_8}$, the Weyl group of 
$E_8 \oplus E_8 \subset \Lambda_{\rm K3}^{(S)}$.

The Weyl group $W_{E_8}$ acts on the 240 roots transitively. We have also 
confirmed that the norm $(-4)$, $(-6)$, $(-10)$ and $(-12)$ points form 
single orbits of $W_{E_8}$ on their own. There are at most two $W_{E_8}$ orbits 
in the norm $(-8)$ points, norm $(-14)$ points and also in the norm $(-16)$ 
points.\footnote{This is done by computing the $\diag(H_4, H_4)$ action on 
$W_{D_8}$ orbits, see \cite{CS-book} for more information. } Thus, it is not 
a particularly bad estimate (for the counting of inequivalent flux vacuum) 
to assume that all the norm $(-2m)$ points in the $E_8$ root lattice form 
a single orbit of $W_{E_8}$ for relatively small $m$ such as $m \leq
12$, which is the range that matters under the tadpole 
condition (\ref{eq:total-tadpole-<=24}). With this crude approximation 
of the modular group $\Gamma$ action on fluxes, the total number of 
``inequivalent'' choices of $X \times S = X_{[a~0~c]} \times
X_{[a'~0~c']}$ and fluxes $G_{\rm tot}$ combined is reduced to 
about $7 \times 10^6$. About 80\% of this ensemble of ``inequivalent''
vacua still come from fluxes on $X \times S = X_{[1~0~1]} \times
X_{[1~0~1]}$, and the totality of the flux choices on the five $X \times S$ 
mentioned above (those with small $a,c,a',c'$) account for 92\%.
Although the precise percentage values should not be taken too seriously 
because of the crude estimate of the modular group action (and
artificial choice of $J_S$), it is trustable at the qualitative
level that the vacuum statistics of  ${\rm K3} \times {\rm K3}$ compactifications of F-theory 
with a $rk_7 = 16$ 7-brane gauge group $W_{\rm unbroken}$ is dominated 
by a small number of $X\times S = X_{[a~b~c]} \times X_{[a'~b'~c']}$ which 
have small values of $a,c,a',c'$ so that the minimum possible value of 
$G_{\rm tot}^2/2$ is small.  
Consequently, the distribution of any observables/moduli parameters 
(such as ${\rm Im}(\rho_H) = \sqrt{ac}$) is determined simply 
by that of $X \times S = X_{[1~0~1]} \times X_{[1~0~1]}$ and a few others. 

Before closing this section, it is worthwhile 
to mention that non-zero flux $G_{\rm tot} = G_1 + G_0$ does not 
imply that all the 18 + 20 complex structure moduli in 
${\cal M}_{{\rm K3};F}^{(X)} \times {\cal M}_{{\rm K3}}^{(S)}$ are stabilized. 
The mass matrix (quadratic part of the superpotential) can be written in the
form of 
\begin{equation}
\propto  \frac{1}{2}
  \left( \delta \Pi^{(1)}, \delta \Pi^{(2)} \right)
  \left( \begin{array}{cc}
      - {\rm Im}(\tau_2)\sqrt{Q_2} \gamma &
      \left. C_I \right) \otimes \left( F_I \right. \\
      \left. F_I \right) \otimes \left( C_I \right. &
      - {\rm Im}(\tau_1)\sqrt{Q_1}  \overline{\gamma}
         \end{array}
      \right)
   \left( \begin{array}{c} \delta \Pi^{(1)} \\ \delta \Pi^{(2)} \end{array}
   \right)\, ,
\end{equation}
where $\delta \Pi^{(X)}$ has 18 components on $W_{\rm frame}$ and 
$\delta \Pi^{(S)}$ 20 components on $S_S$. Mass terms in the diagonal
block are from (\ref{eq:flux-eff-super}), while the off-diagonal block
is due to the $G_0$ type flux. All the moduli would have 
non-zero masses if $G_1 \neq 0$ and $G_0$ were absent
(and similarly, full rank $G_0$ type flux with $rk_7 = 0$ and vanishing 
$G_1$ would stabilize all but two complex structure moduli).
When both are present, however, the diagonal masses from $G_1$ and 
off-diagonal masses from $G_0$ interfere and there may be 
a zero mass-eigenvalue (and an unstabilized direction of the moduli
space) in principle. 

One example is to choose 
$n_{ij}=0$, $m_{11}=m_{12}=m_{22}= -m_{21}=1$ and $F_P = F_Q = 0$ in 
(\ref{eq:G1-nm-parametrized}, \ref{eq:G0-nm-parametrized}) for 
a series of infinite pairs of attractive K3 surfaces 
$X \times S = X_{[k~0~k]} \times X_{[k~0~k]}$ (where $k=1,2 \cdots, \infty$).
Under this choice of the flux, a one parameter ($k$) deformation 
of the complex structure $\Omega_X \wedge \Omega_S = 
[(v_1 + k V_1)+i(v'_1+kV'_1)] \otimes 
[(v_2 + k V_2)+i(v'_2+kV'_2)]$ remains a flat direction, and 
all the $\rho_X = \rho_S = 20$ points $X_{[k~0~k]} \times X_{[k~0~k]}$ 
just sit in this flat direction.

Such flux vacua with an unstabilized direction have been removed from
the vacuum ensemble studied in this section. To be
more precise, our numerical code examined the $4 \times 4$ mass matrix 
on $P_1 \delta \Pi^{(1)}_{P_1} + Q_1 \delta\Pi^{(1)}_{Q_1}$ and 
$P_2 \delta \Pi^{(2)}_{P_2} + Q_2 \delta \Pi^{(2)}_{Q_2}$, and threw
away all the flux configurations for which there is a zero eigenvalue in this 
mass matrix. Certainly, one should examine the eigenvalues of 
a $(18+20) \times (18 + 20)$ mass matrix for each flux 
configuration,\footnote{In the case of
$rk_7=16$, this problem is reduced to that of a $(2+20) \times (2+20)$
matrix, since the mass matrix of the moduli $\delta\Pi^{(1)}$ in
$W_{\rm unbroken}$ does not interfere with the rest.} but we believe
that our short-cut approach does not qualitatively distort the distribution of
observables in the landscape.

\subsection{${\rm U}(1)$ Flux, Heterotic--F-theory Duality, and GUT 7-brane Flux}
\label{ssec:MW-singl-fbr-flux}

The unbroken symmetry $W_{\rm unbroken}$ of interest for a sub-ensemble of 
vacua can be chosen at one's will. We can choose it to be rank-18 as 
in Sections \ref{ssec:review-AK}, \ref{ssec:extended-list} 
and \ref{ssec:sample-stats}, or to be rank-16 as in 
Section \ref{ssec:rank-16}, but other choices such as 
$W_{\rm unbroken} \cong A_4$, a $\SU(5)_{\rm GUT}$ landscape, are just as 
appropriate. 

Studies such as the one in Section \ref{ssec:rank-16} are 
dedicated to a landscape of vacua of individual choices of $W_{\rm unbroken}$, 
which we call the $W_{\rm unbroken}$ landscape.
Beyond such an analysis, however, it is natural to ask such questions as 
how properties of the algebraic information $W_{\rm unbroken}$ generally 
characterize distributions within the $W_{\rm unbroken}$-landscape, or 
how the number of vacua in the $W_{\rm unbroken}$-landscape depends on 
$W_{\rm unbroken}$. As a preparation for such a discussion in 
Section \ref{ssec:orientifold-beyond}, we make a few remarks in this 
Section. 

When we write down the $G_0$ type flux in the 
form (\ref{eq:G0-Wx-Ss-expansion}), we can take the generators 
$\{C_I \}_{I=1,\cdots, 18}$ of $W_{{\rm frame}*}$ to be such that some are from 
$W_{{\rm U}(1)}$, and others from $W_{\rm root}$. We call the $G_0$ flux  
with $C_I$ from $W_{\rm U(1)}$ a ${\rm U}(1)$ flux or Mordell--Weil flux, and 
that with $C_I$ from $W_{\rm root}$ a GUT-brane flux or singular fibre flux.
The GUT 7-brane flux corresponds to the line bundles introduced on 
GUT 7-branes in F-theory (such as those in \cite{Beasley:2008kw}), 
or line bundles on a stack of multiple D7-branes in type IIB 
Calabi--Yau orientifolds. The ${\rm U}(1)$ flux requires \cite{BCV} a special 
choice of complex structure so that there is a non-vertical {\it algebraic} cycle, 
meaning that the Mordell--Weil group is non-trivial; the unavoidable tuning of 
complex structure is an expression equivalent to moduli stabilization. 

\subsubsection{${\rm U}(1)$ flux/Mordell--Weil flux}

Let us first ask what the ${\rm U}(1)$ flux looks like in the light of the
duality between F-theory and heterotic string theory. As discussed in the 
literature (such as \cite{Curio:1998bva, DW-1, Hayashi-1}), 
there must be a component of four-form ${\rm U}(1)$ flux in F-theory whose origin 
in heterotic string is a flux on the spectral surface. Suppose that 
$\pi_Z: Z = T^2 \times S \longrightarrow S$ is the elliptically fibred 
Calabi--Yau threefold for the dual heterotic string compactification, 
and $(C, {\cal N})$ the spectral data describing the vector bundle 
on $Z$. The spectral surface $C$ is a subspace of $Z = {\rm Jac}(T^2) \times S$ 
characterized by the zero points of the elliptic functions 
\begin{equation}
a^{\rm v}_0 + a^{\rm v}_2 x + a^{\rm v}_3 y + \cdots = 0, \qquad 
a^{\rm h}_0 + a^{\rm h}_2 x + \cdots = 0  
\label{eq:Het-spec-surf}
\end{equation}
on $T^2$, and ${\cal N} = {\cal O}_C(\gamma)$ is a line bundle on $C$
specified by a divisor $\gamma$ on $C$. 
{\it For any complex structure} of $Z$, $S$ and $C$
\cite{Friedman:1997yq}, one can always find a divisor
\begin{equation}
 \gamma_{\rm FMW} \propto (n \sigma - \pi_C^*(n K_S + \eta))\, ,
\label{eq:FMW-flux}
\end{equation}
where $\pi_C = \pi_Z|_C$, $\eta$ is a divisor on $S$, $K_S$ the 
canonical divisor of $S$, and we assume $\SU(n) \subset E_8^{\rm vis}$ is the 
structure group of the vector bundle. 
None of the ${\rm U}(1)$ fluxes we talk about here, however, should correspond 
to this Friedman--Morgan--Witten flux in the heterotic ``dual''. 
If this flux were to be dual to the ${\rm U}(1)$ flux in F-theory, we would run 
into a contradiction immediately: an arbitrary complex structure is allowed 
on the heterotic side, while it has to be tuned in F-theory. In fact, 
the generic Friedman--Morgan--Witten flux (\ref{eq:FMW-flux}) always 
vanishes, $\gamma = 0$, when the dual F-theory is a  ${\rm K3} \times {\rm K3}$ compactification.

For finely tuned complex structure of the heterotic string compactification 
data, $(Z, S, C)$, however, there is more variety in the choice of
$\gamma$ on the spectral surface \cite{Friedman:1997yq}. 
Suppose that the base K3 surface $S$ is not the general complex analytic 
one, but has Picard number $\rho_S \geq 2$. This means that we might be
able to find a set of $\{F_A \} \subset [J_S^\perp \subset S_S]$
generating a rank-$k$ sublattice of $S_S$. We only consider vector
bundles with the structure group contained in 
$G^{\rm v}_{\rm str} \times G^{\rm h}_{\rm str} =
\SU(N_v) \times \SU(N_h)$. It then appears that we can think of 
a vector bundle for compactification, 
\begin{equation}
 V = \oplus_{i=1}^{N_v + N_h} 
  {\cal O}_S(D_i) \otimes {\cal O}_{T^2}(p_i-e),  \qquad 
\qquad D_i = \sum_A n^A_{(i)} F_A\,, 
\label{eq:Het-bndl-NvNh}
\end{equation}
where the $\{p_i\}$ and $e$ are the zeros and the pole, respectively, 
of the elliptic functions (\ref{eq:Het-spec-surf}). The spectral surface 
is of the form $C = \cup_i C_{p_i}$ with $C_{p_i} \simeq S$'s corresponding 
to $p_i \in {\rm Jac}(T^2)$, and $\gamma|_{C_{p_i}} = D_i$. 
This vector bundle is poly-stable with respect to the K\"{a}hler form 
$J_Z = J_{T^2} + J_S$. The unbroken symmetry should be 
$E_{9-N_v}^{\rm v} \times E_{9-N_h}^{\rm h}$ within $E_8 \times E_8$. 
Moduli parameters include $\rho_H$ and $\tau_H$ of $T^2$, and 
$(N_v-1) + (N_h-1)$ parameters of the flat bundles 
in (\ref{eq:Het-spec-surf}), and hence,in addition to those from the base $S$, there are $N_v + N_h$ of them.

Dual to this heterotic string compactification should be F-theory
on $Y = X \times S$ where the elliptically fibred K3 surface $X$ has 
the following Neron--Severi and transcendental 
lattices:\footnote{A detailed description of the transcendental lattice 
is found in \cite{Hayashi-4} for the case of $N_v = 3$.}
\begin{equation}
 \Lambda_{\rm K3}^{(X)} \supset S_X \oplus T_X = 
  \left[ U_* \oplus E_{9-N_v} \oplus E_{9-N_h} \right] \oplus 
  \left[ U \oplus U \oplus A_{N_v-1} \oplus A_{N_h-1}\right]\, .
\end{equation}
K3 surfaces $X$ of this form have Picard number $\rho_X = (20-N_v-N_h)$,
so that the moduli space has dimension $(N_v+N_h)$, which agrees 
with the counting above. If we are to find a four-form flux in F-theory 
dual to $\gamma$, that dual flux must be associated with non-$U_*$ cycles 
that are orthogonal to the unbroken symmetry part 
$E_{9-N_v} \oplus E_{9-N_h}$. This means that the flux is not in
an algebraic cycle, and such a configuration is not stable. The complex 
structure moduli dynamically tune themselves (due to the GVW potential) 
so that the associated cycle becomes algebraic. In order to introduce 
$k$ independent ${\rm U}(1)$ fluxes in (\ref{eq:G0-Wx-Ss-expansion}), 
the number of unstabilized directions in the moduli space should be 
reduced by $k$, and the moduli space becomes 
$(N_v+N_h-k)$-dimensional \cite{BCV}.\footnote{On top of
this, $G_1$ type flux is introduced to fix the remaining moduli. But in
the context of heterotic--F-theory duality, we take the $G_1$ component out
of the picture (or assume that it is absent).} 

This reduction of the moduli space should 
also be understandable in the heterotic string description. 
Certainly $V$ in (\ref{eq:Het-bndl-NvNh}) is a holomorphic vector
bundle and satisfies the condition $\int J \wedge J \wedge F = 0$ as 
well as the Bianchi identity of the $B$-field, 
\begin{equation}
 0 = \tr_{\mathfrak{so}(6)} \left[ \left(\frac{R}{2\pi}\right)^2 \right]
   - T_R^{-1}\tr {}_R
                        \left[ \left(\frac{F}{2\pi}\right)^2 \right]
   + 4 \delta^{(4)}_{M5} 
 = 4 \left(-c_2(TZ) - {\rm ch}_2(V) + \delta^{(4)}_{M5} \right)\, , 
\label{eq:B6-Bianchi-Het}
\end{equation}
where $\delta^{(4)}_{M5}$ is the delta-function valued four-form 
associated with 
individual M5-branes wrapped on holomorphic curves in $Z = T^2 \times S$, 
and ${\rm ch}_2(V)$ is the second Chern character of the rank $(N_v+N_h)$ 
vector bundle $V$ (with the structure group $\SU(N_v) \times \SU(N_h)$).
There is a condition that is stronger than (\ref{eq:B6-Bianchi-Het}), 
however: 
\begin{equation}
 0 = H := dB + 
\frac{\alpha'}{4}\left( \tr \left[\omega d\omega + \frac{2}{3}\omega\omega\omega\right] - 
 T_R^{-1} {\rm tr} {}_R \left[A dA - i \frac{2}{3} A A A \right]\right)
+ ({\rm source})_{M5}\, ,
\label{eq:H=0}
\end{equation}
if we are to stick to the framework of heterotic string compactification 
on a K\"{a}hler manifold and constant dilation configuration 
(Chap.16, \cite{GSW}).
This constraint may also be understood as a combination of the 
F-term condition of 
\begin{equation}
 W_{\rm Het} \propto \int_Z \Omega_Z \wedge H + 
   (2\pi)^2\alpha' \int_{\Gamma(M5)} \Omega_Z 
\label{eq:Het-super}
\end{equation}
with respect to the complex structure moduli of $Z$ and 
$\vev{W_{\rm Het}} = 0$ for vanishing cosmological constant.\footnote{
Flux of Friedman--Morgan--Witten type $\gamma$ \eqref{eq:FMW-flux} does not
vanish in the case of a general Calabi--Yau threefold $Z$ with elliptic 
fibration (although it vanishes in the present case). This type of flux, 
however, does not restrict complex structure moduli because it induces a 
vanishing Chern-Simons term.} 
$\Gamma(M5)$ is a real 3-chain in $Z = T^2 \times {\rm K3}$ whose
boundary contains the curves on which M5-branes are wrapped \cite{Witten:1997ep}.
The exterior derivative of this condition (\ref{eq:H=0}) 
reproduces (\ref{eq:B6-Bianchi-Het}). 

The $B$-field Bianchi identity, (\ref{eq:B6-Bianchi-Het}), can be
satisfied by wrapping an appropriate number ($N_{M5}$) of heterotic
NS5-branes (M5-branes in heterotic M-theory) on the fibre $T^2$. 
As is well-known in the literature, 
\begin{equation}
 c_2(TZ) = (24 \; {\rm pt}_{K3}) \otimes 1_{T^2}, \qquad 
  {\rm ch}_2(V) =  \frac{1}{2} \sum_{i=1}^{N_v + N_h}(D_i)^2\, , 
\end{equation}
and the only non-trivial part, which is a four-form on $S$ and scalar on
$T^2$, of the condition (\ref{eq:B6-Bianchi-Het}) gives rise to 
\begin{equation}
 \frac{1}{2} ( G^{(4)}_H )^2 + N_{M5} = 24\, .  
\label{eq:M5-tadpole-S}
\end{equation}
Here, $G^{(4)}_H$ takes values in the lattice\footnote{
Note that we define the $A$--$D$--$E$ lattice to have negative definite symmetric pairing 
in this article, and that $D_i$'s are also negative definite, because 
of the signature $(1, \rho_S-1)$ of $S_S$.} 
$(A_{N_v-1} \oplus A_{N_h-1}) \otimes [J_S^\perp \subset S_S]$, and is
given by
\begin{equation}
  G^{(4)}_H = \sum_{I=1}^{N_v-1} C_I \otimes D_I
            + \sum_{I=1}^{N_h-1} C_I \otimes D_I, \qquad 
   D_I = - (D_{i=1} + \cdots + D_{i=I})\, .  
\label{eq:G4-Het-def}
\end{equation}

The condition (\ref{eq:H=0}) in the two-form component on $S$ and 
one-form on $T^2$, however, contains information that is not captured 
by the Bianchi identity (\ref{eq:B6-Bianchi-Het}). In the presence of 
${\rm U}(1)$ flux ${\cal O}_S(D_i)$ on $S$ and a flat bundle 
${\cal O}_{T^2}(p_i-e)$ (i.e., Wilson line) on $T^2$, there is no
contribution to the Chern character, but the Chern--Simons form 
can be non-zero. Keeping in mind that M5-branes wrapped on $T^2$ as well
as gravitational Chern--Simons form only contribute to (\ref{eq:H=0}) 
in the component purely three-form on $S$, we see that 
\begin{equation}
\left. \left(  dB - \frac{\alpha'}{4} T_R^{-1} \tr {}_R
   \left[ A dA \right]
 \right) \right|_{2\mbox{-}{\rm form~on~}S, \; 1\mbox{-}{\rm form~on~}T^2}= 0\, , 
\label{eq:H=0-fibre-dir}
\end{equation}
where the $A^3$ term in the Chern--Simons form has also been dropped 
for the Cartan flux configuration. Since the Polyakov action remains 
invariant by changing the $B$-field background by $(2\pi)^2\alpha'$ times
an integral two-form on the target space, it is possible that the
$B$-field background configuration expanded by using the same set of $F_A$ as 
in (\ref{eq:Het-bndl-NvNh})
\begin{equation}
  B = \sum_A b_A F_A (2\pi)^2 \alpha' 
\end{equation}
may have scalars $b_A$ varying topologically on $T^2$, so that 
$db_A = \check{\alpha} n^8_A + \check{\beta} n_A^9$  
for some integers $n^8_A$ and $n^9_A$ and a basis 
$\{ \check{\alpha}, \check{\beta} \}$ of $H^1(T^2; \Z)$. 
When $k$ independent $F_A$'s in $[J_S^\perp \subset S_S]$ are involved, 
we find the $k$ independent conditions on the Wilson lines 
$A = A_I C_I$, $A_I = (2\pi) (\check{\alpha} a_I^8 + \check{\beta} a_I^9)$:
\begin{equation}
 \left[  db_A  + \frac{A_I}{2\pi} \left(C_I , C_J \right) n_A^{(J)} 
 \right] \otimes F_A = 0, \qquad 
 2 n_A^8 - a_I^8 q^I_A = 0, \quad 
 2 n_A^9 - a_I^9 q^I_A = 0,   
\end{equation}
where $n_A^{(J)} := - (n^A_{(i=1)} + \cdots + n^A_{(i=J)})$ and 
$q^I_A := - (C_{IJ})n_A^{(J)}$.
Hence $k$ combinations of the Wilson lines $A_I$ are required to be 
torsion points of ${\rm Jac}(T^2)$ (equivalent to the origin of 
${\rm Jac}(T^2)$ when multiplied by some non-zero integer).\footnote{
It is no longer surprising that $a_3^{\rm v} y = 0$ and $a_3^{\rm h} y =
0$ are the spectral surface equation read out from
(\ref{eq:Inose-pencil}). These equations mean that the
$N_v=3$-fold spectral cover (also $N_h=3$) consists of three 
irreducible pieces sitting at the three 2-torsion points of ${\rm Jac}(T^2)$.}
This is equivalent to $k$ conditions imposed on the spectral surface data 
($a^{\rm v}_{0,2,3,\cdots}$ and $a^{\rm h}_{0,2,\cdots}$) and 
$f_0$ and $g_0$ (equivalently $\tau_H$).

Let us now translate this interplay among the Cartan flux, $B$-field
topological configuration and the reduction (stabilization) of the
moduli space in the language of Narain moduli. The Narain moduli covers 
not just the region with parametrically large $[{\rm vol}(T^2)/\ell_s^2]$ 
(where supergravity approximation is good), but also the stringy region. 
When there is a Cartan flux and corresponding topological $dB$
configuration, our observation above---some combinations of $A_I$'s are 
forced to be torsions of ${\rm Jac}(T^2)$---is translated into the
existence of $k$ vectors\footnote{It is worth drawing attention 
to the fact that such vectors cannot always be brought into the 
$[A_{N_v-1} \oplus A_{N_h-1}]$ part of $[U \oplus U \oplus A_{N_v-1}
\oplus A_{N_h-1}]$. The flux vectors that can be brought purely into 
$E_8 \oplus E_8$ may be associated with the flux on a trivial spectral
surface in heterotic language, and are dual to the singular fibre flux (GUT
brane flux) in F-theory. } \footnote{We wonder if we have made 
an error causing the factor of $2$ ? } 
\begin{equation}
 n_A := (2 n^8_A, 0,2 n^9_A, 0, q^{I}_A)^T \in {\rm II}_{2,18}\, , 
\label{eq:Het-flux-data}
\end{equation}
satisfying \footnote{
This condition in heterotic string theory means that 
there is a class of states (vertex operators) with $(k^R, k^L) \in
\R^{2,18}$ satisfying $(\alpha'/4) (k^R)^2 = 0$ and 
$(\alpha'/4) (k^L)^2 = -(n_A)^2/2 = - (q_A)^2/2$. 
As we are considering ${\rm U}(1)$/Mordell--Weil fluxes here, rather than 
a GUT 7-brane/singular fibre flux, $n_A$ should belong to 
$W_{\rm frame} \backslash W_{\rm root}$, meaning that $-(n_A)^2 \geq 4$.
Thus, none of vertex operators for physical states with this 
$(k^R, k^L)$ appear in the massless spectrum.} 
\begin{equation}
  \langle  {\cal Z}^R, n_A \rangle = 0, \qquad  
  {\cal Z}^R \propto \left( -\tau_H, \; - \tilde{\rho}_H, \; 1, 
   \; - \tau_H \tilde{\rho}_H - (a)^2/2, \; a_I  \right);  
\label{eq:Het-ZdotG}
\end{equation}
see Appendix \ref{sec:Het-Narain} for background material 
on Narain moduli in heterotic string theory. 
This mechanism is precisely the same, mathematically, 
as how algebraic cycles emerge for special choice of complex 
structure on the other side of the duality. 

It must be be obvious that one should define a lattice element 
\begin{equation}
 \widetilde{G}^{(4)}_H := \sum_A n_A \otimes F_A  \in 
  [U \oplus U \oplus A_{N_v-1} \oplus A_{N_h-1}] \otimes 
  [J_S^\perp \subset S_S] 
\end{equation}
in heterotic string language in order to formulate the Cartan flux and 
topological $dB$ configuration combined. Within the framework that we have 
assumed in this section so far,\footnote{
Obviously this is not the most general form of $G_{\rm tot}$, in particular
when $G_1 = 0$. For a fully general choice of $G_{\rm tot}$ on the
F-theory side, it is considered that we have to turn on 
$\vev{H} \neq 0$, a non-constant dilaton configuration and non-K\"{a}hler 
metric on the Heterotic string side \cite{Becker:2003yv}.} where $Z = T^2 \times S$ 
is K\"{a}hler, $\vev{\phi}_{\rm Het} = {\rm const}$ and $\vev{H} = 0$ on the heterotic
side, the duality map of the flux is given by
\begin{equation}
 \widetilde{G}^{(4)}_H = G_0 = G_{\rm tot}\, .
\label{eq:G4-H-G4-F-duality}
\end{equation}
Within this class of Cartan flux, the norm of $\widetilde{G}^{(4)}_H$
remains the same as that of $G^{(4)}_H$, because with $w^8 = w^9 = 0$, 
only $n_8^A$ and $n_9^A$ are allowed to be non-zero. 
Therefore, the Bianchi identity (\ref{eq:M5-tadpole-S}) 
becomes dual to the D3-tadpole condition (\ref{eq:total-tadpole-<=24}).

The duality dictionary above is a generalization and refinement of 
the preceding discussion in 
\cite{Friedman:1997yq, Andreas:1997ce, Curio:1998bva, Andreas:1999ng, DW-1, Hayashi-1}, 
in that we can deal with a more general class 
of fluxes on the heterotic spectral surface (by allowing non-trivial 
$\vev{dB}$), keep track also of flux quanta in the $U \oplus U$
components in the form of $\widetilde{G}^{(4)}_H = G_{\rm tot}$ or
$n_A$, maintain a clear distinction between algebraic and 
transcendental cycles on the F-theory side and are able to understand 
how the dimension of moduli space is reduced. 

The conventional dictionary using the stable degeneration limit of K3 
surfaces is understood in the following way from the present perspective. 
Within the framework we have considered so far, the moduli space of 
${\cal Z}^R$ ($\propto \Omega_X$) still maintains a free choice of at
least $\tau_H$ and $\tilde{\rho}_H$, even after maximally possible
Cartan fluxes (rank = $N_v+N_h-2$) are introduced in the 
$\SU(N_v) \times \SU(N_h)$ structure group (no $G_1$ component yet,
in particular).
Thus, from this moduli space we can extract a
family of K3 surfaces parametrized by $\tilde{\rho}_H$. 
This family $\pi: {\cal X} \longrightarrow D$ is defined on a disc 
$D \subset \C$ with $t = 1/\tilde{\rho}_H$ as the coordinate, 
just like the discussion in footnote \ref{fn:degeneration-limit-A} for 
the family defined by (\ref{eq:Weierstrass-degen-lim}). Mathematically, 
this family may be augmented by providing the degeneration limit 
corresponding to $t =1/\tilde{\rho}_H = 0 \in D$, which is called the
central fibre. Each fibre (K3 surface) $\pi^{-1}(\tilde{\rho}_H)$ for
some $1/\tilde{\rho}_H \in D$ contains an algebraic cycle (curve) 
corresponding to $n_A$ in (\ref{eq:Het-flux-data}, \ref{eq:Het-ZdotG}), 
and those algebraic cycles (curves) for various $\tilde{\rho}_H$ are
collectively regarded as an algebraic family of curves, or equivalently 
as a divisor in ${\cal X}$. 
The intersection of this divisor with the central fibre, 
$\pi^{-1}(0) = dP_9 \cup dP_9$, determines an algebraic cycle in 
$dP_9 \cup dP_9$. In this way, the stable degeneration limit of 
$n_A$ is obtained. 

Appendix \ref{sec:Km-E6E6} describes the behaviour of a semistable 
degeneration of a K3 surface and its algebraic cycles using a concrete 
example given by (\ref{eq:Inose-pencil}). We will see there how 
the semistable degeneration limit of an algebraic cycle can have 
an image in only one of the two $dP_9$'s (only in visible/hidden sector 
$dP_9$). 

Just like in the heterotic string picture of F-terms
(\ref{eq:Het-super}) leading to the stabilization of Wilson line moduli 
$\vev{A}_I$, we can also understand\footnote{A closely related discussion
is found in \S7.3 of \cite{Tripathy:2002qw}, see also \cite{Becker:2002sx,Becker:2003yv}.
The presentation here, however, is for $\vev{H}=0$ and K\"{a}hler
metric in the heterotic description.} the stabilization of the D7-brane 
positions in the dual type IIB theory on the orientifold $K3\times T^2/\Z_2$. 
The type IIB version of the superpotential (\ref{eq:Het-super}) is 
\begin{equation}
 W_{\rm IIB} \propto \int_M \Omega_M \wedge G^{(3)}
   +  \int_S i_{X*}\Omega_M \wedge F 
  = \int_M \Omega_M \wedge G^{(3)}
  + \int_S \Omega_{M; ijk} \tr \left[ X^i F_{lm}  \right]\,, \label{iibWD7}
\end{equation}
where $X^i$ is the fluctuation of D7-brane in the transverse direction.
Let us assume that we give a vev $D_i$ to the gauge field strength for
the $i$-th D7-brane. Its contribution can also be described by rewriting the last term above as
$W = \int_{\Gamma(D7)} \Omega_M$. Here, $\Gamma(D7)$ is a 3-chain whose boundary consists of 
a two-cycle Poincar\'e dual (in $S$) to $D_i$ and a reference two-cycle $D_0$. We can expand 
$D_i=n_i^A F_A$. Furthermore, we can introduce a three-form flux $G^{(3)} = F^{(3)} - \phi H^{(3)}$. Here
$\phi$ is the type IIB axiodilaton and $F^{(3)}$ and $H^{(3)}$ are the type IIB R-R and NS-NS fluxes, respectively.
We also expand $G^{(3)} = F^{(3)} - \phi H^{(3)} = (m_F^A - \tau m_H^A)F_A$ for a basis of two-cycles $F_A$ of the $K3$
surface $S$ and some $m_F^A, m_H^A\in H^1(T^2; \mathbb{Z})$. Using that $\Omega_M = \Omega_S\wedge dz$, 
we can evaluate the superpotential to be
\begin{equation}
 W_{\rm IIB} \propto  \sum_A \int_S \Omega_S \wedge F_A \sum_i \left( m_F^A - \phi m_H^A +  \int_0^{X_i} n^i_A \, dz  \right) \, .\label{iibWd7}
\end{equation}
The orientifold involution demands that we supply an image D7-brane at $\hat{X}_i=-X^i$ for any D7-brane at $X^i$. The
${\rm U}(1)$ gauge field surviving the orientifold involution is $\hat{A}^i-A^i$, so that the image brane also carries a 
flux $\hat{D}^i=-D^i$ and the two signs cancel in \eqref{iibWd7}.
The orientifold involution also sends $G^{(3)}\mapsto - G^{(3)}$, so that the only modes that survive are
those expanded into odd three-forms. All three-forms used for the expansion of $G^{(3)}$ are odd, so that this is already
taken care of. We can hence capture the effect of orientifolding by only considering 16 branes at $X^i$ with fluxes
$D_i$ on their worldvolume and putting a factor of two in front of the last term in \eqref{iibWd7}.
The conditions for supersymmetry coming from the superpotential are then written as
\begin{equation}
\sum_i \left( m_F^A - \phi m_H^A +  2\int_0^{X_i} n^i_A \,  dz \right) = 0\, .
\end{equation}
Just as for the Wilson lines, the positions of the branes should add up to a torsion point of $T^2$, so that the above can be 
cancelled by an appropriate choice of $F_3$. Note that $\phi \rightarrow i\infty$ in the weak coupling limit, so that (assuming
$n^i_A$ is finite) we cannot use $H_3$ for the same purpose. This is as expected, as the three-form $dB$, which took part in the
same mechanism on the heterotic side, is turned into $F_3=dC_2$ under the duality map. 

The need to put (some of) the D7-branes at special loci on $T^2/\Z_2$ in order to have supersymmetric world-volume flux
and the connection to the existence of algebraic cycles in the F-theory K3 has also been discussed in \cite{BCV}.

A lesson to be learnt from this is that instead of considering $G_3$ alone,
the combination of $G^{(3)}$ and the contribution from D7-branes we see above
should be of type $(2,1)$ \cite{Lust:2005bd}. This is just 
like the corresponding condition in heterotic string theory, where the vanishing of 
$H$ including both $dB$ and the Chern--Simons terms is required. 
Even for three-form fluxes $H^{(3)}$ and $F^{(3)}$ that cannot be made 
purely type $(2,1)$ and primitive for any choice of $\vev{\phi}$ and 
complex structure of $M$, there may be supersymmetric configurations 
for an appropriate choice of D7/O7-brane configuration. Turning this
argument around, one can also see that once a $G^{(3)}$ configuration 
is found so that the supersymmetric conditions are satisfied for a 
D7/O7-brane configuration, then all the supersymmetric 
configurations consist of a sum of this special combination of flux 
plus any primitive pure $(2,1)$ form.

\subsubsection{GUT 7-brane flux/singular fibre flux}

Let us also make a brief remark on the GUT 7-brane flux (singular fibre
flux), before moving on to the next section. 
Suppose that a flux (\ref{eq:G0-Wx-Ss-expansion}) is introduced for $C_I$ 
(i.e., the corresponding $F_I$ does not vanish) in an irreducible component 
$R$ of $W_{\rm root}$. Although there is a gauge theory on 7-brane 
($\simeq S \times \R^{3,1}$) with the gauge group $R$, the flux turns on 
a non-trivial line bundle on the 7-brane and the symmetry of $R$ is 
broken down to $[C_I^\perp \subset R]$ in the effective theory below the
Kaluza--Klein scale.

The field contents in the effective theory can be described in terms of irreducible 
representations of the unbroken symmetry group $[C_I^\perp \subset R]$. 
Let $\alpha$ be a root of $R$ that does not belong to 
$[C_I^\perp \subset R]$, and $D_\alpha = \sum_I (\alpha, C_I) F_I$ 
a divisor on $S$. Since we have assumed from the outset that 
$F_I \in [J_S^\perp \subset S_S]$, we have that $H^0(S; {\cal O}_S(D_{\pm \alpha})) = 0$, 
as long as the K\"{a}hler form $J_S$ is in ${\rm Amp}_S$ (defined as 
an open set).
If $H^0(S; {\cal O}_S(D_{\alpha}))$ or 
$H^0(S; {\cal O}_S(D_{-\alpha}))$ were non-trivial, then 
either $D_\alpha$ or $D_{-\alpha}$ is a divisor class represented by an
effective curve, which should have a positive intersection number with 
any other divisor (including $J_S$) in the ample cone ${\rm Amp}_S$.
If, however, we allow some $(-2)$-curves to have zero volume under $J_S$ (i.e., 
$J_S \in \overline{{\rm Amp}}_S \backslash {\rm Amp}_S$), some effective 
divisors may have zero volume and 
$H^0(S; {\cal O}_S(D_{\pm \alpha})) \neq \phi$ is not ruled out 
mathematically. This is also true when we cannot rely entirely on the local field 
theory approximation. Just like in toroidal orbifold examples 
where symmetry may sometimes be restored even in the presence of 
symmetry-breaking orbifold twists, stringy effects at zero-volume 
cycles may bring about consequences beyond the local field theory/supergravity 
approximation. Keeping in mind that 
\begin{eqnarray}
 h^0(S; {\cal O}_S(D_{\alpha})) + h^0(S; {\cal O}_S(D_{-\alpha}))
 & = & h^0(S; {\cal O}_S(D_\alpha)) + h^2(S; {\cal O}_S(D_{\alpha})), 
 \nonumber  \\
 & \geq & \chi(S; {\cal O}_S(D_\alpha)) = 2 + \frac{(D_\alpha)^2}{2}  
\end{eqnarray}
on a K3 surface $S$, there is always such a zero-volume $(-2)$ curve  
for a flux such that $(D_\alpha)^2 = -2$. For fluxes where 
$(D_{\alpha})^2 \leq -4$, however, there is no such immediate
consequence.  

As long as we stay within the field theory approximation 
(which means that $J_S \in {\rm Amp}_S$, or at least 
$h^0(S; {\cal O}_S(D_\alpha))=0$), the number of chiral multiplets 
in the irreducible component containing the root $\pm \alpha$ is given by 
\begin{equation}
 h^1(S; {\cal O}_S(D_{ \pm \alpha})) =
 - \chi(S; {\cal O}_S(D_{\pm  \alpha}))
  = - \frac{(D_\alpha)^2}{2} -2. 
\label{eq:nmbr-vctr-like-pair}
\end{equation}
Note that there is no net chirality because of $(D_\alpha, K_S) = 0$.


\subsection{Vacuum Distribution based on continuous approximation}
\label{ssec:orientifold-beyond}

There have been attempts, most notably in \cite{Ashok:2003gk,
Denef:2004ze, DeWolfe:2004ns, Denef:2008wq}, of going beyond a 
case-by-case analysis of flux configuration counting. We initiate an effort 
of generalizing their approach by implementing various concepts that we have already developed in this section, so that 
we can ask statistical questions that are of interest in the context of particle physics, not just in cosmology. 
To do so, we use what we call the {\it restricted complex structure moduli space}, which is the space of complex structure
deformations leaving a chosen set of divisors algebraic. For a ${\rm K3}$ surface (or ${\rm K3} \times {\rm K3}$), we denote 
this by ${\cal M}_*(J_S, W_{\rm noscan})$.
It is the moduli space of complex structures for which the (fixed) K\"ahler form $J_S$ stays purely of type $(1,1)$ and
the divisors spanning $W_{\rm noscan}$ remain algebraic. This means that we only consider complex structures such 
that $\Omega \cdot W_{\rm noscan} = 0$ \footnote{K3 surfaces with this kind of constraint are called \emph{lattice polarized} 
in the mathematics literature.}. For more general fourfolds (where $H^{2,0}=0$) the first condition, $J_Y$ being of type $(1,1)$,
is automatically satisfied, so that we simply use the notation ${\cal M}_*(W_{\rm noscan})$ here.

{\bf I}
An approach of \cite{Ashok:2003gk, Denef:2004ze} is to replace the 
sum over all the flux configurations taking their value in a lattice by 
continuous integration. In a theory where there are flux quanta $N$
specified by $K$ integers, the vacuum index density $d\mu_I$ is defined on 
${\cal M}_*(J_S, W_{\rm noscan})$ (of complex dimension $m$) by 
\begin{eqnarray}
 d\mu_I & = & d^{2m} z \sum_{N} \Theta(L_*-L)
    \delta^{2m}(D_a W,\overline{D_bW})  \; 
  {\rm det}\left( \begin{array}{cc}
       D_c D_d W &
       \partial_c \overline{D_d W} \\
       \bar{\partial}_{\bar{c}}D_dW   &
       \overline{D_cD_d W} \end{array} \right)_{2m \times 2m}
\label{eq:index-density-pre} \\
 & \approx & d^{2m}z \int_N d^K N \;
   \Theta(L_*-L)
    \delta^{2m}(D_a W,\overline{D_bW}) \; 
  {\rm det}\left( \begin{array}{cc}
       D_c D_d W &
       \partial_c \overline{D_dW} \\
       \bar{\partial}_{\bar{c}} D_dW &
       \overline{D_cD_d W} \end{array} \right)_{2m \times 2m} \, . 
\label{eq:index-density}
\end{eqnarray}
Here, the $z^a$ are complex local coordinates on ${\cal M}_*$. The definition 
of this index density $d\mu_I$ is independent of the choice of local 
coordinates on ${\cal M}_*$ and becomes an $(m,m)$-form on 
${\cal M}_*/\Gamma_*$, where $\Gamma_*$ is the modular group. 
The vacuum index density can be used to set a lower bound in the number
of vacua (without the condition $\vev{W_{\rm GVW}} = 0$).

In type IIB compactification on a Calabi--Yau threefold $M_3$ with an 
orientifold projection yielding $n$ irreducible O7-planes, for example, 
${\cal M}_*(D_4^{\oplus n})$ is the same as the moduli space of
complex structures of $M_3$ together with the axi-dilaton. 
We only consider configuration of 7-branes such that they appear 
in the combination of $\SO(8)$ 7-brane gauge groups, i.e. we out four D7-branes on each O7-plane. 
Then the restricted moduli space ${\cal M}_*$ has complex dimension 
$m=(h^{2,1}_{-,{\rm prim.}}(M_3)+1)$ and there are 
$K = 4(h_{-,{\rm prim.}}^{2,1}(M_3) + 1)$ integers for type IIB 
three-form quanta satisfying the primitiveness condition 
$G^{(3)} \wedge J_M = 0$.
The delta functions (F-term conditions) reduce the flux space to 
be integrated from real $K$ dimensions to $K-2m$ dimensions.  
The remaining integral in the flux space (at each point in the moduli
space ${\cal M}_*/\Gamma_*$) is over primitive and (2,1) or (0,3)-type 
(i.e., imaginary self-dual) three-form fluxes $G^{(3)}$ in type IIB 
Calabi--Yau orientifold. It is bounded in region, because 
the contribution of such flux to the D3-tadpole 
\begin{equation}
 \frac{1}{2} G^{(4)}_{\rm tot} \cdot G^{(4)}_{\rm tot} \longrightarrow 
 \frac{i}{2} \frac{G^{(3)} \wedge \overline{G^{(3)}}}{ {\rm Im}(\phi) }
 = H^{(3)} \wedge F^{(3)}  
\end{equation}
is positive definite and is bounded from above by 
$L_*$, the total O3 plane charge in $M_3$.
The vacuum index density thus becomes a finite-valued distribution 
$(m,m)$-form on the moduli space ${\cal M}_*/\Gamma_*$. 

By following the argument of \cite{Ashok:2003gk,
Denef:2004ze}, it is not hard to realize that the integral 
over the finite $(K-2m)$-dimensional region within the flux space 
$d^K N$ yield a factor $L_*^{(K-2m)/2}$ for $z \in {\cal M}_*$. The Jacobian between the 
moduli space coordinates $z_a$ and the remaining real $2m$ coordinates 
of the continuous flux space $d^K N$ gives rise to another factor 
$(L_*)^{2m/2}$. Hence the vacuum index density of the landscape 
is given by 
\begin{equation}
 d\mu_I = \frac{1}{(K/2)! }(2 \pi L_*)^{K/2} \times \rho_{\rm ind.}, 
\label{eq:A-Den-Doug}
\end{equation}
where $\rho_{\rm ind.}$ is an $(m,m)$-form on ${\cal M}_*/\Gamma_*$. 
Note that $K$ and $m$ are distinguished intentionally; although 
the relation $K = 4m$ holds in type IIB Calabi--Yau orientifolds,  
it does not necessarily hold when one pays attention to a restricted 
subset of the full complex structure moduli space of $M_3$ 
(e.g., $K = 3$ and $m=1/2$ model for $M_3 = T^6$ in \cite{DeWolfe:2004ns}), 
or in applications to F-theory. 
The $(m,m)$-form $\rho_{\rm ind.}$ in (\ref{eq:A-Den-Doug}) does not 
depend on $L_*$ primarily\footnote{
It frequently happens, though, that the integral of 
$\rho_{\rm ind.}$ over ${\cal M}_*/\Gamma_*$ is not finite, and/or the
continuous approximation of the flux quanta 
(from (\ref{eq:index-density-pre}) to (\ref{eq:index-density})) 
in some regions of ${\cal M}_*/\Gamma_*$ becomes bad, and/or 
some regions of ${\cal M}_*/\Gamma_*$ correspond to decompactification 
``limits'' in dual theories. Because the cut-off for the region in 
${\cal M}_*/\Gamma_*$ for the continuous approximation depends on 
the value of $L_*$, the integral of $\rho_{\rm ind.}$ with cut-off 
may depend on $L_*$. See \cite{DeWolfe:2004ns} for such an example. }.
For this reason, the factor $(L_*)^{K/2} / (K/2)!$ roughly determines the
overall number of flux vacua in the landscape on ${\cal M}_*/\Gamma_*$ 
and the distribution within the moduli space ${\cal M}_*/\Gamma_*$ is
controlled by $\rho_{\rm ind.}$. 

\begin{figure}
 \begin{center}
\begin{tabular}{ccccc}
     \includegraphics[width=.2\linewidth]{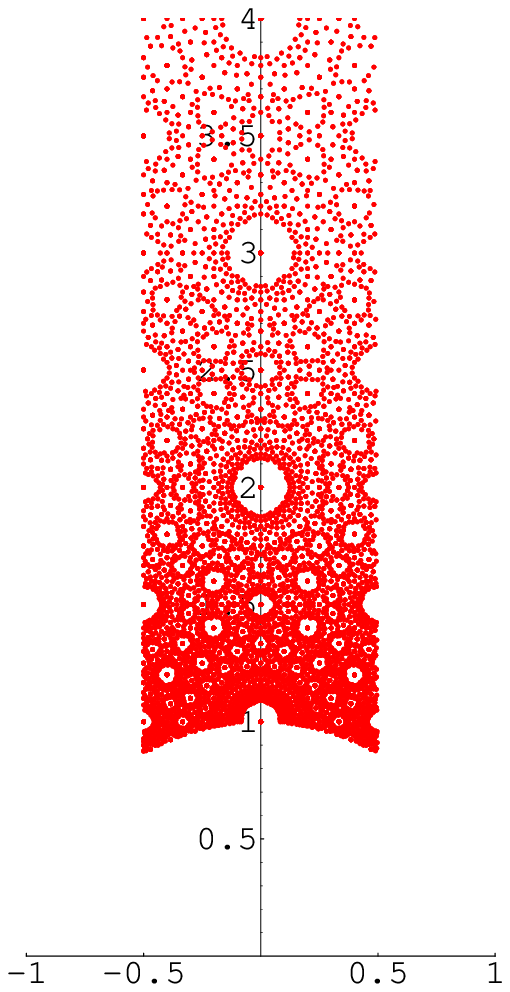}  & $\qquad$ &
     \includegraphics[width=.2\linewidth]{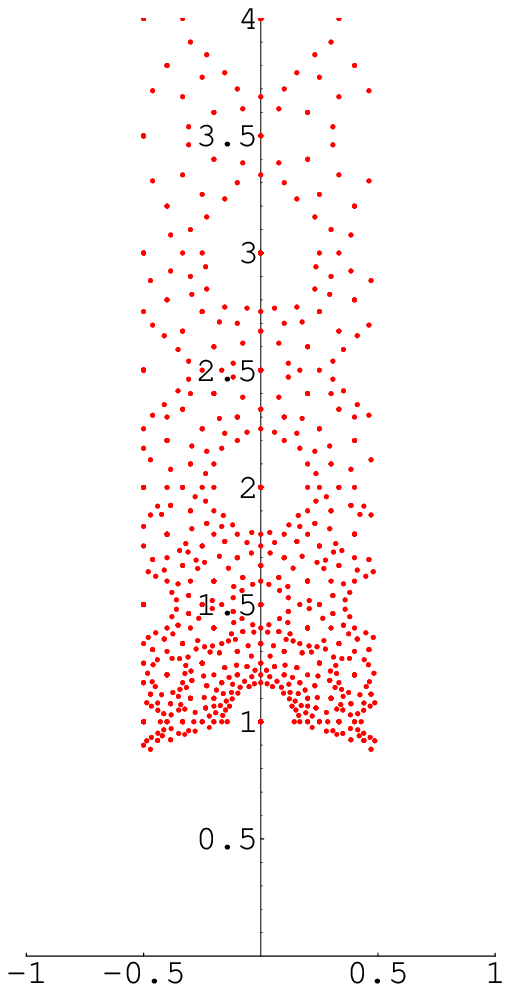}  & $\qquad$ &
     \includegraphics[width=.2\linewidth]{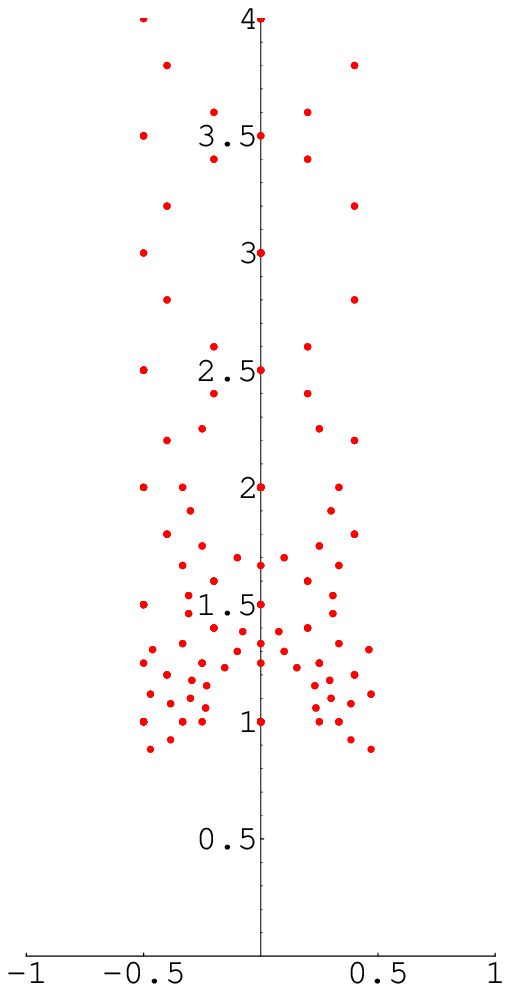}  \\
 $L_* = 150$ & & $L_* = 50$ & & $L_* = 20$
\end{tabular}
\caption{\label{fig:landscape-cont-discr} Vacuum distribution in the
rigid Calabi--Yau model of \cite{Ashok:2003gk,
Denef:2004ze, DeWolfe:2004ns} shown in the fundamental domain of the 
axi-dilaton moduli space. 
Depending on the value of $L_*$, D3-tadpole, the distribution of vacua 
in the moduli space can be either almost continuous or genuinely discrete. }
\end{center}
\end{figure}

Whether the continuous approximation is good or bad crucially 
depends on the value of $L_*$. To take an example, let us consider the 
rigid Calabi--Yau threefold model studied in \cite{Ashok:2003gk,
Denef:2004ze, DeWolfe:2004ns}, where $h^{2,1}(M_3)=0$ and the moduli space
${\cal M}$ is that of the axi-dilaton of type IIB string theory. The
distribution of flux vacua is presented in Figure 
\ref{fig:landscape-cont-discr} without making the continuous
approximation for three different values of $L_*$. 
For the one with $L_* = 150$, which is also found in \cite{Denef:2004ze}, 
the continuous approximation looks reasonably good. 
For cases with small value of $L_*$, however, the continuous
approximation is not very good, as in Figure
\ref{fig:landscape-cont-discr} with $L_*=20$.

When the continuous approximation is not good, it is more appropriate 
to i) specify the set of points in ${\cal M}_*/\Gamma_*$ that admit 
integral fluxes, and ii) describe how many choices of such integral flux 
configurations are available at such points \cite{DeWolfe:2004ns}. 
Suppose the dilaton vev $\vev{\phi}$ takes its value in an algebraic 
extension field ${\cal F}$ over $\Q$ with 
${\cal D} := {\rm dim}_\Q {\cal F} = 2$, and the complex structure 
moduli of $M_3$ are such that $[\Omega_{M_3}] \in {\cal F}\P[H^3(M_3)]$
and $[D_i \Omega_{M_3}] \in {\cal F} \P[H^3(M_3)]$ for 
$i=1, \cdots, h^{2,1}(M_3)$. Then the number of flux quanta at our
disposal (while preserving the F-term conditions) is 
\begin{equation}
 \kappa = 4(h^{2,1}(M)_{-,{\rm prim.}} + 1) - 
         ({\cal D} = 2) \times (h^{2,1}(M)_{-,{\rm prim.}}+1), 
\label{eq:nmbr-flx-quanta-eachpt-IIB}
\end{equation}
which is reduced to 
\begin{equation}
 \kappa' = 4(h^{2,1}(M)_{-,{\rm prim.}} + 1) - 
          ({\cal D} = 2) \times (h^{2,1}(M)_{-,{\rm prim.}}+1) 
        - ({\cal D}=2)  
\label{eq:nmbr-flx-quanta-eachpt-IIB-W0}
\end{equation}
when $\vev{W_{\rm GVW}} = 0$ is required. Consequently, the number 
of flux configuration scales for a given complex structure 
$(\vev{\phi}, \vev{z^a})$ as $(L_*)^{\kappa/2}$ or 
$(L_*)^{\kappa'/2}$, respectively \cite{DeWolfe:2004ns}. 
The overall number of flux vacua, estimated to be 
$(L_*)^{2(h^{2,1}(M)_{-,{\rm prim.}}+1)}$, should be reproduced  
partially from $(L_*)^{\kappa/2}$ times the number of such ${\cal D}=2$
vacua in the fundamental domain of the moduli space, and the rest must
come from similar contributions from vacua with different value of the 
extension degree ${\cal D}$.

The $rk_7 = 16$ landscapes of F-theory on  ${\rm K3} \times {\rm K3}$ in Section \ref{ssec:rank-16} 
fits very well with this discrete picture in two different ways. 
First, $L_*$ in the type IIB language comes from $\chi(Y_4)/24$ in
F-theory, which remains small for $Y_4= {\rm K3} \times {\rm K3}$. 
This is visible most clearly in Figure \ref{fig:gamma-lattice}, 
or in the relatively short list of vacuum points (there are 66) in the moduli space 
in Table \ref{tab:AK-notmorethan24} or merely 170 pairs found in the
analysis of Section \ref{ssec:rank-16}. These results definitely look closer to 
the $L_*=20$ picture in Figure~\ref{fig:landscape-cont-discr} than 
the one with $L_*=150$. 

The second reason is that the number of scanning flux quanta $\kappa$ 
available at each isolated point in ${\cal M}_*(J_S, W_{\rm noscan})$ 
remains the same for all the $W_{\rm noscan}$-landscapes for 
F-theory on ${\rm K3} \times {\rm K3}$ with $rk_7 = {\rm rank}(W_{\rm noscan}) = 16$. 
This is obvious in the case of the 
$W_{\rm noscan} = (D_4^{\oplus 4}); (\Z_2 \times \Z_2)$-landscape,
because this is the type IIB orientifold compactification on 
$M_3 = T^2 \times K3$ with all the O7-planes accompanied by four
D7-branes on top of them. But, even for other choices with 
${\rm rank}(W_{\rm noscan}) = 16$, the number of scanning flux quanta 
$\kappa$, and hence the estimate of the number of ${\cal D}=2$ vacua 
should remain much the same, even if $W_{\rm noscan}$ contains
$E_{6,7,8}$ type gauge group.
%
The number of flux quanta freely scanned over for a given K\"{a}hler 
form (\ref{eq:Kahler-4-F-theory}) and a rank-16 $W_{\rm noscan}$
was
\begin{equation}
 \kappa = 8 + 2 \times 17, \qquad 
 \kappa' = 6 + 2 \times 17  
\end{equation}
in the study in Section \ref{ssec:rank-16}, which agrees with the 
type IIB orientifold value (\ref{eq:nmbr-flx-quanta-eachpt-IIB}, 
\ref{eq:nmbr-flx-quanta-eachpt-IIB-W0}) for 
$h^{2,1}(T^2 \times {\rm K3})_{-,{\rm prim.}} = 1 + 19$.

The action of the modular group $\Gamma_*$ is implemented in the
continuous approximation of \cite{Ashok:2003gk, Denef:2004ze} by simply 
restricting the complex $m$-dimensional restricted moduli space 
${\cal M}_*$ to its fundamental region. If $L_*$ is small and one is in a situation of maintaining 
the discrete approach, one can still restrict the space ${\cal
M}_*$ to its fundamental region under the action of the modular group $\Gamma_*$.
Furthermore, one has to take a quotient of integral flux configurations 
admitted for a given point $[z] \in {\cal M}_*/\Gamma_*$ by 
the residual modular group $\Gamma_*(z)$, the stabilizer subgroup of 
a representative point $z \in {\cal M}_*$ \cite{DeWolfe:2004ns}. 
In the example of type IIB on a rigid Calabi--Yau 
threefold \cite{Ashok:2003gk, Denef:2004ze, DeWolfe:2004ns}, the first few 
axi-dilaton vevs for small flux contribution to the D3-tadpole sit 
at a special point in the axi-dilaton moduli space, $\vev{\phi} = i$,  
where the stabilizer subgroup of the moduli space 
$\Gamma_* = {\rm SL}(2; \Z)$ is non-trivial. Flux configurations  
for $\vev{\phi} = i$ have to be modded out by the non-trivial residual 
modular group.

Exactly the same phenomenon takes place in the case of compactification of F-theory on  ${\rm K3} \times {\rm K3}$. 
For any point $(\omega_X, \omega_S)$ in the moduli space $D^{(X)} \times D^{(S)}$, the stabilizer subgroup 
of $\Gamma$ in (\ref{eq:modular-group-F}) contains
\begin{equation}
  \left(
   \left[ W^{(2)}(S_X) \rtimes {\rm Aut}(X) \right] \times 
   \left[ W^{(2)}(S_S) \rtimes {\rm Aut}(S) \right]
 \right)\, .
\label{eq:temp-2}
\end{equation}
${\cal M}_*(J_S, W_{\rm noscan})$, however, further specifies an 
embedding of $[J_S] \in S_S \otimes \R/\R_{\geq 0}$ and 
$(U_* \oplus W_{\rm noscan}) \subset S_X$. Thus, 
only the subgroup of (\ref{eq:temp-2}) preserving this embedding 
remains as the residual modular group $\Gamma_*$ acting on the flux 
configuration.
None of the reflection subgroup $W^{(2)}(S_S)$ will be left in the 
residual modular group as long as $J_S$ is in the interior 
of the cone $\overline{{\rm Amp}}_S$. In our numerical study in 
Section \ref{ssec:rank-16} $J_S$ is sitting on a boundary 
of $\overline{{\rm Amp}}_S$, so that the $W_{E_8} \times W_{E_8}$ 
Weyl group in $W^{(2)}(S_S)$ is in the residual modular group $\Gamma_*$.
Similarly, most of $W^{(2)}(S_X)$ is also gone from the residual modular 
group because of the embedding of $U_* \oplus W_{\rm noscan}$.
At a point $\omega_X \in D^{(X)}$ where an extra non-Abelian factor 
of $W_{\rm root}$ emerges outside of $W_{\rm noscan}$, however, 
its Weyl group is still a part of $\Gamma_*$.
The automorphisms in ${\rm Aut}(X) \cap \Gamma_*$ at least 
preserve the elliptic fibration morphism and the zero-section, so 
this is a small subgroup of ${\rm Aut}(X)$.
We should also remark that our vacuum counting in 
Section \ref{ssec:sbtl-stat} exploited all possible ${\rm Aut}(S)$
to take a quotient of the set of $G_1$-type fluxes. Since the argument there 
does not refer to the choice of $J_S$, one either has to count 
varieties in $J_S$ or use only the subgroup of ${\rm Aut}(S)$ 
preserving $J_S \in \overline{\rm Amp}_S$ when taking a quotient of the flux
configurations. 

In landscapes where $L_*$ is small, there is a pronounced void structure 
in the vacuum distribution on ${\cal M}_*/\Gamma_*$, and a significant 
fraction of vacua are accumulated at the special points at the 
centre of the voids \cite{Denef:2004ze, DeWolfe:2004ns}. 
Sample statistics in Section \ref{ssec:sample-stats} and, in particular, 
Section \ref{ssec:rank-16} must have been strongly influenced by this effect. 
Especially when it comes to the fraction of symmetric vacua in the
statistics (such as the fraction of CP preserving vacua in 
Section \ref{ssec:sample-stats}), the small value of $L_*$ from 
$\chi(Y_4)/24 = 24$ must have a strong impact. Keeping this in mind, 
one should not view the high fraction of CP symmetric vacua 
as a generic prediction of F-theory.  

\vspace{5mm}

{\bf II} From the perspective of F-theory, there is no reason to 
focus our attention only to $W_{\rm noscan}$-landscapes with 
$rk_7 = {\rm rank}(W_{\rm noscan}) = 16$. It is thus natural to discuss 
flux vacuum distribution on the restricted moduli space 
${\cal M}_*(J_S, W_{\rm noscan})$. In the rest of this section, we treat 
$\chi(Y_4)/24$ (or $L_*$) as if it were a free parameter. 
By doing so, we can get a feeling for how the vacuum distribution depends on such
parameters as $rk_7$---the rank of fixed 7-brane gauge groups---in 
F-theory compactification on general elliptic Calabi--Yau fourfolds 
(not necessarily  ${\rm K3} \times {\rm K3}$). With this in mind, we use the continuous
approximation to the space of flux quanta \cite{Ashok:2003gk,
Denef:2004ze, DeWolfe:2004ns}.

For  ${\rm K3} \times {\rm K3}$, the restricted moduli space ${\cal M}_*(J_S, W_{\rm noscan})$ is of 
complex dimension $m=(18-rk_7)+19$. 
The vacuum index density (\ref{eq:index-density}) of
\cite{Ashok:2003gk, Denef:2004ze} in this set-up becomes an $(m,m)$-form 
on ${\cal M}_*$. Summation over the flux configuration is replaced by an
integral, $d^K N$, and the delta functions (F-term condition) remove 
the integral over the $(3,1)+(1,3)$ Hodge components of the flux.
Since we restrict the flux to be orthogonal to $W_{\rm noscan}$, 
there are only complex $m = (20-2-rk_7) + (20-1)$ dimensions of such 
$(3,1)$ components to begin with, and hence all the integrals of the 
$(3,1)+(1,3)$ components are removed from $\int d^K N$ 
in (\ref{eq:index-density}).
There are still remaining integrals over real 
$K -2m = 2 \times 2 + (18-rk_7) \times 19$ dimensions of the space of 
flux configuration; contribution to the D3-brane tadpole is positive 
definite in this space. 
This argument will lead to the vacuum index distribution that is an 
$(m, m)$-form $\rho_{\rm ind.}$ on 
${\cal M}_*(J_S, W_{\rm noscan})/\Gamma_*$ multiplied by a factor 
$(L_*)^{K/2}$ with $K =[(20-rk_7) \times 21 ]$.
By requiring that $\vev{W_{\rm GVW}} = 0$, $K$ is replaced by $K' = K-2$.
Therefore, the number of vacua in the ensemble of 
a given $W_{\rm noscan}$ and $J_S$ roughly scales as 
$(L_*)^{-(21/2) \times rk_7}$ as a function of $rk_7$.

In order to study the ratio of the number of $\SO(10)$ vacua to 
$\SU(5)$ vacua in F-theory on ${\rm K3} \times {\rm K3}$ flux vacua, one should also address 
K\"{a}hler moduli stabilization, and the vacuum counting on the choice
of $J_S$ need to be used as the weight multiplied on top of the
distribution on ${\cal M}_*(J_S, W_{\rm noscan})/\Gamma_*$.
It will be arguable whether the  ${\rm K3} \times {\rm K3}$ compactification of F-theory 
is phenomenologically interesting enough to motivate such a laborious 
study. Unless the weight factor from K\"{a}hler moduli stabilization 
contains very strong $rk_7$-dependence, it seems very difficult to 
overturn the statistical ratio of $(L_*)^{-10.5}$ of $\SO(10)$ vacua 
to $\SU(5)$ vacua simply coming from the flux statistics for fixed $J_S$. 

In such cases as $W_{\rm noscan} = A_4$ or $D_5$ in $Y = X \times S =
{\rm K3} \times {\rm K3}$ compactification of F-theory, there exists 
a primitive embedding $\phi: U \oplus W_{\rm noscan}
\hookrightarrow \Lambda_{\rm K3}^{(X)} \cong H_2(X; \Z)$.
Furthermore, such an embedding is unique modulo 
${\rm Isom}^+(\Lambda_{\rm K3}^{(X)})$ due to 
Theorem 1.14.4 of \cite{Nik01} and Theorem 2.8 of \cite{Mor} (which is 
also quoted as Thm. $\epsilon$ in \cite{BKW-math}). Thus, the restricted 
moduli space ${\cal M}_*$ can be constructed out of a single
piece\footnote{In principle, it is possible that primitive embeddings 
of $(U \oplus W_{\rm noscan})$ into $\Lambda_{\rm K3}^{(X)}$ are not
unique modulo ${\rm Isom}^+(\Lambda_{\rm K3}^{(X)})$, depending on what 
$W_{\rm noscan}$ is. In such cases, the restricted moduli space ${\cal
M}_*/\Gamma_*$ consists of multiple connected components.} of the
restricted period domain 
\begin{equation}
 \P \left[ \left\{ \left. 
   \Omega_X \in (\Lambda_{\rm K3}^{(X)})^* \otimes \C \; \right| \; 
   \Omega_X \wedge \Omega_X = 0, \;
   \langle \Omega_X, \phi(U \oplus  W_{\rm noscan}) \rangle = 0, \;
   \Omega_X \wedge \overline{\Omega}_X > 0
    \right\} \right] .
\end{equation}
If an attractive K3 surface $X = X_{[a~b~c]}$ admits an elliptic
fibration whose $W_{\rm frame}$ contains a $W_{\rm noscan}$ as above, 
its $\Omega_X$ (equivalently a primitive embedding of its $T_X$
into $[\phi(U \oplus W_{\rm noscan})^\perp \subset \Lambda_{\rm
K3}^{(X)}]$) should be found in this single piece of restricted period 
domain. Such an embedding of $T_X$, however, is no longer unique under 
the subgroup of ${\rm Isom}^+(\Lambda_{\rm K3})$ preserving the embedding 
$\phi: (U \oplus W_{\rm noscan}) \hookrightarrow \Lambda_{\rm
K3}^{(X)}$. In Sections \ref{ssec:J1-J2-phys} and 
\ref{sssec:counting-F-practice}, we discussed multiplicities of elliptic 
fibration on a given attractive K3 surface $X$ and a given isometry 
class of frame lattice. This multiplicity appears as a part of the
non-uniqueness of the embedding of $T_X$ into 
$[\phi(U \oplus W_{\rm noscan})^\perp \subset \Lambda_{\rm K3}^{(X)}]$
modulo the remaining subgroup of ${\rm Isom}^+(\Lambda_{\rm K3})$.

\vspace{5mm} 

{\bf III} Having seen how the vacuum statistics depend on the rank of 
gauge groups, the next question of interest will be how it 
depends on the number of generations of matter fields. Since it is
impossible to generate a non-zero net chirality on non-Abelian 7-branes 
in the  ${\rm K3} \times {\rm K3}$ set-up, we content ourselves with studying the dependence 
on the number of vector-like pairs on a non-Abelian 7-brane. 
As we have seen in Section \ref{ssec:MW-singl-fbr-flux}, the singular 
fibre fluxes generate some vector-like pairs of matter fields, while 
reducing the symmetry group on the non-Abelian 7-brane associated 
with the singular fibre. Thus, we set the problem as follows. 
For some choice of $J_S$ and $W_{\rm noscan} \subset W_{\rm frame}$ as
before, let us specify 
\begin{equation}
   G_{\rm fix} = \sum_A (C_A \otimes F_A), \qquad 
   C_A \in W_{\rm noscan} \cap W_{\rm root}, \qquad
   F_A \in [J_S^\perp \subset S_S]. 
\end{equation}
An ensemble of flux vacua is generated, by allowing the four-form flux 
$G^{(4)}$ to be $G_{\rm scan} + G_{\rm fix}$ for any $G_{\rm scan}$ 
orthogonal to $J_S$ and $W_{\rm noscan}$. The vacuum index density 
distribution of such a landscape is obtained as an $(m,m)$ form 
over the restricted moduli space ${\cal M}_*(J_S, W_{\rm
noscan})/\Gamma_*$. Landscapes with different $G_{\rm fix}$ share 
the same restricted moduli space as long as $W_{\rm noscan}$ and $J_S$ 
remain the same.
When we take $W_{\rm noscan}$ to be $E_7$, $E_6$ or $D_5$, for example, 
and $G_{\rm fix} = (C_A \otimes F_A)$ so that the unbroken symmetry becomes 
$E_6$, $D_5$ or $A_4$, the singular fibre flux $G_{\rm fix}$ determines 
the number of vector-like pairs in the ${\bf 27}+\overline{\bf 27}$, 
${\bf 16}+\overline{\bf 16}$ and ${\bf 10}+\overline{\bf 10}$ 
representations, respectively, through (\ref{eq:nmbr-vctr-like-pair}).
By comparing the numbers of vacua of the ensembles of 
$(J_S, W_{\rm noscan}; G_{\rm fix})$ with the common $(J_S, W_{\rm
noscan})$ and different $G_{\rm fix}$, we can determine the statistical 
cost of leaving vector-like pairs of matter fields in certain classes
of representations in the effective theory.\footnote{The net 
chirality would be proportional to the first power of the four-form flux, 
while the number of vector-like pairs scales as a square of the flux. This 
difference should be kept in mind. Note also that the number of
vector-like pairs of matter fields may also depend on the representation, even 
for a given non-Abelian gauge group. Thus, it is dangerous to extract 
too many lessons out of this result.}

The result is simple. Our estimate of the number of vacua in such an ensemble is given by 
\begin{equation}
 (L_{*,{\rm eff}.})^{K'/2}; \qquad L_{*, {\rm eff.}} = 
   \frac{\chi(X \times S)}{24} - \frac{1}{2}(G_{\rm fix})^2,
\label{eq:def-L-eff-on-Gfix}
\end{equation}
replacing $L_* = \chi(Y)/24$ by the remaining D3-tadpole $L_{*,{\rm
eff.}}$ to be cancelled by the flux other than $G_{\rm fix}$. 
The more the number of vector-like pairs, the less the effective value 
$L_{*,{\rm eff}.}$. 
If $L_* = \chi(Y)/24$ were fairly large (and $K'/2$ is not particularly large), 
then requiring one or two vector-like pairs of the matter field does not 
reduce the number of vacua too much, relatively. If $L_*$ is 
not particularly large, then the number of vacua with a few more 
vector-like pairs of matter fields becomes much smaller. Clearly, this effect is further enhanced
when $K'/2$ is large.

\vspace{5mm}

{\bf IV} Finally, let us study the $(m,m)$-form distribution 
$\rho_{\rm ind.}$ over ${\cal M}_*/\Gamma_*$ in the context of 
F-theory compactification, not just the overall number of flux vacua. 
We begin with a review of what is known about $\rho_{\rm ind}$ 
in \cite{Ashok:2003gk, Denef:2004ze, Denef:2008wq}.  

The most robust result on $\rho_{\rm ind}$ states that, for F-theory compactification without much restriction on 
the space of scanning four-form fluxes $G^{(4)}_{\rm scan}$,
$\rho_{\rm ind}$ is written in the form of the Euler class $e(\nabla)$ 
on ${\cal M}_*$ (which means that it is a differential $2m$-form)
associated with a connection $\nabla$ on a real vector bundle with 
rank $2m$ \cite{Denef:2008wq}.  The formula \cite{Ashok:2003gk, Denef:2004ze}
\begin{equation}
 \rho_{\rm ind} = 
   {\rm det} \left( - \frac{R}{2\pi i} + \frac{\omega}{2\pi} {\bf 1} \right)_{m \times m}
  = e (T{\cal M}_* \otimes {\cal L}^{-1}) = c_n(T{\cal M}_* \otimes {\cal L}^{-1}) , 
\label{eq:ADD-formula}
\end{equation}
is a special case of the more robust result $\rho_{\rm ind} = e(\nabla)$ \cite{Denef:2008wq}.
Here, $\omega$ is the K\"{a}hler form and $R$ the curvature $(1,1)$ form of the holomorphic vector bundle $T{\cal M}_*$,
${\cal L}$ is the line bundle whose first Chern class is $-\omega/(2\pi)$ and $c={\rm det}(-kR/(2\pi i)+{\bf 1})=\sum_k c_k t^k$
defines the Chern classes. It takes an extra effort to find for which real vector bundle 
$\rho_{\rm ind} = e(\nabla)$ for general cases. \\

\noindent
It is known that the formula (\ref{eq:ADD-formula}) can be used at
least for two categories of landscapes. 
\begin{itemize}
\item [A]: type IIB on a Calabi--Yau 
orientifold $M_3$ with full scanning of three-form fluxes $G^{(3)}_{\rm scan}$ and with all the D7-branes appearing 
as an $\SO(8)$ configuration \cite{Denef:2004ze}, i.e. four D7-branes on each O7-plane.
\item [B]: F-theory compactification on a Calabi--Yau fourfold $Y_4$ with the four-form flux 
scanning in a sufficiently large space. 
\end{itemize}
An example of category B is given by \cite{Denef:2008wq}:
\begin{equation}
 G^{(4)}_{\rm scan} \in
    \left[H^{4,0}(Y; \C) + {\rm h.c.} \right]
 \oplus \left[H^{3,1}(Y; \C)_{\rm prim.} + {\rm h.c.} \right]
 \oplus \left[\left( H^{2,2}(Y; \R)_V \right)^\perp \subset H^{2,2}(Y;
	 \R) \right]\, .  
\label{eq:scan-all-nonvertical}
\end{equation}
Here, $H^{(2,2)}(Y; \R)_V$ is the subspace of $H^{2,2}(Y; \R)$ spanned
by the intersection of any pair of divisors of $Y$ (this naive definition is 
made more precise shortly).

Let us explain what we mean by a ``sufficiently large space'' when we introduced category B.
Let $V \subset H^4(Y;\R)$ be the subspace in which the four-form is 
scanned to generate a landscape, and let $\{\psi_{I=1,2,\cdots} \}$ be a basis 
of $V$. The assumption made (implicitly) in \cite{Denef:2008wq} is that 
the vector space $V$ is large enough that one can make a replacement 
\begin{equation}
  \left( \int_Y \psi_I \wedge \varphi \right) (A^{-1})^{IJ} 
  \left( \int_Y \psi_J \wedge \chi \right) \Longrightarrow 
  \int_Y \varphi \wedge \chi, \qquad 
   A_{IJ} := \int_Y \psi_I \wedge \psi_J 
\label{eq:cond-ADD-formula}
\end{equation}
for arbitrary differential forms $\psi$ and $\chi$ that belong to the vector space\footnote{See equations (6.16) 
and (6.17) and compare with (6.33) and (6.34) in \cite{Denef:2008wq}. }
\begin{equation}
 G^{(4)}_{\rm scan} \in
    \left[H^{4,0}(Y_z; \C) + H^{0,4}(Y_z; \C) \right] 
 \oplus \left[H^{3,1}(Y_z; \C)_* + H^{1,3}(Y_z; \C)_* \right]
  \oplus H^{2,2}(Y_z; \R)_{H*}\, .
\label{eq:space-G4-40-31-22H*sect}
\end{equation}
The quantities $[H^{3,1}(Y_z; \C)_*$ and $H^{2,2}(Y_z; \R)_{H*}$ are defined in the next paragraph and
\eqref{eq:H22-horizontal*-def}.

It is a natural generalization of the category B landscapes above to 
require extra divisors corresponding to extra 7-brane gauge groups,
i.e. introduce a $W_{\rm noscan}$ as done earlier in this 
section. The moduli space ${\cal M}_*(J_Y,W_{\rm noscan})$ is then reduced in dimensions from the full space of complex structure moduli 
of $Y$ compatible with a K\"{a}hler form $J_Y$. The $H^{3,1}+H^{1,3}$ components of the flux is then not only required to be primitive, 
$G^{(4)}_{\rm scan} \wedge J_Y = 0$, but the same condition is required with regard to the divisors in $W_{\rm noscan}$. 
Let $[H^{3,1}(Y_z; \C)_* + H^{1,3}(Y_z; \C)_*]$ denote the resulting smaller subspace. 
\label{pg:flx-31-cmp-*} The $(H^{2,2}_V)^\perp$ component
of the flux space is also reduced, because $H^{2,2}(Y_z; \R)_V$ is
larger in dimensions. This new $H^{2,2}(Y_z; \R)_V$ is denoted by 
$H^{2,2}(Y_z; \R)_{V*}$. We then have an ensemble of vacua with the four-form flux scanning over the vector space 
\begin{equation}
 G^{(4)}_{\rm scan} \in
    \left[H^{4,0}(Y; \C) + {\rm h.c.} \right]
 \oplus \left[H^{3,1}(Y; \C)_{*} + {\rm h.c.} \right]
 \oplus \left[\left( H^{2,2}(Y; \R)_{V*} \right)^\perp \subset H^{2,2}(Y;
	 \R) \right];  
\label{eq:scan-all-nonvertical-gen}
\end{equation}
Let us refer to these landscapes as category B'.

In order to state the relation between the category A landscapes and 
category B' landscapes, we need to prepare the following language.
Consider a family of Calabi--Yau fourfolds  
$\pi: {\cal Y} \longrightarrow {\cal M}_*$, where $Y_z: = \pi^{-1}(z)$
is the Calabi--Yau fourfold corresponding to $z \in {\cal M}_*$.
We have in mind a restricted moduli space ${\cal M}_*$ for a specific 
choice of K\"{a}hler form $J_Y$ and some set of divisors 
$W_{\rm noscan}$ corresponding to the 7-brane gauge groups.
The $H^{2,2}(Y_z; \R)$ vector space over $\R$ for each $z \in {\cal
M}_*$ is decomposed as follows. First, $H^{2,2}(Y_z; \R)_{V*}$ is 
defined\footnote{This (more precise) definition of $H^{2,2}(Y_z; \R)_V$ 
is not the same as the naive ``definition'' right 
after (\ref{eq:scan-all-nonvertical}), when $z \in {\cal M}_*$ 
is in a special locus of ${\cal M}_*$ so that there are more divisors
in $Y_z$ than in generic points of ${\cal M}_*$. Contributions from such 
extra divisors are not included in $H^{2,2}(Y_z; \R)_V$ in the precise 
definition.} 
as the subspace spanned by intersection of $Y_z$ with any pair of
divisors of ${\cal Y}$. Secondly, another subspace $H^{2,2}(Y_z; \R)_{H*} \subset H^{2,2}(Y_z; \R)$ 
is defined as 
\begin{equation}
 {\rm Span}_\C \left\{ (D_a D_b \Omega_Y), \overline{(D_cD_d \Omega_Y)}  
               \right\}|_{a, b,c,d \in 1, \cdots, m} \cap H^{2,2}(Y_z; \R)\, , \hspace{.5cm}m={\rm dim}\, {\cal M}_*\, . 
\label{eq:H22-horizontal*-def}
\end{equation}
It is known that \cite{Denef:2008wq}
\begin{itemize}
 \item [i)] the vector space of (2,2) Hodge components is 
divided into $H^{2,2}_{\ell=0} \oplus H^{2,2}_{\ell=1} \oplus H^{2,2}_{\ell=2}$
 \item [ii)] the intersection form is positive definite on 
$H^{2,2}_{\ell=0} \oplus H^{2,2}_{\ell=2}$ and negative definite on $H^{2,2}_{\ell=1}$
 \item [iii)]  $H^{2,2}_{\ell=1} \oplus H^{2,2}_{\ell=2}$ is contained in 
$H^{2,2}(Y_z; \R)_{V*}$
\end{itemize}
The differential forms in the first component, $H^{2,2}_{\ell=0}$ are the primitive $(2,2)$ 
forms. Thus, the intersection form is positive definite on the orthogonal complement 
$[(H^{2,2}(Y_z; \R)_{V*})^\perp \subset H^{2,2}(Y_z; \R)]$, and there is a 
well-defined orthogonal decomposition of the $H^{2,2}(Y_z; \R)$ vector space:
\begin{equation}
 H^{2,2}(Y_z; \R) = H^{2,2}(Y_z; \R)_{H*} \oplus H^{2,2}(Y_z ; \R)_{RM}
  \oplus H^{2,2}(Y_z; \R)_{V*}. 
\label{eq:dcmp-22-HVrm}
\end{equation}
Note that the remnant subspace $H^{2,2}(Y_z; \R)_{RM}$ is not empty 
in general \cite{Mor_com}. In particular, as explained in detail in Appendix \ref{sec:ADD},
this happens already in the case of $Y={\rm K3} \times {\rm K3}$. 
As discussed in Appendix \ref{sec:ADD}, the landscape of category A 
(roughly speaking, type IIB orientifolds with three-form 
scanning (\ref{eq:scan-3form-in-Fdescr})) 
corresponds to scanning the four-form in the $H^{2,2}(Y;\R)_{H*}$ subspace 
in addition to the first two components of (\ref{eq:scan-all-nonvertical}), 
i.e., the space (\ref{eq:space-G4-40-31-22H*}).
This is a smaller subspace than (\ref{eq:scan-all-nonvertical-gen}), 
in principle, in the sense that $H^{2,2}(Y;\R)_{RM}$ can be non-empty. 
This indicates that the scanning space of the flux can be 
sufficiently large, for the condition (\ref{eq:cond-ADD-formula}) to hold
(and the formula (\ref{eq:ADD-formula}) also holds consequently), even 
when the scanning space is smaller than (\ref{eq:scan-all-nonvertical-gen}). 

As discussed in Appendix \ref{sec:ADD}, the four-form scanning we have 
introduced in this section also corresponds to the vector space (\ref{eq:space-G4-40-31-22H*}). 
It is thus reasonable 
to think of one more category of landscapes in F-theory which contains 
both the category A landscapes and the ones on $Y={\rm K3} \times {\rm K3}$ 
with arbitrary $rk_7$ we have considered. Let us call category A' 
any landscapes generated by scanning four-form flux in the vector space 
(\ref{eq:space-G4-40-31-22H*}) for an elliptic fibred Calabi--Yau 
fourfold\footnote{A more appropriate way to phrase this is a family of 
elliptically fibred Calabi--Yau fourfolds $\pi: {\cal Y} \longrightarrow {\cal M}_*$ for
which a generic fibre is not necessarily in the form of  ${\rm K3} \times {\rm K3}$ or a 
K3-fibration.} $\pi_Y: Y \rightarrow B_3$.
This is to be contrasted with those in category B' where the four-form 
flux is scanned in (\ref{eq:scan-all-nonvertical-gen}).

It must now be obvious that the formula (\ref{eq:ADD-formula}) holds, 
not just for the landscapes in category B', but also for the landscapes in 
category A'. Since the condition (\ref{eq:cond-ADD-formula}) requires 
that $\psi_I (A^{-1})^{IJ} \psi_J$ be an insertion of a complete system only
for $\varphi, \chi$ in (\ref{eq:space-G4-40-31-22H*}), it is equivalent to 
say that the entire vector space (\ref{eq:space-G4-40-31-22H*}) is contained 
in the space of four-form scanning (or not).
Appendix \ref{sec:ADD} also provides an alternative (more down to earth)
derivation of the formula (\ref{eq:ADD-formula}) following the line of 
argument in one of the original articles \cite{Denef:2004ze}), rather than 
the refined version of \cite{Denef:2008wq}, with an assumption 
(\ref{eq:self-real-cond-22horizontal-4fold}) and a little more concrete 
consequence (\ref{eq:def-Riemann-curvature-analog-mirror}). We understand 
that the argument here is enough to justify the formula (\ref{eq:ADD-formula}) 
for landscapes in category A'. 

There is an immediate consequence of this observation. 
Any landscape (an ensemble of vacua) in category B' is decomposed into 
multiple landscapes (ensembles of vacua) in category A' labelled by 
$G^{(4)}_{\rm fix}$ in $H^{2,2}(Y; \R)_{\rm RM}$. Since all of these landscapes 
in category A' share the same restricted moduli space ${\cal M}_*$, 
and since the formula (\ref{eq:ADD-formula}) determines $\rho_{\rm ind}$ 
only in terms of geometry of ${\cal M}_*$, the vacuum index distribution 
$d\mu_I$ for these landscapes in category A' are the same apart form the 
$G^{(4)}_{\rm fix}$-dependent overall normalization (\ref{eq:def-L-eff-on-Gfix}).

Let us use (\ref{eq:ADD-formula}) to derive an explicit result on the
distribution of moduli parameters. Consider a landscape 
of F-theory compactification on $Y=X \times S = {\rm K3} \times {\rm K3}$ 
with some rank-16 $W_{\rm noscan}$ and $G^{(4)}_{\rm fix}$.   
Then the restricted moduli space is of the form 
${\cal M}_* = {\cal M}_{\rho_1} \times {\cal M}_{\rho_2} \times 
{\cal M}_{\rm K3}(S; J_S)$. This is because the lattice 
$[(U_* \oplus W_{\rm noscan})^\perp \subset H^2(X; \Z)]$ is always signature 
$(2, 2)$, and the intersection form can be made $U \oplus U$ for some 
$\R$-coefficient basis of the vector space obtained by tensoring $\R$ with  
the signature $(2, 2)$ lattice above. The K\"{a}hler forms of the moduli spaces 
${\cal M}_{\rho_1}$ and ${\cal M}_{\rho_2}$ are 
$K^{(\rho_1)} = - \ln [ (\rho_1-\bar{\rho}_1)/i]$ and 
$K^{(\rho_2)} = - \ln [ (\rho_2-\bar{\rho}_2)/i]$, respectively. 
The modulus $\rho_1$ is interpreted as the axi-dilaton of type IIB orientifold  
in the case of $W_{\rm noscan} = D_4^{\oplus 4}; (\Z_2 \times \Z_2)$. 
By closely following the computation of Section 3.1.2 in \cite{Denef:2004ze}, 
we obtain  
\begin{eqnarray}
 \rho_{\rm ind} & = &
   \left( -\frac{R^{(\rho_1)}}{2\pi i} + \frac{\omega^{(S)}}{2\pi}+\frac{\omega^{(\rho_1)}}{2\pi} + \frac{\omega^{(\rho_2)}}{2\pi}
   \right)
   \wedge 
  \left( -\frac{R^{(\rho_2)}}{2\pi i} + \frac{\omega^{(S)}}{2\pi}+\frac{\omega^{(\rho_1)}}{2\pi} + \frac{\omega^{(\rho_2)}}{2\pi}
  \right) \wedge \nonumber \\
&&  \qquad 
  {\rm det}\left( -\frac{R^{(S)}}{2\pi i} + 
            \left(\frac{\omega^{(S)}}{2\pi}+\frac{\omega^{(\rho_1)}}{2\pi} + \frac{\omega^{(\rho_2)}}{2\pi}\right){\bf 1} \right),  \\
  & = & 2 \frac{\omega^{(\rho_1)}}{2\pi} \wedge \frac{\omega^{(\rho_2)}}{2\pi} \wedge 
   \left\{ c_{m-2}^{(S)} + \left(\frac{\omega^{(S)}}{2\pi}\right)^2 \wedge c_{m-4}^{(S)} \right\}, 
\end{eqnarray}
where $\omega^{(\rho_1)}$, $\omega^{(\rho_2)}$ and $\omega^{(S)}$ are
the K\"{a}hler forms on the moduli spaces ${\cal M}_{\rho_1}$, 
${\cal M}_{\rho_2}$ and ${\cal M}_{\rm K3}(S; J_S)$, respectively, 
and $R^{(S)}$, $R^{(\rho_2)}$ and $R^{(\rho_1)}$ the curvature 
$(1,1)$-forms of the tangent bundles of those moduli spaces \footnote{They satisfy $R^{(\rho_i)}=2i\omega^{(\rho_i)}$.}.
$c_r^{(S)}$'s are the $r$-th Chern class of the holomorphic rank-$(m-2)$ vector 
bundle $T{\cal M}_{\rm K3}(S; J_S) \otimes ({\cal L}^{(S)})^{-1}$, where 
${\cal L}^{(S)}$ is the line bundle whose first Chern class is $-\omega^{(S)}/2\pi$.
Thus, in these set-ups the vacuum index density distribution $\rho_{\rm ind.}$ is 
factorized for the three pieces of the moduli space. 
If one is interested only in the distribution of any one among
$\rho_1$, $\rho_2$ and moduli of the K3 surface $S$, but not
altogether, then the distribution $\rho_{\rm ind}$ can be integrated 
over the irrelevant coordinates first. 
For the moduli $\rho_1$ and $\rho_2$, in particular, 
\begin{equation}
  \rho_{\rm ind.} (\rho_1) \propto
     d [{\rm Re}\rho_1] d [{\rm Im}\rho_1] \frac{1}{({\rm Im}(\rho_1))^2}, 
 \qquad \qquad 
  \rho_{\rm ind.} (\rho_2) \propto
     d [{\rm Re}\rho_2] d [{\rm Im}\rho_2] \frac{1}{({\rm
     Im}(\rho_2))^2}.  
\label{eq:ADD-predict-phi-rho2-(K3)}
\end{equation} 
Applying this result to the $W_{\rm noscan} = E_8 \oplus
E_8$-landscape, in particular, the prediction in the continuous
approximation (\ref{eq:ADD-predict-phi-rho2-(K3)}) for $\rho_2$ can also 
be read as that of $[{\rm vol}(T^2)/\ell_s^2]_{\rm Het}$.
Our numerical results (not relying on continuous approximation) 
in Section \ref{ssec:rank-16} agree with this prediction based on the continuous 
approximation qualitatively in that large 
${\rm Im}(\tilde{\rho}_H) = [{\rm vol}(T^2)/\ell_s^2]_{\rm Het}$ is 
statistically disfavoured. Note that this happens although we should not expect the continuous 
approximation to be very good because $\chi(Y)/24=24$ is not particularly large.

The result (\ref{eq:ADD-predict-phi-rho2-(K3)}) also indicates 
that there is no correlation between the distribution of $\rho_1$,
$\rho_2$ and the moduli of K3 surface $S$, if we ignore the difference 
between the vacuum distribution and the vacuum index distribution. 
Figure \ref{fig:vsA} (iii) in Section \ref{ssec:rank-16} may be regarded 
as a manifestation of the absence of any correlation between $\rho_1$ and $\rho_2$.

It is an interesting question how $\rho_{\rm ind}$ depends on 
${\rm Im}(\tilde{\rho}_H)$ for different values of $rk_7$. $\rho_{\rm ind}$
is not expected to have a factorized form as 
(\ref{eq:ADD-predict-phi-rho2-(K3)}), but it will be in the form of 
\begin{equation}
 \rho_{\rm ind} = c_{18-rk_7}^{(X)} c_{19}^{(S)} 
   + \omega^{(X)} \omega^{(S)} c_{17-rk_7}^{(X)} c_{18}^{(S)} 
   + (\omega^{(X)} \omega^{(S)} )^2 c_{16-rk_7}^{(X)} c_{17}^{(S)} + \cdots, 
\end{equation}
where $c_r^{(X)}$'s are the $r$-th Chern class of the rank-$(18-rk_7)$ 
holomorphic vector bundle $T{\cal M}_{\rm K3}(X; U_* \oplus W_{\rm noscan}) \otimes {\cal L}_{(X)}^{-1}$.
It must still be possible to extract the leading power-law behaviour in 
${\rm Im}(\tilde{\rho}_H)$ in the large ${\rm Im}(\tilde{\rho}_H)$ region 
of the moduli space, using the parametrization (\ref{eq:Narain-parametrization-full}). 
We leave this as an open problem in this article.

\subsection*{Acknowledgements}

We thank
Kazuo Hosomichi for useful comments, and 
Tetsuji Shioda for giving a lecture on Mordell--Weil lattice on K3 surface, 
in January 2010 at IPMU, Tokyo.
A.~P.~B. likes to thank the IPMU Tokyo for kind hospitality during his
visit in summer of 2012, which initiated this collaboration.
The work of A.~P.~B. was supported by a JSPS postdoctoral fellowship 
under grant PE 12530, and by the STFC under grant ST/J002798/1
and that of T.~W. by WPI Initiative, MEXT, Japan 
and a Grant-in-Aid for Scientific Research on Innovative Areas 2303. 

\appendix

\section{A Note on Heterotic String Narain Moduli}
\label{sec:Het-Narain}

Although Narain compactificaiton of heterotic string theory is a well-known
subject, we leave a brief note here for summary of conventions used 
in the main text of this article. 

A compactification of heterotic string theory on $T^2$ is specified by embedding  
an even self-dual lattice ${\rm II}_{2,18}$ in a space 
\begin{equation}
 \R^{2,18} = \left\{ \sqrt{\frac{\alpha'}{2}} 
   (k^R_8, k^R_9, k^L_8, k^L_9, k_{i=11, \cdots, 26})^T \right\}
\end{equation}
where the metric on $\R^{2,18}$ is
$\diag({\bf 1}_{2\times 2}, {\bf 1}_{18 \times 18})$. Let 
$\{ e_{K8}, e_{\bar{w}8}, e_{K9}, e_{\bar{w}9}, e_{I=11, \cdots, 26} \}$
be a set of generators of ${\rm II}_{2,18}$ (as well as its image in $\R^{2,18}$)
where $U = {\rm Span}_Z\{e_{K8}, e_{\bar{w}8}\}$, 
$U={\rm Span}_\Z\{e_{K9}, e_{\bar{w}9}\}$, and $E_8 \oplus E_8$ is generated by 
the rest, $\{e_{I=11, \cdots,26}\}$. Thus, the data of this compactification 
is written in the form of a $(2+18) \times (2+2+16)$ matrix, 
\begin{equation}
 {\cal Z} = \left[ 
      e_{K8}, e_{\bar{w}8}, e_{K9}, e_{\bar{w}9}, e_{I=11}, \cdots, e_{I=26} \right], 
  \qquad  
{\cal Z}^T \cdot 
  \left( \begin{array}{cc} {\bf 1}_{2\times 2} & \\ & - {\bf 1}_{18 \times 18}
         \end{array} \right) \cdot 
{\cal Z} =
 \left( \begin{array}{cccc} 
     U &   &        &    \\
       & U &        &    \\
       &   & C_{E_8} &    \\
       &   &        & C_{E_8} 
   \end{array} \right),  
\label{eq:constraint-Narain-data}
\end{equation}
where the $U$ in \eqref{eq:constraint-Narain-data} denotes the matrix 
$\left[ \begin{array}{cc} & 1 \\ 1 & \end{array} \right]$ and 
$C_{E_8}$ the negative of the Cartan matrix of $E_8$, which is also the 
(negative definite) intersection form of the $E_8$ root lattice. 
General elements of ${\rm II}_{2,18}$ are written in the form of 
$n_8 e_{K8} + (-w^8) e_{\bar{w}8} + n_9 e_{K9} + (-w^9) e_{\bar{w}9} + 
\sum_I q^I e_I$, which is also denoted by $(n_8, -w^8, n_9, (-w^9), q^I)^T$ 
in the component description. The generators $e_{K8, K9}$ correspond to 
states with elementary Kaluza--Klein excitation in $T^2$, and 
$e_{\bar{w}8, \bar{w}9}$ to states winding $T^2$ once. 

The first two rows of the matrix ${\cal Z}$ are denoted by ${\cal Z}^R_m$ 
with $m=8,9$. Introducing ${\cal Z}^R := {\cal Z}^R_8 + i {\cal Z}^R_9$, it follows
from relation (\ref{eq:constraint-Narain-data}) that 
\begin{equation}
 {\cal Z}^R \cdot
 \left( \begin{array}{cccc} 
   U^{-1} & & & \\ & U^{-1} & & \\ & & (C_{E_8})^{-1} & \\ & & & (C_{E_8})^{-1}
 \end{array} \right) \cdot \left({\cal Z}^{R} \right)^T = 0\, .
\end{equation}
Because ${\cal Z}^R$ can be regarded as an element of 
${\rm Hom}({\rm II}_{2,18}, \C)$, 
\begin{equation}
 {\cal Z}^R: (n_8, -w^8, n_9, -w^9, q^I)^T \longmapsto 
  \sqrt{\frac{\alpha'}{2}} (k^R_8 + i k^R_9) \; {\rm component~of} \; 
   (e_{K8} n^8 + e_{\bar{w}8} (-w^8) + \cdots + \sum_I e_I q^I), 
 \nonumber 
\end{equation}
we see that the relation above can also be written as 
$({\cal Z}^R, {\cal Z}^R) = 0$ using the symmetric pairing of 
the dual lattice ${\rm II}_{2,18}^*$ naturally extended bilinearly 
to ${\rm II}_{2,18}^* \otimes \C$. With the same notation, it also follows 
that 
\begin{equation}
 ( {\cal Z}^R, \overline{{\cal Z}^R}) = +2.
\end{equation}

The moduli space of this Narain compactification is parametrized by 
18 complex numbers. With a parametrization that is understood intuitively 
in the supergravity approximation of heterotic $E_8 \times E_8$ string theory, 
$k^R_{m=8,9}$ in ${\cal Z}^R$ are written as follows \cite{polchinski}:
\begin{equation}
 k^R_m = \frac{n_m}{R_m} - \frac{R_m w^m}{\alpha'} - \sum_I q^I A_{I, m} 
   + \frac{1}{2}\sum_{I,J} A_{I, m} C^{IJ} (\sum_n w^n R_n A_{J,n}); 
 \qquad ({\rm no~summation~in~}m)
\end{equation}
$A_{I,m}$'s ($m=8,9$ and $I=11, \cdots, 26$) are the Wilson lines on $T^2$ 
for the simple roots of $E_8 \times E_8$, and $R_{m=8,9}$ the radii of 
the two directions of $T^2$. Together with two more $\R$-valued parameters 
that we have omitted here ($\vev{B} = 0$, and $T^2$ is rectangular, just 
for better readability), there are $2 \times 18$ parameters in total. 
$C^{IJ}$ is the inverse of the matrix $C_{E_8 \times E_8}$. Therefore, 
${\cal Z}^R$ is parametrized by 18 complex numbers. In the component description of 
$[U \oplus U \oplus E_8 \oplus E_8]^* \otimes \C$, it is given by
\begin{equation}
 {\cal Z}^R = \frac{i}{ \sqrt{
   2 {\rm Im}(\tau_H) {\rm Im}(\tilde{\rho}_H)  
 + {\rm Im}(a_P) C^{PQ} {\rm Im}(a_Q)} }
   ( - \tau_H, \; - \tilde{\rho}_H , \; 1, \;
        -\tau_H \tilde{\rho}_H - a_K C^{KL} a_L/2 , \; a_I ),
\label{eq:Narain-parametrization-full}
\end{equation}
where (under the condition that $\vev{B}=0$ and $T^2$ is rectangular)
\begin{equation}
 \tau_H = i \frac{R_9}{R_8}, \quad 
 \rho_H = i \frac{R_8R_9}{\alpha'}, \quad 
 a_I = iR_9 (A_{I,8}+iA_{I,9}) = 
  \sqrt{\alpha' \tau_H \rho_H} (A_{I,8} + i A_{I,9}), 
\end{equation}
and 
\begin{equation}
 \qquad 
 \tilde{\rho}_H = \rho_H - 
   \frac{a_I C^{IJ}(a-\bar{a})_J}{4i\,{\rm Im}(\tau_H)}. 
\end{equation}
%

\section{Heterotic Cartan Flux and Semistable Degeneration}
\label{sec:Km-E6E6}

\subsection{Set-up}

In this appendix, the discussion on heterotic--F-theory duality of Cartan/Mordell--Weil flux 
of Section \ref{ssec:MW-singl-fbr-flux} is made explicit by using the 
two-parameter family of K3 surfaces (\ref{eq:Inose-pencil}, \ref{eq:Inose-pencil-2}) for F-theory 
(and its heterotic dual) as an example. 

If we are to require a pair of ${\rm IV}^*$ singular fibres, 
this specifies a family of $\rho_X = 12 + 2=14$ K3 surfaces, whose 
moduli space has dimension 6. Let 
\begin{equation}
 T_X^{\rho=14} = U \oplus U \oplus A_{2} \oplus A_2, \qquad 
 W_{\rm frame}^{\rho=14} = E_6 \oplus E_6.
\label{eq:SU(3)xSU(3)-Het-rho14-TandS}
\end{equation}
The two parameter family of K3 surfaces, 
$X = {\rm Km}(E_{\rho_1} \times E_{\rho_2})$, has four more independent 
algebraic cycles, $\rho_X = 18$, and the transcendental lattice is only
of rank 4, $T_X^{\rho=18} = U[2] \oplus U[2]$. Such a special family of 
K3 surfaces can be identified by an embedding $T_X^{\rho=18}
\longrightarrow T_X^{\rho=14}$:
\begin{equation}
 (\overline{C}_{32}, \overline{C}_{14}, \overline{C}_{12}, \overline{C}_{43})
  \longmapsto  
 (v, V, v', V',\alpha^{\rm v}_1, \alpha^{\rm v}_2, 
    \alpha^{\rm h}_1, \alpha^{\rm h}_2) 
   \left[ \begin{array}{cc|cc}
        1 & 1 & & \\
          & 2 & & \\
       \hline 
          &   & 1 & 1 \\
          &   &   & 2 \\
       \hline
          & 1 &   &   \\
          &   &   & 1 \\
       \hline
          & -1&   &   \\
          &   &   & 1 
\end{array}
   \right],
\label{eq:KmEF-embed-rho14Tx}
\end{equation}
where $\overline{C}_{32,14,12,43}$ are generators of 
$T_X^{\rho=18} = U[2] \oplus U[2]$, $\{v, V,v', V'\}$ those of 
$U \oplus U \subset T_X^{\rho=14}$, and $\{\alpha^{\rm v}_{1,2}\}$ and 
$\{\alpha^{\rm h}_{1,2}\}$ those of the structure group $A_2 = \SU(3)$
in the visible and hidden sectors, respectively. 

In this two parameter family of K3 surfaces (Kummer surfaces), 
the four extra independent algebraic cycles are the generators of the 
orthogonal complement of the image of $T_X^{\rho=18}$ within
$T_X^{\rho=14}$. This lattice is also regarded as the orthogonal
complement of the $E_6 \oplus E_6$ lattice within the frame lattice 
$W^{\rho=18}$ for $\rho=18$. Hence this is also the essential
lattice $L(X) = W_{{\rm U}(1)}$ of this elliptic fibration. 
$L(X)=W_{{\rm U}(1)}$ is generated by 
\begin{equation}
 (P, Q, P', Q')   
 = (v, V, v', V',\alpha^{\rm v}_1, \alpha^{\rm v}_2, 
    \alpha^{\rm h}_1, \alpha^{\rm h}_2) 
   \left[ \begin{array}{cc|cc}
        & 2 &   & -1 \\  & & & \\ \hline
     -1 &   & 2 &    \\  & & & \\ \hline
      1 & 1 &   &    \\
        &   & 1 & 1  \\    \hline
      1 &-1 &   &    \\
        &   & 1 & -1       
   \end{array}  \right].
\label{eq:Het-flux-data-Km-E6E6}
\end{equation}
The period integral is therefore in the form of 
\begin{equation}
 {\cal Z}^R \propto \Omega_X = \left( -\tau_H, -\tilde{\rho}_H; 
    1, -\tau_H \tilde{\rho}_H + \frac{\tau^2_H+1}{2};  
    \tau_H+\frac{1}{2}, -\left(\frac{\tau_H}{2}+1\right), 0^6; 
    -\left(\tau_H-\frac{1}{2}\right), \frac{\tau_H}{2}-1,0^6 \right)
\label{eq:Het-Narain-Km-E6E6}
\end{equation}
in the component description of 
$[U \oplus U \oplus E_8 \oplus E_8]^* \otimes \C$.  
In the component description using $(T_X^{\rho=18})^* \otimes \C$ 
(and irrelevant $(W_{{\rm U}(1)} \oplus E_6 \oplus E_6)^* \otimes \C$), 
this becomes 
\begin{equation}
 \left( -\tau_H, -(2\tilde{\rho}_H - \tau_H), 
        1, - \tau_H (2\tilde{\rho}_H - \tau_H) \right).
\end{equation}
Thus, in the parametrization of $\Omega_X$ in terms of $\tau_H$ and 
$\tilde{\rho}_H$, the K3 surface 
$X = {\rm Km}(E_{\rho_1} \times E_{\rho_2})$ varies as 
\begin{equation}
 \rho_1 = \tau_H, \qquad \rho_2 = 2 \tilde{\rho}_H - \tau_H.
\label{eq:Het-F-moduli-map-Km-E6E6}
\end{equation}
Obviously the parametrization (\ref{eq:Het-Narain-Km-E6E6}) follows 
the convention of Narain moduli for heterotic string compactification, 
and the correspondence (\ref{eq:Het-F-moduli-map-Km-E6E6}) should be 
read as the duality map between the coordinates of the heterotic and 
F-theory moduli spaces. 

In the heterotic string language, the component description of 
${\cal Z}^R$ in (\ref{eq:Het-Narain-Km-E6E6}) is a manifest consequence 
of the condition (\ref{eq:Het-ZdotG}), with the flux data $n_A$'s in 
(\ref{eq:Het-flux-data}) given by the $k=4$ column vectors in 
(\ref{eq:Het-flux-data-Km-E6E6}). The Wilson lines are constrained to be 
torsion points in ${\rm Jac}(T^2)$ of the heterotic compactification:
\begin{equation}
 \diag \left(\frac{\tau_H}{2}, - \frac{\tau_H+1}{2}, \frac{1}{2}
       \right) \subset \mathfrak{su}(3)^{\rm vis.}, \qquad 
 \diag \left(- \frac{\tau_H}{2},\frac{\tau_H-1}{2}, \frac{1}{2} \right)
        \subset \mathfrak{su}(3)^{\rm hid.}.
\end{equation}
Wilson lines in the dimension-3 representations of the structure group 
$\SU(3)^{\rm vis.} \times \SU(3)^{\rm hid.}$ take values in 2-torsion 
points in this case, which is also the direct consequence of the spectral 
surface equations $a^{\rm v}_3 y = 0$ for the visible sector and 
$a^{\rm h}_3 y = 0$ for the hidden sector.

\subsection{Mordell--Weil Group and Narrow Mordell--Weil Group}

In the two parameter family of K3 surfaces $X = {\rm Km}(E_{\rho_1}
\times E_{\rho_2})$ with an elliptic fibration of type 
$2{\rm IV}^* + 8I_1$, the essential lattice is 
\begin{equation}
 L(X)= W_{{\rm U}(1)} \cong A_2[2] \oplus A_2[2] \cong 
  {\rm Span}_\Z\{P, P'\} \oplus {\rm Span}_\Z \{Q, Q' \}.
\end{equation}
$E_6 \oplus E_6$ forms the root lattice $W_{\rm root}$ of the frame
lattice $W_{\rm frame}$ for generic $\rho_1$ and $\rho_2$, and 
the frame lattice is obtained by adding, to 
$W_{\rm root} \oplus W_{{\rm U}(1)}$, such glue vectors as 
\begin{eqnarray}
 \frac{2P+P'}{3} + \omega^{\rm v}_{\bf 27} + \omega^{\rm h}_{\bf 27}, \qquad 
 \frac{P+2P'}{3} + \omega^{\rm v}_{\overline{\bf 27}}
                 + \omega^{\rm h}_{\overline{\bf 27}} \\
 \frac{2Q+Q'}{3} + \omega^{\rm v}_{\bf 27}
                 + \omega^{\rm h}_{\overline{\bf 27}}, \qquad 
 \frac{Q+2Q'}{3} + \omega^{\rm v}_{\overline{\bf 27}}
                 + \omega^{\rm h}_{{\bf 27}}, 
\label{eq:MW-Km-E6E6-4generators-lattice}
\end{eqnarray}
where $\omega_{\bf 27}^{\rm v,h}$ [resp. $\omega_{\overline{\bf
27}}^{\rm ,h}$] are the weights of the ${\bf 27}$ 
[resp. $\overline{\bf 27}$] representation of the visible/hidden sector 
$E_6$ symmetry. The quotient space $W_{\rm frame}/W_{\rm root} \cong
MW(X) \cong \Z^{\oplus 4}$ is generated by those four elements and 
the height pairing of the Mordell--Weil lattice is 
$A_2^*[-2] \oplus A_2^*[-2]$ in this basis (as is well-known 
in the literature). 

Reference \cite{KuwShio} provides explicitly expressions of four generators 
of $MW(X) \cong \Z^{\oplus 4}$ for the $2{\rm IV}^*+8I_1$-type elliptic 
fibration on $X = {\rm Km}(E_{\rho_1} \times E_{\rho_2})$. 
The Weierstrass model of $X$ given by (\ref{eq:Inose-pencil}) has 
four independent non-zero sections, two of which---denoted by 
$\bar{P}_4$ and $\bar{P}_8$---are given by 
\begin{eqnarray}
 (X_4, Y_4) &=& (-4 \lambda_1 \lambda_2 z^2,
      -4z^2 (\lambda_1(\lambda_1-1)z^2 + \lambda_2(\lambda_2-1))), \\ 
 (X_8, Y_8) &=& (-4 z^2, 4 z^2 ( \lambda_1(\lambda_1-1)z^2 
                              + \lambda_2(\lambda_2-1) ) ) \, .
\end{eqnarray}
Readers interested in the expressions for the two other sections 
$\bar{P}_7$ or $\bar{P}_5$ are referred to \cite{KuwShio}.
We follow \cite{Sch} and denote rational points of the elliptic curve, as 
well as the corresponding divisors and elements of the Neron-Severi lattice or the 
Mordell-Weil group, by $P$ and the corresponding curves by $\bar{P}$. 

Modulo $U_* = {\rm Span}_\Z\{ [F], [\sigma + F]\}$ 
and $W_{\rm root}$, these four sections generate the Mordell--Weil group $MW(X)$, 
$P_4$ and $P_8$ for one $A_2^*[-2]$, and 
$P_7$ and $P_5$ for the other $A_2^*[-2]$.
All four of those sections, however, meet the $E_6$ singularities at 
$z=0$, $(X, Y) = (0,0)$ and at $z=\infty$, $(X/z^4, Y/z^6) = (0,0)$.
Hence they are not contained within the narrow Mordell--Weil group, $MW(X)^0$, which 
is consistent with the fact that the generators (\ref{eq:MW-Km-E6E6-4generators-lattice}) contain 
weights of non-singlet representations of $E_6^{\rm vis}$ and $E_6^{\rm hid.}$.

Just like the generators of the essential lattice $L(X) \cong
W_{{\rm U}(1)}$, $\{P, P', Q, Q'\}$, are obtained as $\Z$-linear 
combinations of the generators in
(\ref{eq:MW-Km-E6E6-4generators-lattice}), the corresponding sections for
the generators of the narrow Mordell--Weil lattice $MW(X)^0$ should 
also be obtained through the group-law sum of the sections
$\bar{P}_{4,8}$ and $\bar{P}_{7,5}$. Sections corresponding to 
$(2P_4-P_8) \in MW(X)$ and $(2P_8-P_4) \in MW(X)$ are given---by using the 
ordinary group law sum on elliptic curves---by 
\begin{eqnarray}
 X_{2P_4-P_8} & = & \frac{4\{ \lambda_2^2(\lambda_2-1)^2
     + z^2 (\lambda_2^2(\lambda_1^2-1) -\lambda_1^2)
     + z^4 \lambda_1^2(\lambda_1-1)^2\}} 
   {(\lambda_1+\lambda_2 - \lambda_1\lambda_2)^2}, \\
 Y_{2P_4-P_8} & = & \frac{4 \left\{
   \begin{array}{rl}
        2\lambda_2^3(\lambda_2-1)^3 &
      + z^2 (\lambda_2^5 (\lambda_1-1)^2(\lambda_1+1)+O(\lambda_2^4)) \\
      + 2 z^6 \lambda_1^3(\lambda_1-1)^3 &  
      + z^4 (\lambda_2^3 (\lambda_1-1)^3\lambda_1(\lambda_1+1)
                   + O(\lambda_2^2))

    \end{array}
   \right\}  }
  { (\lambda_1+\lambda_2 - \lambda_1\lambda_2)^3        }, \\
 X_{2P_8-P_4} & = & \frac{ 4\{
     \lambda_2^2(\lambda_2-1)^2
  - z^2 \lambda_1 \lambda_2 (\lambda_2^2+\lambda_1^2-1) 
  + z^4 \lambda_1^2(\lambda_1-1)^2 \} }
    {(1-\lambda_1-\lambda_2)^2},  \\
 Y_{2P_8-P_4} & = & \frac{4 \left\{
  \begin{array}{rl}
     2 \lambda_2^3(\lambda_2-1)^3 &
   - z^2 (\lambda_2^5(2\lambda_1-1) + O(\lambda_2^4)) \\
   + 2 z^6 \lambda_1^3(\lambda_1-1)^3 & 
   - z^4 (\lambda_2^3 \lambda_1(\lambda_1-1)(2\lambda_1-1)
             + O(\lambda_2^2) ) 
  \end{array}
\right\}} 
  { (1-\lambda_1-\lambda_2)^3 }. 
\end{eqnarray}
They belong to the narrow Mordell--Weil group $MW(X)^0$. Indeed, 
\begin{equation}
 (X_{2P_4-P_8}, Y_{2P_4-P_8})|_{z=0} = \left( 
    \left( \frac{2\lambda_2(\lambda_2-1)}
                {\lambda_1+\lambda_2-\lambda_1\lambda_2}\right)^2 , 
    \left( \frac{2\lambda_2(\lambda_2-1)}
                {\lambda_1+\lambda_2-\lambda_1\lambda_2}\right)^3  
    \right) \neq (0,0),  
\end{equation}
\begin{equation}
 \left.
 \left( \frac{X_{2P_4-P_8}}{z^4}, \frac{Y_{2P_4-P_8}}{z^6}
 \right)\right|_{z=\infty} = 
 \left(
    \left( \frac{ 2\lambda_1(\lambda_1-1)}
                { (\lambda_1+\lambda_2-\lambda_1\lambda_2)^2 } \right)^2 ,
    \left( \frac{ 2\lambda_1(\lambda_1-1)}
                { (\lambda_1+\lambda_2-\lambda_1\lambda_2)^2 } \right)^3 
\right) \neq (0,0) ,
\end{equation}
and a similar calculation proves that the section $(2P_8-P_4)$ also 
stays away from the two $E_6$ singularities at $z=0$ and $z=\infty$.
The height pairing is $A_2[-2]$ on the basis of 
$\{ 2P_4-P_8, 2P_8-P_4 \}$, the opposite of 
$A_2[2] \subset L(X) = W_{{\rm U}(1)}$, as expected. 

Similarly, we can construct section contained in the narrow Mordell--Weil group from the sections $\bar{P}_7$ and $\bar{P}_5$. 
We have computed the sections corresponding to $(2P_7-P_5)$ and $-(P_7+P_5)$ in $MW(X)$ 
and confirmed that they indeed belong in $MW(X)^0$. The height pairing on the basis $\{2P_7-P_5, -(P_7+P_5)\}$ is $A_2[-2]$. 
Since we use the explicit expressions of these sections later, we leave them here as a record:
\begin{eqnarray}
 X_{2P_7-P_5} & = & \frac{ 4 \{
     \lambda_2^2(\lambda_2-1)^2
   - z^2 \lambda_1 (\lambda_2^2(\lambda_1^2-1) + 1) 
   + z^4 \lambda_1^2(\lambda_1-1)^2 \}}
   { ( \lambda_2 - \lambda_1 \lambda_2 -1)^2 }, \\
 Y_{2P_7-P_5} & = & \frac{ 4 \left\{
   \begin{array}{rl}
      2 \lambda_2^3(\lambda_2-1)^3 & 
    - z^2 ( \lambda_2^5 \lambda_1^2 (\lambda_1-2) + O(\lambda_2^4) ) \\
    + 2 z^6 \lambda_1^3 (\lambda_1-1)^3 & 
    + z^4 ( \lambda_2^3 \lambda_1^3 (\lambda_1-1)(\lambda_1-2) + 
              O(\lambda_2^2) )
    \end{array}
\right\}}
    { ( \lambda_2 - \lambda_1 \lambda_2 -1)^3 }, \\
 X_{-(P_7+P_5)} & = & \frac{ 4 \{ 
   \lambda_2^2(\lambda_2-1)^2 
 - z^2 (\lambda_2^3 \lambda_1 + O(\lambda_2^2) )
 + z^4 \lambda_1^2(\lambda_1-1)^2 \} }
   {(\lambda_1-\lambda_2)^2}, \\
 Y_{-(P_7+P_5)} & = & \frac{ 4 \left\{ 
    \begin{array}{rl}
        2 \lambda_2^3(\lambda_2-1)^3 &
      + z^2 ( \lambda_2^5 (2\lambda_1-1) + O(\lambda_2^4) ) \\
      - 2 z^6 \lambda_1^3 (\lambda_1-1)^3 & 
      - z^4 ( \lambda_2^3 (2\lambda_1-1) \lambda_1(1-\lambda_1) 
                + O(\lambda_2^2) )
    \end{array}
\right\}}
    { (\lambda_1-\lambda_2)^3 }\, .
\end{eqnarray}  

None of the four sections corresponding to $(2P_4-P_8)$, $(2P_8-P_4)$, 
$(2P_7-P_5)$ and $-(P_7+P_5)$ in $MW(X)^0$ meet the zero section,
$\sigma$, of the Weierstrass model given by (\ref{eq:Inose-pencil}).
Therefore, $\overline{(2P_4-P_8)}-\sigma$ and 
$\overline{(2P_8-P_4)}-\sigma$ generate $A_2[2] \subset W_{{\rm U}(1)}$, and 
$\overline{(2P_7-P_5)}-\sigma$ and $\overline{- (P_7 + P_5)} - \sigma$
generate the other $A_2[2] \subset W_{{\rm U}(1)}$. Dual to the rank-$k=4$
Cartan flux in the heterotic string $\SU(N_v=3) \times \SU(N_h=3)$
bundle compactification should be four-form fluxes involving 
the Poincar\'e dual of these algebraic cycles in $X$. 

\subsection{Semistable Degeneration}

In the large ${\rm Im}(\tilde{\rho}_H)$ region of moduli space, 
where the supergravity description is a good approximation of heterotic string theory, 
it is more intuitive to choose $\{P+Q, P'+Q'\}$ and $\{P-Q, P'-Q'\}$ 
as the basis of the rank-$k=4$ lattice of Cartan flux quanta. The first two 
generate the $A_2[4]$ sublattice for the visible sector $\SU(N_v=3)$ structure 
group and the last two another $A_2[4]$ sublattice for the hidden sector 
$\SU(N_h=3)$. In F-theory language, the Poincar\'e dual of the algebraic cycles $\{ P+ Q, P' + Q'\}$ and 
$\{P-Q, P'-Q'\}$ should thus be interpreted as those for the visible and hidden sectors, respectively. 

Let us take one step further and identify the equivalent of the 
visible and hidden sector basis $\{P \pm Q, P' \pm Q'\}$ not just 
in terms of the heterotic string, or in algebraic (lattice) language for 
F-theory, but also in terms of the geometry of the K3 surface of F-theory. 
We have identified four independent algebraic cycles in $W_{{\rm U}(1)}$, which 
are also in one-to-one correspondence with elements in $MW(X)^0$. 
$\{ (2P_4-P_8), (2P_8-P_4), (2P_7-P_5), -(P_7+P_5) \} \subset MW(X)^0$ are 
generators of $W_{{\rm U}(1)} \cong A_2[2] \oplus A_2[2]$ and are equivalent 
to $\{P,P',Q,Q'\}$. We claim that the visible and hidden sector basis 
is given by 
\begin{eqnarray}
({\rm visible~sect.}) & 
\overline{(2P_4-P_8)} - \overline{-(2P_7-P_5)}, & \qquad 
\overline{(2P_8-P_4)} - \overline{(P_7+P_5)},  
     \label{eq:id-K3-RESRES-vis} \\
({\rm hidden~sect.}) & 
\overline{(2P_4-P_8)} - \overline{(2P_7-P_5)}, & \qquad 
\overline{(2P_8-P_4)} - \overline{-(P_7+P_5)} .
     \label{eq:id-K3-RESRES-hid}
\end{eqnarray}
This idea comes from the following observations in geometry.

As we have already made clear, the coordinate rescaling 
in footnote \ref{fn:Inose-rescale} and the coordinate redefinition in 
footnote \ref{fn:degeneration-limit-A} allow us to 
see (\ref{eq:Inose-pencil-2}) as a family of elliptic K3 surface showing 
semistable degeneration. In one of the affine patches, the set of equations 
\begin{equation}
\left\{
 \begin{array}{l}
 \tilde{\eta}^2 =
   \left(\xi + \frac{4}{\lambda_2}\right)
   \left(\xi + 4\left(1 + \frac{\lambda_1}{\lambda_2}\right) \right)
   \left(\xi + 4\lambda_1 \right) + 2^3
       \left( (1-1/\lambda_2) u + \lambda_1(\lambda_1-1) v \right) \; 
       \tilde{\eta} , \\
  u v = 1/\lambda_2  
  \end{array}
\right.
\label{eq:E6E6-family}
\end{equation}
defines a family of K3 surfaces elliptically fibred over a curve 
$\{u v = t | (u,v) \in \C^2 \}$ parametrized by 
$t := 1/\lambda_2 \in D \subset \C$. In the large $\lambda_2$ limit, 
$t=0$, the base curve splits into two irreducible pieces, and the K3 surface 
also splits into two rational elliptic surfaces (a.k.a $dP_9$) glued 
together at one common fibre elliptic curve. 
\begin{eqnarray}
 \tilde{\eta}^2 & = & \xi \left(\xi + 4 \right)
     \left(\xi + 4\lambda_1 \right) + 2^3 u \; \tilde{\eta} \, , \qquad v = 0\, , \\
 \tilde{\eta}^2 & = & \xi \left(\xi + 4 \right)
     \left(\xi + 4\lambda_1 \right) + 2^3
       \lambda_1(\lambda_1-1) v \; \tilde{\eta} \, , \qquad u = 0\, , 
\end{eqnarray}
are the visible and hidden sector $dP_9$'s, respectively. The
$E_6$ singularities are at $u=\infty$ in the visible sector $dP_9$ and 
at $v = \infty$ in the hidden sector $dP_9$. The common fibre at $u=v=0$ 
is given by 
\begin{equation}
  \tilde{\eta}^2 = \xi (\xi + 4) (\xi + 4 \lambda_1)\, .
\end{equation}
The two $dP_9$'s (rational elliptic surfaces) should be ``type No.27'' 
in the classification in \cite{OgShio}.

The sections $\overline{(2P_4-P_8)}$, $\overline{(2P_8-P_4)}$, 
$\overline{(2P_7-P_5)}$ and $\overline{-(P_7+P_5)}$, as well as the sections 
corresponding to their inverse elements in $MW(X)^0$, such as 
$\overline{(P_7+P_5)}$, define divisors in the threefold given by 
(\ref{eq:E6E6-family}). Intersection of those divisors with the 
$t = 1/\lambda_2 = 0$ divisor---$dP_9 \cup dP_9$---defines their 
semistable degeneration limits mathematically  (whatever this means in physics).  
Working this out explicitly, we found that the limit of both sections 
$\overline{(2P_4-P_8)}$ and $\overline{(2P_7-P_5)}$ are precisely 
the same in the visible sector $dP_9$, 
\begin{eqnarray}
  \xi = \left(\frac{2}{1-\lambda_1} \right)^2 u^2, \qquad 
  \tilde{\eta} = \left(\frac{2}{1-\lambda_1}\right)^3
     \left\{ u^3 + \frac{1}{2} (\lambda_1-1)^2(\lambda_1+1) u \right\}.
\end{eqnarray}
This common limit in the visible sector passes through one of the 2-torsion 
point $(\xi, \tilde{\eta}) = (0,0)$ in the common elliptic fibre.  
The semistable degeneration limit of the two sections, however, 
remain different in the hidden sector $dP_9$. In the fibre in $v=\infty$, 
for example, 
\begin{equation}
 \left( \frac{\xi}{v^2}, \frac{\tilde{\eta}}{v^3} \right)
   \rightarrow  
   \left( (- 2\lambda_1)^2, (-2\lambda_1)^3 \right)
  \qquad {\rm v.s.} \qquad 
   \rightarrow 
   \left( (2\lambda_1)^2, (2\lambda_1)^3 \right)
\end{equation}
for $\overline{(2P_4-P_8)}$ and $\overline{(2P_7-P_5)}$, respectively.
They are inverse elements under the group law of the elliptic curve. 
This is why the algebraic cycle 
$\overline{(2P_4-P_8)} - \overline{(2P_7-P_5)}$ is considered to be 
purely in the hidden sector $dP_9$. It must also be easy to see 
that the algebraic cycle $\overline{(2P_4-P_8)} - \overline{-(2P_7-P_5)}$ 
is purely in the visible sector $dP_9$. 
A similar story holds also for the pair of sections $\overline{(2P_8-P_4)}$ and 
$\overline{-(P_7+P_5)}$. We do not present details here, except noting that 
those sections pass through another 2-torsion point in the common elliptic fibre: $(\xi, \tilde{\eta}) = (-4\lambda_1,0)$.  

Back in the regime of finite $|\lambda_2|$, the two sections 
$\overline{(2P_4-P_8)}$ and $\overline{(2P_7-P_5)}$ both cover 
the entire base $\P^1$, from the visible sector 7-brane at $z=0$ to 
the hidden sector 7-brane at $z=\infty$. These two sections are distinct,
but they remain very close in the small $z$ region (near the visible 
sector), with the difference scaling as 
$1/ \lambda_2 \sim e^{ 2 \pi i \tilde{\rho}_H  }$. It is thus reasonable 
to understand this as a stringy effect. When we ignore differences of order 
${\cal O}(1/\lambda_2)$ to restore the supergravity approximation, the geometric 
picture described above (using $dP_9 \cup dP_9$) is a reasonably good 
approximation for large $|\lambda_2|$ and fits perfectly with our intuitive 
understanding of Cartan fluxes in the visible as well as hidden sector structure group. 
This is how we were led to the claim (\ref{eq:id-K3-RESRES-vis}, 
\ref{eq:id-K3-RESRES-hid}), and it is this interesting behaviour of 
sections under the semistable degeneration of K3 surfaces that reconciles the notion of
having Cartan flux purely in the visible/hidden sector with considering sections 
of the elliptic K3 surface. 

Before closing this section, let us try to place the observations 
based on the example characterized by (\ref{eq:SU(3)xSU(3)-Het-rho14-TandS}) 
and (\ref{eq:KmEF-embed-rho14Tx}) (or equivalently by (\ref{eq:SU(3)xSU(3)-Het-rho14-TandS}) 
and (\ref{eq:Het-flux-data-Km-E6E6})). It is more natural from the perspective of heterotic string theory to take 
(\ref{eq:Het-flux-data-Km-E6E6}) as input data for compactification because 
they are flux data of the gauge fields and $B$-field. In F-theory language, 
the essential lattice $L(X) = W_{{\rm U}(1)}$ of an elliptic fibration is specified 
by (\ref{eq:Het-flux-data-Km-E6E6}), while the embedding 
(\ref{eq:KmEF-embed-rho14Tx}) determines the transcendental lattice of a 
$\rho=18$ (two parameter) family of K3 surfaces. When we replace 
(\ref{eq:KmEF-embed-rho14Tx}, \ref{eq:Het-flux-data-Km-E6E6}) by some other choice, 
this means we take different flux quanta for the rank-$k=4$ Cartan flux 
in the $\SU(3) \times \SU(3)$ bundle compactification of heterotic string 
theory, or to use a $\rho=18$ family of K3 surfaces different from
$X = {\rm Km}(E_{\rho_1} \times E_{\rho_2})$. 
There is nothing wrong in doing so. For all different choices of 
(\ref{eq:KmEF-embed-rho14Tx}, \ref{eq:Het-flux-data-Km-E6E6}), one can 
construct a two parameter family of K3 surfaces with four independent 
sections in the narrow Mordell--Weil group. 

Since we are not interested in literally taking ${\rm Im}(\rho_H) = 
[{\rm vol}(T^2)/\ell_s^2]_{\rm Het}$ to infinity for practical applications, 
we do not need to study the semistable degeneration limit of K3 surfaces, but
rather want to consider ${\rm Im}(\rho_H)$ very large, but finite.

Given the fact that literature referring to the heterotic--F-theory duality 
dictionary on the flux has often relied on the stable degeneration limit, 
however, it is not uninteresting to ask whether the $dP_9 \cup dP_9$ picture 
loses some information.  
When compactification data is given in terms of a one parameter family of 
$dP_9 \cup dP_9$ both in the ``type No.27'' of \cite{OgShio}, 
along with $G_H^{(4)}$ in (\ref{eq:G4-Het-def}), one has to make sure 
that the sections pass through some torsion points in the common elliptic fibre. 
Using these torsion points and the Cartan flux quanta $G^{(4)}_H$, 
the $B$-field flux quanta on $T^2$ must be reproduced at least to some extent. 
Thus, apart from how far one should go back from the semistable degeneration 
limit (e.g., the value of $\lambda_2$), a great deal of information may be
recovered from the description using ($dP_9 \cup dP_9$, $G^{(4)}_H$) by 
paying attention to such subtleties. We remain inconclusive about this 
question, however. 

\section{Ashok--Denef--Douglas Formula for F-theory}
\label{sec:ADD}

In this section, we begin with a review of the derivation of 
the vacuum index density distribution (\ref{eq:ADD-formula}) 
in \cite{Ashok:2003gk, Denef:2004ze} for type IIB Calabi--Yau
orientifolds, and then generalize its derivation for more general 
landscapes based on F-theory compactifications, where the four-form 
fluxes are scanned within the subspace $H^{2,2}(Y_z; \R)_H$. 
We largely follow the presentation in \cite{Denef:2004ze}, which 
maintains more intuitive control over what is being done than the 
sophisticated and polished-up style of \cite{Denef:2008wq}. 
Along the way, we will see that the three-form scanning in type IIB
orientifolds and the four-form scanning considered in 
Section \ref{ssec:orientifold-beyond} correspond to scanning only in 
$H^{2,2}(Y_z; \R)_H$ rather than the entire orthogonal complement 
$[(H^{2,2}(Y_z; \R)_V)^\perp \subset H^{2,2}(Y_z; \R)] = 
H^{2,2}(Y_z)_H \oplus H^{2,2}(Y_z)_{RM}$. 

\vspace{5mm}

The vacuum index density for F-theory flux vacua is defined by 
\cite{Ashok:2003gk, Denef:2004ze}\footnote{It is worth noting that the diagonal blocks
$D_aD_bW$ and $\overline{D_cD_dW}$ are the same as the fermion mass matrix of the low energy effective
field theory below the Kaluza-Klein scale or below the moduli mass scale $M_{KK}^3/M_{\rm Str}^2$. 
Fluctuations in the directions tangential to ${\cal M}_*$ are just as heavy as those in the 
directions normal to the restricted moduli space generically. The determinant of the $2m$ x $2m$
matrix just makes sure that each topological flux $N$ contributes to $\int d\mu_I$ by 1 (TW thanks
T.~Eguchi and Y.~Tachikawa for discussion).}
\begin{eqnarray}
 d\mu_I = d^{2m} z \sum_N \Theta(L_* - L)
     \delta^{2m}(D_aW,\overline{D_aW}) 
     {\rm det} \left( \begin{array}{cc}
        D_aD_bW & \partial_a \overline{D_d W} \\
        \bar{\partial}_{\bar{c}} D_bW & \overline{D_cD_dW}
               \end{array} \right)_{2m \times 2m}, \label{vacdensapp}
\end{eqnarray}
where $a,b,c,d \in \{1, \cdots, m\}$ label $m$ local complex coordinates 
of some restricted moduli space ${\cal M}_*$ 
(see Section \ref{ssec:orientifold-beyond} for various ${\cal M}_*$ of 
interest). In dealing with such integrals, we have adopted the conventions of
\cite{polchinski}, where $\int d^2 z \delta^2(z,\bar{z}) = 1$. The tadpole $L$ and the superpotential
are given by
\begin{equation}
 W \propto \int_Y G^{(4)}_{\rm scan} \wedge \Omega_Y, \qquad 
 L = \frac{1}{2} \int_Y G^{(4)}_{\rm scan} \wedge G^{(4)}_{\rm scan}\, .
\end{equation}
This $d\mu_I$ is a distribution function over the space ${\cal M}_*$ and captures all the flux vacua 
for which the D3-tadpole from the flux configuration $L$ is not more than $L_*$. The sum over flux 
quanta $\sum_N$ is replaced by its continuous approximation $\int d^K N$.
This expression can be rewritten as \cite{Ashok:2003gk}
\begin{eqnarray}
 d\mu_I & = & \frac{(\alpha_0 L_*)^{K/2} }{(K/2)!} \rho_{\rm ind}(\alpha_0), \\
 \rho_{\rm ind}(\alpha_0) & := &  
d^{2m}z\int d^K N e^{-\alpha_0 L}
     \delta^{2m}(D_aW,\overline{D_aW}) 
     {\rm det} \left( \begin{array}{cc}
        D_aD_bW & \partial_a \overline{D_d W} \\
        \bar{\partial}_{\bar{c}} D_bW & \overline{D_cD_dW}
               \end{array} \right); 
\label{eq:rho-ind-near-def-exprs}
\end{eqnarray}
since $\rho_{\rm ind}(\alpha_0)$ scales as $(\alpha_0)^{-K/2}$, the vacuum 
index density $d\mu_I$ does not depend on the choice of $\alpha_0$. 
By setting $\alpha_0 = L_*^{-1}$, one can see where (and how) the $L_*$ dependence arises
in the expression of $\rho_{\rm ind}(L_*^{-1})$.
In contrast, by setting $\alpha_0 = 2\pi$, the $L_*$-dependence of 
the overall number of vacua in this landscape is seen clearly. 
We take $\alpha_0 = 2\pi$ throughout this article (as in
\cite{Ashok:2003gk, Denef:2004ze}), and $\rho_{\rm ind}(\alpha_0=2\pi)$ is 
simply denoted by $\rho_{\rm ind}$.
The distribution $\rho_{\rm ind}$ can be rewritten in a more useful form
in some cases, and that is the subject of the following.

The formulation in \cite{Denef:2004ze} accommodates scanning four-form
fluxes in 
\begin{equation}
 G^{(4)}_{\rm scan} \in
    \left[H^{4,0}(Y_z; \C) + H^{0,4}(Y_z; \C) \right] 
 \oplus \left[H^{3,1}(Y_z; \C)_* + H^{1,3}(Y_z; \C)_* \right]
  \oplus H^{2,2}(Y_z; \R)_{H*}, 
\label{eq:space-G4-40-31-22H*}
\end{equation}
where $H^{3,1}(Y_z; \C)_*$ has been introduced in 
p.~\pageref{pg:flx-31-cmp-*}, and $H^{2,2}(Y_z; \R)_{H*}$ was 
defined in (\ref{eq:H22-horizontal*-def}). The first two components of 
$G^{(4)}_{\rm scan}$ is parametrized as follows, by $1 + m$ complex
numbers $\{ N_X, N_Y^a\}|_{a=1,\cdots, m}$ ($m={\rm dim}_\C {\cal M}_*$): 
\begin{equation}
 \Delta G^{(4)}_{\rm scan} =
  \left[ N_X \Omega_Y + \bar{N}_X \overline{\Omega}_Y \right]+ 
  \left[ N_Y^a (D_a \Omega_Y) +
 \bar{N}_Y^{\bar{b}} (\bar{D}_{\bar{b}} \overline{\Omega}_Y) \right] , 
\label{eq:G4-expand-horizontal-1}
\end{equation}
using $\Omega_Y$ and $\{ (D_a\Omega_Y) \}_{a=1,\cdots, m}$ as the basis 
of $H^{4,0}(Y_z; \C)$ and $H^{3,1}(Y_z; \C)_*$, respectively. Here, 
\begin{equation}
 D_a \Omega_Y = \partial_a \Omega_Y + K_a \Omega_Y, \qquad 
 D_a D_b \Omega_Y = (\partial_a + K_a) D_b\Omega_Y - \Gamma_{ba}^c D_c \Omega_Y
\end{equation}
\begin{equation}
 K = - \ln \left[ \int_Y \Omega_Y \wedge \overline{\Omega}_Y \right], 
 \qquad 
 K_a := \partial_a K.
\end{equation}
The last component, $H^{2,2}(Y_z; \R)_{H*}$, is parametrized 
by 
\begin{equation}
 \Delta G_{\rm scan} = \sum_{I=1}^{\widetilde{K}} 
     \tilde{N}_I \Omega^{(2,2)}_I, \qquad 
  A_{IJ} := \int_Y \Omega^{(2,2)}_I \wedge \Omega^{(2,2)}_J, 
\label{eq:G4-expand-horizontal-2}
\end{equation}
by using a basis $\{ \Omega^{(2,2)}_I \}|_{I=1,\cdots, \widetilde{K}}$ 
of the vector space $H^{2,2}(Y_z; \R)_{H*}$ over $\R$; 
the generators $(D_aD_b\Omega_Y)$ and $\overline{(D_cD_d\Omega_Y)}$ 
of $H^{2,2}(Y_z; \R)_{H*}$ are not necessarily linearly independent.
Thus, the continuous approximation $\int d^K N$ of the flux
configuration $\sum_N$ is given by 
\begin{equation}
  \int dN_X d\bar{N}_X e^{-K} \; \int \prod_{a=1}^m [dN_Y^a
   d\bar{N}_Y^{\bar{a}}] e^{-m K} {\rm det}(K_{c\bar{d}})_{m \times m} \;
  \int d^{\widetilde{K}} \tilde{N} \sqrt{A_{IJ}}.
\label{eq:flux-integrate-measure}
\end{equation}
We expect very little confusion to arise from the fact that we use $A_{IJ}$ as the 
intersection form on the vector space $H^{2,2}(Y_z; \R)_{H*}$ here, 
while it is the intersection form on $H^4(Y_z; \R)$ in (\ref{eq:cond-ADD-formula}).

In the case of a landscape based on a type IIB orientifold using a 
Calabi--Yau threefold $M_3$ with 7-branes in the $\SO(8)$ configuration 
and scanning three-form fluxes $F^{(3)}_{\rm scan}$ and $H^{(3)}_{\rm scan}$, 
the four-form is given by 
\begin{equation}
 G^{(4)}_{\rm scan} = \frac{1}{\phi - \bar{\phi}}
    \left[  \overline{G^{(3)}}_{\rm scan} \wedge \Omega_{T^2}
          - G^{(3)}_{\rm scan} \wedge \overline{\Omega}_{T^2}
    \right], \qquad 
 G^{(3)}_{\rm scan} := F^{(3)}_{\rm scan} - \phi H^{(3)}_{\rm scan}, 
\label{eq:scan-3form-in-Fdescr}
\end{equation}
we can take \cite{Denef:2004ze}
\begin{equation}
 N_Z^{\phi i}(D_\phi D_i \Omega_Y) + {\rm h.c.} \qquad \qquad 
  \left(i=1,\cdots, m-1
 =  h^{2,1}_{-,{\rm prim.}}(M_3) \right)
\end{equation}
with $N_Z^{\phi i} \in \C$ as a non-redundant parametrization of 
$H^{2,2}(Y_z; \R)_{H*}$. This is due to a relation 
\begin{equation}
 (D_j D_k \Omega_M) \wedge \Omega_{T^2} =
   - {\cal F}_{ljk} {\cal F}_\phi K^{\bar{i}l} K^{\bar{\phi}\phi}
     e^{K}
     \overline{ ( D_i D_\phi (\Omega_M \wedge \Omega_{T^2}) ) }  
\label{eq:IIB-real-constraint}
\end{equation}
that follows from\footnote{For $T^2$, 
${\cal F}_\phi := \int_{T^2} \Omega_{T^2} \wedge (D_\phi \Omega_{T^2})$.   
For a Calabi--Yau threefold $M$, ${\cal F}_{ijk} := 
\int_M \Omega_M \wedge (D_iD_jD_k \Omega_M)$.
For $T^2$, there is the relation 
$K_{\phi\bar{\phi}}= |{\cal F}_\phi|^2 e^{2K^{(T^2)}}$.
When we choose the normalization $\Omega_{T^2} = dx + \phi dy$, we have that ${\cal F}=1$.} 
\begin{equation}
 D_\phi \Omega_{T^2} = i {\cal F}_{\phi} e^{K^{(T^2)}} \overline{\Omega}_{T^2},  \qquad 
 D_a D_b \Omega_M = i {\cal F}_{d a b} K^{\bar{c}d}
    e^{K^{(M)}} \overline{D_c \Omega_M}.  
\end{equation} 

Let us now
consider a more general cases of F-theory
compactifications where the moduli space ${\cal M}_*$ is not necessarily 
in the form of ${\cal M}_{\rm cpx}(M_3) \times {\cal M}_\phi$, or 
7-branes are not necessarily in an $\SO(8)$ configuration. We consider 
a class of landscapes where the restricted moduli space ${\cal M}_*$
of a Calabi--Yau fourfold $Y$ is specified by divisors $J_Y$ and 
$W_{\rm noscan}$ such that there is a relation
\cite{Greene:1993vm}\footnote{
$\bar{D}_{\bar{g}}(D_aD_b\Omega_Y)$ has only $(3, 1)$ Hodge 
components \cite{Greene:1993vm}, although $\overline{D_g(D_eD_f \Omega_Y)}$
may also have $(2, 2)$ components in addition. 
In this sense, ${\cal F}_{abcd} \tilde{B}^{cd,\bar{e}\bar{f}}$
plays the role of $S^{(2)}$ in eq. (2.20) of \cite{Greene:1993vm}. 
$\tilde{B}^{cd,\bar{e}\bar{f}} e^K$ in this article corresponds to 
$B^{cd,\bar{e}\bar{f}}$ in \cite{Greene:1993vm}.} 
among differential forms\footnote{
For a Calabi--Yau fourfold $Y$, ${\cal F}_{abcd}:= \int_Y \Omega_Y
\wedge (D_aD_bD_cD_d \Omega_Y)$. Similarly, for a K3 surface $X$, 
${\cal F}_{\alpha\beta} := 
\int_X \Omega_X \wedge (D_\alpha D_\beta \Omega_X)$.} 
\begin{equation}
 (D_a D_b \Omega_Y) = {\cal F}_{abcd} \tilde{B}^{cd,\bar{e}\bar{f}} e^K
    \overline{(D_e D_f \Omega_Y)}, \qquad 
 \overline{(D_aD_b \Omega_Y)} = 
    \overline{\cal F}_{\bar{a}\bar{b}\bar{c}\bar{d}} 
    \tilde{B}^{cd,\bar{c}\bar{d}}  e^K
    (D_cD_d\Omega_Y)
\label{eq:self-real-cond-22horizontal-4fold}
\end{equation}
for some ${}^\exists \tilde{B}^{cd,\bar{e}\bar{f}}$ over the moduli 
space ${\cal M}_*$ ($a,b,c,d,e,f \in \{1, \cdots, m\}$). 

Obviously this is a generalization of (\ref{eq:IIB-real-constraint}).
It is not hard also to see that 
$Y = X \times S = {\rm K3} \times {\rm K3}$ also has this property. 
Using the relation 
\begin{equation}
 D_\alpha \Omega_X = {\cal F}_{\alpha\beta}^{(X)} K^{\bar{\gamma}\beta}
    e^{K^{(X)}} \overline{D_\gamma \Omega_X} 
\label{eq:self-real-cond-11form-K3}
\end{equation}
for a K3 surface $X$ and the fact that 
${\cal F}_{\alpha\beta \kappa\lambda} = {\cal F}_{\alpha\beta}^{(X)} 
{\cal F}_{\kappa\lambda}^{(S)}$, one can see that 
$\tilde{B}^{\beta\lambda, \bar{\gamma}\bar{\mu}} = 
 K^{\bar{\gamma}\beta}_{(X)} K^{\bar{\mu}\lambda}_{(S)}$ do the job.
We will comment on $\tilde{B}^{\alpha\beta, \bar{\kappa}\bar{\lambda}}$ 
later. 

Under the condition that $\tilde{B}^{cd, \bar{c}\bar{d}}$ exists, 
one can choose $(D_a D_b \Omega_Y)$'s or $\overline{(D_cD_d \Omega_Y)}$'s 
as a (still possibly redundant) set of $\C$-coefficient generators of 
$H^{2,2}(Y_z; \R)_{H*}$. Thus, the $H^{2,2}(Y_z; \R)$ component 
(\ref{eq:G4-expand-horizontal-2}) may be written as 
\begin{equation}
\Delta G^{(4)}_{\rm scan} =    N_Z^{ab} (D_aD_b\Omega_Y) 
 = \bar{N}_Z^{\bar{c}\bar{d}} \overline{(D_cD_d\Omega_Y)} 
\end{equation}
for some complex valued $N_Z^{ab}$ or their complex conjugates 
$\bar{N}_Z^{\bar{a}\bar{b}}$. The following reality condition must be 
satisfied by the (in-principle) complex valued $N_Z^{ab}$, so that 
the two expressions agree: 
\begin{equation}
 \bar{N}_Z^{\bar{c}\bar{d}} = N_Z^{ab} 
   {\cal F}_{abcd}\tilde{B}^{cd,\bar{c}\bar{d}} e^K, \qquad 
 N_Z^{cd} = \bar{N}_Z^{\bar{a}\bar{b}}
    \overline{\cal F}_{\bar{a}\bar{b}\bar{c}\bar{d}} 
    \tilde{B}^{cd, \bar{c}\bar{d}} e^K . 
\label{eq:real-cond}
\end{equation}

In the following, we closely follow the presentation in
\cite{Denef:2004ze}, and see that the formula (\ref{eq:ADD-formula}) 
holds also in this case. 
The integration measure (\ref{eq:flux-integrate-measure}) is used 
as it is. 
The D3-tadpole contribution from the flux is written as\footnote{
The following relation is used \cite{Strominger:1990pd, Greene:1993vm}: 
\begin{equation}
 \int (D_aD_b\Omega_Y) \wedge \overline{(D_cD_d\Omega_Y)} =
 - e^{-K} \left[R_{a\bar{c}b\bar{d}} -
    K_{a\bar{c}} K_{b\bar{d}} - K_{a\bar{d}}K_{b\bar{c}} \right]
 =     e^{K} \tilde{B}^{ef,\bar{e}\bar{f}}
      {\cal F}_{abef} \overline{\cal F}_{\bar{c}\bar{d}\bar{e}\bar{f}}.
\end{equation}
} 
\begin{equation}
 L = e^{-K} \left( 
   |N_X|^2 - K_{a\bar{b}} N_Y^a \bar{N}_Y^{\bar{b}} 
   + \frac{1}{2} e^{2K} \tilde{B}^{ef,\bar{e}\bar{f}}
      {\cal F}_{abef} \overline{\cal F}_{\bar{a}\bar{b}\bar{e}\bar{f}}
        N_Z^{ab} \bar{N}_Z^{\bar{c}\bar{d}}
            \right);
\end{equation}
the last term is of type $(2, 2)$ and hence is 
positive definite. 
%
%
%
The F-term conditions (delta-functions) 
\begin{equation}
\delta^{2m}(D_aW,\overline{D_bW}) = \delta^{2m}(N_Y,\bar{N}_Y)
  \;  \left( e^{- m K} \; {\rm det}(K_{a\bar{b}}) \right)^{-2}
\end{equation}
eliminate the flux space integral over the $(3,1)+(1,3)$ components 
from the measure (\ref{eq:flux-integrate-measure}), and all the
remaining directions in the flux space have positive definite 
contributions to the D3-tadpole \cite{Denef:2004ze}. 

The parametrization of the $(2, 2)$ flux component in terms of the
$N_Z^{ab}$ satisfying (\ref{eq:real-cond}) may be redundant in general
($Y={\rm K3} \times {\rm K3}$ is an example; see the discussion later).
Thus, a set of independent flux space coordinates $\tilde{N}_I \in \R$ 
($I=1, \cdots, \widetilde{K}$) is introduced and we parametrize 
\begin{equation}
 N_Z^{ab} = \sum_I Z_I^{ab} \tilde{N}_I, \qquad
 \bar{N}_Z^{\bar{a}\bar{b}} = \sum_I \bar{Z}_I^{\bar{a}\bar{b}} \tilde{N}_I 
\end{equation}
without redundancy. The integration measure (\ref{eq:flux-integrate-measure}) 
is still used, 
but now there is an alternative expression for $A_{IJ}$:
\begin{eqnarray}
 A_{IJ} & = & Z_I^{ab} \int_Y (D_aD_b\Omega_Y) \wedge 
          \overline{(D_aD_b\Omega_Y)} \bar{Z}_J^{\bar{a}\bar{b}} 
  = Z_I^{ab} e^K \tilde{B}^{cd,\bar{c}\bar{d}} 
      {\cal F}_{abcd} \overline{\cal F}_{\bar{a}\bar{b}\bar{c}\bar{d}}
    \bar{Z}_J^{\bar{a}\bar{b}}, \nonumber \\
  & = & \bar{Z}_I^{\bar{c}\bar{d}}
          \overline{\cal F}_{\bar{a}\bar{b}\bar{c}\bar{d}} 
          \bar{Z}_J^{\bar{a}\bar{b}} = 
        Z_I^{ab} {\cal F}_{abcd} Z_J^{cd}.
\end{eqnarray}
The last term in the D3-tadpole contribution is also written as 
$\Delta L = \tilde{N}_I\tilde{N}_J A_{IJ} / 2$. 

In this case, we case, we can write the determinant in \eqref{vacdensapp} as follows:
\begin{eqnarray}
(-1)^{m} {\rm det}\left( \begin{array}{cc}
	   D_aD_bW & \partial_a \overline{D_d W} \\
           \bar{\partial}_{\bar{c}} (D_bW) & \overline{D_cD_dW }
		 \end{array}\right) & = &
   \int d^m \theta d^m\bar{\theta} d^m\psi d^m\bar{\psi} \\
  & & \exp \left[ \quad 
     \theta^a \psi^b {\cal F}_{abdf} Z_I^{ef} \tilde{N}_I 
      + \bar{\theta}^{\bar{c}} \bar{\psi}^{\bar{d}}
          \overline{\cal F}_{\bar{c}\bar{d}\bar{a}\bar{b}}
          \bar{Z}_J^{\bar{a}\bar{b}} \tilde{N}_I
        \right. \nonumber \\
  & & \qquad \left.
    + \theta^a \bar{\psi}^{\bar{d}} N_X e^{-K} K_{a\bar{d}}
    + \bar{\theta}^{\bar{c}} \psi^b \bar{N}_X e^{-K} K_{b\bar{c}} \right].
     \nonumber 
\end{eqnarray}
Carrying out Gaussian integrals over the complex $N_X$ and real
$\tilde{N}_I$ coordinates, we obtain the following formula: 
\begin{eqnarray}
 \rho_{\rm ind} & = & (-1)^{m}\frac{d^{2m} z}{(2\pi)^m}\; 
  e^{m K} [{\rm det}(K_{a\bar{b}})]^{-1} \; 
  \int d^m \theta d^m \bar{\theta} d^m\psi d^m\bar{\psi} 
      \label{eq:ADD-formula-generalized-a}  \\
 & & \!\!\!\!\!\!\!\!\!\!
  \exp \left[ \quad e^{-K} (\theta^a \bar{\psi}^{\bar{d}} K_{a\bar{d}}) 
                        (\bar{\theta}^{\bar{c}} \psi^b K_{b\bar{c}}) 
         \right. \nonumber \\
 & &  \!\!\!\!\!\!\!\!\!\!
  \qquad \left.
   + \left( \theta^a \psi^b {\cal F}_{abcd} Z_I^{cd} + 
            \bar{Z}_I^{\bar{a}\bar{b}} \overline{\cal F}_{\bar{a}\bar{b}\bar{c}\bar{d}}
                \bar{\theta}^{\bar{c}}\bar{\psi}^{\bar{d}} \right) 
       \frac{(A^{-1})^{IJ}}{2} 
     \left( \theta^p \psi^q {\cal F}_{pqrs} Z_J^{rs} + 
             \bar{Z}_J^{\bar{p}\bar{q}} 
               \overline{{\cal F}}_{\bar{p}\bar{q}\bar{r}\bar{s}}
               \bar{\theta}^{\bar{r}} \bar{\psi}^{\bar{s}}
      \right)
   \right].   \nonumber
\end{eqnarray}

In fact, his expression can be further simplified to (\ref{eq:ADD-formula}). 
To see this, note that possibly redundant set of generators 
$\{(D_aD_b\Omega_Y) \}$ or $\{\overline{(D_cD_d\Omega_Y)} \}$ can be written 
as 
\begin{equation}
 (D_aD_b\Omega_Y) = e_{ab}^{\; I} \Omega^{(2,2)}_I, \qquad 
 \overline{(D_cD_d\Omega_Y)} = \bar{e}_{\bar{c}\bar{d}}^{\; I} 
    \Omega^{(2,2)}_I,
\end{equation}
using a basis $\{ \Omega^{(2,2)}_I \}_{I=1,\cdots, \tilde{K}}$ of the 
vector space $H^{2,2}(Y_z; \R)_{H*}$ over $\R$.
The complex valued coefficients $e_{ab}^{\; I}$ and 
$\bar{e}_{\bar{c}\bar{d}}^{\; I}$ should satisfy 
\begin{equation}
 Z_I^{\; ab} e_{ab}^{\; J} = \delta_{I}^{\; J}, \qquad 
 \bar{Z}_I^{\bar{a}\bar{b}} \bar{e}_{\bar{a}\bar{b}}^{\; J} =
 \delta_I^{\; J}.
\end{equation}
From this, we obtain ${\cal F}_{abcd} = e_{ab}^{\; I} Z_{I}^{\; ef} {\cal F}_{efcd}$.

Using this relation, the $\theta^2\psi^2$ term in the exponent of 
(\ref{eq:ADD-formula-generalized-a}) can be rewritten as 
\begin{equation}
 \frac{1}{2} (\theta^a \psi^b e_{ab}^{\; I}) (A_{IK}) (A^{-1})^{KL} (A_{LJ})
    (\theta^p \psi^q e_{pq}^{\; J})
 = \frac{1}{2} (\theta^a \psi^b e_{ab}^{\; I}) (A_{IJ}) 
    (\theta^p \psi^q e_{pq}^{\; J})
 = \frac{1}{2} \theta^a \psi^b {\cal F}_{abpq} \theta^p \psi^q.
   \nonumber 
\end{equation}
This vanishes because of the totally symmetric nature of ${\cal F}_{abpq}$ and 
Grassmann nature of the $\theta^a \theta^p$. 
The $\theta\bar{\theta}\psi\bar{\psi}$ terms in the exponent, on the 
other hand, become 
\begin{equation}
 (\theta^a \psi^b e_{ab}^{\; I}) (A_{IK}) (A^{-1})^{KL} (A_{LJ}) 
 (\bar{\theta}^{\bar{p}} \bar{\psi}^{\bar{q}} \bar{e}_{\bar{p}\bar{q}}^{\; J})
 =  (\theta^a \psi^b e_{ab}^{\; I}) A_{IJ} 
 (\bar{\theta}^{\bar{p}} \bar{\psi}^{\bar{q}} \bar{e}_{\bar{p}\bar{q}}^{\; J})
 = \theta^a \psi^b {\cal F}_{abef} \tilde{B}^{ef,\bar{e}\bar{f}} e^K 
   \overline{\cal F}_{\bar{c}\bar{d}\bar{e}\bar{f}}
   \bar{\theta}^{\bar{c}}\bar{\psi}^{\bar{d}}.
  \nonumber 
\end{equation}
Using all these relations above, one arrives at the expression 
\begin{eqnarray}\label{rhodetapp}
\rho_{\rm ind.} & = & (-1)^{\frac{3m^2-m}{2}} [{\rm det}(K_{a\bar{b}})]^{-1}
    \int d^m\theta d^m\bar{\theta}
   \exp \left[ \theta^a \bar{\theta}^{\bar{b}}
     \left(K_{a\bar{d}} K_{c\bar{b}} - e^{2K} {\cal F}_{acef} \tilde{B}^{ef,\bar{e}\bar{f}}
                 \overline{\cal F}_{\bar{b}\bar{d}\bar{e}\bar{f}} \right)
 \frac{dz^c\wedge d\bar{z}^{\bar{d}}}{2\pi i} \right] \nonumber \\
 & = & {\rm det}\left(-\frac{R^b_{\hspace{1ex} a}}{2\pi i}+\frac{\omega}{2\pi}\delta^b_a \right),
\end{eqnarray}
Here
\begin{eqnarray}
 - R_{\bar{b} a c \bar{d}} = R_{a \bar{b} c \bar{d}}
      = K_{a\bar{d}} K_{c\bar{d}} + K_{a\bar{b}} K_{c\bar{d}}
           - e^{2K} {\cal F}_{acef} \tilde{B}^{ef,\bar{e}\bar{f}}
               \overline{\cal F}_{\bar{b}\bar{d}\bar{e}\bar{f}}
\label{eq:def-Riemann-curvature-analog-mirror}\, , \\
 R^b_{\hspace{1ex} a} = R^b_{ \hspace{1ex} a c \bar{d}} 
 \hspace{1ex}  dz^c \wedge dz^{\bar{d}} \, , \hspace{1cm} R^b_{\hspace{1ex}a c \bar{d}}= K^{\bar{b}b}R_{\bar{b}ac\bar{d}}\, .
\end{eqnarray}
$R^b_{\hspace{1ex}a}$ is the curvature $(1,1)$ form of the holomorphic tangent bundle $T{\cal M}_*$
and $\omega = i K_{c\bar{d}}  \hspace{1ex}  dz^c \wedge dz^{\bar{d}} $ the K\"ahler form on
${\cal M}_*$. The determinant in \eqref{rhodetapp} is computed with respect 
to the $a,b$ indices, so that the result is a $2m$-form on moduli space.

Finally, let us work out detailed descriptions of the vector space $H^{2,2}(Y;\R)_{H*}$ as well as 
the decomposition (\ref{eq:dcmp-22-HVrm}) in the case of $Y = X \times S$. There are 
$K= (20-rk_7) \times 21$ scanning (real-valued) flux quanta of $G^{(4)}_{\rm scan}$ 
introduced in the discussion of \ref{ssec:orientifold-beyond}. Among them, 
two correspond to the $(4,0)+(0,4)$  components 
$N_X (\Omega_X \otimes \Omega_S) + {\rm h.c}$, and $2m$ to 
the $(3,1)+(1,3)$ component 
\begin{equation}
N_Y^\alpha \; [(D_\alpha \Omega_X) \otimes \Omega_S] + 
 N_Y^\kappa \; [\Omega_X \otimes (D_\kappa \Omega_S)] + {\rm h.c.}
\end{equation}
in (\ref{eq:G4-expand-horizontal-1}), where 
$\alpha = 1,\cdots, (18-rk_7)$ and $\kappa = 1,\cdots, 19$. 
The remaining $2 + (18-rk_7) \times 19$ real-valued flux quanta correspond 
to the coefficients of these differential forms: 
\begin{equation}
 \Omega_X \otimes \overline{\Omega}_S, \qquad 
 \overline{\Omega}_X \otimes \Omega_S, \qquad 
 (D_\alpha \Omega_X) \otimes (D_\kappa \Omega_S).
\end{equation}
Noting that there is a relation 
$(D_\alpha D_\beta \Omega_X) = {\cal F}_{\alpha \beta} e^{K^{(X)}} \overline{\Omega}_X$
for a K3 surface $X$, one finds that 
i) all of $(D_\alpha D_\beta \Omega_X) \otimes \Omega_S$ for 
$\alpha, \beta = 1,\cdots, (18-rk_7)$ are the same 
as differential forms on $Y=X \times S$ up to normalization, at each 
given point in the moduli space ${\cal M}_*$, 
ii) all of the $2 \times (18-rk_7) \times 19$ differential forms 
above belong to $H^{2,2}(Y;\R)_{H*}$, and are furthermore 
linearly independent; iii) this is even a basis of $H^{2,2}(Y; \R)_{H*}$,  
because all the differential forms in the form of 
$D_aD_b(\Omega_X \otimes \Omega_S)$ have already been exploited, 
given the relation (\ref{eq:self-real-cond-11form-K3}).
All of these observations combined indicate that the vector space 
of scanning four-form flux considered in Section \ref{ssec:orientifold-beyond}
corresponds precisely to the space (\ref{eq:space-G4-40-31-22H*}).

In the case of $Y=X \times S = {\rm K3} \times {\rm K3}$, 
another vector subspace $H^{2,2}(Y; \R)_{V*} \subset H^{2,2}(Y; \R)$ 
is generated, on the other hand, by 
\begin{equation}
 H^4(X; \R) \otimes 1_S, \qquad 1_X \otimes H^4(S; \R), \qquad 
 (U_* \oplus W_{\rm noscan}) \otimes J_S \otimes \R.
\end{equation}
Thus, the remaining component, consisting of cycles which are neither ``horizontal'' or ``vertical'', is given by 
\begin{equation}
H^{2,2}(Y_z; \R)_{RM} \cong  
 \left( U_* \oplus W_{\rm noscan}\right) \otimes 
 \left[J_S^\perp \subset H^{1,1}(S;\R) \right] \oplus 
 \left[ (U_* \oplus W_{\rm noscan})^\perp \subset H^{1,1}(X;\R) \right]
 \otimes [J_S] .
\end{equation}
%
This is not empty, and in fact, the first component is where the 
singular fibre flux (GUT 7-brane flux) $G^{(4)}_{\rm fix}$ in 
Section \ref{ssec:MW-singl-fbr-flux}-II and \ref{ssec:orientifold-beyond}-III
is introduced.

For ${\rm K3} \times {\rm K3} = X \times S$, the Riemann curvature tensor 
should become block-diagonal, which is verified as in 
\begin{equation}
 R_{\alpha\bar{\beta} \kappa \bar{\lambda}} =
   K_{\alpha \bar{\beta}}^{(X)} K_{\kappa \bar{\lambda}}^{(S)} - 
   {\cal F}_{\alpha \gamma}^{(X)} {\cal F}_{\kappa \mu}^{(S)} \; 
   K^{\bar{\delta} \gamma}_{(X)} K^{\bar{\nu}\mu}_{(S)} \; 
   e^{2(K^{(X)}+K^{(S)})} \;
   \overline{\cal F}^{(X)}_{\bar{\beta} \bar{\delta}} 
   \overline{\cal F}^{(S)}_{\bar{\lambda} \bar{\nu}}
 =    K_{\alpha \bar{\beta}}^{(X)} K_{\kappa \bar{\lambda}}^{(S)} - 
      K_{\alpha \bar{\beta}}^{(X)} K_{\kappa \bar{\lambda}}^{(S)} =0.
\end{equation}
Diagonal blocks are given by 
\begin{equation}
 R_{\alpha \bar{\beta} \gamma \bar{\delta}} = 
   K_{\alpha \bar{\beta}}^{(X)} K_{\gamma \bar{\delta}}^{(X)} + 
   K_{\alpha \bar{\delta}}^{(X)} K_{\gamma \bar{\beta}}^{(X)} - 
   e^{2K^{(X)}} 
  {\cal F}_{\alpha \gamma}^{(X)}
  \overline{\cal F}_{\bar{\beta}\bar{\delta}}^{(X)},
\end{equation}
where we used 
\begin{equation}
 \tilde{B}^{\alpha \beta, \bar{\alpha}\bar{\beta}} = 
   \frac{ K^{\bar{\alpha} \alpha}_{(X)} K^{\bar{\beta}\beta}_{(X)} }
        { {\rm dim}_\C {\cal M}_{\rm cpx}(X;U_*\oplus W_{\rm noscan}) }, \qquad 
 \tilde{B}^{\alpha \beta, \bar{\kappa} \bar{\lambda} } = 
 \tilde{B}^{\kappa \lambda, \bar{\alpha} \bar{\beta}} = 0.
\end{equation}
%


\end{document}